\documentclass{jfm}
\usepackage[dvipsnames]{xcolor}
\usepackage{amsmath}
\usepackage{mathtools}
\usepackage{amssymb}
\usepackage{bm}
\usepackage{subfig}
\usepackage{graphicx}
\usepackage{tikz}
\begin{document}

\newtheorem{lemma}{Lemma}
\newtheorem{corollary}{Corollary}
\newcommand{\rom}[1]
    {\MakeUppercase{\romannumeral #1}}
\shorttitle{Hydrodynamic stability of compressible boundary layers} 
\shortauthor{B. Sharma and S. S. Girimaji} 

\title{Prandtl number effects on the hydrodynamic stability of compressible boundary layers: flow-thermodynamic interactions}

\author
 {
 Bajrang Sharma\aff{1}
  \corresp{\email{bajrangsharma@tamu.edu}}
  \and 
  Sharath S. Girimaji\aff{1, 2}
  }

\affiliation
{
\aff{1}
Department of Aerospace Engineering, Texas A\&M University, College Station, TX 77843, USA
\aff{2}
Department of Ocean Engineering, Texas A\&M University, College Station, TX 77843, USA
}

\maketitle

\begin{abstract}
Hydrodynamic stability of compressible boundary layers is strongly influenced by Mach number ($M$), Prandtl number ($Pr$) and thermal wall boundary condition. These effects manifest on the flow stability via the flow-thermodynamic interactions. Comprehensive understanding of stability flow physics is of fundamental interest and important for developing predictive tools and closure models for integrated transition-to-turbulence computations. The flow-thermodynamic interactions are examined using linear analysis and direct numerical simulations (DNS) in the following parameter regime:  $0.5 \leq M \leq 8$; and, $0.5 \leq Pr \leq 1.3$.  
For adiabatic wall boundary condition, increasing Prandtl number has a destabilizing effect.
In this work, we characterize the behavior of production, pressure-strain correlation and pressure-dilatation as functions of Mach and Prandtl numbers. First and second instability modes exhibit similar stability trends but the underlying flow physics is shown to be diametrically opposite. The Prandtl-number influence on instability is explicated in terms of the base flow profile with respect to the different perturbation mode shapes.
\end{abstract}

\section{Introduction}

High Reynolds number flows tend to become hydrodynamically unstable wherein a small perturbation of the velocity field grows rapidly resulting in a transformation of the base flow field in finite time \citep{drazin2002introduction}. In general, instability draws energy from organized base flow and deposits into a less-organized perturbation field. When the velocity perturbations grow to a threshold magnitude relative to the base flow, nonlinear effects set in \citep{reshotko1976boundary,morkovin1994transition} initiating breakdown of the base flow toward a turbulent state which is characterized by chaotic velocity fluctuations. As a flow transitions from a base laminar flow to a chaotic turbulent state, there is a significant change in the overall mass, momentum and energy transport characteristics \citep{pope2001turbulent,monin2013statistical}. The study of instabilities is therefore of much importance for flows in nature and engineering.

In incompressible flows, all of the kinetic energy extracted from the mean flow by the instability-enabled production mechanism goes toward energizing the perturbation velocity field \citep{pope2001turbulent,george2013lectures}. Pressure is merely a Lagrange multiplier with the sole function of preserving a dilatation-free velocity field and hence does no work (energy transfer) on the velocity field \citep{pope2001turbulent}. As a result, instability and ensuing turbulence analyses do not entail thermodynamic or internal energy considerations.

In compressible flows, the role of pressure reverts to that of a thermodynamic state variable \citep{anderson1990modern}. Pressure field now evolves according to a wave equation derived from internal energy balance and equation of state \citep{landau1987fluid,lele1994compressibility}. The change in the fundamental nature of pressure action triggers important flow-thermodynamics interactions. Most importantly, the velocity field develops a dilatational component allowing for pressure to perform work on the velocity field (or vice versa) via the pressure-dilatation mechanism. Thus additional degrees of freedom and a new component of energy (internal) enter into instability analysis. From the perspective of energetics, the kinetic energy extracted from the mean flow can be diverted away from perturbation kinetic energy to perturbation internal energy by the pressure-dilatation mechanism \citep{sarkar1991analysis,sarkar1992pressure,praturi2019effect,mittal2020nonlinear}. 

The initial growth/decay of small perturbations is described by linear stability theory. Linear stability analysis (LSA) of incompressible boundary layer shows the emergence of Tollmien-Schlichting instability (also termed first mode) beyond a critical Reynolds number \citep{schmid2002stability}. 
Akin to incompressible flows, instability in compressible boundary layers has been studied in literature using linear stability analysis \citep{lees1946investigation, mack1984boundary, reed1996linear,criminale2018theory}. \cite{lees1946investigation} extended the Rayleigh stability criterion \citep{rayleigh1880stability} to compressible flows, and established that an extremum of mean angular momentum ($D(\overline{\rho}D\overline{U})=0$) is necessary for inviscid instability. \cite{mack1984boundary} developed a more complete theory for boundary layers by performing extensive stability calculations. 
At subsonic Mach numbers, compressibilty is known to have a stabilizing effect. Unlike subsonic flows, beyond $M=1$, oblique first modes are more unstable than their 2D counterparts. In addition to the first mode, at high Mach numbers a new family of instability modes coexist along with the first mode \citep{mack1984boundary}. These additional modes belong to the family of trapped acoustic waves and exist whenever there is a relative supersonic region in the flow, i.e, the relative Mach number is greater than 1. The first of these additional modes termed the second or Mack mode becomes the dominant instability \citep{mack1984boundary} at $M\geq4$ for an adiabatic flat plate.
\cite{Gushchin1990ExcitationAD} show that the second mode instability occurs in a region where two modes of the discrete spectrum are synchronized leading to the branching of discrete spectrum. These discrete modes were categorized as fast ($F$) and slow ($S$) by \cite{fedorov2011transition} based on their asymptotic behaviour near the leading edge. The branching pattern of the discrete spectrum is dependent on the Mach number at a fixed Reynolds number \citep{fedorov2011high}. As a result, depending on the flow parameters the second mode can be associated with the fast or slow mode.
Extensive studies have been conducted on the effect of wall cooling for these modes \citep{lees1946investigation,mack1984boundary,malik1989prediction,masad1992effect,mack1993effect}. In general, cooling stabilizes the first mode, while destabilizing the second mode. Consequently for cold walls, the second mode becomes the dominant instability at even lower Mach numbers. Recently, \cite{bitter2015stability} have shown the existence of unstable supersonic modes at very high levels of cooling causing the flow to become unstable over much wider range of frequencies. \cite{malik1991real} investigated real gas effects on the stability of hypersonic boundary layers by considering disassociation of air and conclude that real gas effects stabilize the first mode, while destabilizing the second mode. 

At high temperatures excitation of internal modes and disassociation can lead to large deviation of the effective Prandtl number from it's baseline value \citep{hansen1958approximations,capitelli2000transport}. The Prandtl number for air at atmospheric pressure and extremely high temperatures can be $0.9$ or higher, whereas at low pressures and high temperature the Prandtl number can be as low as $0.3$ \citep{hansen1958approximations,capitelli2000transport}. 
Such combinations of extremely high temperature and low pressure can be experienced during hypersonic reentry in the Jovian atmosphere \citep{seiff1998thermal}. The effect of flow parameters such as Mach number, wall temperature and Reynolds number on compressible boundary layer stability have been studied extensively in literature \citep{mack1984boundary,malik1989prediction,masad1992effect,fedorov2011high}. However, studies examining the effect of Prandtl number have been limited. 
\cite{ramachandran2015linear} investigate the effect of Prandtl number on the eigenspectrum of hypersonic boundary layers. They observe destabilization of both first and second mode with increasing Prandtl number. Moreover, their findings also suggest that the discrete spectrum branching pattern is dependent on Prandtl number. Although the effect of Prandtl number on instability trends and eigenspectrum branching has been discussed, the underlying physics leading to the destabilization has not been clearly explained. 


Comprehensive understanding of instability at different Mach and Prandtl numbers is of much value for many engineering flows for developing predictive tools. Specifically, there is much interest in a unified reduced-order computational tool capable of accurately simulating the entire transition-to-turbulence process. This entails developing closure models for various flow mechanisms and processes contributing toward instability.

In this work, we seek to understand the flow physics underlying the effect of Prandtl number on boundary layer instability with adiabatic walls. These effects manifest on the flow stability via the flow-thermodynamic interactions. Thus we investigate perturbation internal energy, kinetic energy and pressure-velocity interactions. We establish kinetic and internal energy levels of first and second mode at different Mach and Prandtl numbers. 
The various flow thermodynamic interactions and turbulence mechanisms contributing to first and second instability modes are also examined.
We explicate the observed instability behavior by characterizing production, pressure-strain correlation and pressure-dilatation at different Mach and Prandtl numbers.
The profiles of the base-flow and the stresses are analyzed to explain the different trends shown by first and second mode.
The results obtained by LSA are corroborated by direct numerical simulations performed using the gas kinetic method \citep{Xu2001}. Thus the work also leads to the validation of the kinetic-theory based numerical scheme and computational code.

\section{\label{sec:GovEquation}Governing equations and linear analysis}
The compressible Navier-Stokes equations for an ideal fluid are as follows
\begin{subequations}\label{Eq:full_eq}
\begin{align}
    \begin{split}
    \frac{\partial \rho^*}{\partial t^*} + \frac{\partial}{\partial x_j^*} (\rho^* u_j^*) = & 0, 
    \label{Eq:cont} 
    \end{split}\\
    \begin{split}
    \frac{ \partial (\rho^* u_i^*)}{\partial t^*} + \frac{\partial (\rho^* u_i^* u_j^*)}{\partial x_j^*} = & -\frac{\partial p^*}{\partial x_i^*} + \frac{\partial \tau_{ij}^*}{\partial x_j^*}, 
    \label{Eq:mom} 
    \end{split} \\
    \begin{split}
        \frac{\partial}{\partial t^*} \bigg(\frac{p^*}{\gamma-1}\bigg) + \frac{\partial }{\partial x_j^*}\bigg(\frac{p^* u_j^*}{\gamma-1}\bigg) = & \frac{\partial}{\partial x_j^*} \bigg( \kappa^* \frac{\partial T^*}{\partial x_j^*} \bigg) 
         - p^* \frac{\partial u_k^*}{\partial x_k^*} + \tau_{ij}^*\frac{\partial u_i^*}{\partial x_j^*}, 
    \label{Eq:energy}
    \end{split} \\
    \begin{split}
    p^* = & \rho^* R T^*, \label{Eq:State}
    \end{split}
\end{align}
\end{subequations}
where, the superscript $^{*}$ is used to denote the dimensional variables. The density of the fluid is denoted by $\rho^*$, velocity by $u_i^*$, temperature by $T^*$ and pressure by $p^*$. $\gamma$ is the specific heat ratio, $\kappa^*$ is the coefficient of thermal conductivity, $R$ is the universal gas constant and $\tau_{ij}^*$ is the viscous stress tensor given by:
\begin{equation}
    \label{eq:stress_tensor}
    \tau_{ij}^*=\mu^*\left(\frac{\partial u_i^*}{\partial x_j^*}+\frac{\partial u_j^*}{\partial x_i^*}\right)-\frac{2}{3}\mu^*\frac{\partial u_k^*}{\partial x_k^*}\delta_{ij}.
\end{equation}
The coefficient of viscosity $\mu^*$ is dependent on the local temperature as dictated by the Sutherland's law \citep{lii1893viscosity}. 

The evolution of total kinetic energy ($\rho^*u_i^*u_i^*/2$) as obtained from the momentum equations \eqref{Eq:mom} is given by
\begin{equation}
    \label{eq:TotalKineticEnergy}
    \begin{split}
        \frac{ \partial (\rho^* u_i^*u_i^*/2)}{\partial t^*} + \frac{\partial (u_j^*\rho^* u_i^*u_i^*/2 )}{\partial x_j^*}  =  \underbrace{p^*\frac{\partial u_i^*}{\partial x_i}}_{\Pi} \underbrace{-\tau_{ij}^*\frac{\partial u_i^*}{\partial x_j}}_{\epsilon}  +\underbrace{\frac{\partial}{\partial x_j}\left[\tau_{ij}^*u_i^*-p^*u_i^*\delta_{ij}\right]}_{\mathcal{T}}.
    \end{split}
\end{equation}
Here $\Pi$ represents pressure-dilatation, $\epsilon$ is viscous dissipation of kinetic energy and $\mathcal{T}$ is the kinetic energy transport term. From equation \eqref{Eq:energy} it is evident that the pressure-dilatation and dissipation terms couple the kinetic and internal/pressure modes. Pressure-dilatation enables a reversible exchange between the internal and kinetic energies. On the other hand, the dissipation of kinetic energy to the internal mode is irreversible.

\subsection{Linear Stability Analysis}
\label{sec:LSA}
The dimensional variables are normalized as follows,
\begin{equation}
\begin{split}
     & u_i = \frac{u_i^*}{U_\infty}, \hspace{2mm} \rho = \frac{\rho^*}{\rho_\infty}, \hspace{2mm} T = \frac{T^*}{T_\infty}, \hspace{2mm} p = \frac{p^*}{\rho_\infty U_\infty^2}, \\
     & x = \frac{x_i^*}{L_r}, \hspace{2mm} t = \frac{t^*U_\infty}{L_r} , \hspace{2mm} \mu = \frac{\mu^*}{\mu_\infty}, \hspace{2mm}  \kappa = \frac{\kappa^*}{\kappa_\infty}, 
\end{split}
\end{equation}
where, $U_\infty$ is the freestream velocity, $\rho_\infty$ is the freestream density, $T_\infty$ is the freestream temperature, $\mu_\infty$ and $\kappa_\infty$ are the freestream viscosity and thermal conductivity respectively. The spatial co-ordinate $x_i^*$ is normalized by the Blassius length scale $L_r=\sqrt{\mu_\infty x^*/\rho_\infty U_\infty}$.

\begin{figure}
    \centering
    \includegraphics[width=0.7\textwidth, keepaspectratio]{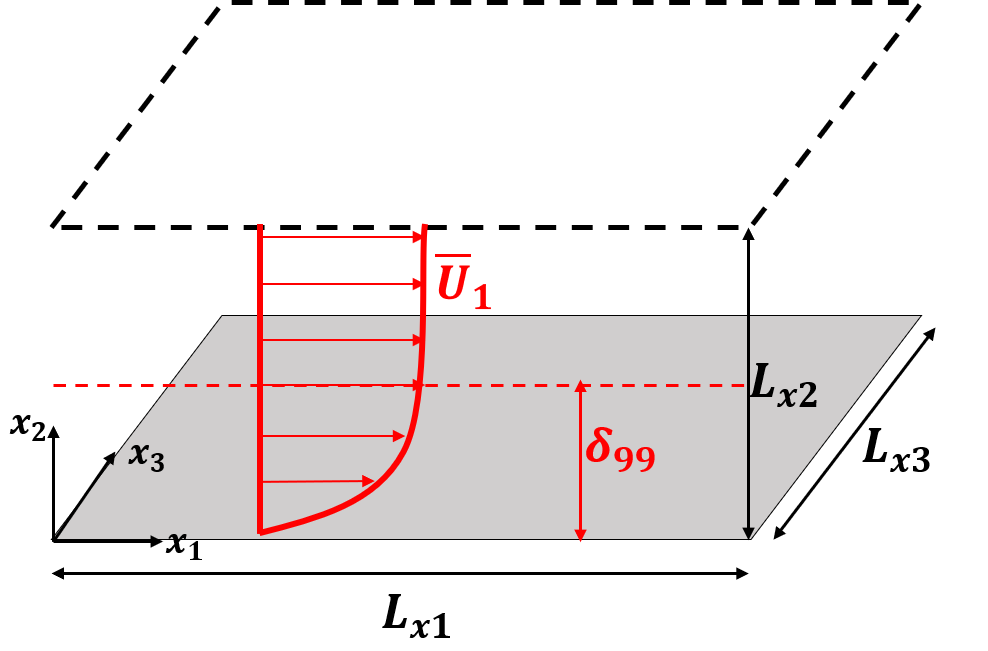}
    \caption{Schematic of the basic state for LSA and the problem setup for DNS. $L_{x1}$, $L_{x2}$ and $L_{x3}$ represent the domain sizes in the streamwise, wall-normal and spanwise directions respectively. $\delta_{99}$ denotes the 99\% boundary layer thickness.}
    \label{fig:Setup}
\end{figure}

The flow variables are then decomposed into a basic state and perturbations.
\begin{equation}
    \label{eq:linear_decomp}
    A=\overline{A}+A'.
\end{equation}
 Here, $A$ represents the flow variables ($u_i, \rho, p, T$). We assume a two dimensional locally parallel basic state wherein the wall-normal and spanwise base velocities are zero. Moreover, the basic state properties only vary along the wall-normal direction $x_2$. A schematic of the base flow is shown in figure \ref{fig:Setup}. The fluid properties, viscosity and thermal conductivity are also decomposed into a base state and perturbations. The base viscosity ($\overline{\mu}$) is obtained from the Sutherland's law of viscosity \citep{lii1893viscosity} while the base thermal conductivity ($\overline{\kappa}$) is varied to ensure a constant Prandtl number across the boundary layer. 
 The viscosity and thermal conductivity perturbations are expressed in terms of temperature fluctuations as:
\begin{equation}
    \label{eq:linear_mupert}
    \mu'=\frac{d\overline{\mu}}{d\overline{T}}T';\quad\quad\kappa'=\frac{d\overline{\kappa}}{d\overline{T}}T',
\end{equation}
where the derivatives $d\overline{\mu}/d\overline{T}$ and $d\overline{\kappa}/d\overline{T}$ are also computed from Sutherland's law.

The basic state is obtained by solving the two dimensional compressible laminar flat plate boundary layer equations using the Levy-Lees similarity transformation \citep{rogers1992laminar}. The physical coordinates ($x_1^*, x_2^*$) are transformed to the ($\xi-\eta$) space using the following relations.
\begin{equation}
    \label{eq:Levy_Lees}
    \begin{split}
        \xi=\int_0^{x_1^*} \rho_\infty\mu_\infty U_\infty dx_1^*, \\
        \eta=\frac{U_\infty}{\sqrt{2\xi}}\int_0^{x_2^*} \overline{\rho} dx_2^*.
    \end{split}
\end{equation}
The basic state equations in the transformed space reduces to a system of ordinary differential equation (ODE) given by \citep{rogers1992laminar} :
\begin{subequations}\label{eq:BaseFlow}
    \begin{gather}
        (Cf'')' + ff''=0,\\
        \left(\frac{C}{Pr}g'\right)' + fg' + (\gamma-1)M^2Cf''^2=0.
    \end{gather}
\end{subequations}
Here, the similarity variable $f'=\overline{U}_1$ is the non-dimensional streamwise velocity, $g=\overline{T}$ is the non-dimensional temperature, $C=\overline{\rho}\overline{\mu}$ is the Chapman-Rubesin factor, $M$ is the free stream Mach number and  $Pr$ is the Prandtl number. 
For an adiabatic flat plate, the set of ODEs \eqref{eq:BaseFlow} are subjected to the following boundary conditions.
\begin{equation}
    \label{eq:BaseFLowBc}
    \begin{split}
        & f(0)=0;\quad f'(0)=0;\quad g'(0)=0; \\
        & f'(\eta\to\infty)=1;\quad g(\eta\to\infty)=1.
    \end{split}
\end{equation}
The resulting boundary value problem is solved using the Nachtsheim-Swigert iteration technique \citep{nachtsheim1965statisfaction}. 

The basic state equations are subtracted from the full Navier-Stokes equations \eqref{Eq:full_eq} and  the higher order terms are neglected to obtain the linearized perturbation equations. The full form of the linearized perturbation equations is detailed in appendix \ref{app:A1}. 
The non-dimensional parameters in the perturbation equations are defined below.
\begin{equation}
    \label{eq:nondparam}
    Re=\frac{\rho_\infty U_\infty L_r}{\mu_\infty}\quad M=\frac{U_\infty}{\sqrt{\gamma R T_\infty}}\quad Pr=\frac{\mu_\infty C_p}{\kappa_\infty},
\end{equation}
where, $C_P=\gamma R/(\gamma-1)$ is the specific heat at constant pressure and the specific heat ratio $\gamma=1.4$.

The perturbations are then expressed in the normal mode form as,
\begin{equation}
\label{eq:linear_mode}
    A'=\hat{A}(x_2)e^{\iota(\alpha x_1+\beta x_3-\omega t)},
\end{equation}
where $\alpha$, $\beta$ are the wavenumbers in the streamwise and spanwise wavenumbers respectively, $\omega$ is the temporal frequency and $\hat{A}$ is the amplitude of perturbation varying  in the wall-normal direction. For temporal stability analysis, $\alpha$ and $\beta$ are assumed to be real and specified apriori, while, $\omega$ is the complex eigenvalue obtained from analysis. The sign of imaginary part of $\omega$ ($\omega_i$) determines the stability: perturbations grow if $\omega_i>0$ and decay if $\omega_i<0$. 

Substituting the modal form of perturbations \eqref{eq:linear_mode} into the linearized perturbation equation (\ref{eq:PerturbationContinuity}-\ref{eq:PerturbationEnergy}) yields the following eigenvalue problem:
\begin{equation}
    \label{eq:linear_eig}
    \omega\Phi =\mathbf{A}^{-1}\mathbf{B}(\alpha,\beta,Re,M,Pr)\Phi.
\end{equation}
Here, $\Phi=[\hat{u}_1,\hat{u}_2,\hat{u}_3,\hat{T},\hat{p}]$ are the eigenmode shapes corresponding to the eigenvalue $\omega$. The elements of the $5^{th}$ order coefficient matrix $\mathbf{A}$ and $\mathbf{B}$ are listed in appendix \ref{app:A2}.
The eigenvalue problem is solved by discretizing equation \eqref{eq:linear_eig} using Chebyshev polynomials \citep{malik1990numerical} on collocation points. The Chebyshev polynomials are defined on the following Gauss-Labato points ($\xi_i$) in the interval $[-1,1]$.
\begin{equation}
    \label{eq:gauss_labato}
    \xi_i=\cos{\frac{\pi i}{N}}\quad\quad i=0,1...N,
\end{equation}
where, $N$ is the number of collocation points. The physical domain ($x_2 \in [0,L_{x2}]$) is mapped to the computational domain using an algebraic stretching function \citep{malik1990numerical}.
\begin{equation}
    \label{eq:algebraic_stretching}
    x_2=a\frac{1+\xi}{b-\xi};\quad\mbox{where } b=1+\frac{2a}{L_{x2}};\mbox{ \& } a=\frac{y_lL_{x2}}{L_{x2}-2y_l}
\end{equation}
Here, $L_{x2}$ is the edge of physical domain and half of the grid points lie between the wall and the parameter $y_l$. The parameter $y_l$ is selected to be half of the 99\% boundary layer thickness ($\delta_{99}$) in all the stability calculations. No slip and zero thermal perturbation boundary conditions are used for velocity and temperature, while, a Neumann boundary condition for pressure is obtained by solving the wall-normal momentum equation. The global eigenvalue problem is solved using the QZ algorithm \citep{moler1973algorithm} at 199 collocation points. 

We validate the results of linear stability analysis by comparison against the multi domain spectral method (MDSP) of \cite{malik1990numerical}. The eigenvalue $\omega$ of the most unstable mode for two different cases are listed in table \ref{table:EigVerify}. The eigenvalues obtained from the code used in the current work are in excellent agreement with the MDSP results from \cite{malik1990numerical}.  

\begin{table}
	\begin{center}
	  \begin{tabular}{ccccccc}
 	 \hspace{1mm} $\mathbf{M}$ \hspace{1mm} & \hspace{1mm} $\mathbf{Re}$ \hspace{1mm} & \hspace{1mm} $\mathbf{T_0 (K)}$ \hspace{1mm}  
 	& \hspace{1mm} $\mathbf{\alpha}$ \hspace{1mm} & \hspace{1mm} $\mathbf{\beta}$ \hspace{1mm} & \hspace{1mm} $\mathbf{\omega}$ (\textbf{MDSP} \citep{malik1990numerical}) \hspace{1mm} & \hspace{1mm} $\mathbf{\omega}$ (\textbf{Current}) \hspace{1mm}  \\ \hline 
  	$0.5$ & $2000$ & $278$ & $0.1$& $0$ & $0.0290817+0.0022441\iota$ & $0.0290829+0.0022441\iota$ \\
  	$2.5$ & $3000$ & $333$ & $0.06$& $0.1$ & $0.0367340+0.0005840\iota$ & $0.0367379+0.0005875\iota$ \\ 
    \end{tabular}  
	\end{center}
    \caption{Comparison of eigenvalues of the most unstable mode.}
    \label{table:EigVerify}
\end{table}

\subsection{\label{sec:FlowProcesses}Flow processes in the linear limit}
In this subsection, we discuss the role of key turbulent processes in the linear limit. Starting from the perturbation momentum equation \eqref{eq:PerturbationMomentum} the instantaneous perturbation kinetic energy ($k=\overline{\rho}u_i'u_i'/2$) equation can be derived as,
\begin{equation}
   \label{eq:KeEquation}
    \begin{split}
        \frac{\partial k}{\partial t}+ 
        \overline{U}_i\frac{\partial k}{\partial x_i} = -\overline{\rho}u_i'u_k'\frac{\partial \overline{U}_i}{\partial x_k} 
        + p'\frac{\partial u_i'}{\partial x_i} 
         -\frac{1}{Re}\tau_{ik}'\frac{\partial u_i'}{\partial x_k}
        +\frac{\partial}{\partial x_k}\left[\frac{1}{Re}\tau_{ik}'u_i'-p'u_i'\delta_{ik}\right].
    \end{split}
\end{equation}
The instantaneous kinetic energy equation \eqref{eq:KeEquation} is averaged in the homogeneous $x_1$ and $x_3$ directions to derive the average kinetic energy equation.
\begin{equation}
   \label{eq:KeBudget}
    \begin{split}
        \frac{\partial \langle k\rangle }{\partial t}+ 
        \overline{U}_i\frac{\partial\langle k\rangle }{\partial x_i} = 
        - \underbrace{\langle\overline{\rho}u_i'u_k'\rangle \frac{\partial \overline{U}_i}{\partial x_k}}_{P_k} 
        +   \underbrace{\left\langle p'\frac{\partial u_i'}{\partial x_i}\right\rangle }_{\Pi_k} 
          \underbrace{-\frac{1}{Re}\left\langle\tau_{ik}'\frac{\partial u_i'}{\partial x_k}\right\rangle }_{\epsilon_k} 
        +  \underbrace{\frac{\partial}{\partial x_k}\left[\frac{1}{Re}\langle \tau_{ik}'u_i'\rangle -\langle p'u_i'\rangle \delta_{ik}\right]}_{\mathcal{T}_k},
    \end{split}
\end{equation}
where the notation $\langle \rangle $ denotes the averaging operator in the homogeneous directions and is defined as follows,
\begin{equation}
    \langle k \rangle = \frac{1}{L_{x1}L_{x3}}\int_0^{L_{x3}}\int_0^{L_{x1}}k dx_1dx_3.
\end{equation}
The key turbulent processes are defined in equation \eqref{eq:KeBudget}. $P_k$ denotes the production of kinetic energy, $\Pi_k$ is pressure-dilatation, $\epsilon_k$ is dissipation of kinetic energy and $\mathcal{T}_k$ is the transport term. The role of the aforementioned processes is well known in the context of turbulence. A brief overview in the current context is presented here. The perturbation velocity field extracts energy from the basic state via production. Pressure-dilatation quantifies the amount of pressure work on the velocity field. 
In incompressible flows, the net work done by pressure on the velocity field is zero at each point in the flow field due to the solenoidal nature of velocity field.
On the other hand, flow-thermodynamic interactions become important for compressible flows as $\Pi_k$ becomes significant. The energy transfer enabled by pressure-dilatation is reversible.
The dissipation process irreversibly transfers energy from the perturbation velocity field to the mean flow internal energy in both compressible and incompressible flows.
 The transport terms merely redistributes energy in space. It must be noted that the transport terms are zero in the streamwise and spanwise directions due to spatial homogeneity.  
 In the linear limit of small perturbation the basic state remains unaltered. Therefore the basic state can be considered an infinite source/sink of energy. 

We now derive the perturbation internal energy equation. In the linear limit, pressure variance can be approximated as the internal energy \citep{sarkar1991analysis, mittal2019mathematical}. The instantaneous perturbation internal energy is defined as,
\begin{equation}
    \label{eq:InternalEnergyDef}
    e=\frac{p'p'}{2\gamma\overline{P}}.
\end{equation}
The governing equation for the averaged internal energy ($\langle e\rangle$) in pressure fluctuations is,
\begin{equation}
    \label{eq:InternalEnergyBudget}
        \begin{split}
         \frac{\partial \langle e\rangle }{\partial t} + \overline{U}_i\frac{\partial\langle e\rangle }{\partial x_i}  = & -\underbrace{\left\langle p'\frac{\partial u_k'}{\partial x_k}\right\rangle }_{\Pi_k}
         \underbrace{-\frac{1}{\gamma RePrM^2\overline{P}}\left\langle p'\frac{\partial q_k'}{\partial x_k}\right\rangle }_{T_s} \\ 
          & + \underbrace{\frac{\gamma-1}{\gamma Re\overline{P}}\left[\langle p'\tau_{ij}'\rangle \frac{\partial \overline{U}_i}{\partial x_j}+\overline{\tau}_{ij}\left\langle p'\frac{\partial u_i'}{\partial x_j}\right\rangle \right]}_{\epsilon_s}.   
    \end{split}
\end{equation}
Here, $T_s$ and $\epsilon_s$ denotes the thermal flux and viscous contribution to internal energy respectively. It is evident from \eqref{eq:InternalEnergyBudget} that pressure-dilatation couples the internal and kinetic modes of the perturbation field. The thermal flux and viscous flux terms represent the interaction of fluctuating internal field with the mean internal field via heat conduction and viscous action respectively.

Finally, the evolution equation for the stress components $R_{ij}=-\overline{\rho}u_i'u_j'$ are derived. The evolution of averaged stresses $\langle R_{ij}\rangle $ is given by the following equation.
\begin{equation}
   \label{eq:ReynoldsStressBudget}
    \begin{split}
        \frac{\partial \langle R_{ij}\rangle }{\partial t} & + 
        \overline{U}_k\frac{\partial \langle R_{ij}\rangle }{\partial x_k} =  \underbrace{-\langle\overline{\rho}u_j'u_k'\rangle\frac{\partial \overline{U}_i}{\partial x_k}-\langle\overline{\rho}u_i'u_k'\rangle\frac{d\overline{U}_j}{dx_k}}_{P_{ij}} 
         + \underbrace{\left\langle p'\left(\frac{\partial u_i'}{\partial x_j}+\frac{\partial u_j'}{\partial x_i}\right)\right\rangle}_{\Pi_{ij}} \\
        & \underbrace{-\frac{1}{Re}\left(\left\langle\tau_{ik}'\frac{\partial u_j'}{\partial x_k}\right\rangle+\left\langle\tau_{jk}'\frac{\partial u_i'}{\partial x_k}\right\rangle\right)}_{\epsilon_{ij}}
         +\underbrace{\frac{\partial}{\partial x_k}\left[\left\langle\frac{1}{Re}\tau_{ik}'u_j'+\frac{1}{Re}\tau_{jk}'u_i'-p'u_i'\delta_{jk}-p'u_j'\delta_{ik}\right\rangle\right]}_{\mathcal{T}_{ij}},
    \end{split}
\end{equation}
where, $P_{ij}$ are the components of production tensor for stresses, $\Pi_{ij}$ denotes the components of pressure-strain correlation, $\epsilon_{ij}$ is the dissipation tensor and $\mathcal{T}_{ij}$ denotes the diffusion term. The trace of $P_{ij}$ and $\Pi_{ij}$ is equal to twice the production and pressure-dilatation respectively. The pressure-strain correlation redistributes energy among different stress components.

In a recent work by \cite{weder2015decomposition} a balance equation for the total disturbance energy is derived and the temporal growth rate is decomposed into production and dissipation components. Such a decomposition \citep{weder2015decomposition} is aimed at isolating the contribution of processes facilitating an exchange between the base and perturbation field. Consequently, flow-thermodynamic interactions in the perturbation field cannot be analyzed within this framework. In this work, the budget equations \eqref{eq:KeBudget}-\eqref{eq:ReynoldsStressBudget} are examined to highlight both base-perturbation interactions and the energy exchanges within the perturbation field. 

\subsection{Dependence of base flow on Prandtl number}
\begin{figure}
    \centering
    \subfloat[$\overline{U}_1$]{\includegraphics[width=0.45\textwidth, keepaspectratio]{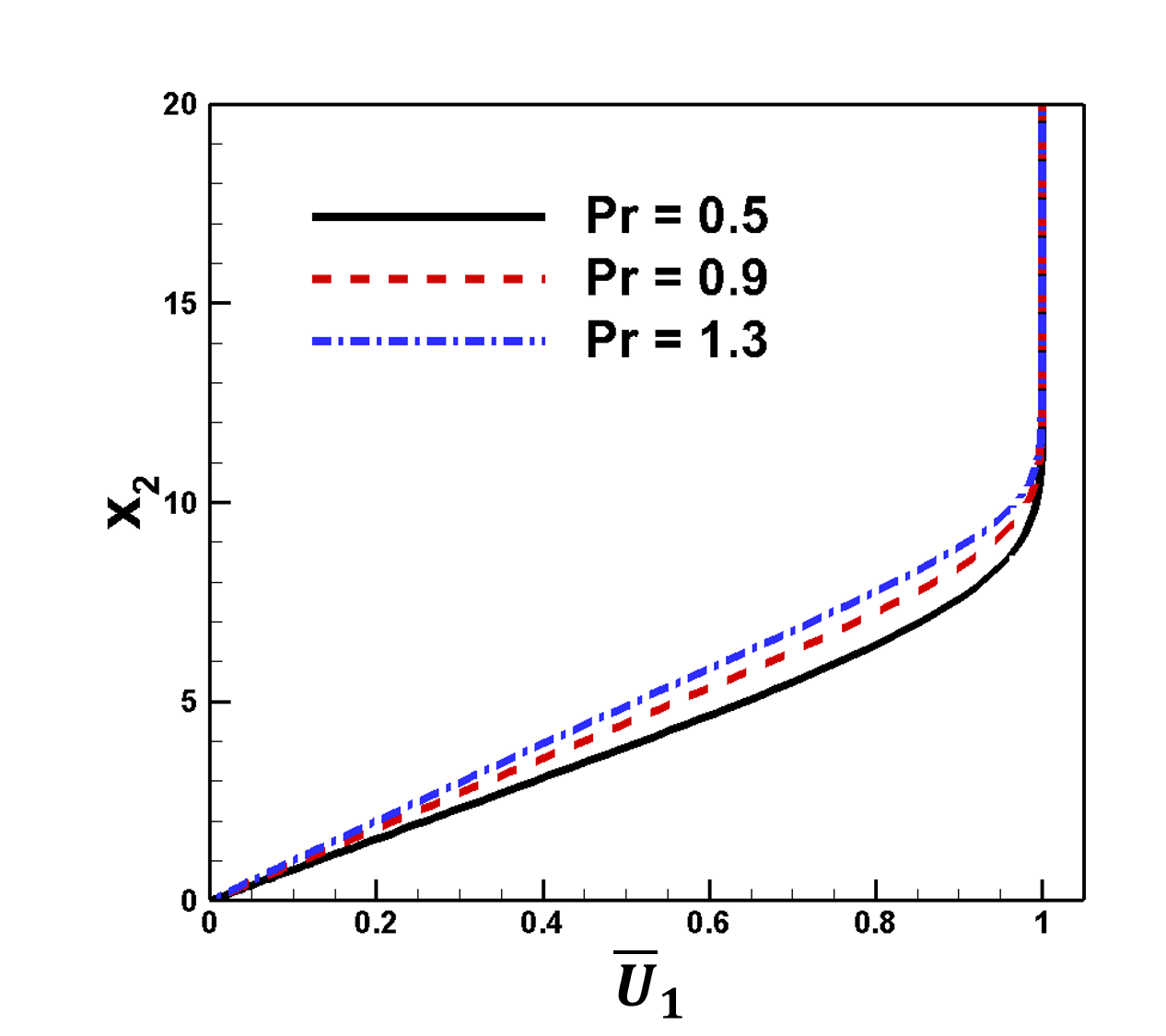}}
    \subfloat[$\overline{T}$]{\includegraphics[width=0.45\textwidth, keepaspectratio]{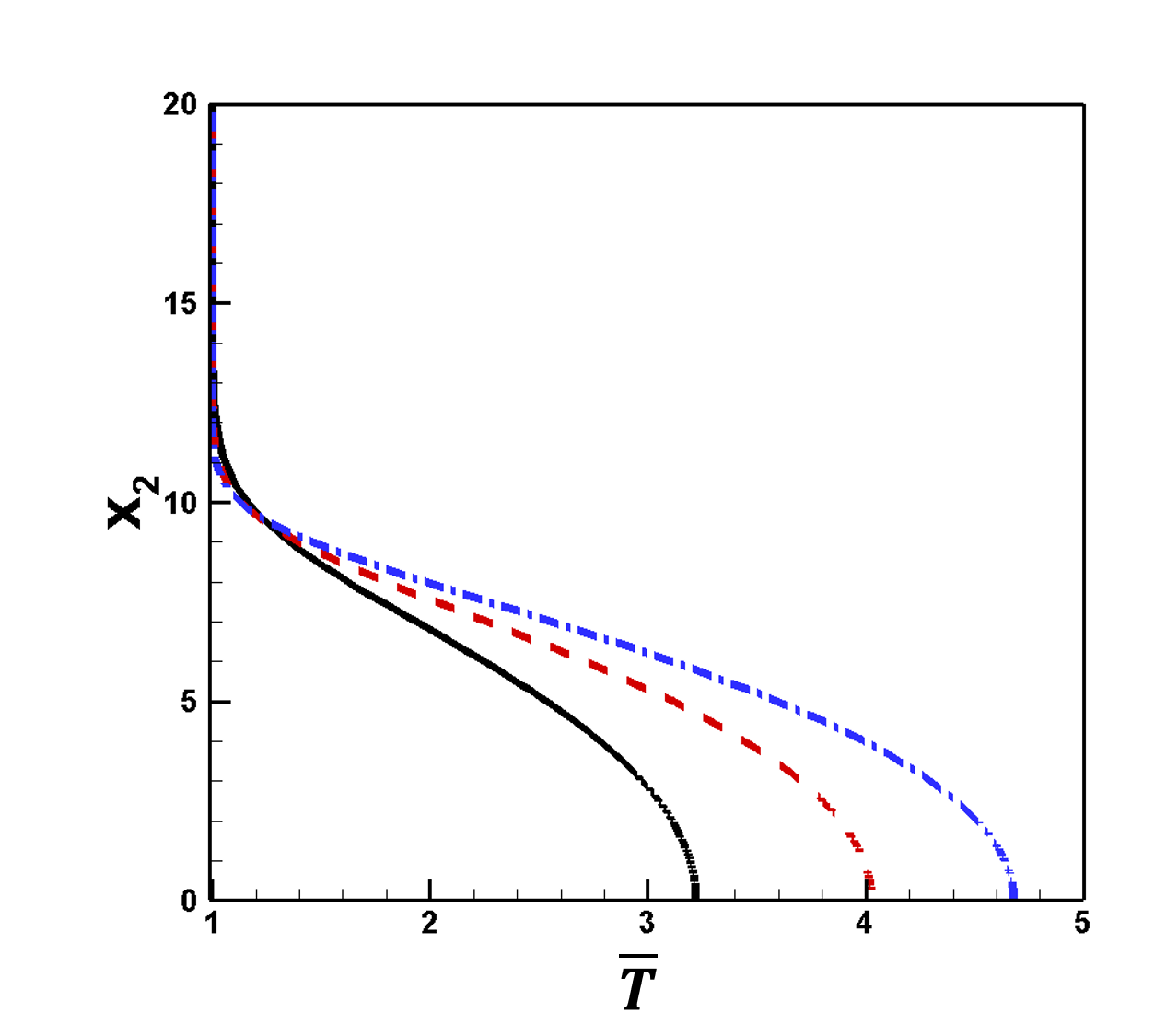}} \\
    \subfloat[$d \overline{U}_1/d x_2$]{\includegraphics[width=0.45\textwidth, keepaspectratio]{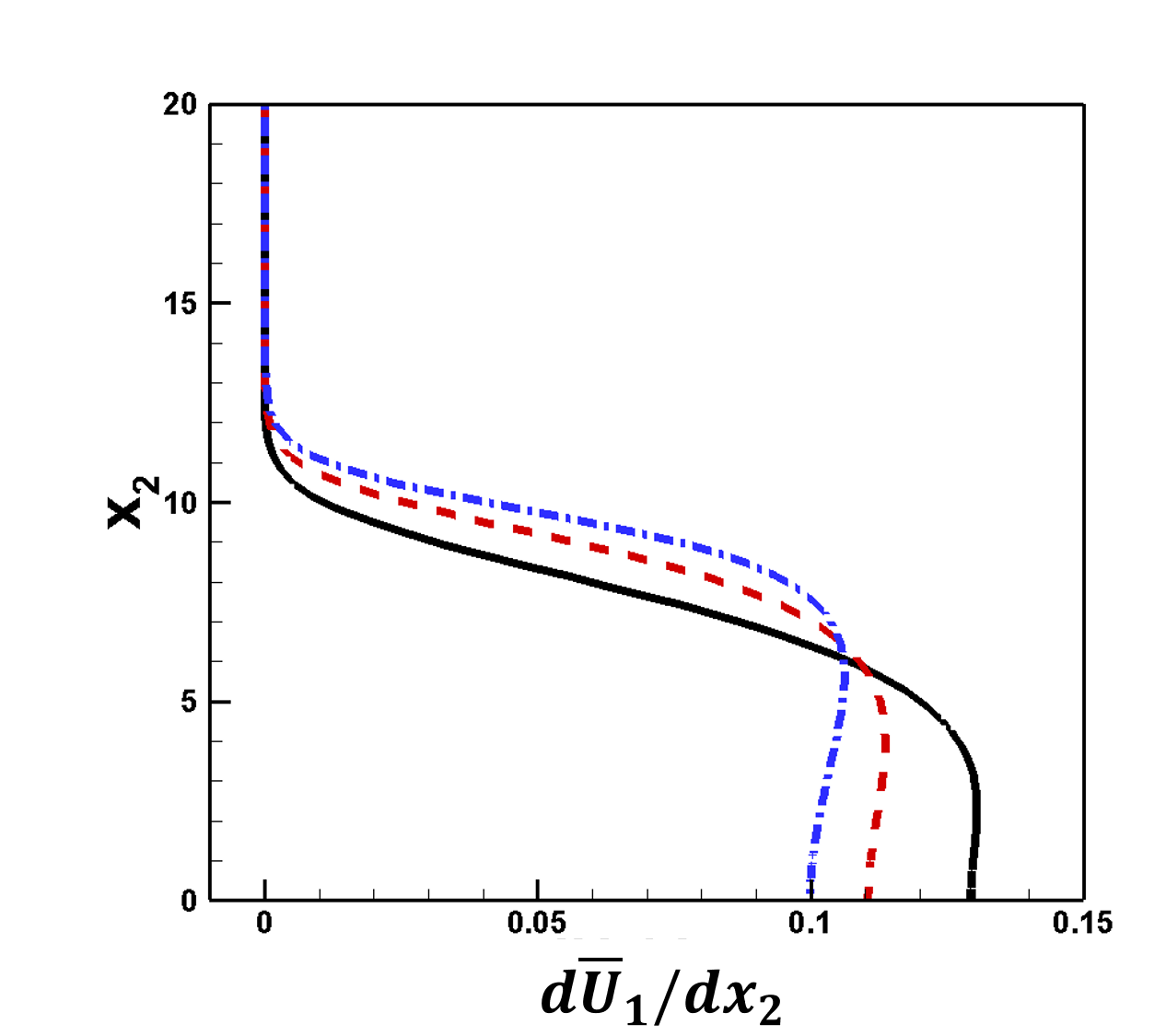}}
    \subfloat[$d \overline{T}/d x_2$]{\includegraphics[width=0.45\textwidth, keepaspectratio]{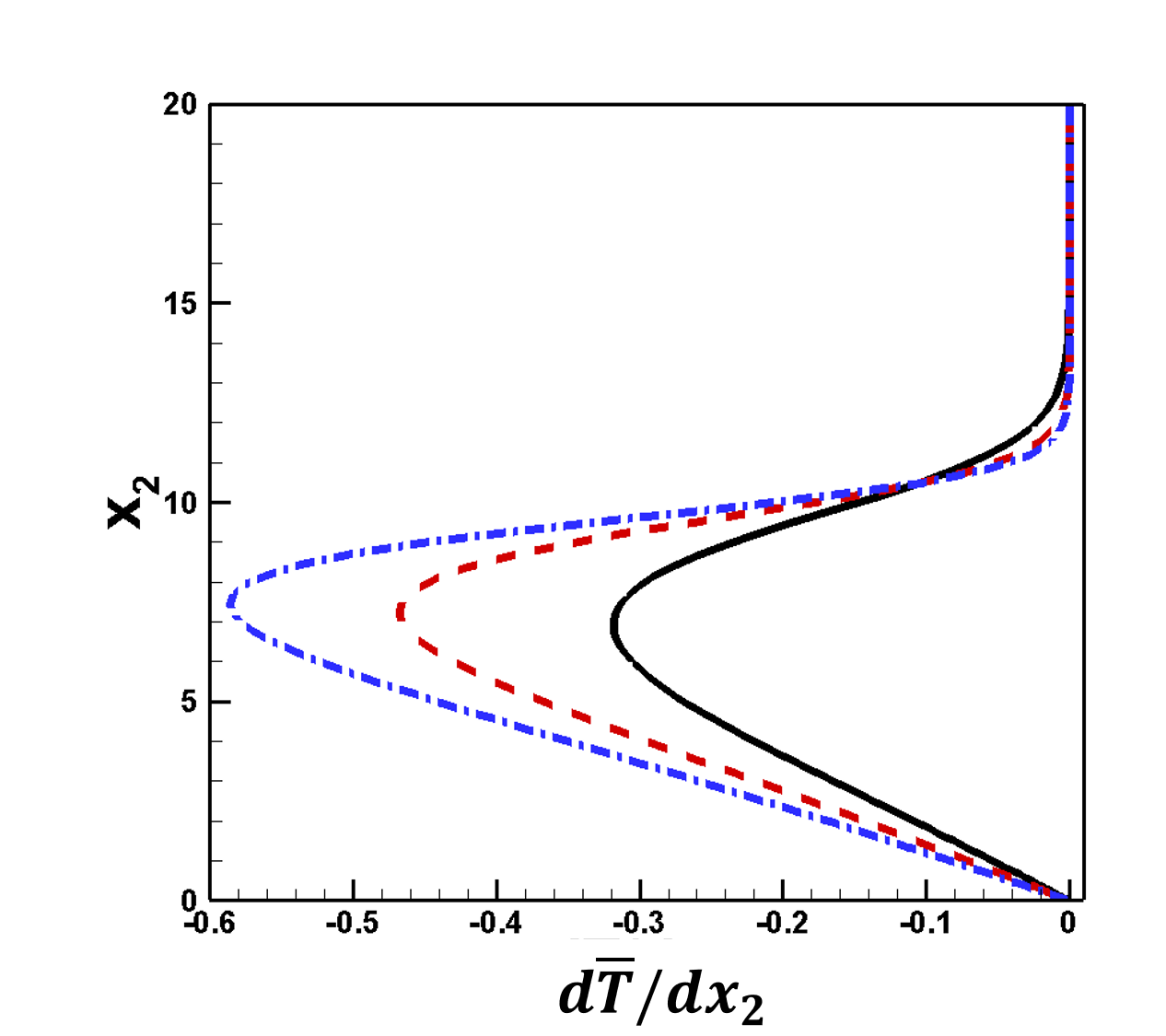}}
    \caption{Profiles of base (a) velocity ($\overline{U}_1$), (b) temperature ($\overline{T}$), (c) velocity gradient ($d\overline{U}_1/dx_2$) and, (d) temperature gradient ($d\overline{T}/dx_2$) at $M=4$ for three different Prandtl numbers.}
    \label{fig:BasicState}
\end{figure}

 The basic state plays a key role in the instability dynamics. It directly influences several key processes such as production, dissipation and thermal flux, and it varies substantially with Prandtl number. Figures \ref{fig:BasicState}(a)-(b) plots the base velocity and temperature profiles for an adiabatic flat plate boundary layer at $M=4$ at three different Prandtl numbers. The base velocity and temperature gradient are also plotted in figures \ref{fig:BasicState}(c)-(d). The adiabatic wall temperature ($\overline{T}_{aw}$) is dependent on $M$ and $Pr$ according to the following relation \citep{rogers1992laminar,dorrance2017viscous}.
\begin{equation}
    \overline{T}_{aw}=1+\frac{\gamma-1}{2}M^2\sqrt{Pr}
\end{equation}
It is evident from figure \ref{fig:BasicState} that the temperature at the wall increases with Prandtl number. Consequently, the peak temperature gradient inside the boundary layer is stronger at higher Prandtl numbers. Increasing the Prandtl number leads to stronger viscous transport compared to thermal diffusion. As a result the boundary layer thickness ($\delta_{99}$) increases while the thermal boundary layer thickness decreases with increasing Prandtl number. The velocity gradient is stronger near the wall at lower Prandtl numbers. The velocity gradient weakens toward the boundary layer edge. Beyond $x_2\approx 6$ ($0.55\delta_{99}$) the velocity gradient is stronger at $Pr=1.3$ compared to the lower Prandtl number cases.      

\section{Methodology for direct numerical simulations}
Although the linear analysis employs Navier-Stokes equations, direct numerical simulations of a temporally evolving boundary layer are performed using a finite volume solver based on the gas kinetic method \citep{Xu2001} (GKM). The GKM solver is capable of accommodating non-equilibrium thermodynamic effects. The GKM-DNS results will be compared against linear theory for validation of the numerical method.

A brief overview of the gas kinetic method is provided here, and for more details the reader is referred to \cite{Xu2001}.
The GKM solves the Boltzmann equation 
\begin{equation}
    \frac{\partial f}{\partial t} + \vec{c} \cdot \nabla f + \vec{a}\cdot\nabla_c f = \bigg(\frac{\partial f}{\partial t}\bigg)_{collisions}, \label{Eq:bolt}
\end{equation}
describing the evolution of single particle probability density function, $f (\vec{x},\vec{c},t)$, defined as a function of physical space, velocity space and time \citep{Xu2001}. Here, $\vec{a}$ is the particle acceleration. Solving the more fundamental Boltzmann equation allows  applicability over a wider range of flow conditions for addressing non-equilibrium and non-continuum effects encountered in high speed flows. The collision terms in the Boltzmann equation are modeled using the Bhatnagar-Gross-Krook (BGK) model resulting in the following Boltzmann-BGK (BBGK) equation.
\begin{equation}
    \frac{\partial f}{\partial t} + \vec{c} \cdot \nabla f + \vec{a}\cdot\nabla_c f = \frac{g-f}{\tau}, \label{Eq:BGK}
\end{equation}
where, $g$ is the equilibrium (i.e. Maxwellian) particle distribution function and $\tau$ is the characteristic relaxation time.

The macroscopic variables, $U = [\rho, \rho u_i , E]^T$, are obtained from the distribution function, $f$, using,
\begin{equation}
    U = \int_{-\infty}^{\infty} \psi f d\Xi.
\end{equation}
Here $\rho$ is fluid density, $u_i$ is macroscopic velocity, $E$ is the sum of kinetic and thermal energy densities, $\psi = [1,c_i,\frac{1}{2}(c_i^2 + \xi^2)]^T$, $\xi$ is an internal variable with $K=(5-3\gamma)/(\gamma-1)$ degrees of freedom and $d\Xi = dc_i d\xi$ is a volume element in phase space. Being a finite-volume based solver the GKM is governed by,
\begin{equation}
    \frac{\partial}{\partial t} \int_\Omega U dx + \oint_A \vec{F}.d\vec{A} = 0,
    \label{Eq:FVM}
\end{equation}
where $\Omega$ is the control volume, $A$ is the surface of control volume and $\vec{F}$ is the flux of macroscopic variables. Equation \eqref{Eq:FVM} is integrated in time and discretized in space. The solution update $U$ at time step $n+1$ and at the cell centre $(i,j,k)$ is obtained as,
\begin{equation}\label{Eq:FVM_disc}
\begin{split}
    U_{i,j,k}^{n+1} = & U_{i,j,k}^n - \frac{1}{\Delta x} \int_0^t F_{i+1/2,j,k}(t) - F_{i-1/2,j,k}(t) dt \\
    & - \frac{1}{\Delta y} \int_0^t G_{i,j+1/2,k}(t) - G_{i,j-1/2,k}(t) dt \\
    &  - \frac{1}{\Delta z} \int_0^t H_{i,j,k-1/2}(t) - H_{i,j,k-1/2}(t) dt ,
\end{split}
\end{equation}
where $\mathbf{F_i} = [F,G,H]$ are the fluxes for the conservative variables. The flux at the cell interface $(i+1/2,j,k)$ are then calculated from the distribution function using the following relation.
\begin{equation}
    \mathbf{F_i} = [F_\rho , F_{\rho u_i} , F_E]^T = \int_{-\infty}^{\infty} c_i \psi f_{i+1/2,j,k} (c, t , \xi) d\Xi, \label{Eq:Flux}
\end{equation}
Here, $F_\rho$, $F_{\rho u_i}$ and $F_{E}$ represent the density, momentum and energy flux respectively. The flux calculations at the cell interface require the interpolation of conservative variables from the cell center. The interpolation is performed by a $5^{th}$ order weighted essentially non-oscillatory scheme \citep{kumar2013weno}.

The GKM solver used in the current work has already been validated for various compressible flows: channel flows \citep{mittal2020nonlinear}, decaying and homogeneous shear turbulence \citep{kumar2013weno,Kumar2014} and mixing layers with Kelvin-Helmholtz instability \citep{MONA_2016}. In this work, the DNS results will be compared against linear stability analysis for the case of high speed boundary layers.

Temporal simulations \citep{adams1992method,adams1993numerical,adams1996subharmonic} of a flat plate adiabatic boundary layer is considered. The problem setup is shown in figure \ref{fig:Setup}. The temporal approach allows for the use of periodic boundary conditions in both streamwise ($x_1$) and spanwise ($x_3$) directions. A forcing term \citep{adams1996subharmonic} is added to the governing equation to ensure that the boundary layer is locally parallel and the basic state is independent of $x_1$. 
It must be noted that the effects of boundary layer growth has not been accounted for in the current computations. As a result, the basic state stays invariant allowing for a direct comparison with temporal stability analysis described in subsection \ref{sec:LSA}. 
At the wall, no slip boundary conditions for velocity is employed, while temperature is set to the adiabatic wall temperature. The Dirichlet boundary condition for temperature ensures consistency with linear analysis, wherein the temperature perturbation vanishes at the wall. A zero gradient boundary condition is used for density at the wall. At the top boundary, all the variables are set to their respective freestream values.
The simulations are initialized with laminar basic state superposed with low intensity perturbations. The basic state solution is the same as used earlier in the linear stability analysis. 

The non-dimensional parameters and the grid sizes for the simulations are listed in table \ref{table:ParametersBasic}.  The simulations $C_1-C_3$ are initialized with the most unstable first mode and $C_4-C_6$ are initialized by the most unstable second mode. The domain size in the streamwise direction ($L_{x1}$) is set to twice the wavelength of the instability. 
The computational grid is uniform in the streamwise and spanwise direction while a stretched grid with a cell to cell grading of $r =1.015$ is employed in the wall normal direction. The DNS results are validated in \S\ref{sec:Results} by comparing the growth rate of kinetic energy and other statistics against linear stability analysis. 

 \begin{table}
\begin{center}
\begin{tabular}{cccccccccccc}
 	  \textbf{Case} & $\mathbf{Re}$ & $\mathbf{Pr}$ & $\mathbf{M}$ & $\mathbf{\rho_\infty (kg\cdot m^{-3})}$ & $\mathbf{T_\infty (K)}$ & $\mathbf{L_{x1}}$ & $\mathbf{L_{x2}}$ & $\mathbf{L_{x3}}$ & $\mathbf{N_{x1}}$ & $\mathbf{N_{x2}}$ & $\mathbf{N_{x3}}$   \\ \hline 
  	$C_1$ & $4000$ & $0.5$ & $0.5$ & $1.0$ & $353$ & $144.4$ & $40$ & $5.8$ & $100$ & $400$ & $4$ \\
  	$C_2$ & $4000$ & $0.9$ & $6.0$ & $1.0$ & $353$ & $193.3$ & $124$ & $127.6$ & $100$ & $200$ & $66$ \\
  	$C_3$ & $4000$ & $1.3$ & $6.0$ & $1.0$ & $353$ & $104.7$ & $124$ & $98.4$ & $100$ & $200$ & $94$ \\
  	$C_4$ & $4000$ & $0.5$ & $4.0$ & $1.0$ & $353$ & $37.2$ & $76$ & $1.5$ & $100$ & $400$ & $4$ \\
  	$C_5$ & $4000$ & $1.3$ & $4.0$ & $1.0$ & $353$ & $35.0$ & $76$ & $1.4$ & $100$ & $300$ & $4$ \\
  	$C_6$ & $4000$ & $1.3$ & $6.0$ & $1.0$ & $353$ & $65.7$ & $124$ & $2.63$ & $100$ & $200$ & $4$ \\
\end{tabular}
\end{center}
	\caption{Non-dimensional parameters, freestream properties and grid sizes for DNS. The domain sizes are normalized by the Blassius length scales $L_R$. $N_{x1}$, $N_{x2}$ and $N_{x3}$ denote the number of grid points in $x_1$, $x_2$ and $x_3$ directions respectively. The grid resolutions are selected after conducting appropriate grid convergence studies.}
	\label{table:ParametersBasic}
\end{table}

\section{Neutral stability curves and eigenspectrum}
\begin{figure}
    \centering\subfloat[$M=0.5$]{\includegraphics[trim=0 10 10 0, clip,width=0.45\textwidth, keepaspectratio]{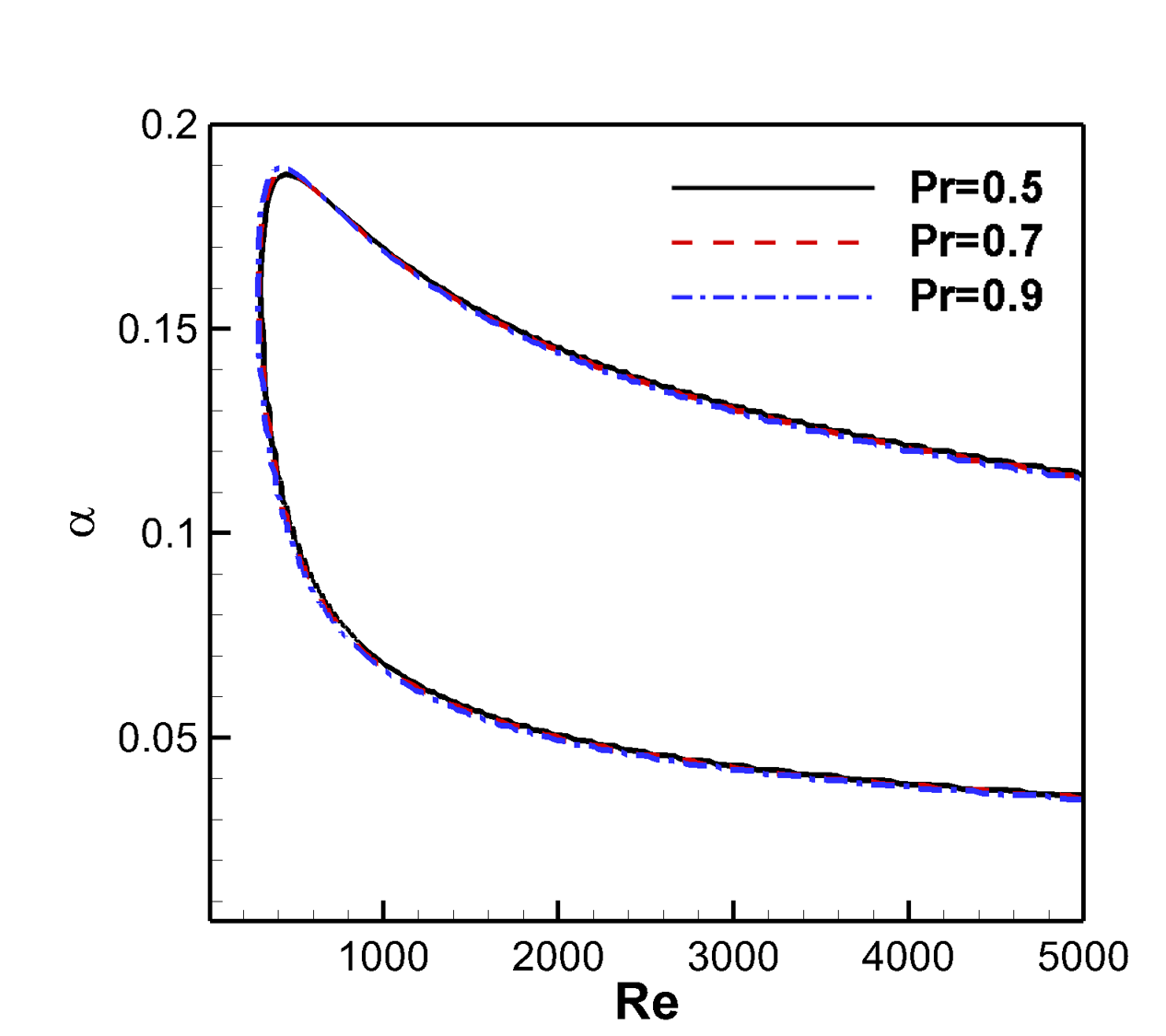}} 
    \subfloat[$M=1$]{\includegraphics[trim=0 10 10 0, clip, width=0.45\textwidth, keepaspectratio]{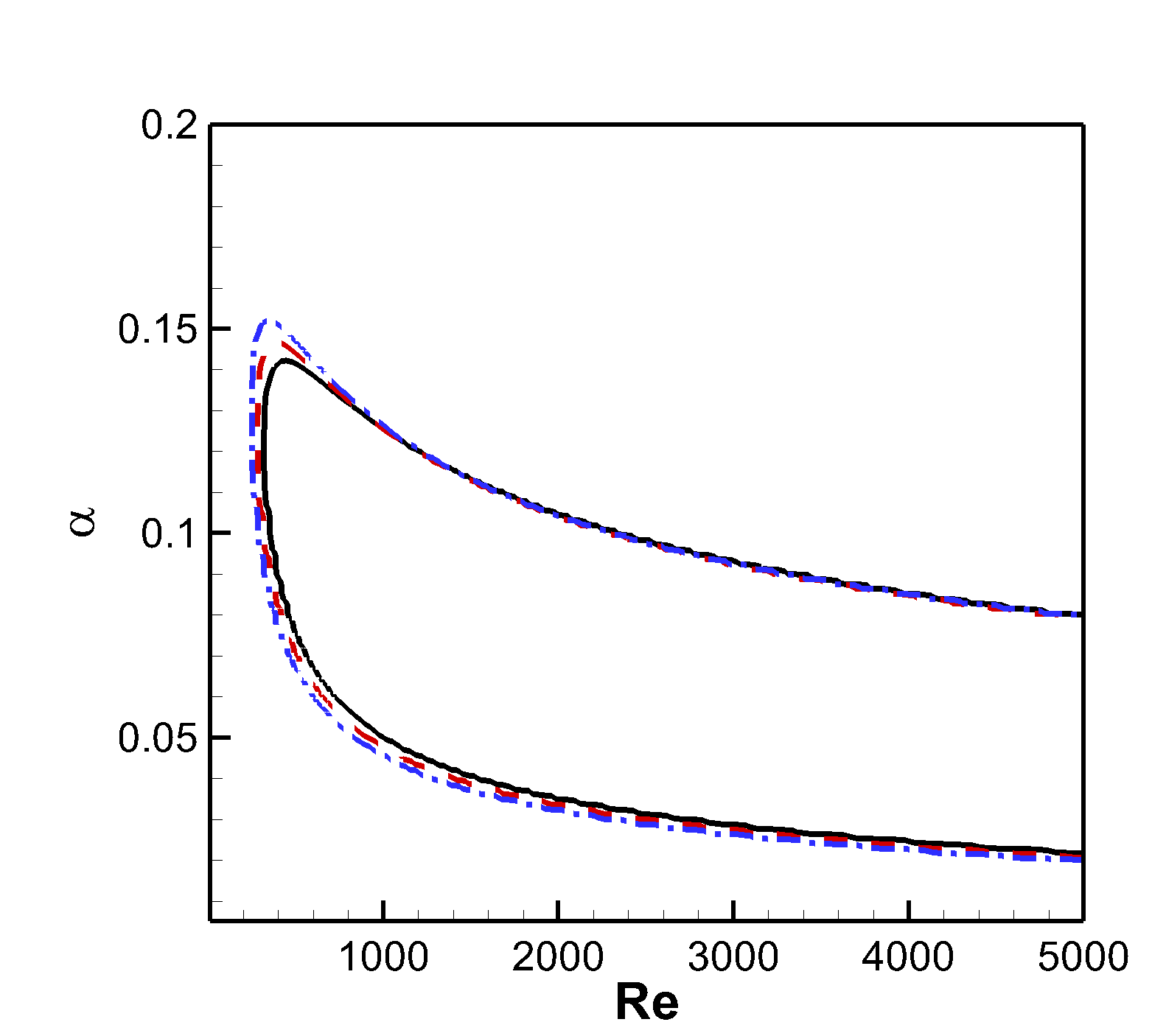}} \\
    \subfloat[$M=4$]{\includegraphics[trim=0 10 10 0, clip, width=0.45\textwidth, keepaspectratio]{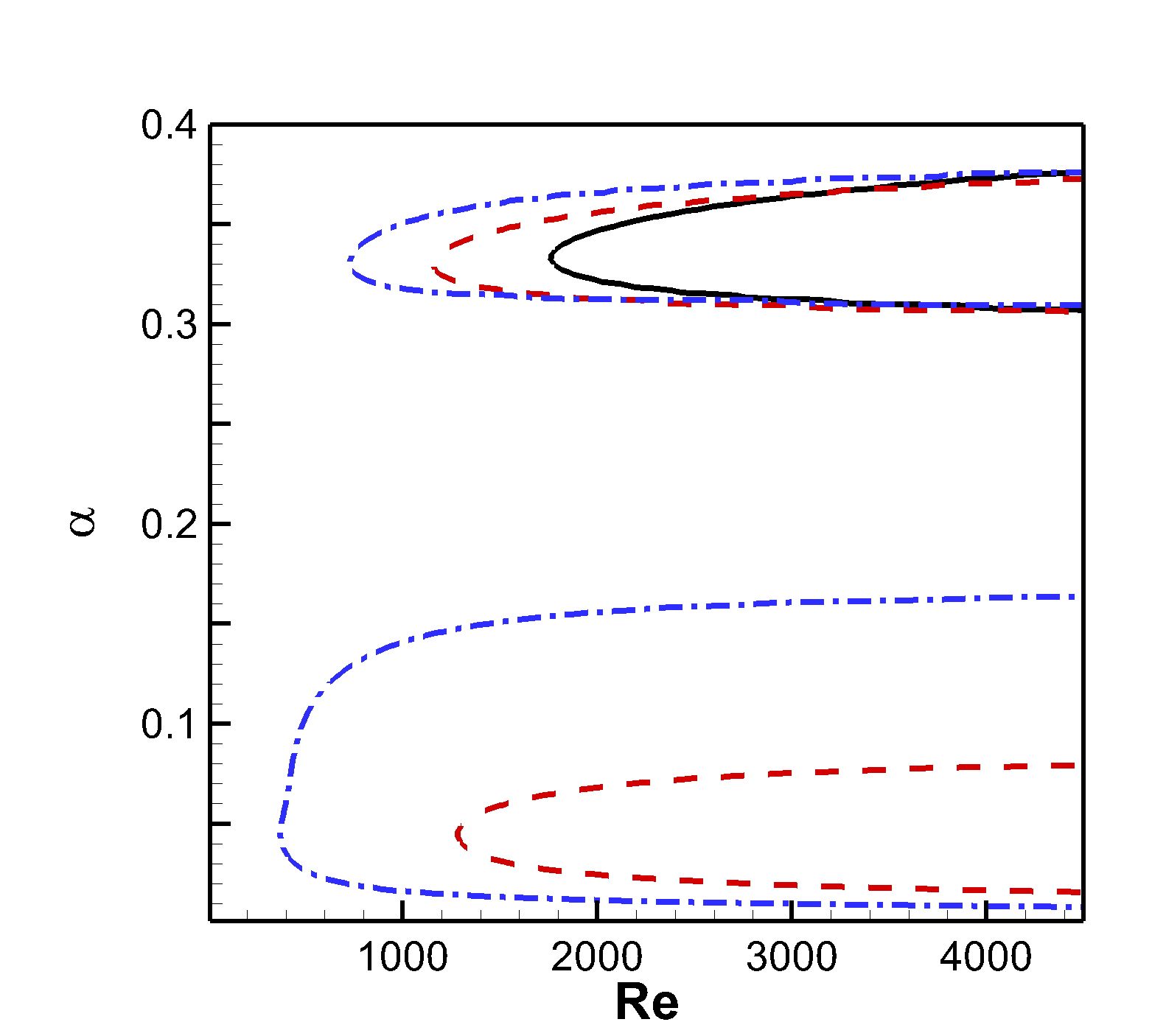}} 
    \subfloat[$M=6$]{\includegraphics[trim=0 10 10 0, clip, width=0.45\textwidth, keepaspectratio]{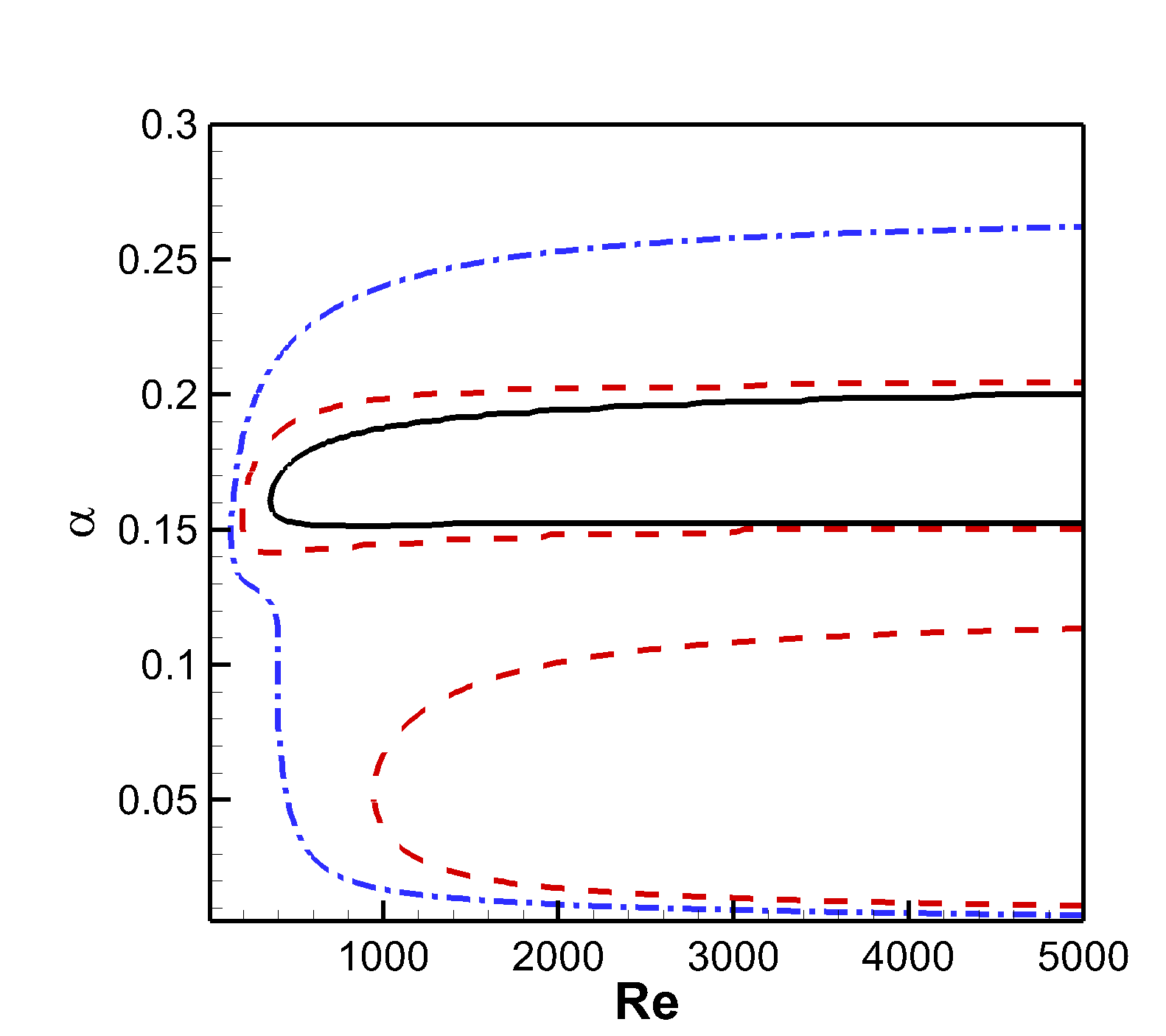}}
    \caption{Neutral stability curves of $2D$ disturbances for different Prandtl numbers at (a) $M=0.5$, (b) $M=1$, (c) $M=4$ and (d) $M=6$.}
    \label{fig:NeutralStability2D}
\end{figure}
\begin{figure}
    \centering
    \subfloat[$M=2$]{\includegraphics[trim=0 10 10 0, clip,width=0.45\textwidth, keepaspectratio]{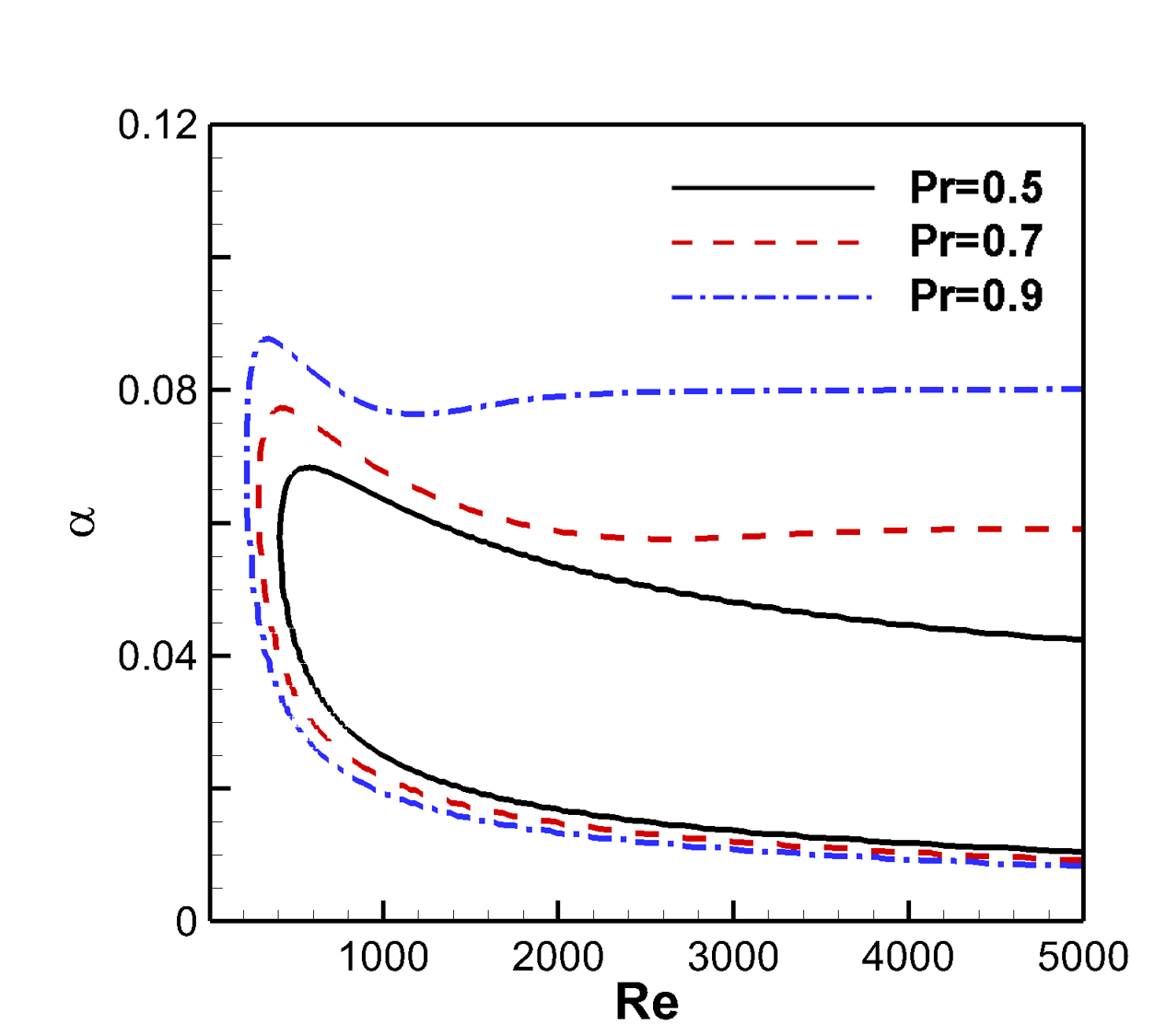}}
    \subfloat[$M=3$]{\includegraphics[trim=0 10 10 0, clip,width=0.45\textwidth, keepaspectratio]{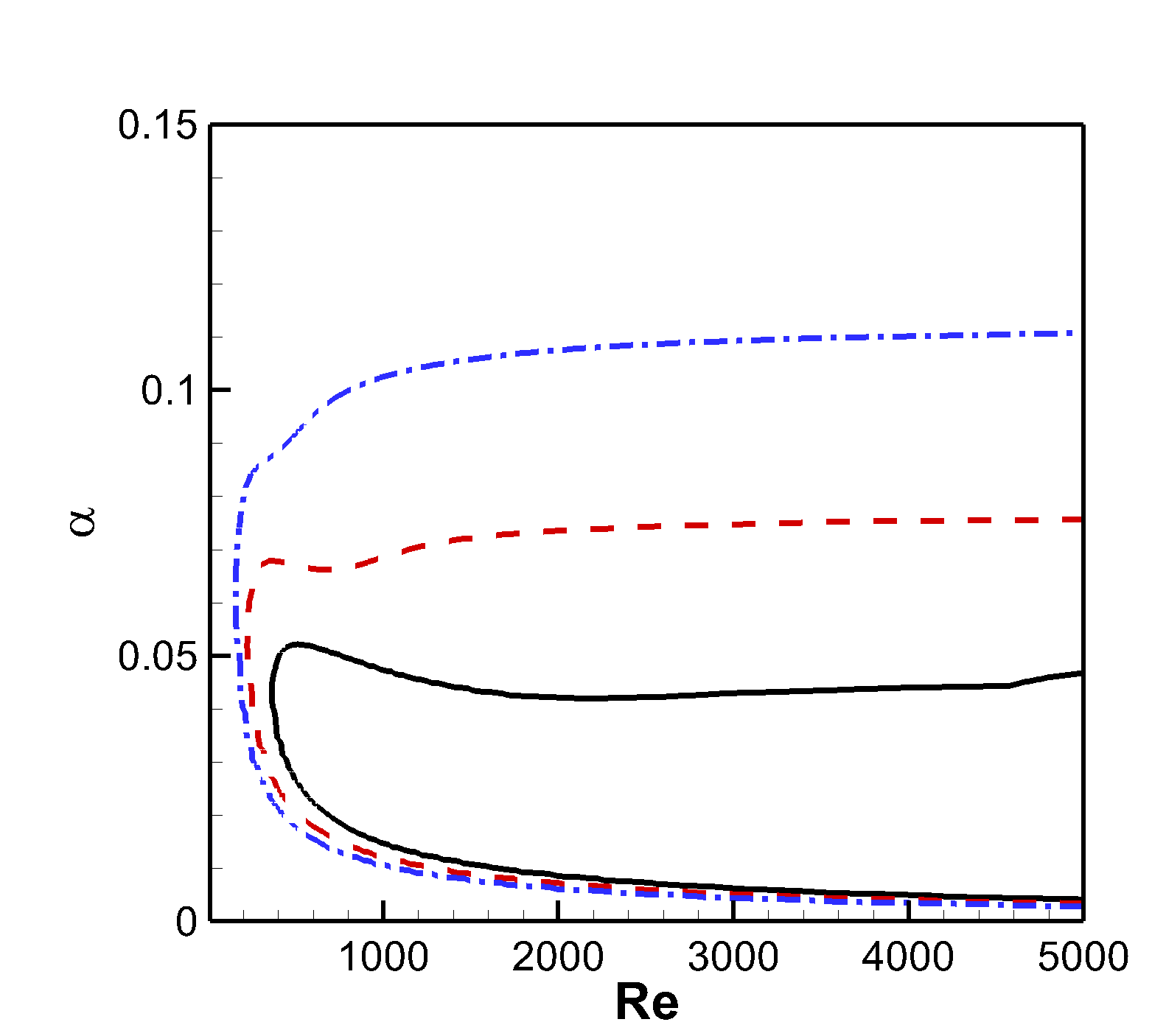}}
    \caption{Neutral stability curves of $3D$ disturbances ($\Psi=60^o$) for different Prandtl numbers at (a) $M=2$ and (b) $M=3$. }
    \label{fig:NeutralStability3D}
\end{figure}

The neutral stability curves for two dimensional disturbances at different Mach and Prandtl numbers are displayed in figure \ref{fig:NeutralStability2D}. The neutral stability curves represent contours of zero growth rate in the $Re-\alpha$ plane. The curves shown in figure \ref{fig:NeutralStability2D} are computed for $Re\in[10,5000]$.  At low Mach numbers, the stability curves are reasonably invariant with Prandtl number. 
This is not surprising as the base flow for the low Mach number cases is more or less unaltered in the Prandtl number regime considered. 
For $M\geq4$, there are two loops of instability in the $\alpha-Re$ plane. The loop at low wavenumbers corresponds to the first mode instability while the second mode is unstable at higher wavenumbers. The streamwise first mode is stable at $M=4$ and $M=6$ for $Pr=0.5$ over the range of Reynolds number considered. As the Prandtl number is increased the first mode becomes unstable over a wider range of wavenumbers and the critical Reynolds number ($Re_{cr}$) for the first mode decreases. The instability region of the second mode also expands with increasing Prandtl number. For $Pr\leq0.7$ at $M=4$, the first mode destabilizes at a higher Reynolds number than the second mode. However at $Pr=0.9$, $Re_{cr}$ for the first mode is lower than the second mode. The loops corresponding to first and second mode fuse at $M=6$ for $Pr=0.9$ as destabilization increases with Prandtl number. \cite{ramachandran2015linear} also observed a similar merger of the loops of first and second mode. 
Figure \ref{fig:NeutralStability3D} shows the effect of Prandtl number on the stability characteristics of 3D disturbances. The wave angle for the oblique waves is defined by the following relation  
\begin{equation}
    \psi=\tan^{-1}\left(\frac{\beta}{\alpha}\right).
\end{equation}
The stability curves shown in figure \ref{fig:NeutralStability3D} correspond to $\Psi=60^o$. Much like the 2D disturbances oblique waves are also destabilized at high Prandtl number. The critical Reynolds number for 3D disturbances also decreases with increasing Prandtl number.   

\begin{figure}
    \centering 
    \subfloat[$C_r$]{\includegraphics[width=0.45\textwidth, keepaspectratio]{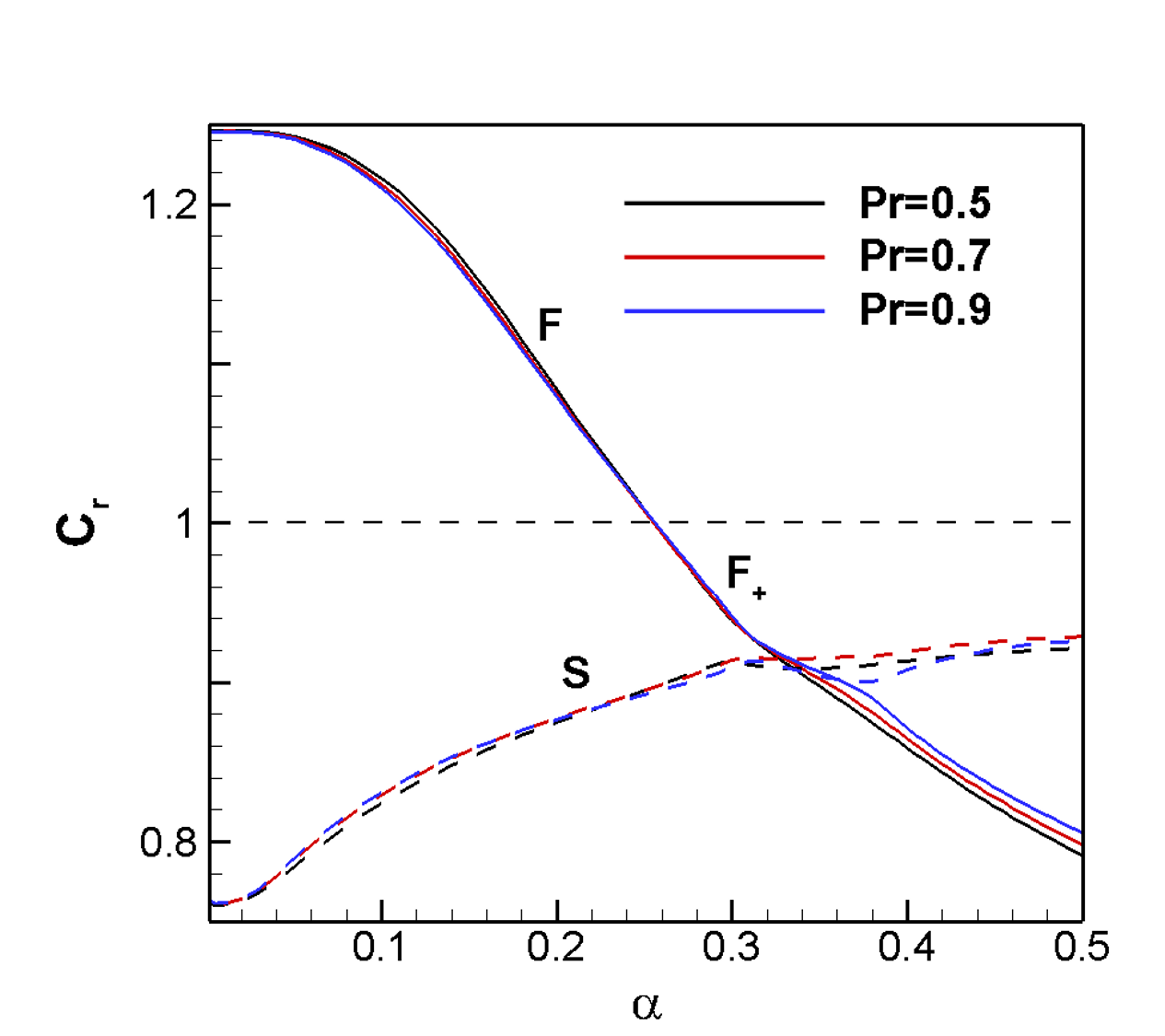}}
    \subfloat[$C_i$]{\includegraphics[width=0.45\textwidth, keepaspectratio]{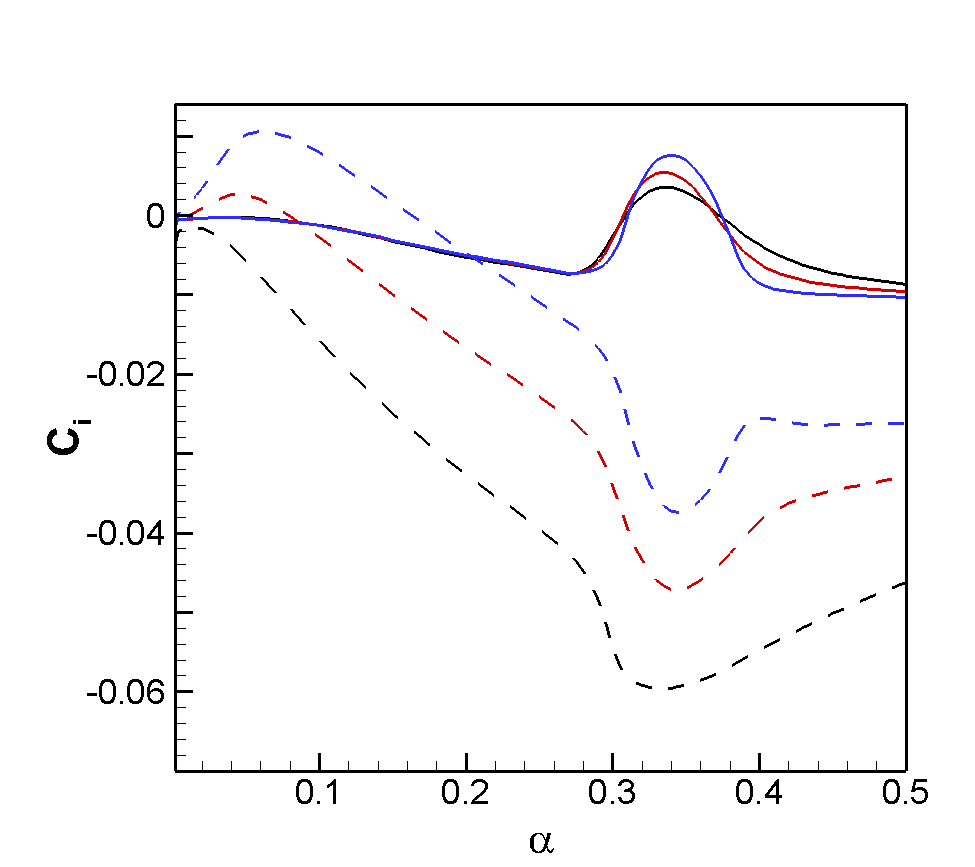}}
    \caption{Variation of (a) phase speed and (b) growth rate for fast and slow modes with wavenumber at $M=4$, $Re=4000$ for three different $Pr$. Solid lines correspond to fast mode and dashed lines represent slow mode. }
    \label{fig:EigenSpectraM4}
\end{figure}

\begin{figure}
    \centering
    \subfloat[$C_r$]{\includegraphics[width=0.45\textwidth, keepaspectratio]{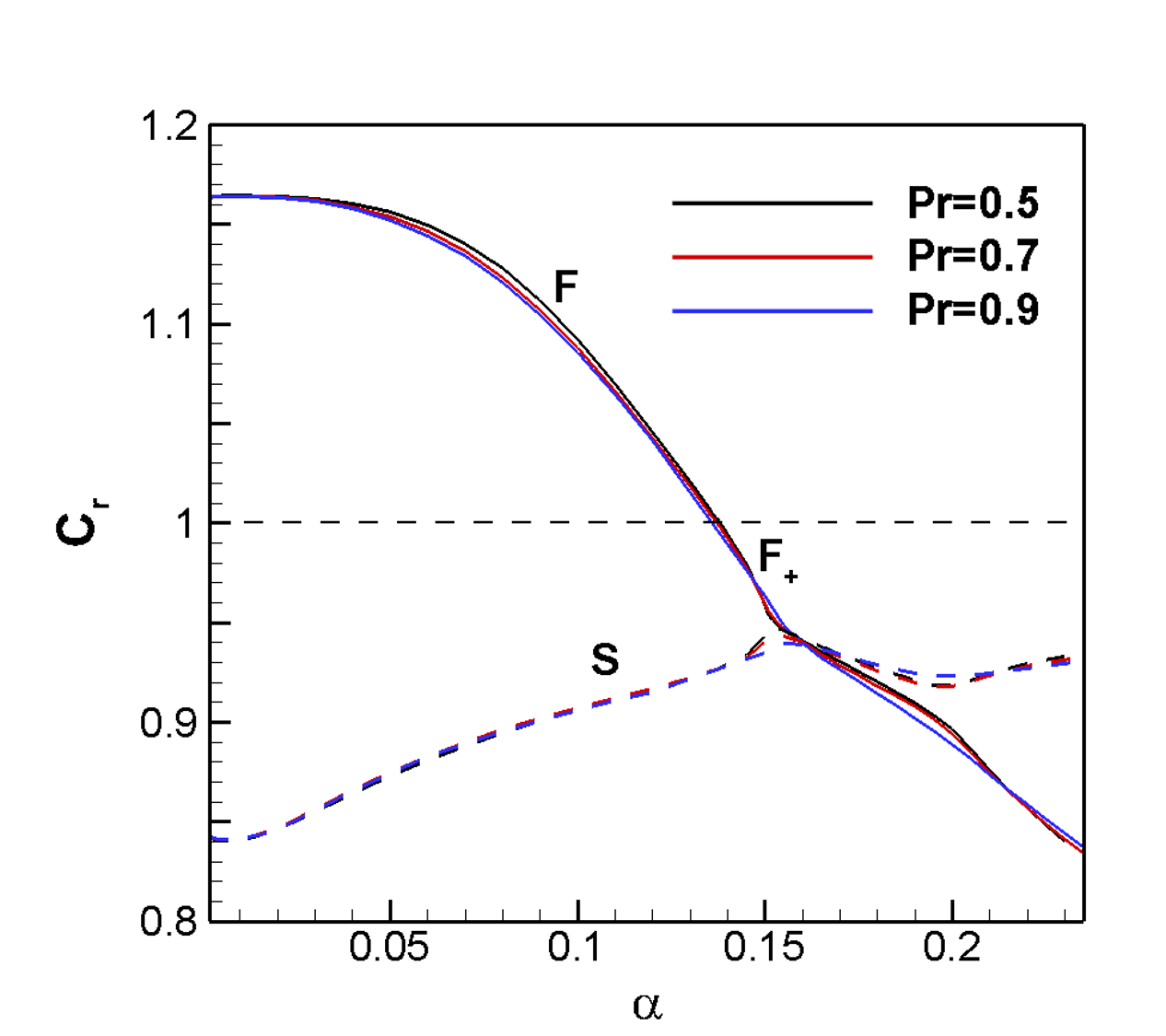}}
    \subfloat[$C_i$]{\includegraphics[width=0.45\textwidth, keepaspectratio]{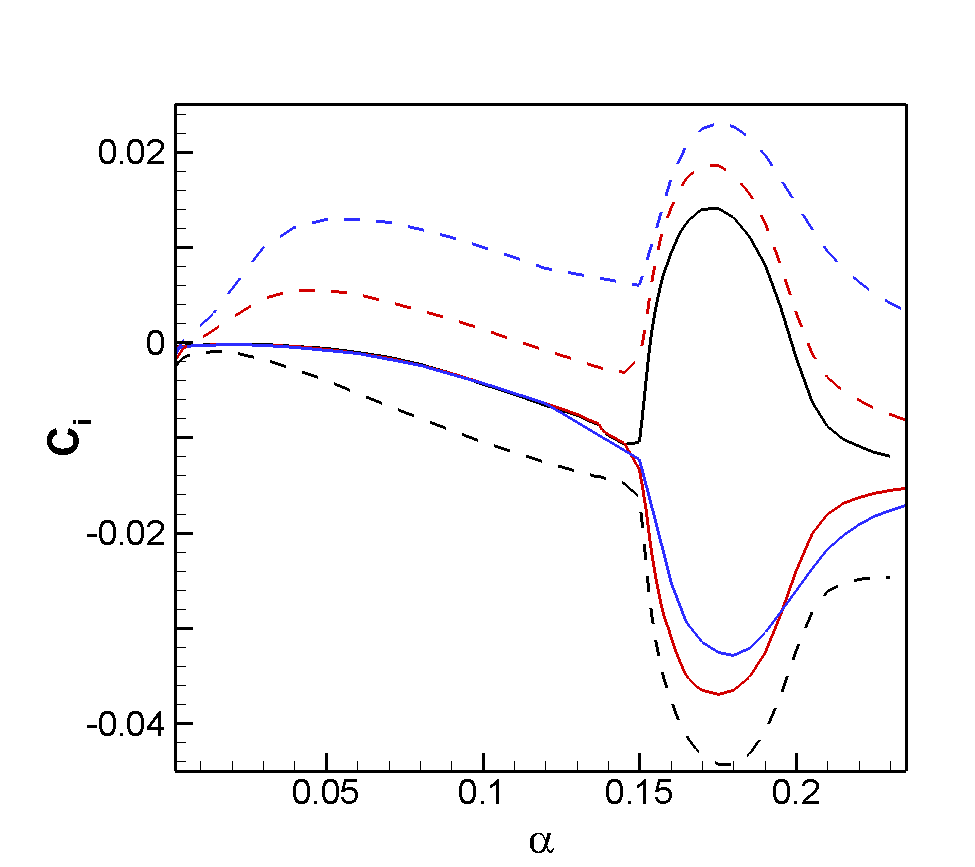}}
    \caption{Variation of (a) phase speed and (b) growth rate for fast and slow modes with wavenumber at $M=6$, $Re=4000$ for three different $Pr$. Solid lines correspond to fast mode and dashed lines represent slow mode. }
    \label{fig:EigenSpectraM6}
\end{figure}

We now investigate the effect of Prandtl number on the eigenspectrum by examining the variation of phase speed and growth rate for the fast and slow modes \citep{fedorov2011high}. The phase speed and growth rate variation for different $Pr$ at $M=4$ are shown in figure \ref{fig:EigenSpectraM4}. 
In the limit of $\alpha\to0$ the fast and slow modes are synchronized with the acoustic wave ($c_{a\pm}=1\pm1/M$). The phase speed of the fast mode decreases with increasing wavenumber and synchronizes with the continuous spectrum branch corresponding to entropy and vorticity modes ($C_r=1$). The fast mode after synchronization with the vorticity/entropy modes is termed as the mode $F_+$ \citep{fedorov2011high}. The phase speed of the fast mode decreases further and, it synchronizes with the slow mode. Due to this synchronization the growth rates of the fast and slow mode exhibit a peak and trough \citep{fedorov2011high}. The phase speed evolution for the fast and slow modes shown in figure \ref{fig:EigenSpectraM4}(a) are similar for all three Prandtl numbers considered. The synchronization point between the fast and entropy/vorticity mode and the location of discrete spectrum branching is weakly dependent on Prandtl number. At $M=4$, the slow mode is unstable at low wavenumbers while the fast mode ($F_+$) becomes unstable at high $\alpha$ for all three Prandtl numbers considered. Figure \ref{fig:EigenSpectraM6} plots the phase speed and growth rates for fast and slow modes at $M=6$. Similar to the case at $M=4$, the phase speed evolution and the synchronization wavenumbers do not have a strong dependence on Prandtl number for $M=6$ as well. However, the branching pattern of the eigenspectrum is dependent on Prandtl number. For low Prandtl number, the mode $F_+$ becomes unstable at high wavenumbers and exhibits a strong peak. On the other hand for $Pr\geq0.7$ the slow mode after synchronization with the fast mode becomes the dominant instability. \cite{fedorov2011high} and \cite{ramachandran2015linear} also report similar branching pattern of the eigenspectrum depending on Mach and Prandtl numbers.

\begin{figure}
    \centering
    \subfloat[$\hat{u}_1$]{\includegraphics[trim=0 50 10 0, clip, width=0.31\textwidth, keepaspectratio]{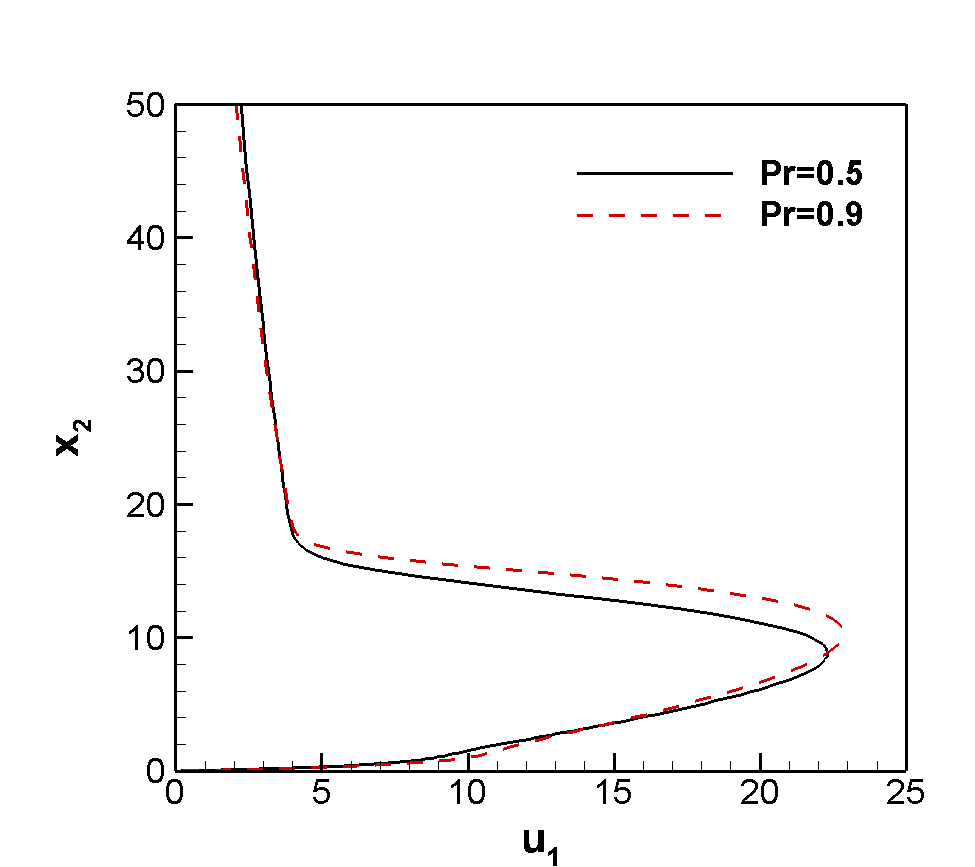}}
    \subfloat[$\hat{T}$]{\includegraphics[trim=0 50 10 0, clip, width=0.31\textwidth, keepaspectratio]{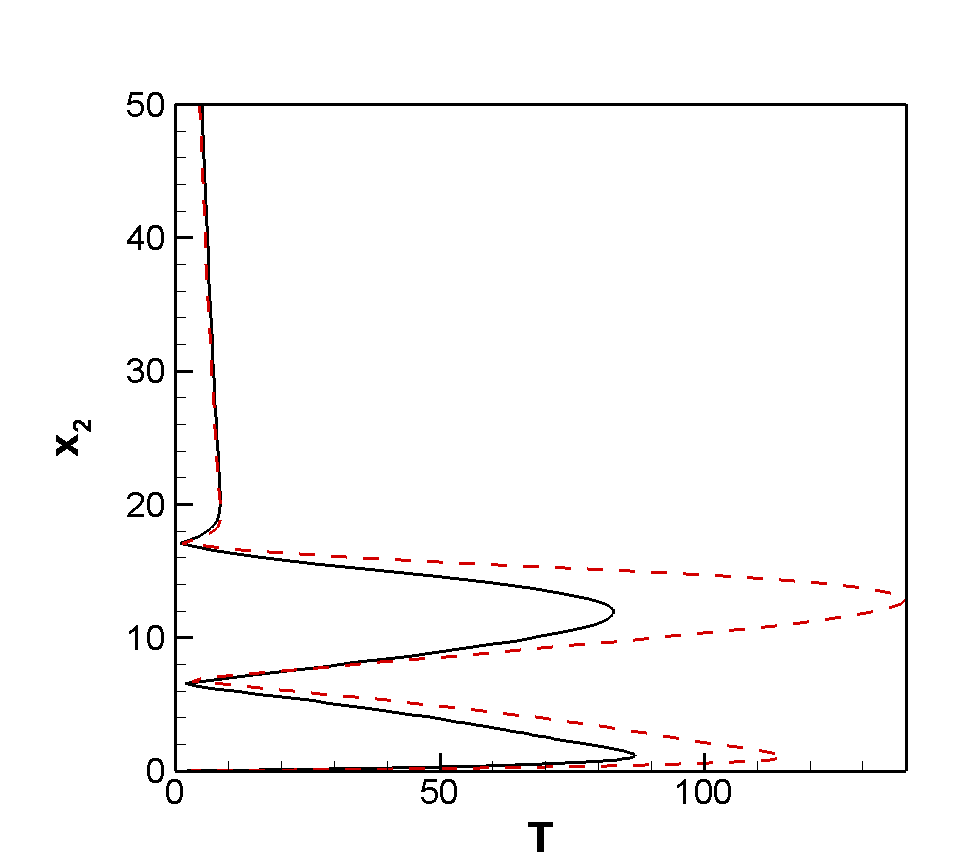}}
    \subfloat[$\hat{p}$]{\includegraphics[trim=0 50 10 0, clip, width=0.31\textwidth, keepaspectratio]{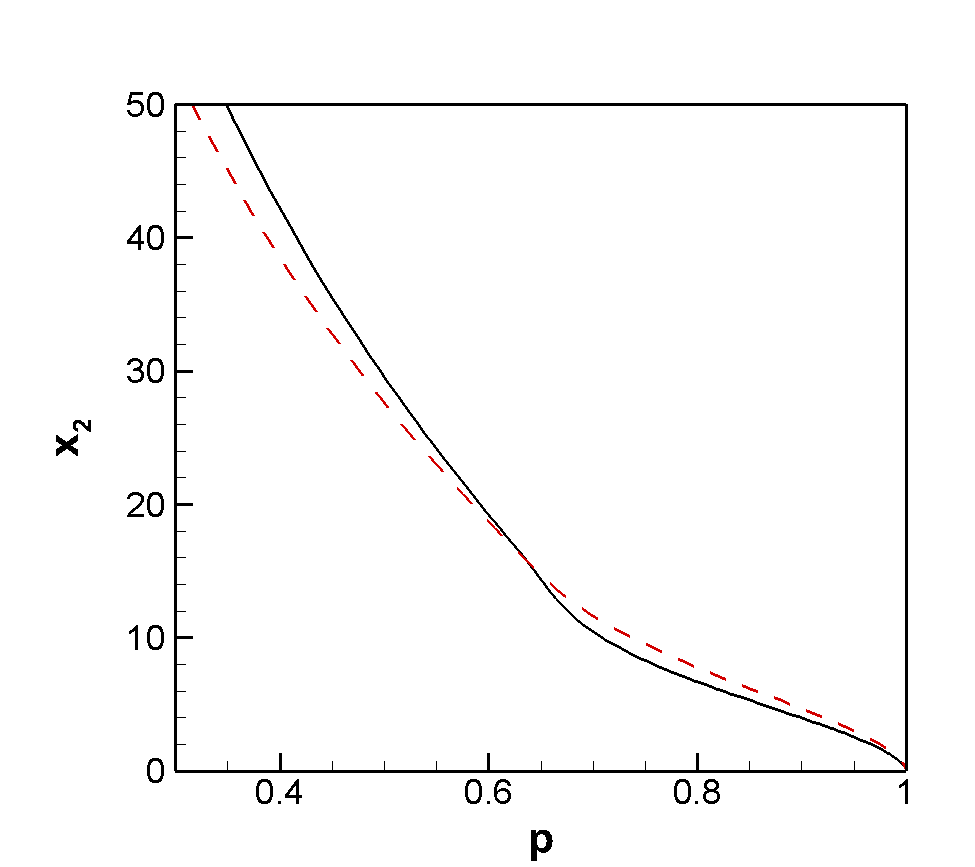}} \\
    \subfloat[$\hat{u}_1$]{\includegraphics[trim=0 50 10 0, clip, width=0.31\textwidth, keepaspectratio]{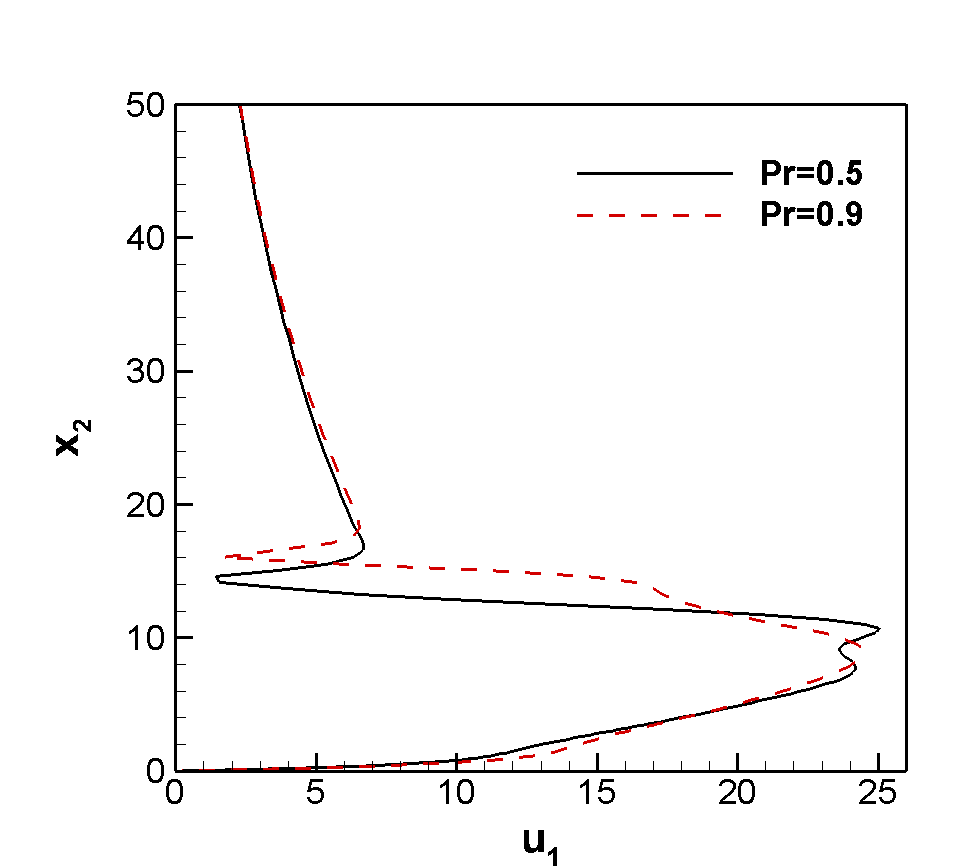}}
    \subfloat[$\hat{T}$]{\includegraphics[trim=0 50 10 0, clip, width=0.31\textwidth, keepaspectratio]{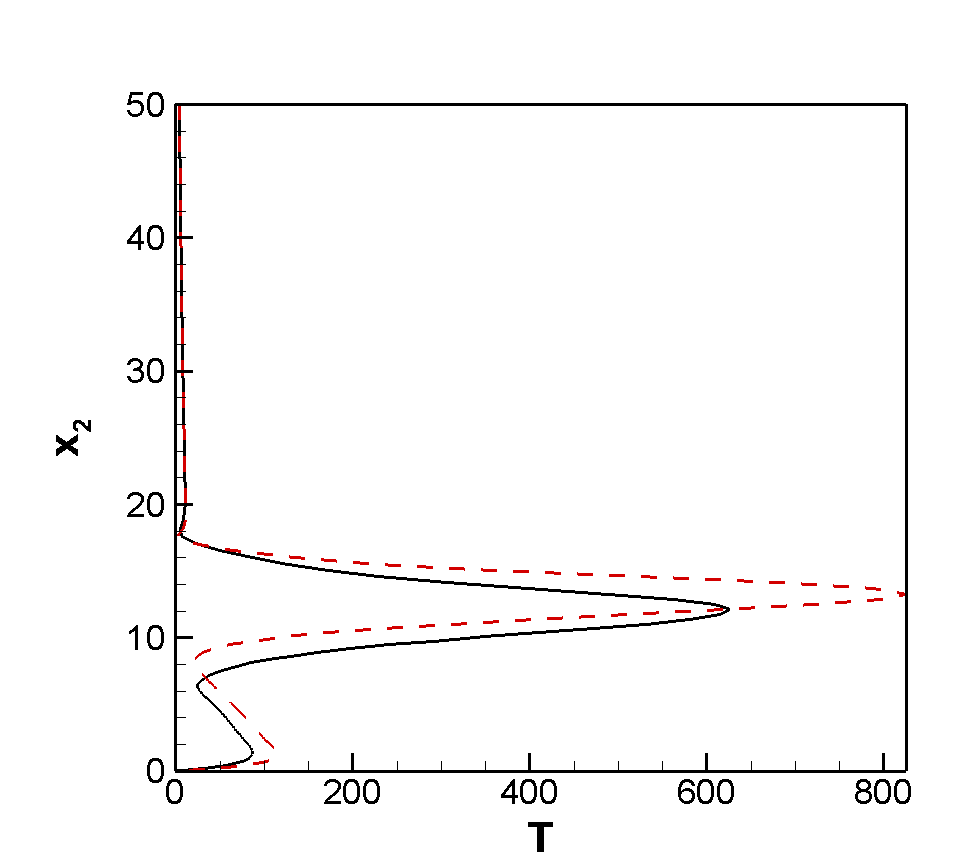}}
    \subfloat[$\hat{p}$]{\includegraphics[trim=0 50 10 0, clip, width=0.31\textwidth, keepaspectratio]{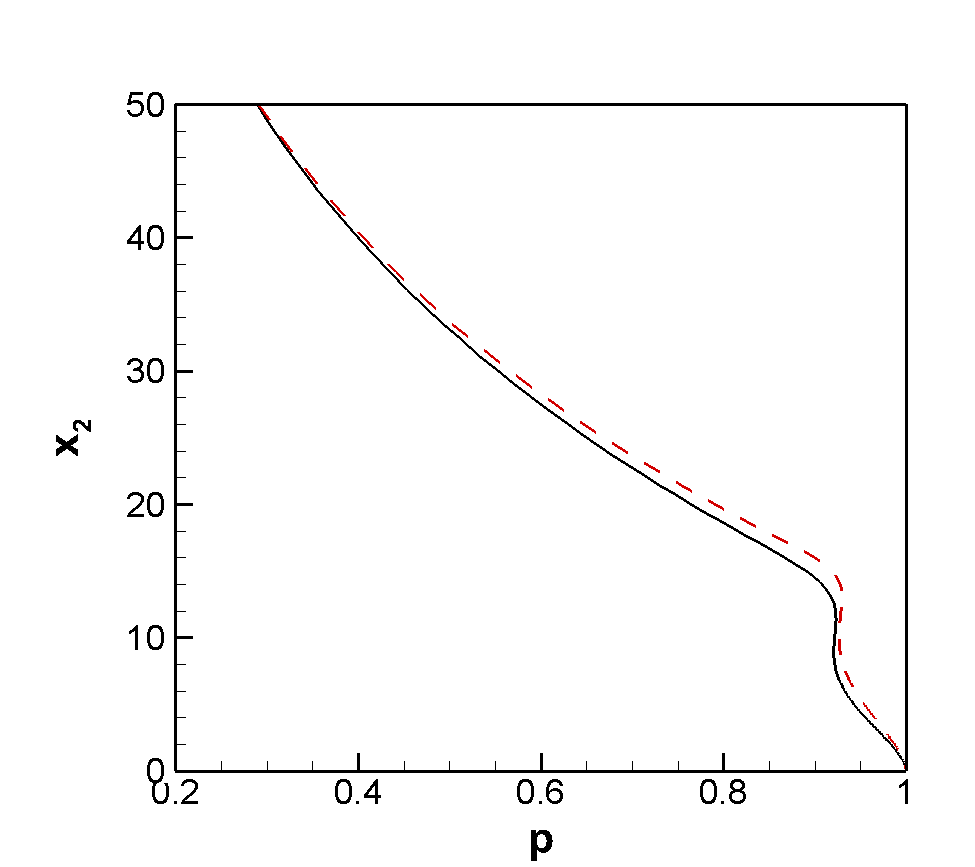}} \\
    \caption{Eigenmode shapes of the (a)-(c) fast ($F$) and (d)-(f) slow ($S$) modes at $M=6$, $Re=4000$, $\alpha=0.05$, $\beta=0$ for two different Prandtl numbers. }
    \label{fig:ModeShapes_FS}
\end{figure}

The effect of Prandtl number on the eigenfunctions of the fast and slow mode at $M=6$, $\alpha=0.05$ is shown in figure \ref{fig:ModeShapes_FS}. The eigenfunctions are normalized by the magnitude of pressure perturbation at the wall. 
In the low wavenumber limit, the eigenfunctions of velocity and pressure for both the fast and slow modes do not have a strong dependence on Prandtl number. The eigenfunctions of temperature for the slow mode peaks near the critical layer. The critical layer ($y_{cl}$) is the location in the flow where the phase speed of the instability ($C_r$) equals the base velocity \citep{mack1984boundary}. In general, $y_{cl}$ increases with Prandtl number as a result there is a moderate shift in the location of peak temperature at high Prandtl number. The peak value of temperature eigenfunction is also larger at $Pr=0.9$ compared to $Pr=0.5$.   

\begin{figure}
    \centering
    \subfloat[$\hat{u}_1$]{\includegraphics[trim=0 50 10 0, clip, width=0.31\textwidth, keepaspectratio]{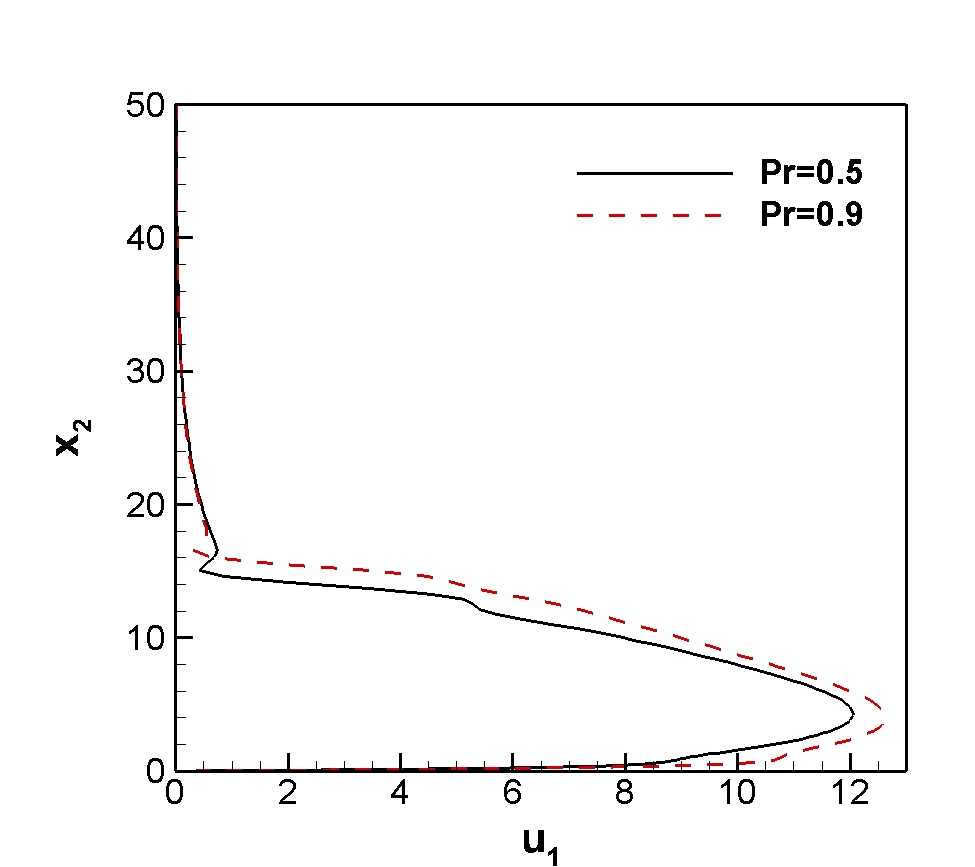}}
    \subfloat[$\hat{T}$]{\includegraphics[trim=0 50 10 0, clip, width=0.31\textwidth, keepaspectratio]{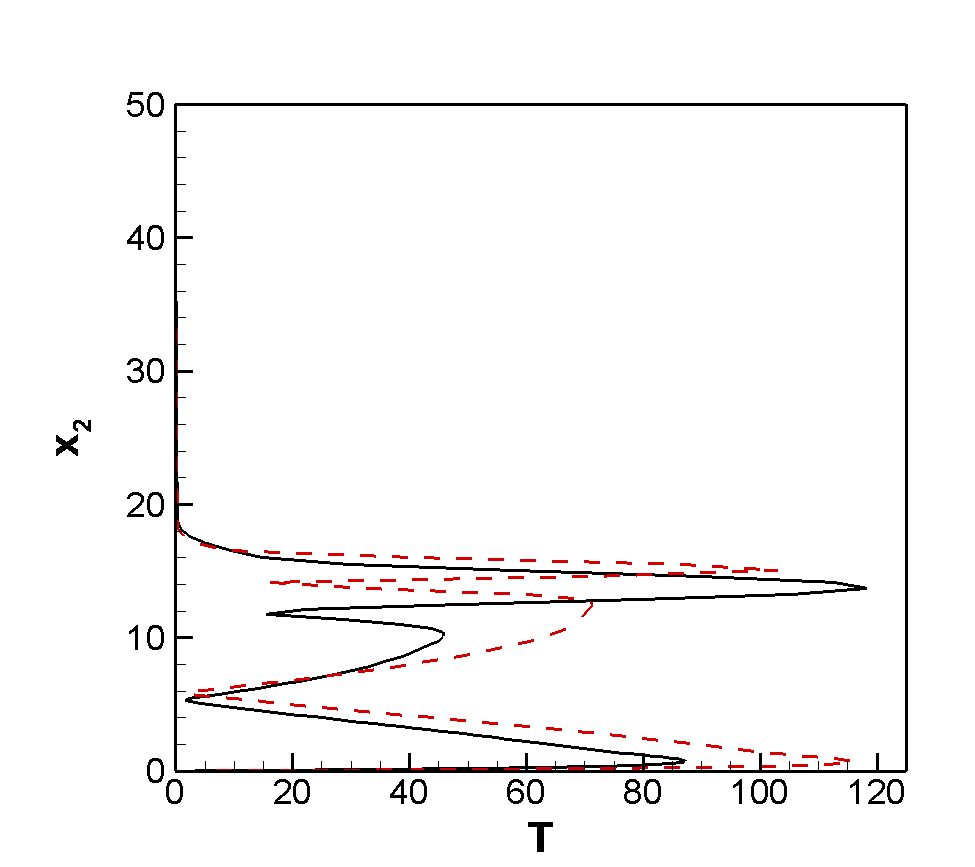}}
    \subfloat[$\hat{p}$]{\includegraphics[trim=0 50 10 0, clip, width=0.31\textwidth, keepaspectratio]{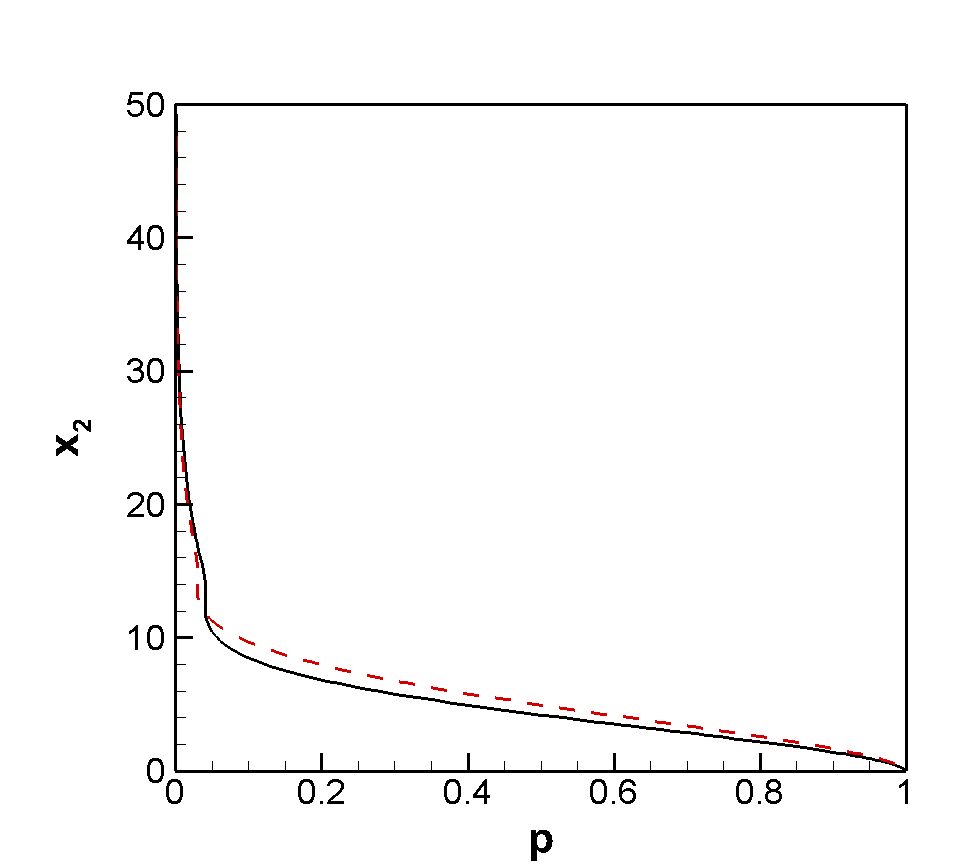}} \\
    \subfloat[$\hat{u}_1$]{\includegraphics[trim=0 50 10 0, clip, width=0.31\textwidth, keepaspectratio]{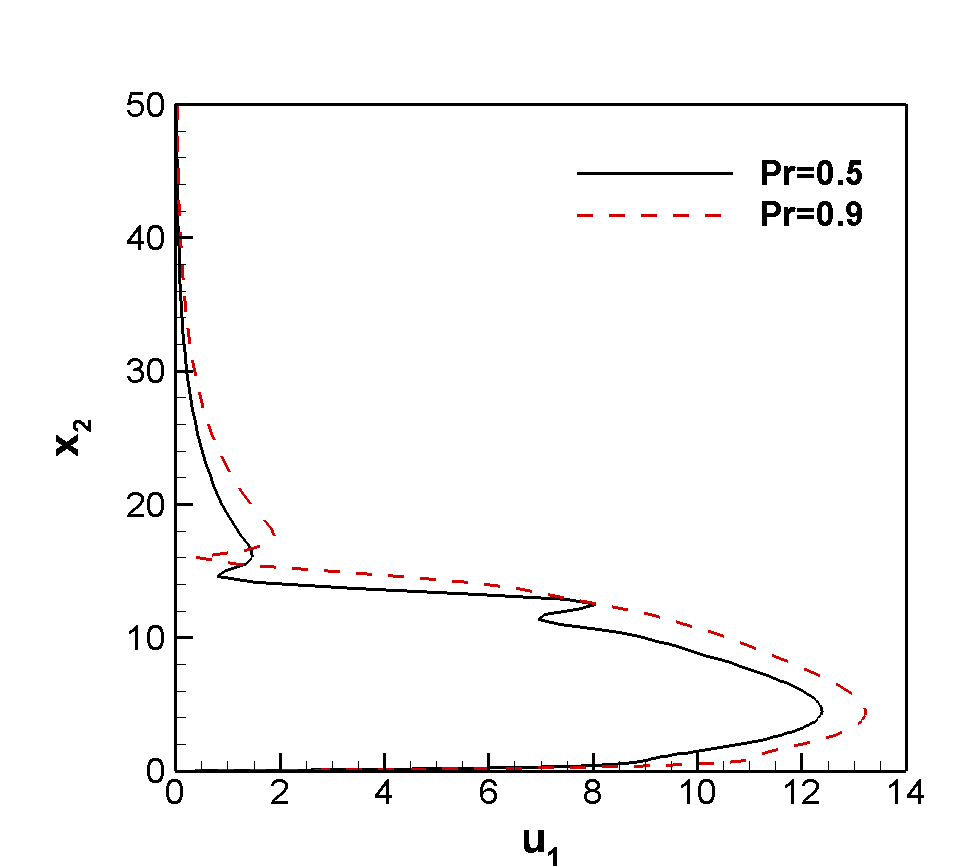}}
    \subfloat[$\hat{T}$]{\includegraphics[trim=0 50 10 0, clip, width=0.31\textwidth, keepaspectratio]{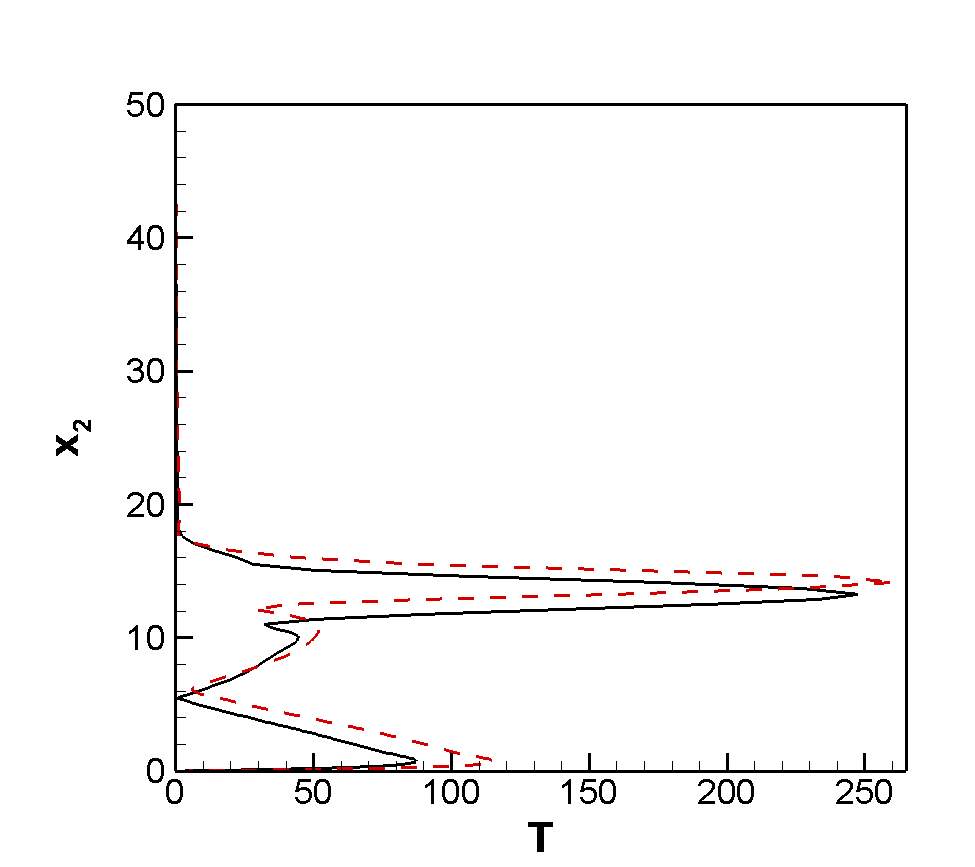}}
    \subfloat[$\hat{p}$]{\includegraphics[trim=0 50 10 0, clip, width=0.31\textwidth, keepaspectratio]{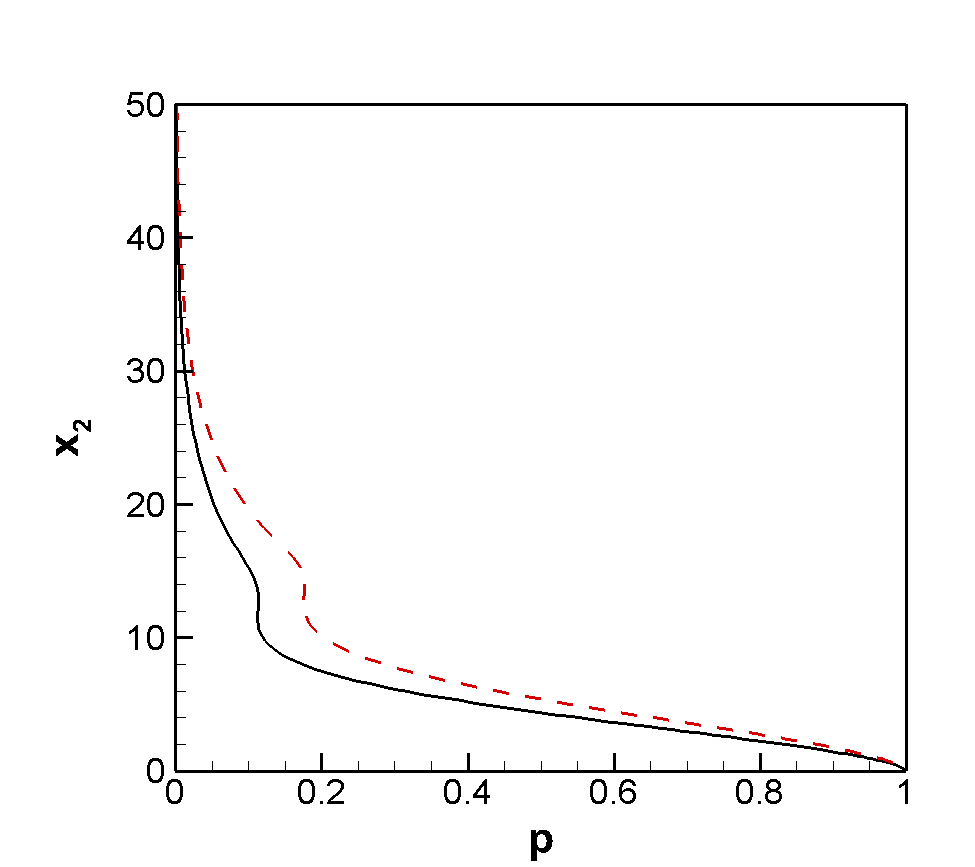}} \\
    \caption{Eigenmode shapes of the (a)-(c) fast ($F_+$) and (d)-(f) slow ($S$) modes before the branch point at $M=6$, $Re=4000$, $\alpha=0.15$, $\beta=0$ for two different Prandtl numbers. }
    \label{fig:ModeShapes_FS1}
\end{figure}

\begin{figure}
    \centering
    \subfloat[$\hat{u}_1$]{\includegraphics[trim=0 50 10 0, clip, width=0.31\textwidth, keepaspectratio]{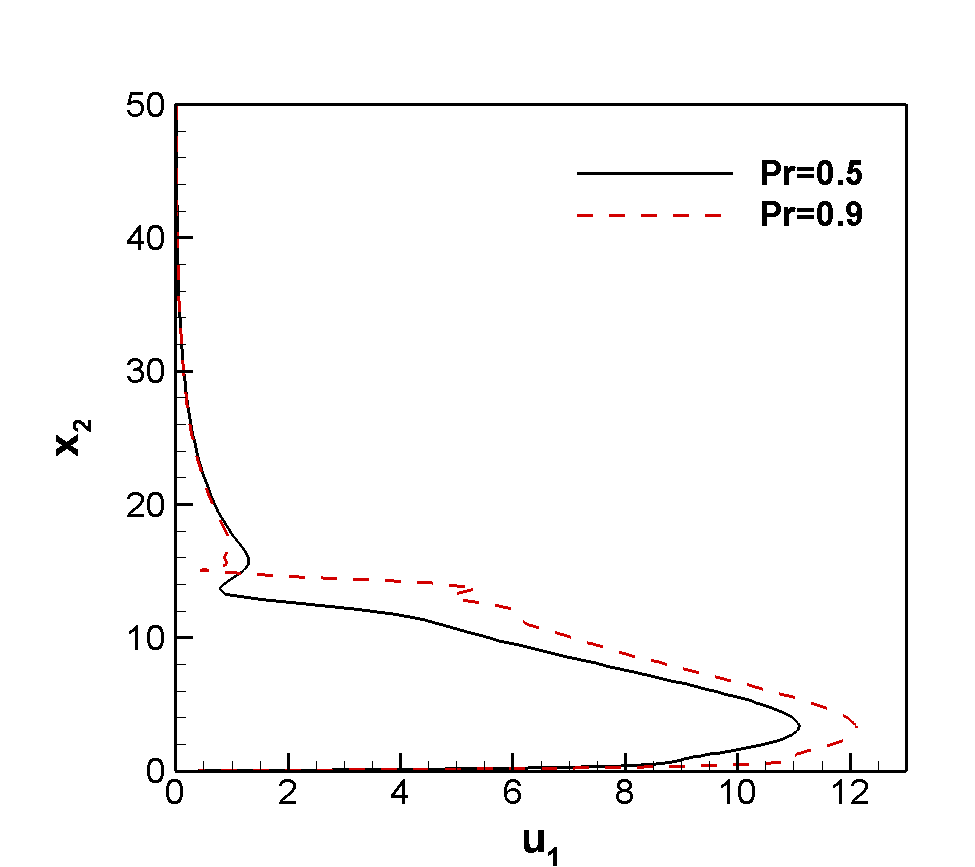}}
    \subfloat[$\hat{T}$]{\includegraphics[trim=0 50 10 0, clip, width=0.31\textwidth, keepaspectratio]{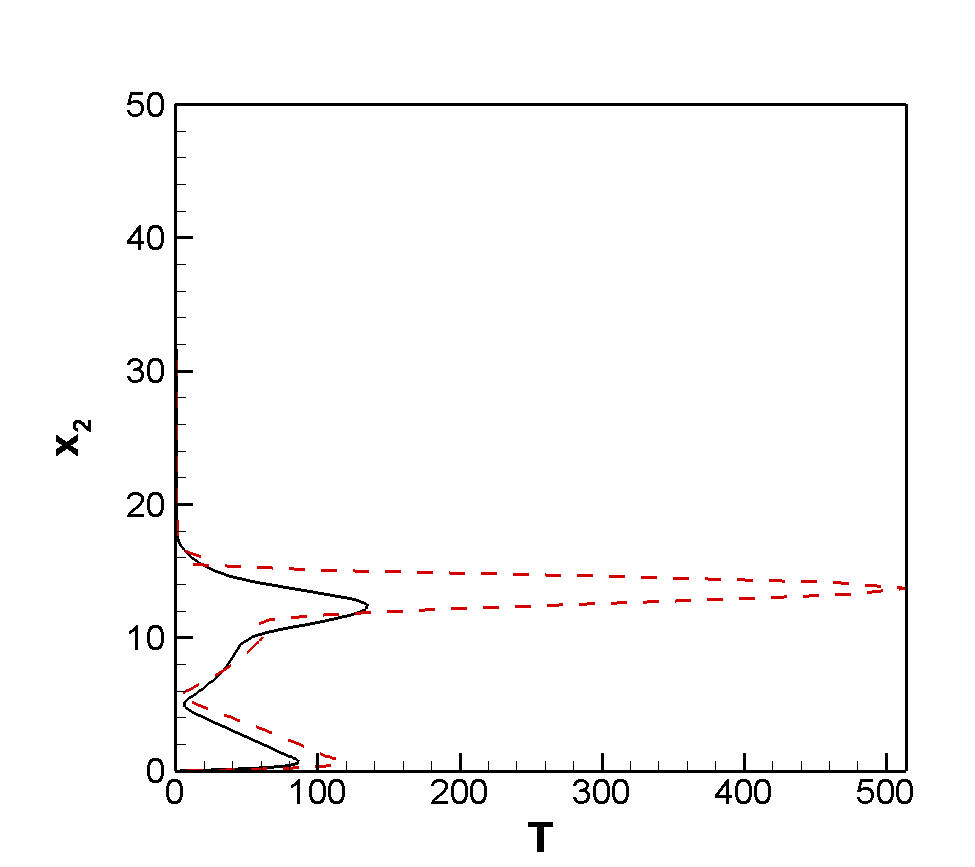}}
    \subfloat[$\hat{p}$]{\includegraphics[trim=0 50 10 0, clip, width=0.31\textwidth, keepaspectratio]{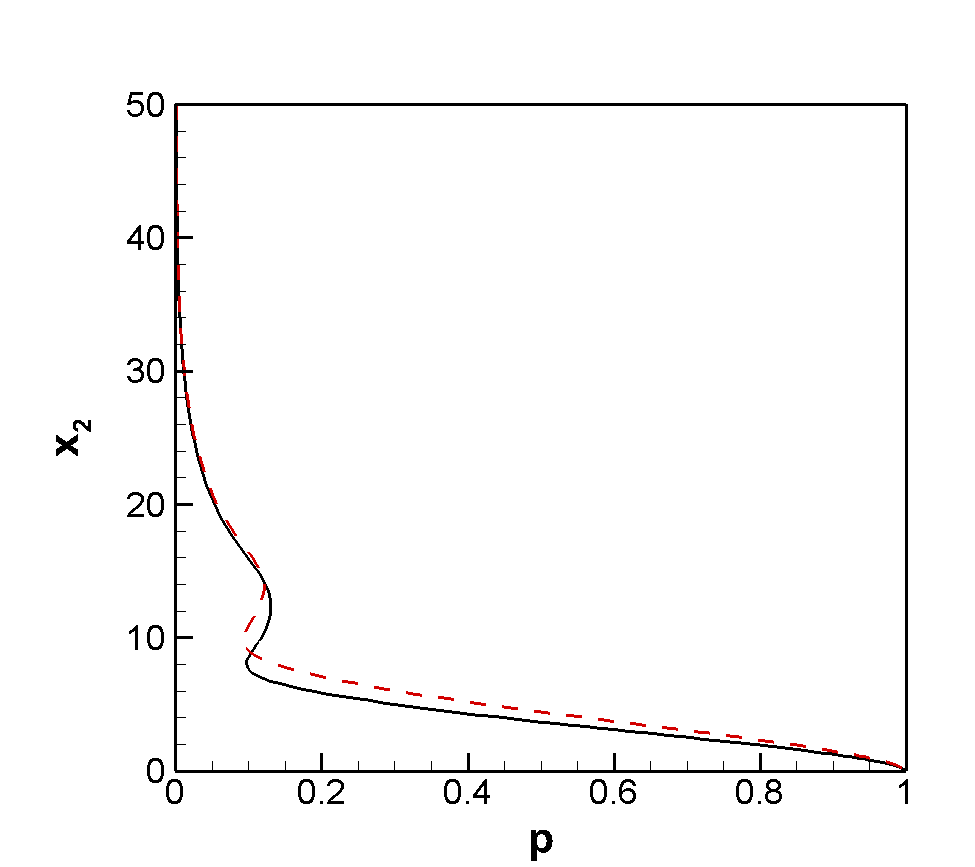}} \\
    \subfloat[$\hat{u}_1$]{\includegraphics[trim=0 50 10 0, clip, width=0.31\textwidth, keepaspectratio]{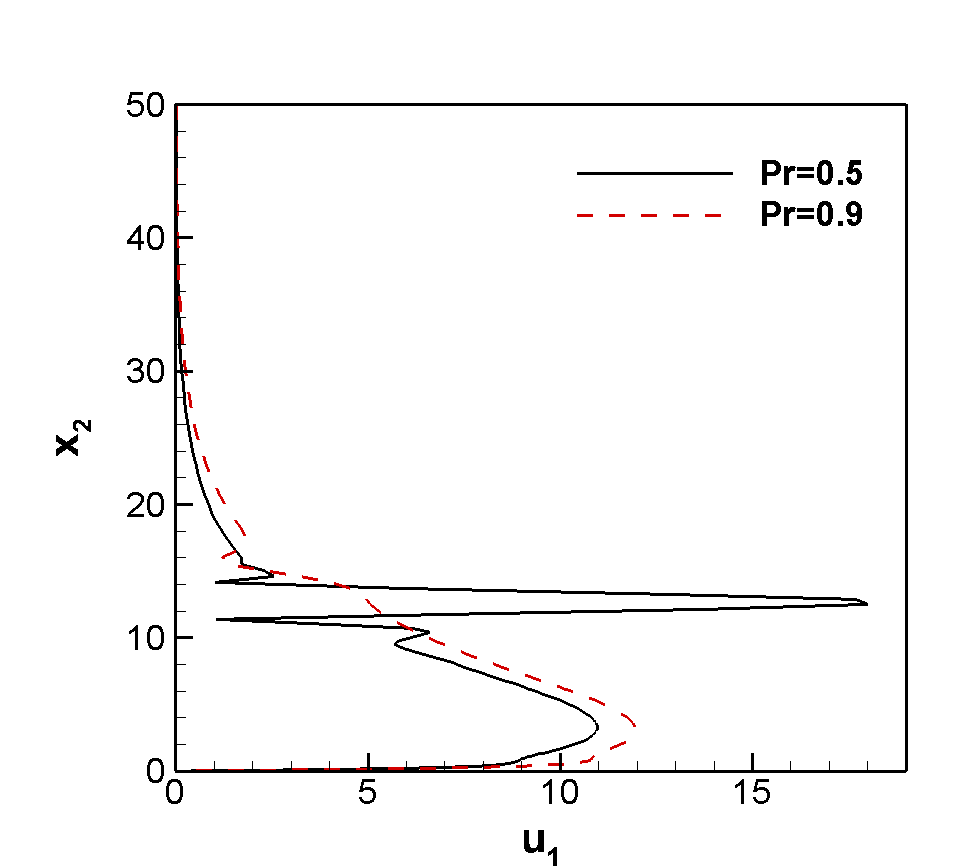}}
    \subfloat[$\hat{T}$]{\includegraphics[trim=0 50 10 0, clip, width=0.31\textwidth, keepaspectratio]{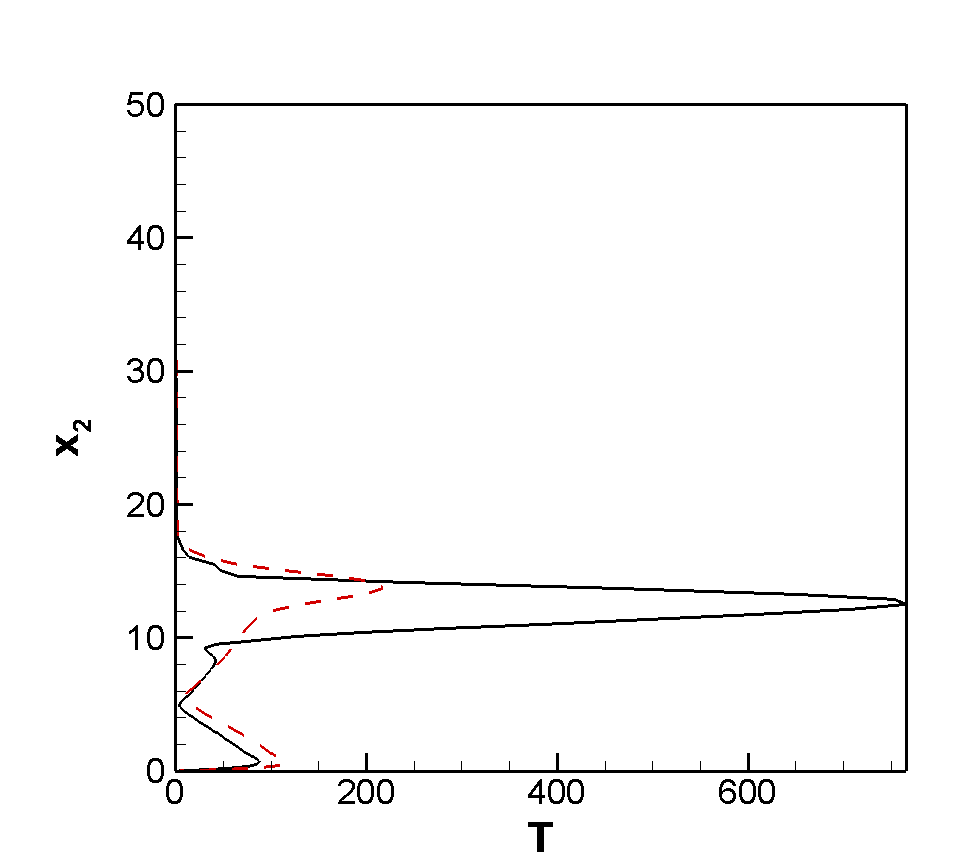}}
    \subfloat[$\hat{p}$]{\includegraphics[trim=0 50 10 0, clip, width=0.31\textwidth, keepaspectratio]{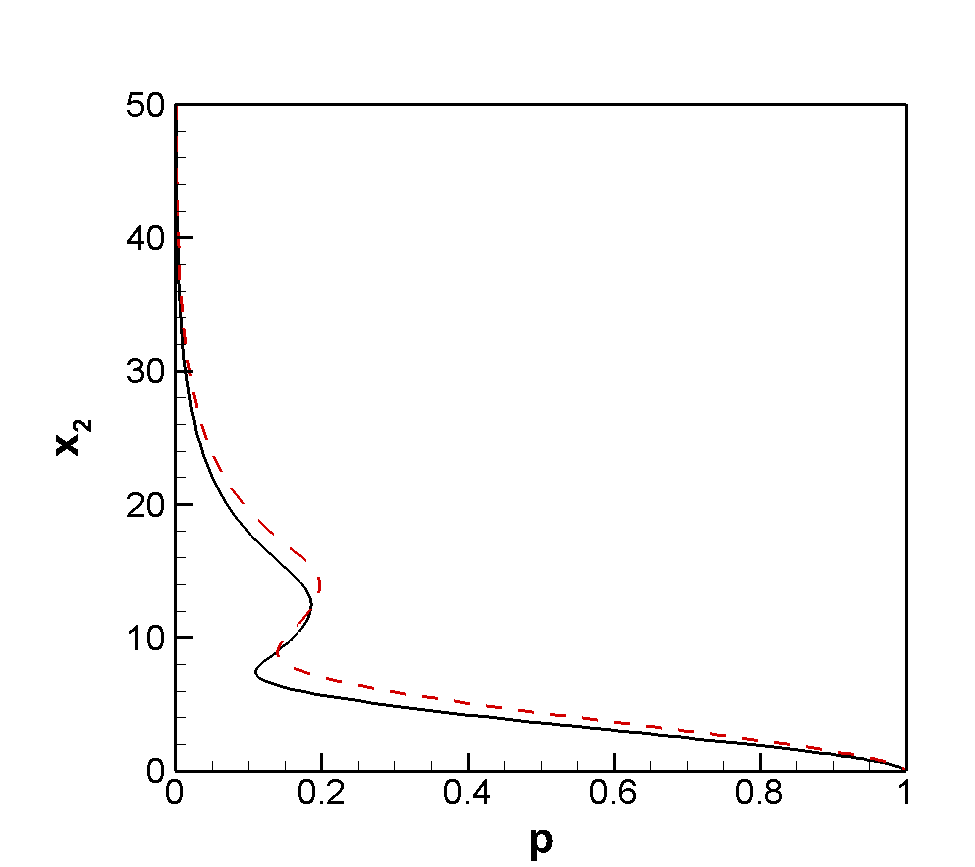}} \\
    \caption{Eigenmode shapes of the (a)-(c) fast ($F_+$) and (d)-(f) slow ($S$) modes near peak/trough in growth rate at $M=6$, $Re=4000$, $\alpha=0.175$, $\beta=0$ for two different Prandtl numbers.}
    \label{fig:ModeShapes_FS2}
\end{figure}

The eigenfunctions of the fast ($F_+$) and slow ($S$) modes before the branching of discrete spectrum are presented in figure \ref{fig:ModeShapes_FS1}. Before the branching of discrete spectrum the fast mode is more unstable at $Pr=0.5$ (figure \ref{fig:EigenSpectraM6}) while the slow mode is dominant instability at $Pr=0.9$. The eigenfunctions for the fast and slow mode at both Prandtl numbers are similar before the synchronization point. Figure \ref{fig:ModeShapes_FS2} displays the eigenfunctions of the fast and slow mode near the peak/trough in growth rates. The pressure eigenfunctions for both $F_+$ and $S$ modes are reasonably invariant with $Pr$. The temperature eigenfunction for the $F_+$ mode exhibits a stronger peak at $Pr=0.9$, while the slow mode has higher peak temperature at $Pr=0.5$. This can be attributed to the different branching pattern observed for $Pr=0.5$ and $Pr=0.9$.

\section{Prandtl number effects on flow-thermodynamic interactions}
\label{sec:Results}
\begin{figure}
    \centering
    \subfloat[First mode]{\includegraphics[width=0.45\textwidth, keepaspectratio]{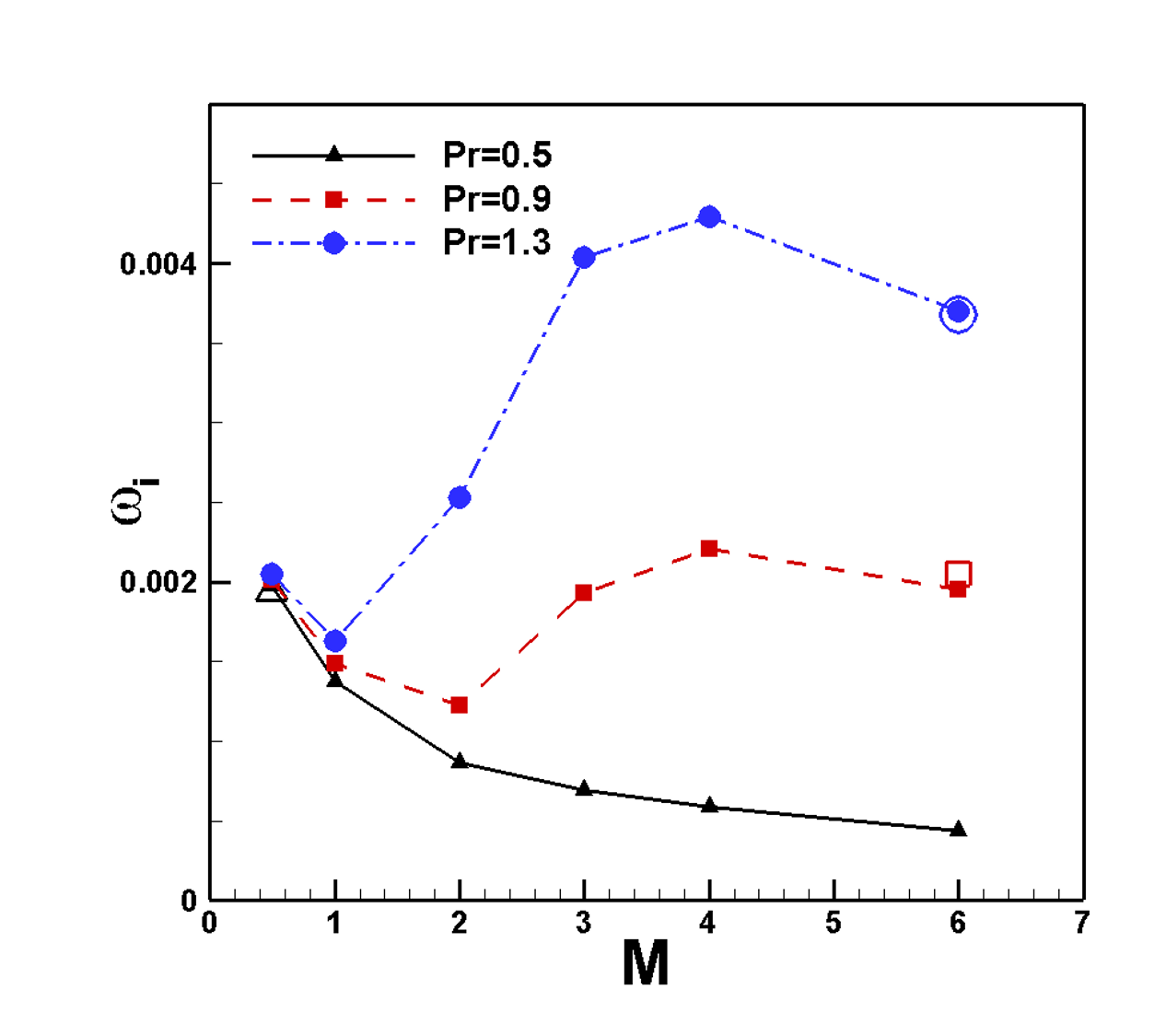}}
    \subfloat[Second mode]{\includegraphics[width=0.45\textwidth, keepaspectratio]{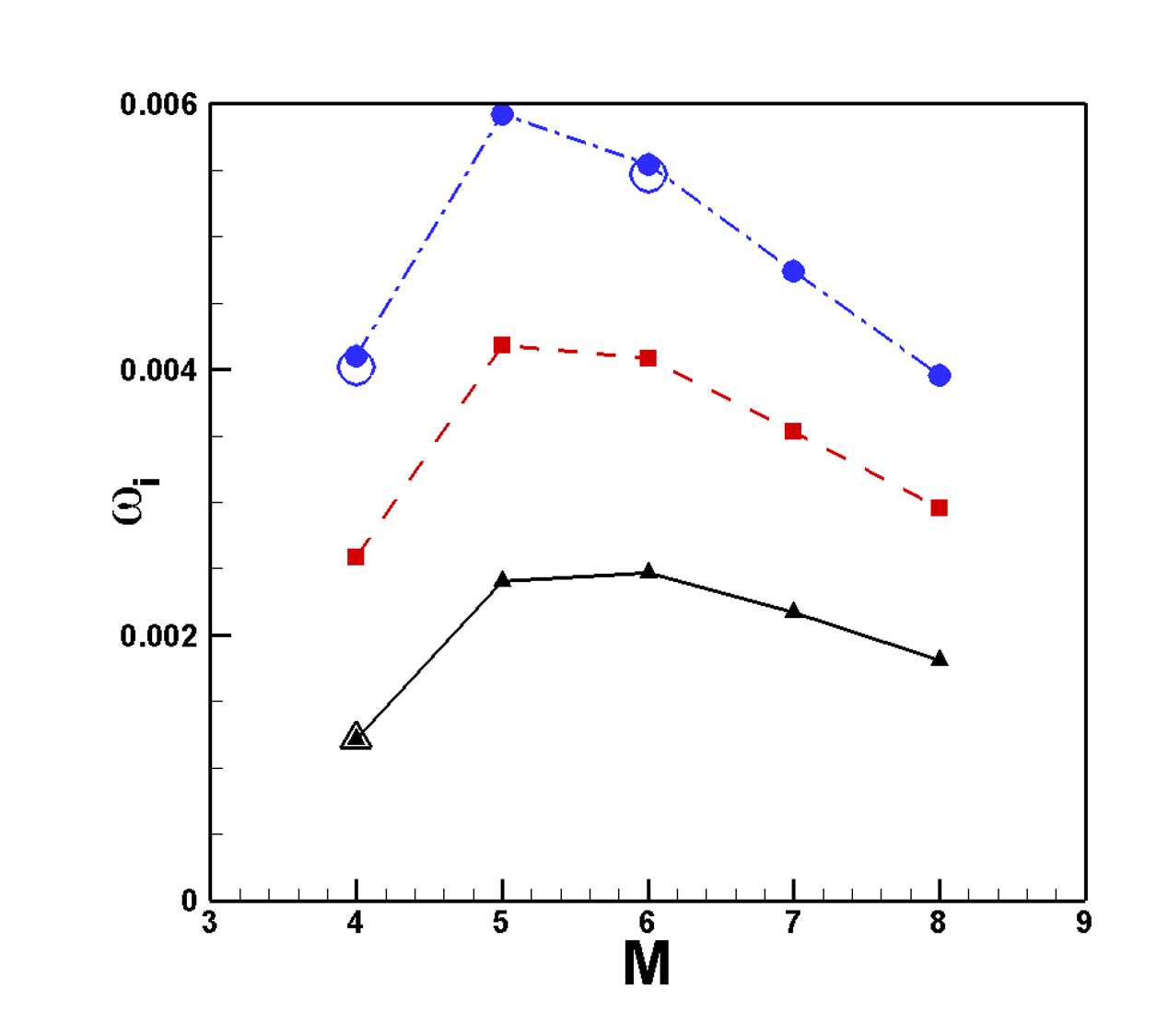}}
    \caption{Growth rates for the most unstable (a) first mode and (b) second mode. Filled symbols are from LSA computations and unfilled symbols correspond to results of GKM-DNS cases outlined in table \ref{table:ParametersBasic}.}
    \label{fig:Grate}
\end{figure}
  The effect of Prandtl number on the flow-thermodynamic interactions is investigated in this section. For simplicity we only consider the most unstable first/second mode for a given $(Re,Pr,M)$ combination. The most unstable mode is obtained by sweeping over a range of the streamwise-spanwise wavenumber pair ($\alpha, \beta$). The Reynolds number for all the cases considered here is maintained at $Re=4000$. At each Prandtl number, the most unstable first mode is obtained for $M=\{0.5,1,2,3,4,6\}$, while the most unstable second mode is computed for $M=\{4,5,6,7,8\}$. 

We first analyze the effect of Mach number and Prandtl number on the instability growth rate. The growth rates for the most unstable first and second modes are shown in figure \ref{fig:Grate}. The mean growth rates obtained from GKM-DNS for cases $C1-C6$ outlined in table \ref{table:ParametersBasic} are also plotted in figure \ref{fig:Grate}. The growth rates predicted by GKM-DNS are in excellent agreement with linear analysis for both first and second mode cases. A more rigorous validation by comparing the mode shapes of perturbations for cases $C_{3}$ and $C_{6}$ is provided in appendix \ref{app:B}. 

The most unstable first mode is streamwise for the subsonic Mach numbers, and oblique for the supersonic and hypersonic Mach numbers. The obliqueness angle for the most unstable mode decreases with Prandtl number at a given Mach number.  
At $Pr=0.5$, the growth rate for the first mode decreases monotonically with Mach number. The high Mach number ($M\geq2$) cases are destabilized with increasing Prandtl number while the low Mach number cases are unaffected by Prandtl number changes. 
The instability growth rates at high Mach numbers increases tenfold as the Prandtl number is increased from $0.5$ to $1.3$. This is consistent with the findings of \cite{ramachandran2015linear} wherein a similar destabilization of the streamwise first mode is observed for $M=4$.

The most unstable second mode is always aligned along the streamwise direction as the relative supersonic region is of maximum extent for 2D waves \citep{mack1984boundary}. Much like the first mode the second mode is also destabilized with increasing Prandtl number, although the destabilization is not as strong as the first mode. As shown in figure \ref{fig:Grate}(b), the growth rate for all Mach numbers considered at $Pr=1.3$ is more than double the growth rate at $Pr=0.5$. A similar destabilization of the second mode was also observed by \citep{ramachandran2015linear}. The main novelty of the present work is to examine the physics underlying the destabilization with increasing Prandtl number.

\subsection{Flow thermodynamic interactions for the first Mode}
\begin{figure}
    \centering
    \subfloat[$e^g$]{\includegraphics[width=0.45\textwidth, keepaspectratio]{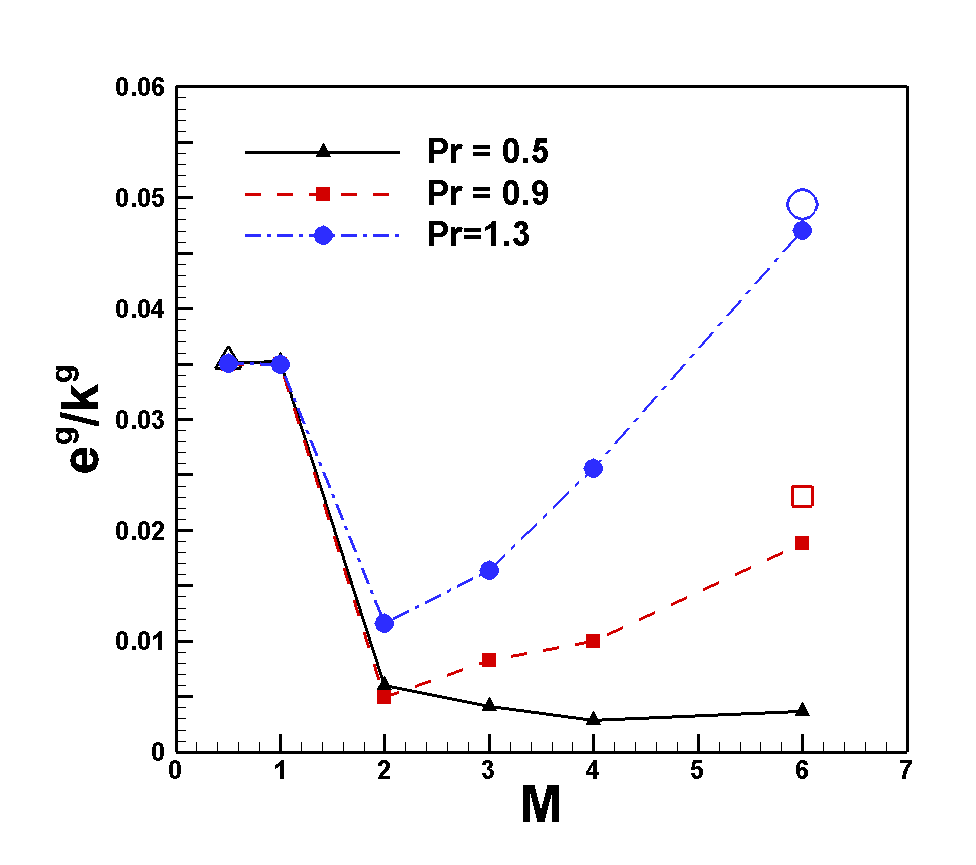}} 
    \subfloat[$\Pi_k^g/P_k^g$]{\includegraphics[width=0.45\textwidth, keepaspectratio]{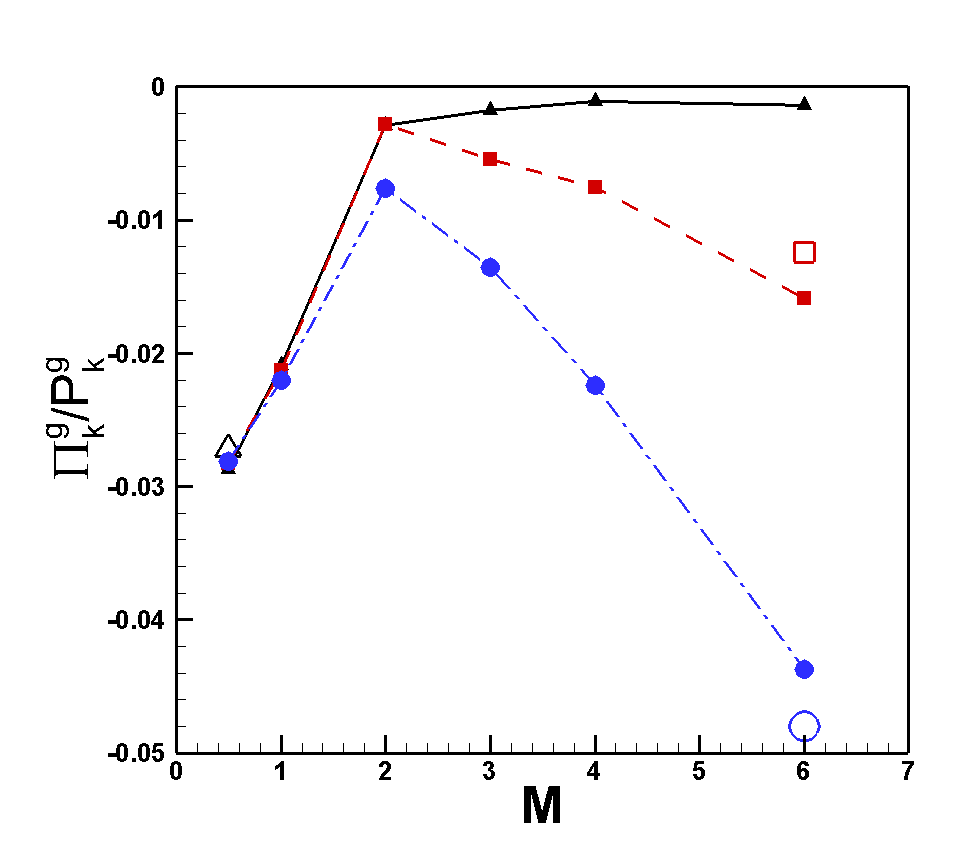}}
    \caption{Global averaged (a) internal energy fraction ($e^g/k^g$) and (b) pressure-dilatation to production ratio for the most unstable first mode. The symbols are same as figure \ref{fig:Grate}.}
    \label{fig:FmodeInternalEnergy}
\end{figure}
The influence of Prandtl number on the growth rate can be best understood by examining the flow-thermodynamic interactions in the flow.
Toward this end, the internal-kinetic energy exchange for the first mode instability is analyzed. For this analysis we define the global average $Q^g$ as:
\begin{equation}
    \label{eq:GlobalAverage}
    Q^g=\frac{1}{L_{x2}}\int_0^{L_{x2}}\langle Q(x_1,x_2,x_3) \rangle dx_2.
\end{equation}
The global averaged perturbation internal energy is obtained from linear stability analysis by integrating the amplitude of pressure perturbations in the wall-normal direction.
\begin{equation}
    \label{eq:GlobalInternalLSA}
    e^g = \frac{1}{L_{x2}}\int_0^{L_{x2}}\frac{\hat{p}(x_2)\hat{p}^c(x_2)}{2\gamma\overline{P}}  dx_2,
\end{equation}
where $\hat{p}$ is the mode shape of pressure perturbation obtained from linear theory and $\hat{p}^c$ is the complex conjugate of $\hat{p}$. Similarly, the global averaged perturbation kinetic energy $k^g$ is determined by the following expression.
\begin{equation}
    \label{eq:GlobalKineticLSA}
    k^g = \frac{1}{2L_{x2}}\int_0^{L_{x2}}\overline{\rho}(x_2)\left[\hat{u}_1(x_2)\hat{u}_1^c(x_2)+\hat{u}_2(x_2)\hat{u}_2^c(x_2)+\hat{u}_3(x_2)\hat{u}_3^c(x_2)\right]dx_2.
\end{equation}

The global averaged perturbation internal energy normalized by the global averaged perturbation kinetic energy at different $M$ and $Pr$ is presented in figure \ref{fig:FmodeInternalEnergy}(a). It is evident from figure \ref{fig:FmodeInternalEnergy}(a) that the internal energy content increases with increasing Prandtl number at high Mach numbers, suggesting thermodynamic effects are stronger in high Prandtl number fluids. For the first mode, the perturbation internal energy content is at least 20 times smaller than the kinetic energy.  
As mentioned previously, the internal and kinetic modes are coupled via pressure-dilatation. The perturbation velocity field interacts with the mean flow and the perturbation internal field via production and pressure-dilatation, respectively. Therefore, the ratio of pressure-dilatation to production is key for quantifying internal-kinetic energy exchange. The ratio of globally averaged pressure-dilatation to production is shown in figure \ref{fig:FmodeInternalEnergy}(b). The ratio is always negative indicating energy is transferred from kinetic to the internal mode. The plots also indicate that production is an order of magnitude greater than pressure-dilatation for all cases. As pressure-dilatation is small compared to production the internal-kinetic energy exchange is not significant for the first mode. 

The globally averaged internal energy and pressure-dilatation to production ratio obtained from GKM-DNS are also shown in figures \ref{fig:FmodeInternalEnergy}(a)-(b). As the DNS simulations are initialized by the mode computed from linear stability analysis, the mean values of $e^g$ and $\Pi_k^g/P_k^g$ in time are presented. The spatial derivatives in the $\Pi_k^g$ computations are obtained with spectral accuracy in the homogeneous directions while a fourth order central difference scheme is employed in the wall-normal direction. The mean $e^g$ and pressure-dilatation to production ratio obtained from GKM-DNS are in good agreement with linear analysis. 
\begin{figure*}
    \centering
    \subfloat[$P_k^g$]{\includegraphics[width=0.33\textwidth, keepaspectratio]{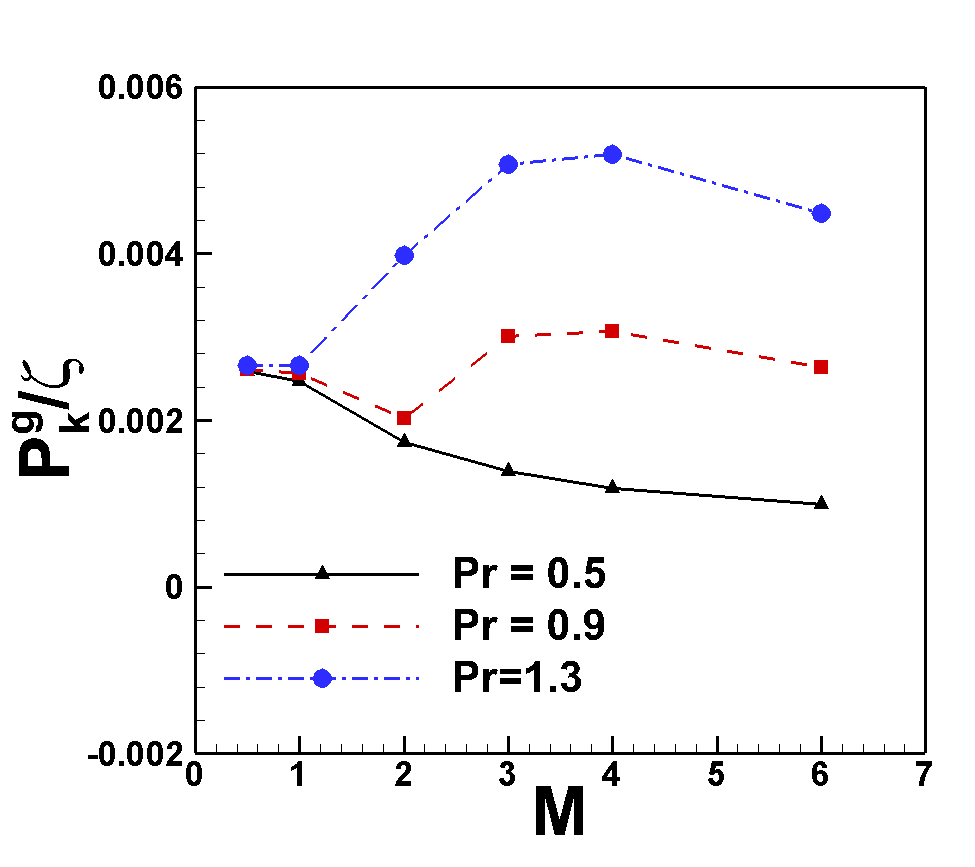}}
    \subfloat[$\Pi_k^g$]{\includegraphics[width=0.33\textwidth, keepaspectratio]{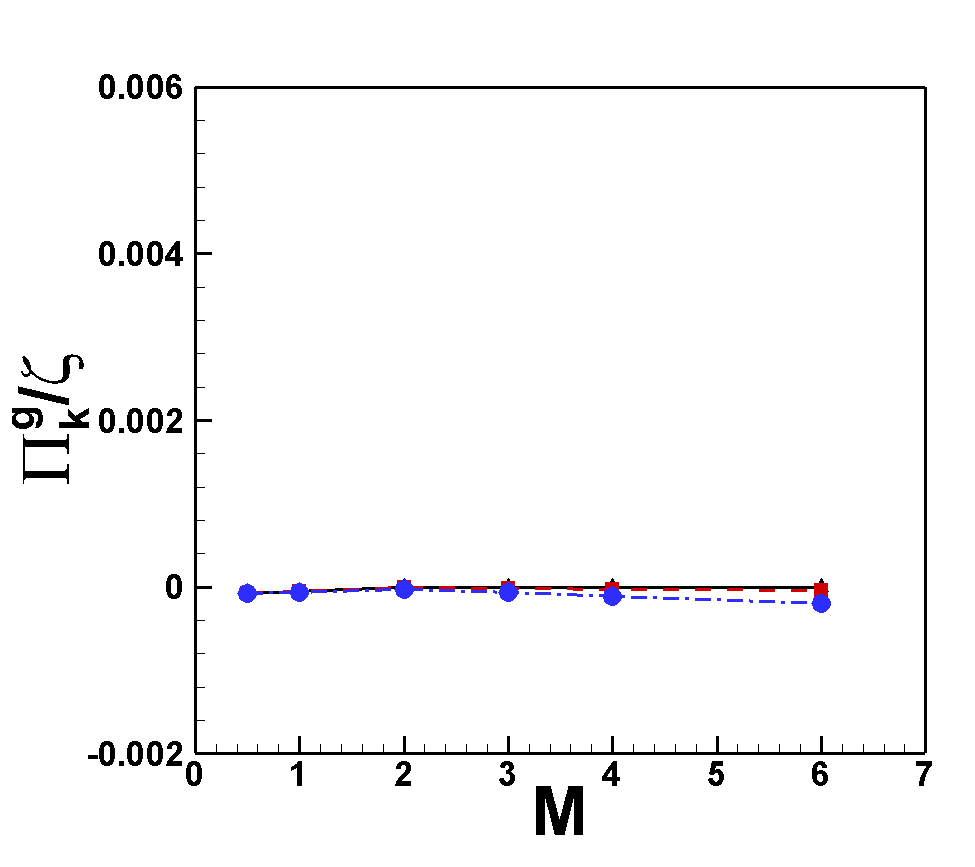}}
     \subfloat[$\epsilon_k^g$]{\includegraphics[width=0.33\textwidth, keepaspectratio]{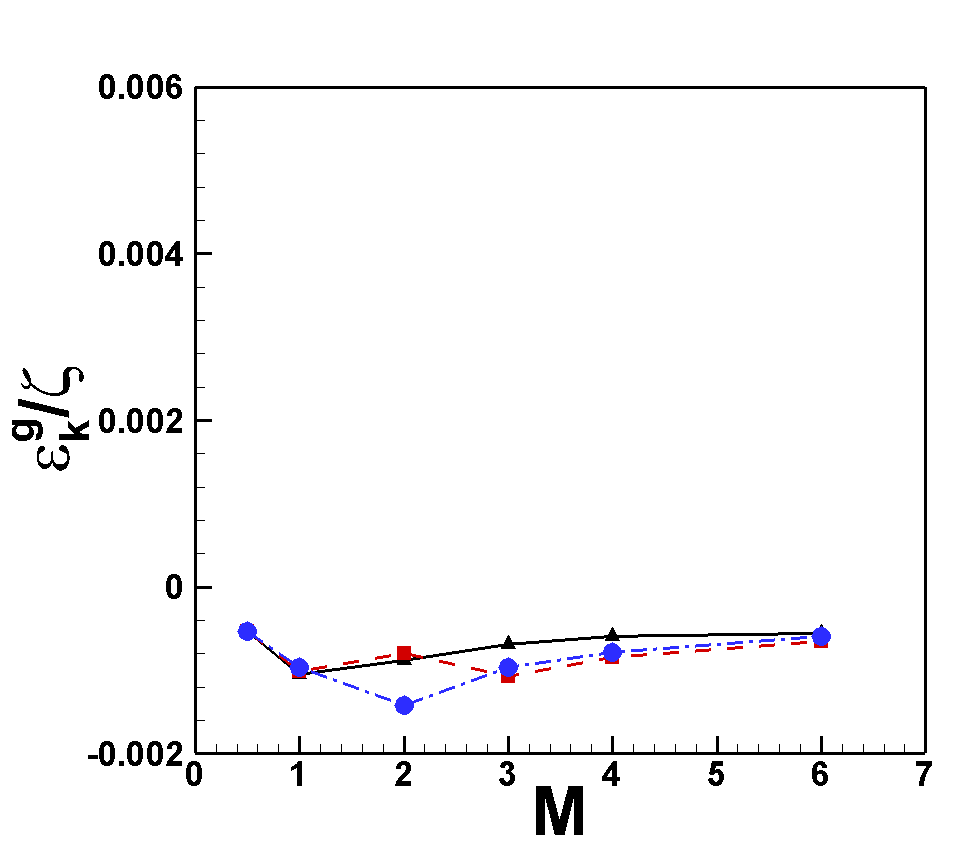}}
    \caption{Global averaged terms in the kinetic energy budget: (a) production $P_k^g$, (b) pressure-dilatation $\Pi_k^g$ and (c) dissipation $\epsilon_k^g$ for the most unstable first mode. All budget terms are normalized by $\zeta=2k^g U_\infty/L_r$. }
    \label{fig:FmodeKineticEnergyBudget}
\end{figure*}

We now consider the effect of Prandtl number on the various processes in the kinetic energy budget detailed in equation \eqref{eq:KeBudget}. Once again the statistics are averaged in the wall-normal direction. The global averaged terms in the kinetic energy budget are normalized by $\zeta=2k^g U_\infty/L_r$. The normalization ensures that the same level of perturbation kinetic energy is maintained allowing for a valid comparison across different cases.
The global average of the transport term is identically zero as the velocity perturbations vanish at the top and bottom boundaries.
Figure \ref{fig:FmodeKineticEnergyBudget} plots global averaged production ($P_k^g$), pressure-dilatation ($\Pi_k^g$) and dissipation ($\epsilon_k^g$) for the first mode at three different Prandtl numbers. At all $Pr-M$ combinations, production is the dominant process and pressure-dilatation is considerably smaller. The kinetic energy budget is essentially a balance between production and dissipation. At $Pr=0.5$, production is highest for $M=0.5$ and decreases monotonically with Mach number. Dissipation is fairly constant across all $M$ at this Prandtl number. The production decrease leads to stabilization of the first mode with Mach number at $Pr=0.5$. As the Prandtl number is increased, production increases almost tenfold for the high Mach number cases. On the other hand, production levels are unchanged for the low Mach number ($M\leq1$) cases. Dissipation exhibits only a marginal increase with Prandtl number. Thus, the main cause of increased destabilization with Pr is the enhancement of production.    
\begin{figure*}
    \centering
    \subfloat[$P_{11}^g$]{\includegraphics[width=0.33\textwidth, keepaspectratio]{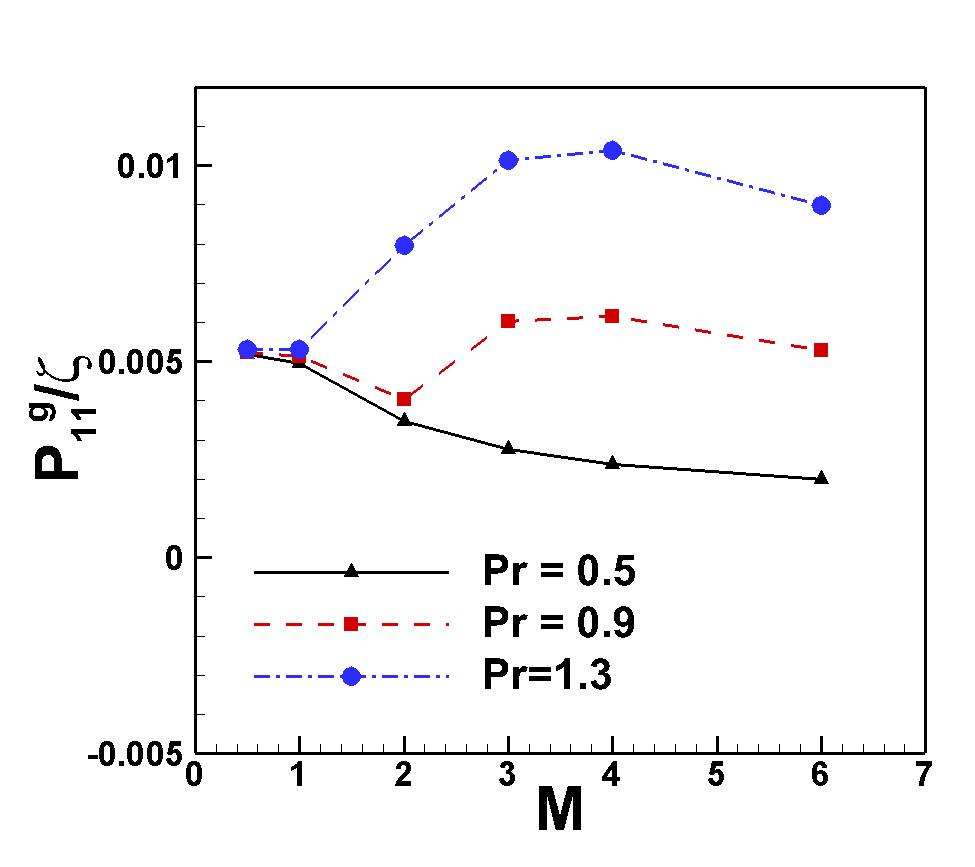}}
    \subfloat[$\Pi_{11}^g$]{\includegraphics[width=0.33\textwidth, keepaspectratio]{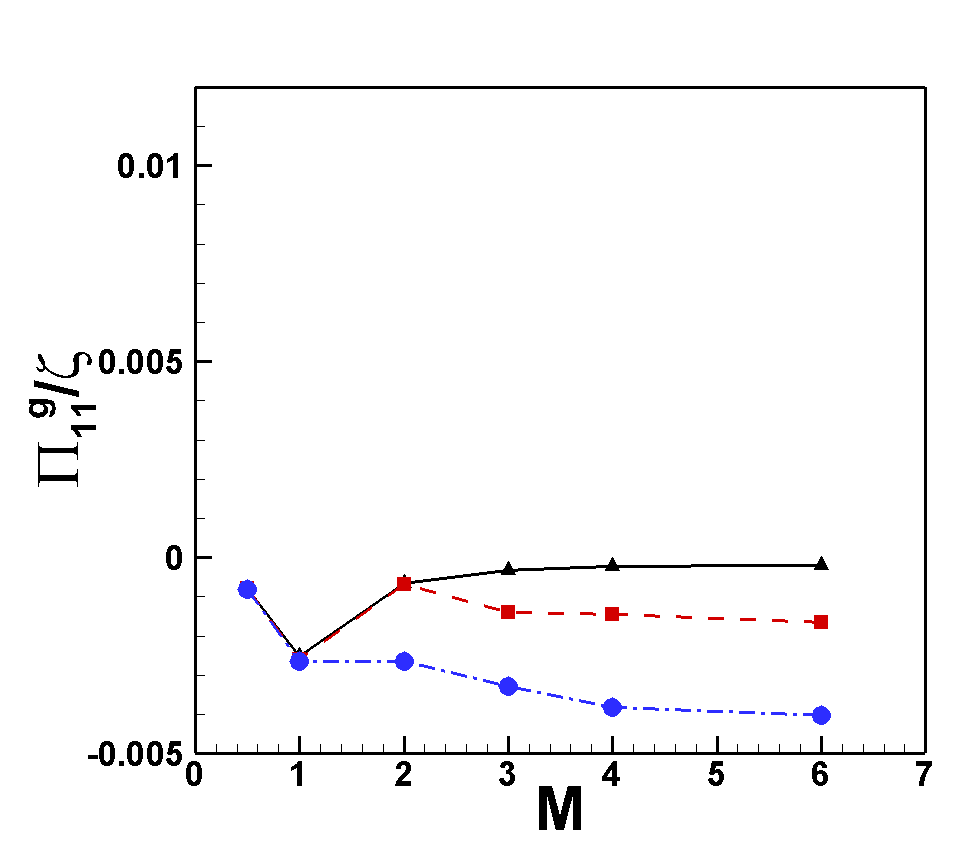}}
     \subfloat[$\epsilon_{11}^g$]{\includegraphics[width=0.33\textwidth, keepaspectratio]{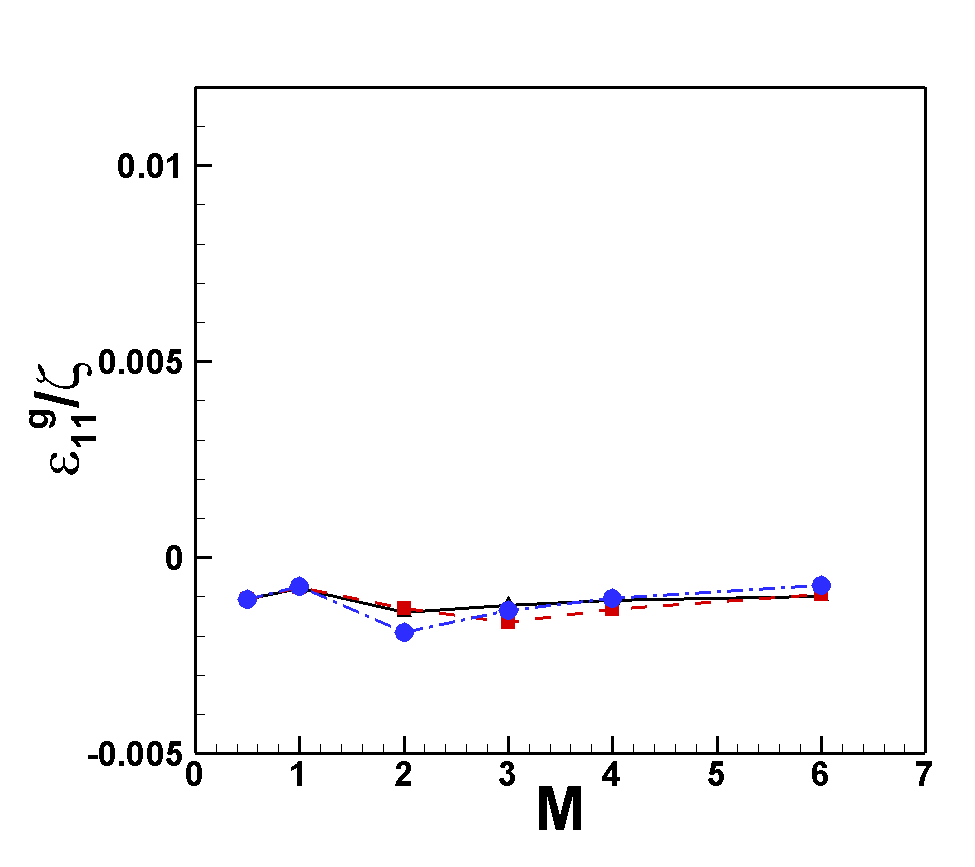}}
    \caption{Global averaged terms in the streamwise kinetic energy ($R_{11}$) budget: (a) production $P_{11}^g$, (b) pressure-strain correlation $\Pi_{11}^g$ and (c) dissipation $\epsilon_{11}^g$ for the most unstable first mode. All budget terms are normalized by $\zeta=2k^g U_\infty/L_r$.}
    \label{fig:FmodeStreamEnergyBudget}
\end{figure*}

Delving further, we examine the inter-component exchange of energy amongst the three kinetic modes by considering the diagonal components of the stress budget \eqref{eq:ReynoldsStressBudget}. The globally averaged terms on the right hand side (RHS) of the streamwise component of the stress ($R_{11}^g$) budget are shown in figures \ref{fig:FmodeStreamEnergyBudget}(a-c). Since, the base flow is two dimensional and parallel, the streamwise energy production ($P_{11}^g$) is the only non-zero diagonal component of the production tensor. Consequently, $P_{11}^g$ is equal to twice the total production of kinetic energy. The streamwise component of pressure-strain correlation ($\Pi_{11}^g$) is of the order of $P_{11}^g$ even though pressure-dilatation is negligible. It is evident form figure \ref{fig:FmodeStreamEnergyBudget}(b) that the sign of $\Pi_{11}^g$ is always negative, indicating that energy is extracted from the streamwise mode. The amount of energy extracted by pressure-strain correlation from the streamwise mode increases with Prandtl number. The streamwise component of the dissipation tensor ($\epsilon_{11}^g$) is fairly constant with Prandtl number and is the least significant of the three processes. Overall, $P_{11}^g+\Pi_{11}^g$ increases with Pr for a given Mach number. 

\begin{figure*}
    \centering
    \subfloat[$\Pi_{22}^g$]{\includegraphics[width=0.33\textwidth, keepaspectratio]{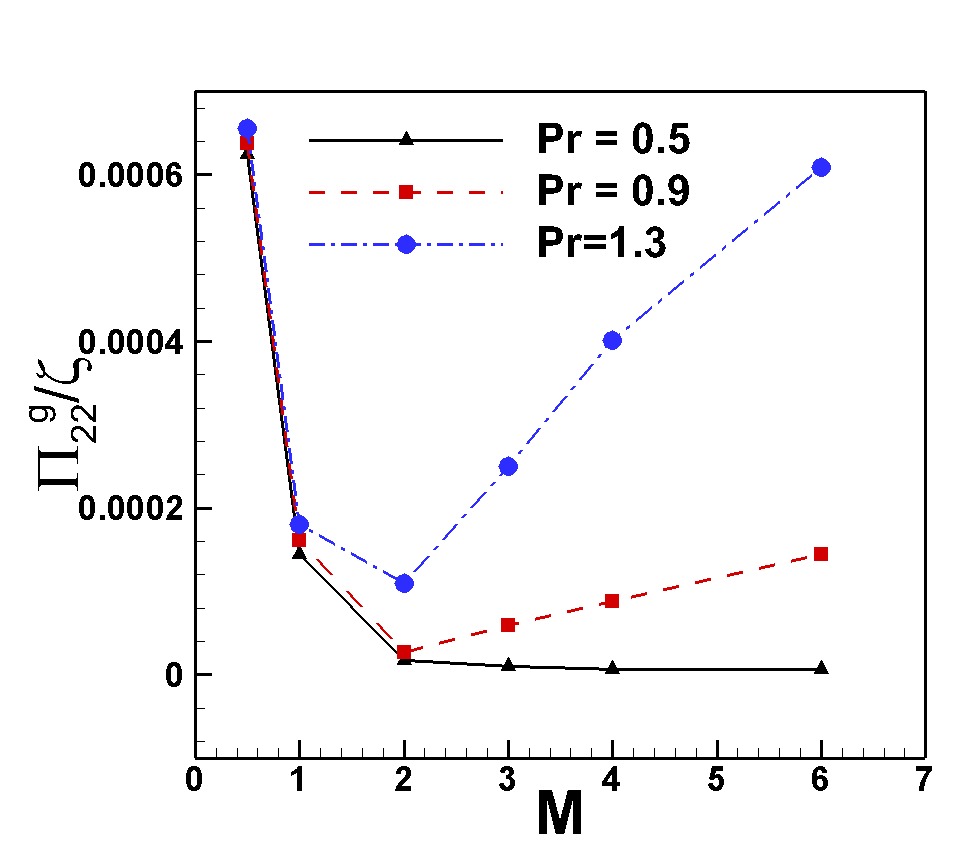}}
    \subfloat[$\epsilon_{22}^g$]{\includegraphics[width=0.33\textwidth, keepaspectratio]{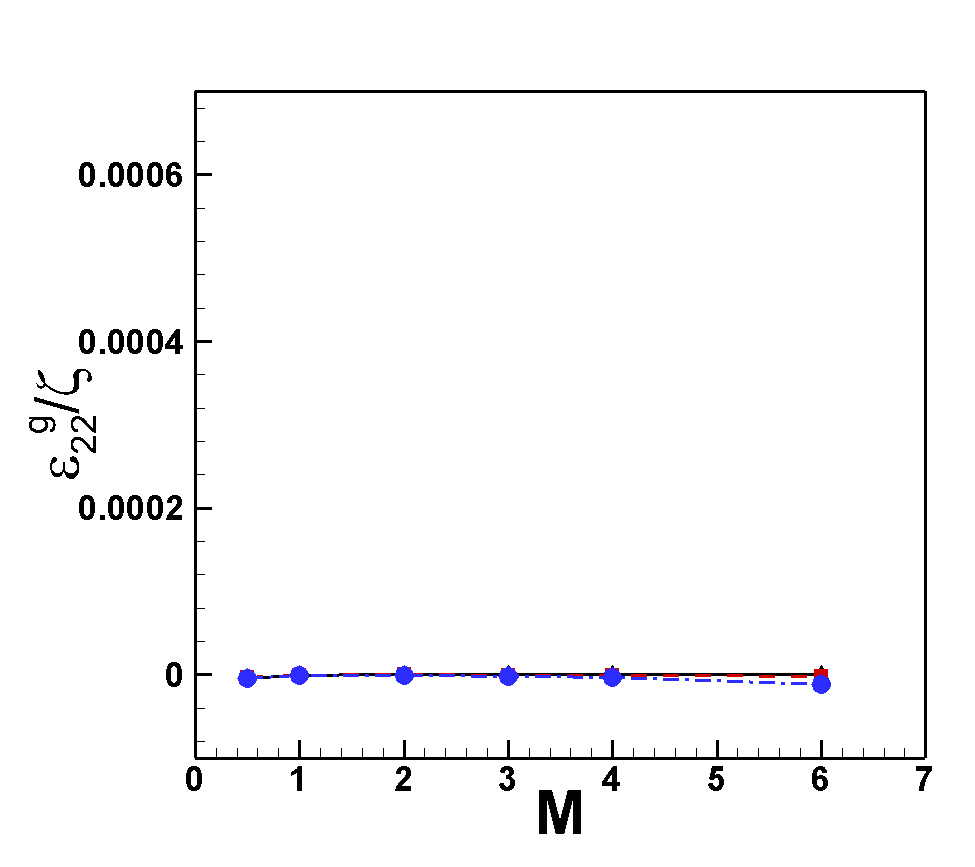}}
     \subfloat[$R_{22}^g$]{\includegraphics[width=0.33\textwidth, keepaspectratio]{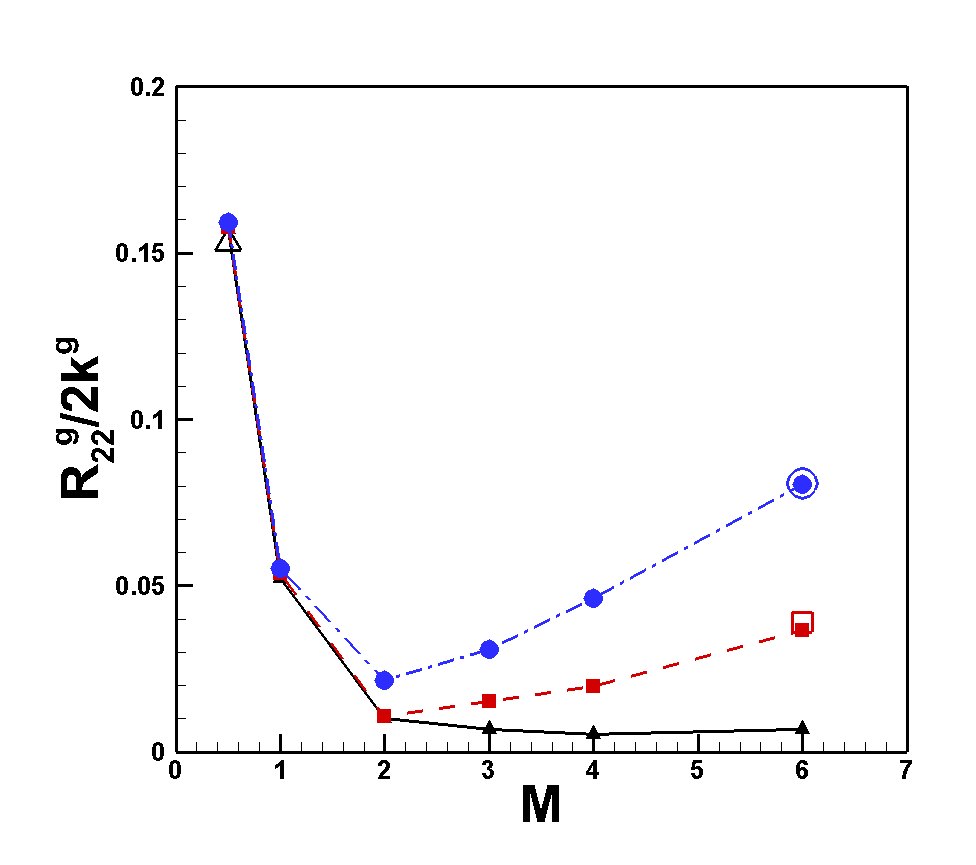}}
    \caption{Global averaged terms in the wall-normal kinetic energy ($R_{22}$) budget: (a) pressure-strain correlation $\Pi_{22}^g$ and (b) dissipation $\epsilon_{22}^g$ for the most unstable first mode. (c) Wall-normal kinetic energy fraction ($R_{22}^g/2k^g$) for the most unstable first mode. All budget terms are normalized by $\zeta=2k^g U_\infty/L_r$. The symbols are same as figure \ref{fig:Grate}.}
    \label{fig:FmodeWallEnergyBudget}
\end{figure*}

The averaged terms on the RHS of the wall-normal kinetic energy budget are shown in figures \ref{fig:FmodeWallEnergyBudget}(a-b). $P_{22}^g$ is identically zero due to the absence of mean velocity in the wall-normal direction. Energy extracted from the streamwise mode is transferred to the wall-normal mode via $\Pi_{22}^g$. The wall-normal component of the dissipation tensor is negligible indicating almost all the energy transferred to the wall-normal mode is retained. $\Pi_{22}^g$ increases with Prandtl number resulting in higher kinetic energy in the wall-normal direction. This is shown in figure \ref{fig:FmodeWallEnergyBudget}(c) wherein the wall-normal kinetic energy fraction is plotted. The wall-normal kinetic energy content ($R_{22}^g$) increases with Prandtl number due to increase in $\Pi_{22}^g$. The fraction of wall-normal kinetic energy as obtained from GKM-DNS is also presented in figure \ref{fig:FmodeWallEnergyBudget}(c). The results from DNS are in excellent agreement with linear analysis. 
\begin{figure*}
    \centering
    \subfloat[$\Pi_{33}^g$]{\includegraphics[width=0.33\textwidth, keepaspectratio]{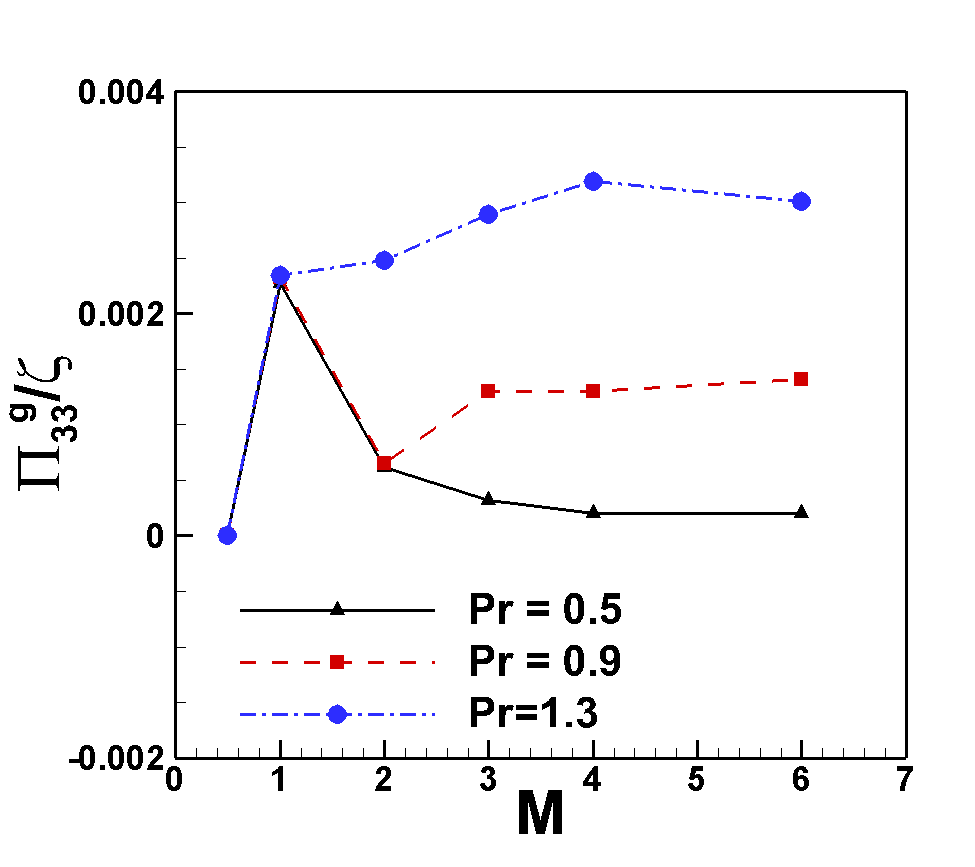}}
    \subfloat[$\epsilon_{33}^g$]{\includegraphics[width=0.33\textwidth, keepaspectratio]{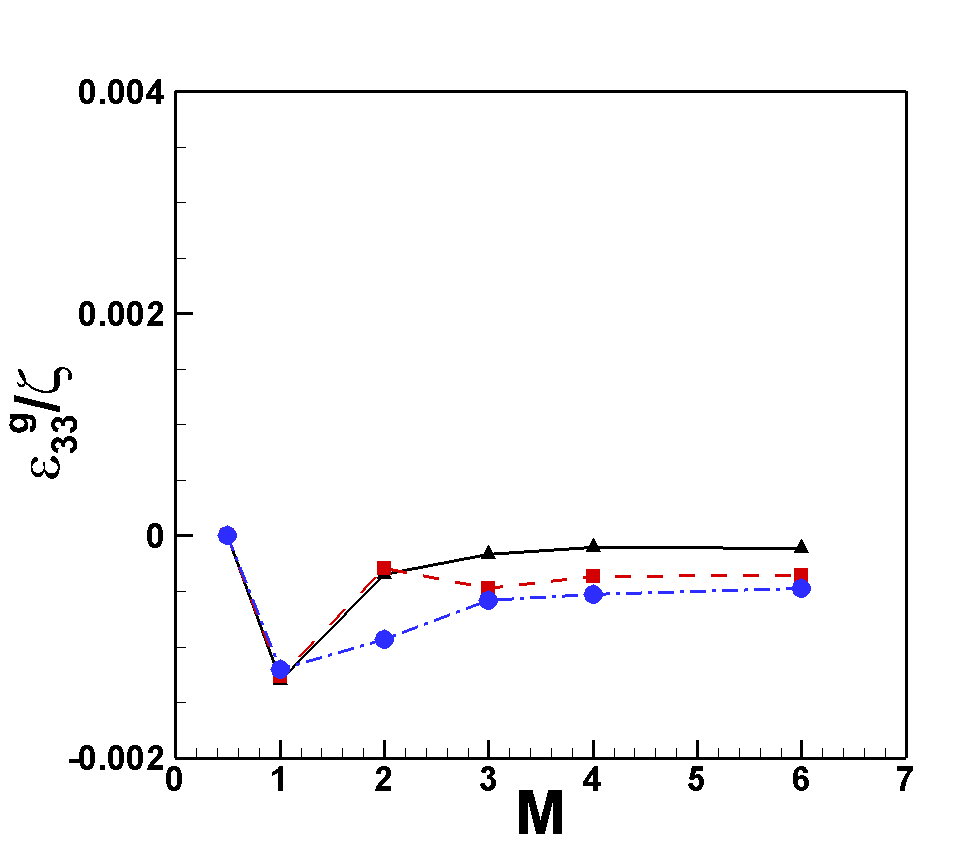}}
     \subfloat[$R_{33}^g$]{\includegraphics[width=0.33\textwidth, keepaspectratio]{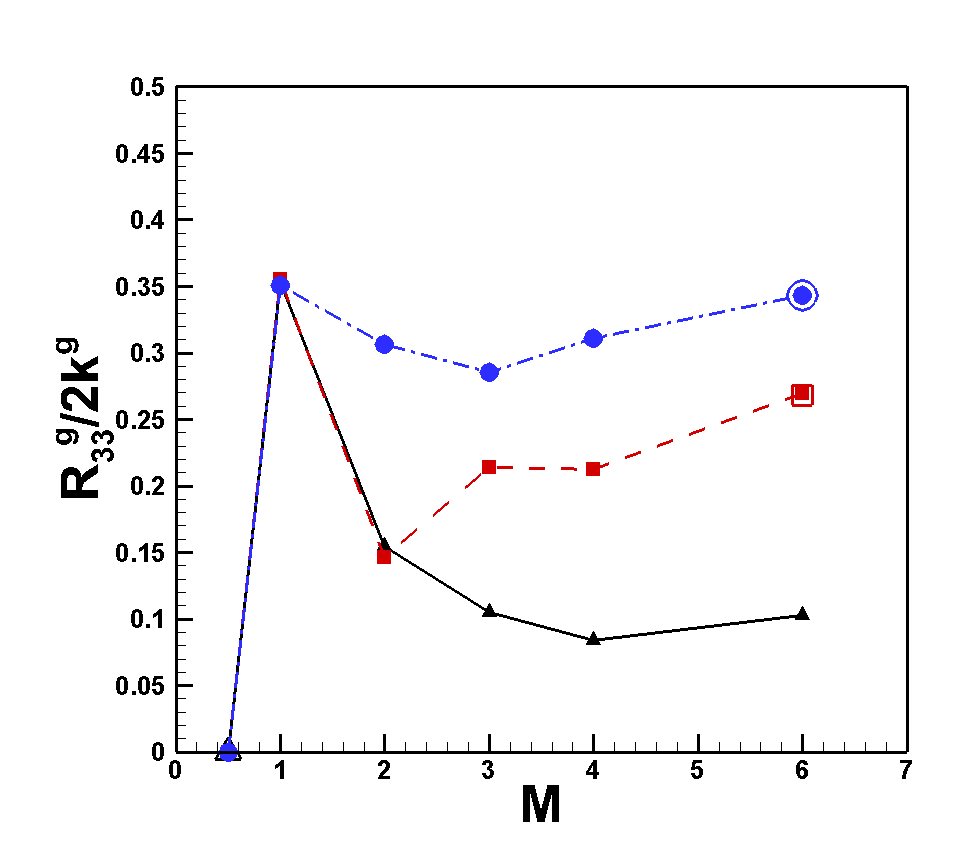}}
    \caption{Global averaged terms in the spanwise kinetic energy ($R_{33}$) budget: (a) pressure-strain correlation $\Pi_{33}^g$ and (b) dissipation $\epsilon_{33}^g$ for the most unstable first mode. (c) Spanwise kinetic energy fraction ($R_{33}^g/2k^g$) for the most unstable first mode. All budget terms are normalized by $\zeta=2k^g U_\infty/L_r$. The symbols are same as figure \ref{fig:Grate}.}
    \label{fig:FmodeSpanEnergyBudget}
\end{figure*}

Figures \ref{fig:FmodeSpanEnergyBudget}(a-b) presents the global averaged terms on the RHS of spanwise kinetic energy budget. As the most unstable first mode at $M=0.5$ is aligned in the streamwise direction, the spanwise kinetic energy is negligible. For the oblique modes ($M\geq1$), $\Pi_{33}^g$ provides a source of spanwise energy. Some of this energy is dissipated via the spanwise component of the dissipation tensor $\epsilon_{33}^g$. The dissipated energy is not significant compared to $\Pi_{33}^g$ at high Prandtl numbers. $\Pi_{33}^g$ increases with Prandtl number resulting in increased energy in the spanwise mode as shown in figure \ref{fig:FmodeSpanEnergyBudget}(c). The DNS results for the spanwise energy fraction are also shown in figure \ref{fig:FmodeSpanEnergyBudget}(c), and once again the agreement with linear analysis is very good. 
\begin{figure*}
    \centering
        \subfloat[$P_{12}^g$]{\includegraphics[width=0.33\textwidth, keepaspectratio]{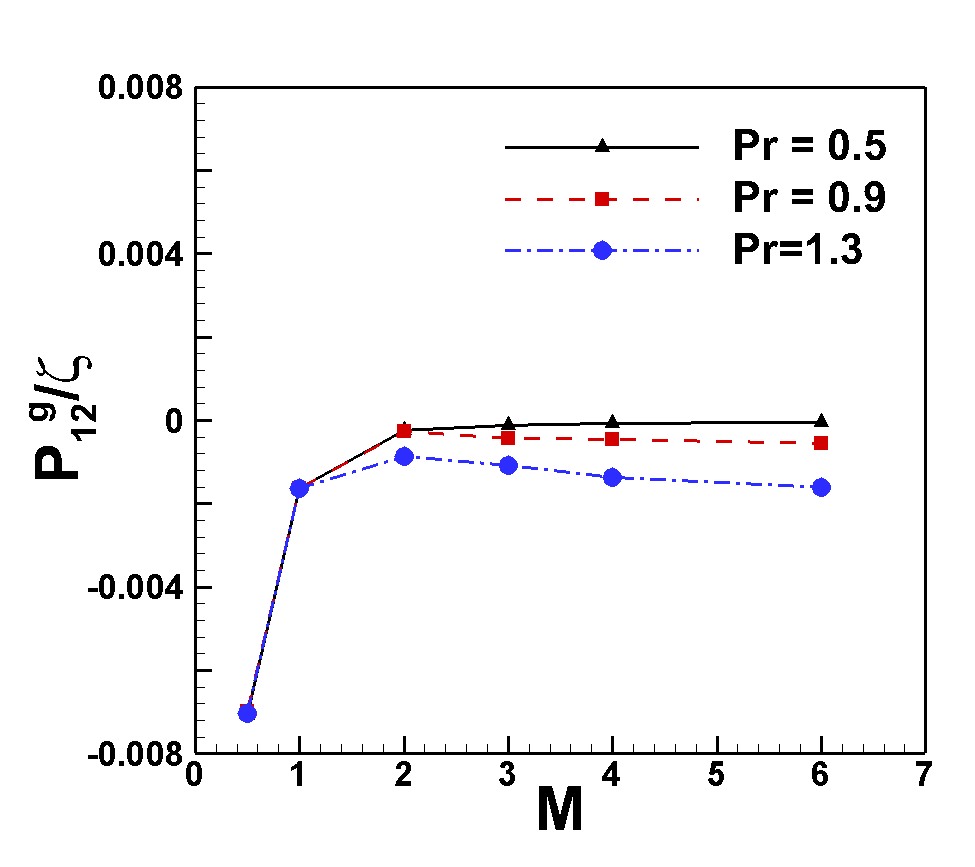}}
    \subfloat[$|\Pi_{12}^g/P_{12}^g|$]{\includegraphics[width=0.33\textwidth, keepaspectratio]{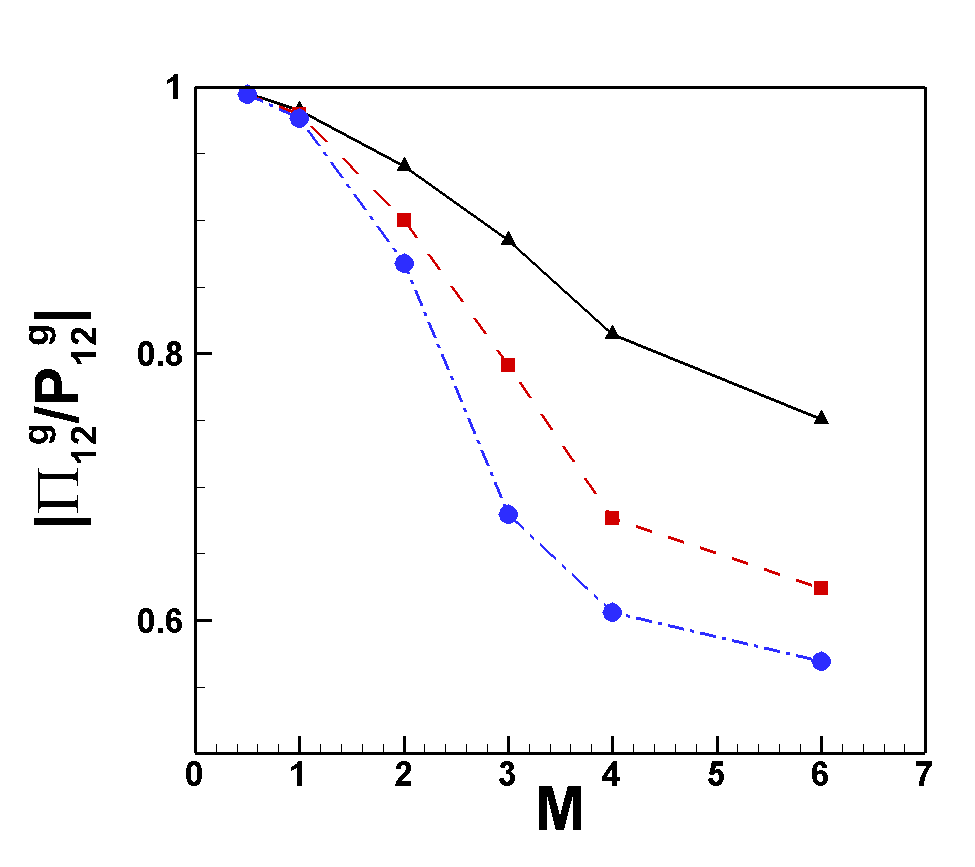}}
     \subfloat[$\Pi_{12}^g+P_{12}^g+\epsilon_{12}^g$]{\includegraphics[width=0.33\textwidth, keepaspectratio]{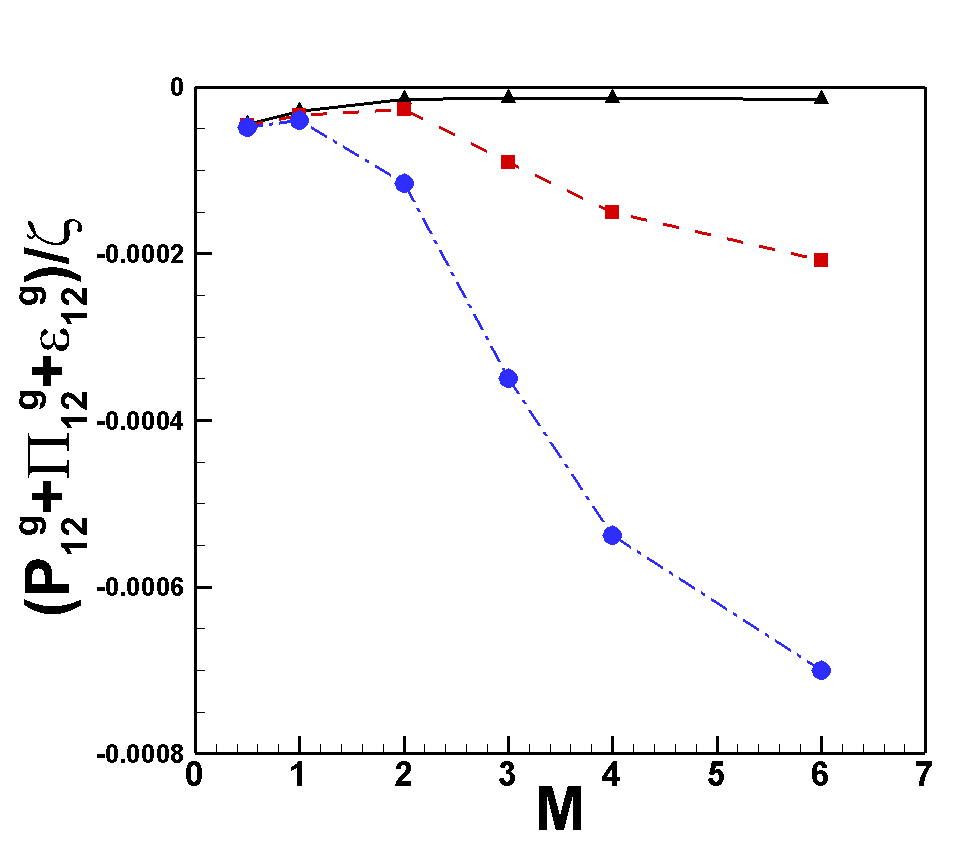}}
    \caption{(a) Global averaged shear stress production $P_{12}^g$, (b) absolute value of ratio of pressure-strain correlation $\Pi_{12}^g$ to shear stress production for the most unstable first mode. (c) Sum of all the terms on the right hand side of the shear stress budget equation \eqref{eq:ReynoldsStressBudget}. All budget terms are normalized by $\zeta=2k^g U_\infty/L_r$.}
    \label{fig:Fmodeu1u2EnergyBudget}
\end{figure*}

The above results indicate that the destabilization of the first mode with Prandtl number is due to increased production. For a parallel flow, production is dependent on the shear stress anisotropy $\langle R_{12} \rangle$. The shear stress budget is examined in figures \ref{fig:Fmodeu1u2EnergyBudget}(a-c). The shear stress production ($P_{12}$) and the shear component of pressure-strain correlation tensor ($\Pi_{12}$) are given by the following relation.
\begin{equation}
    P_{12} =-\langle\overline{\rho}u_2'u_2'\rangle\frac{d\overline{U}_1}{dx_2}; \quad \Pi_{12} =\left\langle p'\left(\frac{\partial u_1'}{\partial x_2}+\frac{\partial u_2'}{\partial x_1}\right)\right\rangle
\end{equation}
$P_{12}$ is dependent on the wall-normal kinetic energy and the mean velocity gradient. As mentioned previously, the wall-normal energy content increases with Prandtl number. Consequently, the global averaged production of shear stress ($P_{12}^g$) increases in magnitude as shown in figure \ref{fig:Fmodeu1u2EnergyBudget}(a). 
The absolute value of the ratio $\Pi_{12}^g/P_{12}^g$ is presented in figure \ref{fig:Fmodeu1u2EnergyBudget}(b). $\Pi_{12}^g$ is positive at all $(M,Pr)$ combination. The ratio $|\Pi_{12}^g/P_{12}^g|$ is approximately unity at low Mach numbers and decreases monotonically with increasing Mach number. Furthermore, as the Prandtl number is increased the ratio $|\Pi_{12}^g/P_{12}^g|$ decreases. It must be noted that $\Pi_{12}^g$ increases with Prandtl number in such a way that the ratio of pressure-strain correlation to shear production decreases. This results in a net increase in the magnitude of the shear stress budget (figure \ref{fig:Fmodeu1u2EnergyBudget}(c)). 
\begin{figure*}
    \centering
    \subfloat[$R_{12}^g$]{\includegraphics[width=0.45\textwidth, keepaspectratio]{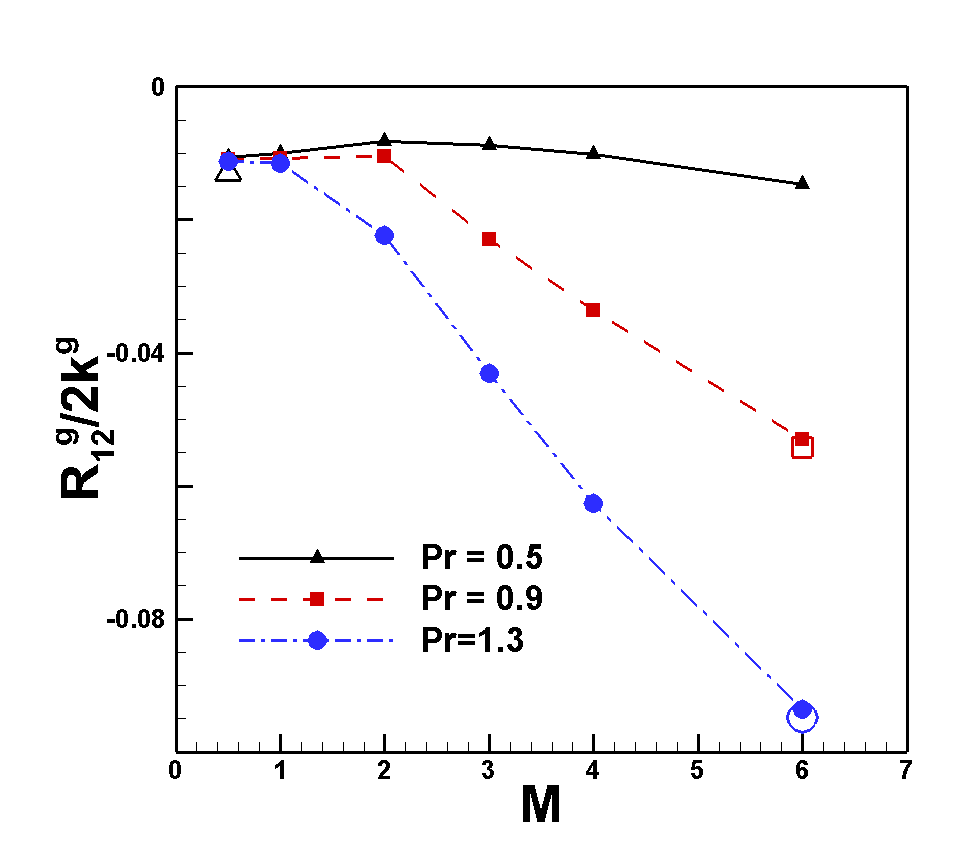}} 
    \subfloat[$d\overline{U}_1/dx_2$]{\includegraphics[width=0.45\textwidth, keepaspectratio]{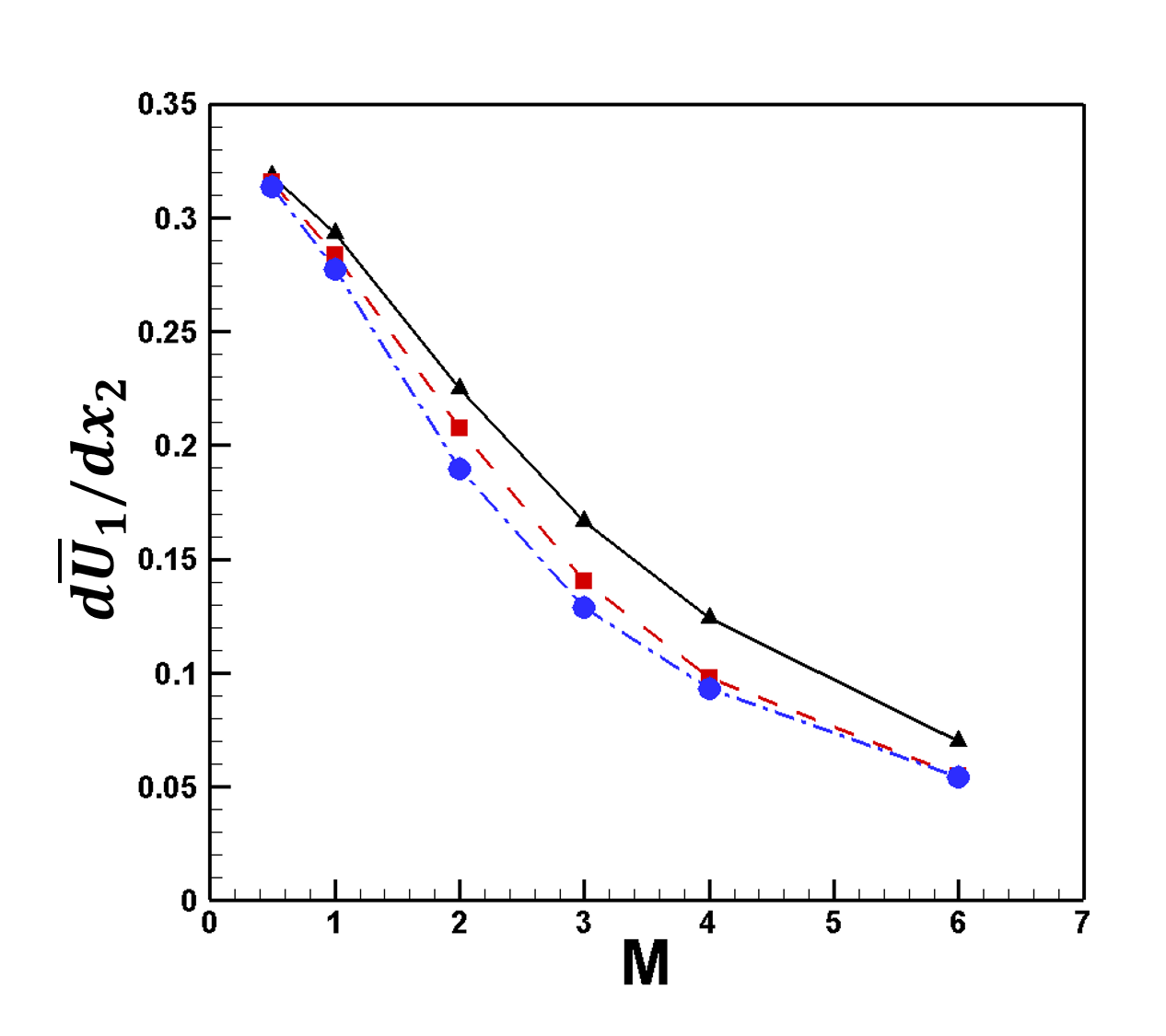}}
    \caption{(a) Shear stress anisotropy ($R_{12}^g/2k^g$) for the most unstable first mode. (b) Mean velocity gradient ($d\overline{U}_1/dx_2$) at the critical layer for the most unstable first mode. The symbols are same as figure \ref{fig:Grate}.}
    \label{fig:Fmodeu1u2}
\end{figure*}

Seeking the reason for increased production with Pr, we now examine shear stress and mean velocity gradients. The shear stress is shown in figure \ref{fig:Fmodeu1u2}(a). The shear stress becomes more negative with increasing Prandtl number. This is also confirmed by DNS results presented in figure \ref{fig:Fmodeu1u2}(a). The base velocity gradient at the critical layer in the flow is presented in figure \ref{fig:Fmodeu1u2}(b). 
The location of the critical layer corresponds to maximum production in the flow. Although the base velocity gradient at the critical layer decreases slightly with Prandtl number, the significant increase in the shear stress anisotropy allows for higher production. At $Pr=1.3$, the magnitude of shear stress anisotropy increases monotonically with Mach number. The mean velocity gradient on the other hand decreases monotonically with the Mach number. The competing trends for shear stress anisotropy and mean velocity gradient leads to a production peak (figure \ref{fig:FmodeKineticEnergyBudget}) at $M=4$ for $Pr=1.3$. 

\subsection{Flow-thermodynamic interactions for the second mode}
\begin{figure}
    \centering
    \subfloat[$e^g$]{\includegraphics[width=0.45\textwidth, keepaspectratio]{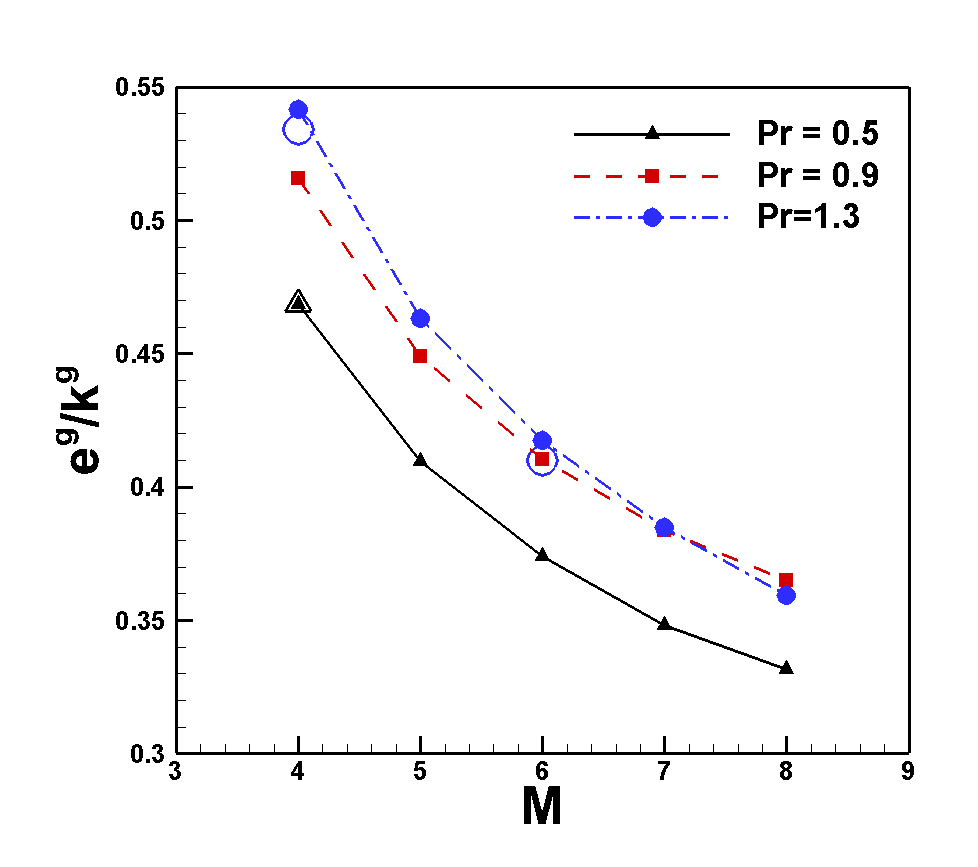}} 
    \subfloat[$\Pi_k^g/P_k^g$]{\includegraphics[width=0.45\textwidth, keepaspectratio]{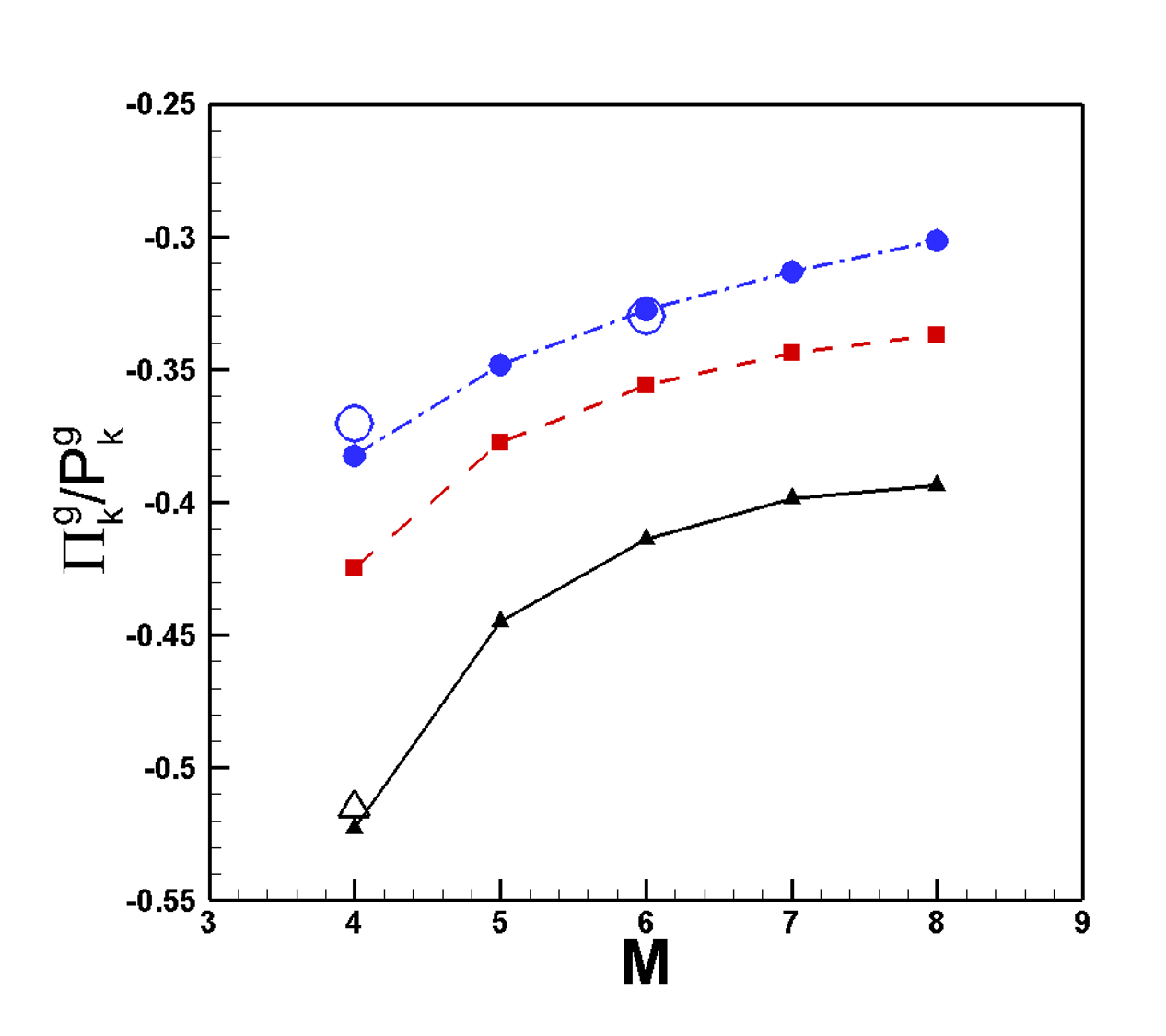}}
    \caption{Global averaged (a) internal energy fraction ($e^g/k^g$) and (b) pressure-dilatation to production ratio for the most unstable second mode. The symbols are same as figure \ref{fig:Grate}.}
    \label{fig:SmodeInternalEnergy}
\end{figure}

In this subsection, we analyze the effect of Prandtl number on the flow thermodynamics interaction of the most unstable second mode. The global averaged perturbation internal energy normalized by $k^g$ for the most unstable second mode is shown in figure \ref{fig:SmodeInternalEnergy}(a). The internal energy content increases with Prandtl number as thermodynamic effects become stronger with increasing Prandtl number. Unlike the first mode, internal energy for the second mode is of the same order as the perturbation kinetic energy. The internal-kinetic energy interactions can be quantified by the ratio of pressure-dilatation to production shown in figure \ref{fig:SmodeInternalEnergy}(b). The ratio $\Pi_k^g/P_k^g\sim\mathcal{O}(1)$, indicating pressure work is significant for the second mode. Energy is transferred from the kinetic to the internal mode as $\Pi_k^g/P_k^g$ is always negative. For a given Mach number, the ratio of pressure-dilatation to production decreases in magnitude with Prandtl number, and yet the internal energy fraction increases with Prandtl number. The mean values of $e^g$ and $\Pi_k^g/P_k^g$ computed from GKM-DNS also shown in figures \ref{fig:SmodeInternalEnergy}(a)-(b) are in excellent agreement with linear analysis.

\begin{figure*}
    \centering
    \subfloat[$P_k^g$]{\includegraphics[width=0.33\textwidth, keepaspectratio]{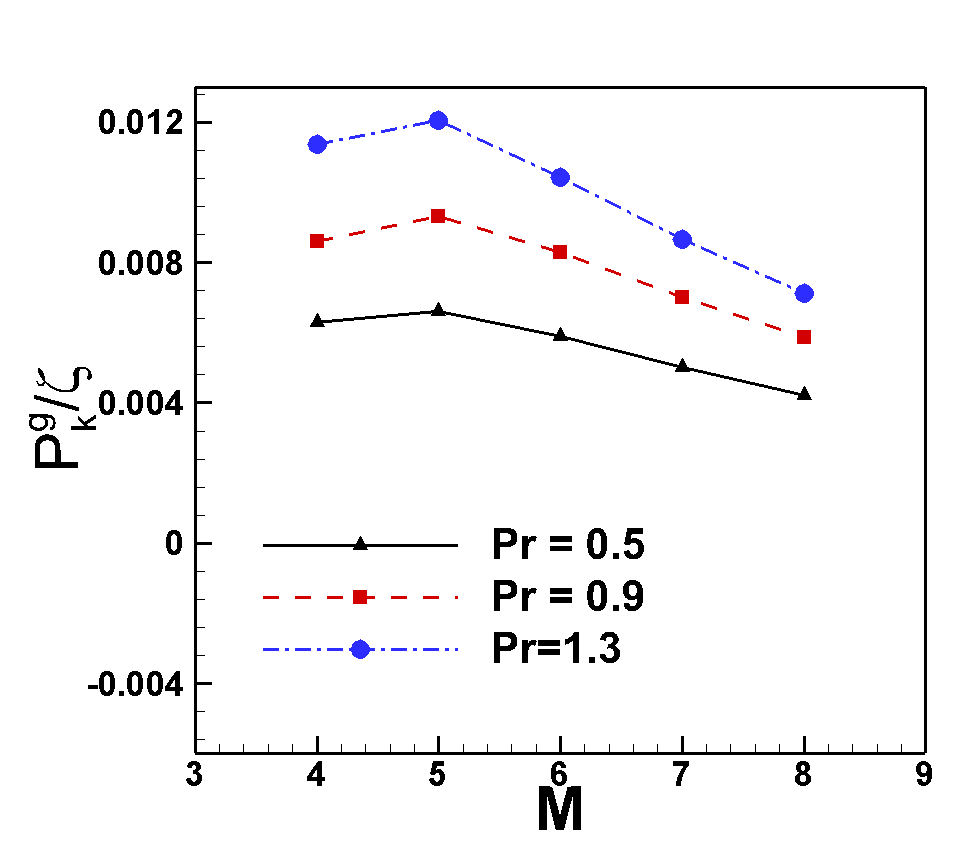}}
    \subfloat[$\Pi_k^g$]{\includegraphics[width=0.33\textwidth, keepaspectratio]{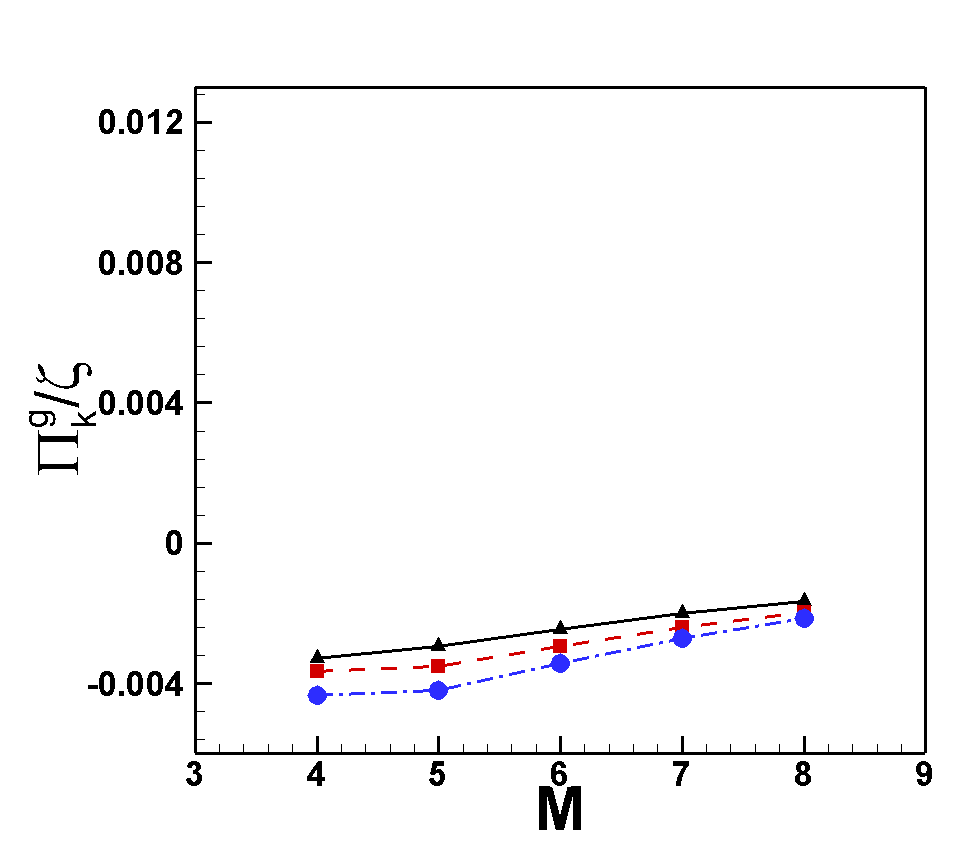}}
     \subfloat[$\epsilon_k^g$]{\includegraphics[width=0.33\textwidth, keepaspectratio]{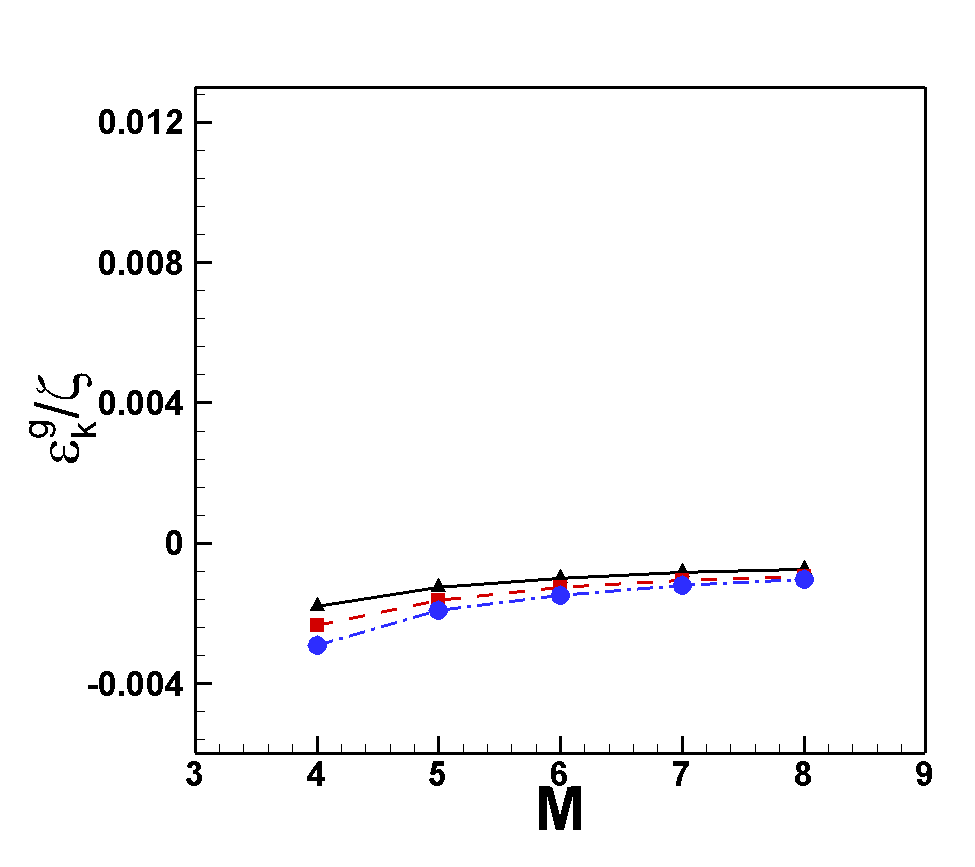}}
    \caption{Global averaged terms in the kinetic energy budget: (a) production $P_k^g$, (b) pressure-dilatation $\Pi_k^g$ and (c) dissipation $\epsilon_k^g$ for the most unstable second mode. All terms in the budget are normalized by $\zeta=2k^g U_\infty/L_r$. }
    \label{fig:SmodeKineticEnergyBudget}
\end{figure*}

We now analyze the effect of Prandtl number on the flow processes. The global average of the terms on the RHS of the kinetic energy budget \eqref{eq:KeBudget} are shown in figures \ref{fig:SmodeKineticEnergyBudget}(a-c). As in the case of the first mode, production is the dominant process and increases with Prandtl number. Dissipation also increases with Prandtl number, although it is small compared to production at high Prandtl numbers. Pressure-dilatation for the second mode is non-negligible as seen before. This kinetic to internal energy transfer becomes stronger as the Prandtl number of the fluid increases. Nonetheless, the increase in $P_k^g$ is larger compared to the other two processes resulting in higher kinetic energy growth rates with increasing Prandtl number.

\begin{figure*}
    \centering
    \subfloat[$P_{11}^g$]{\includegraphics[width=0.33\textwidth, keepaspectratio]{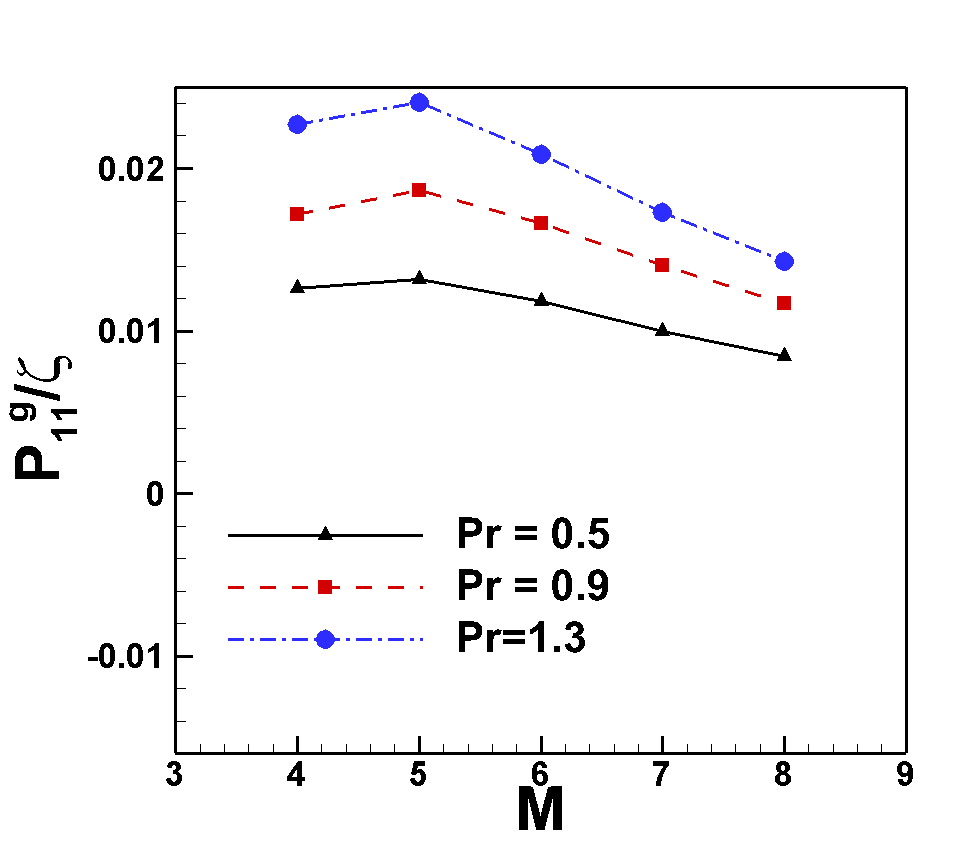}}
    \subfloat[$\Pi_{11}^g$]{\includegraphics[width=0.33\textwidth, keepaspectratio]{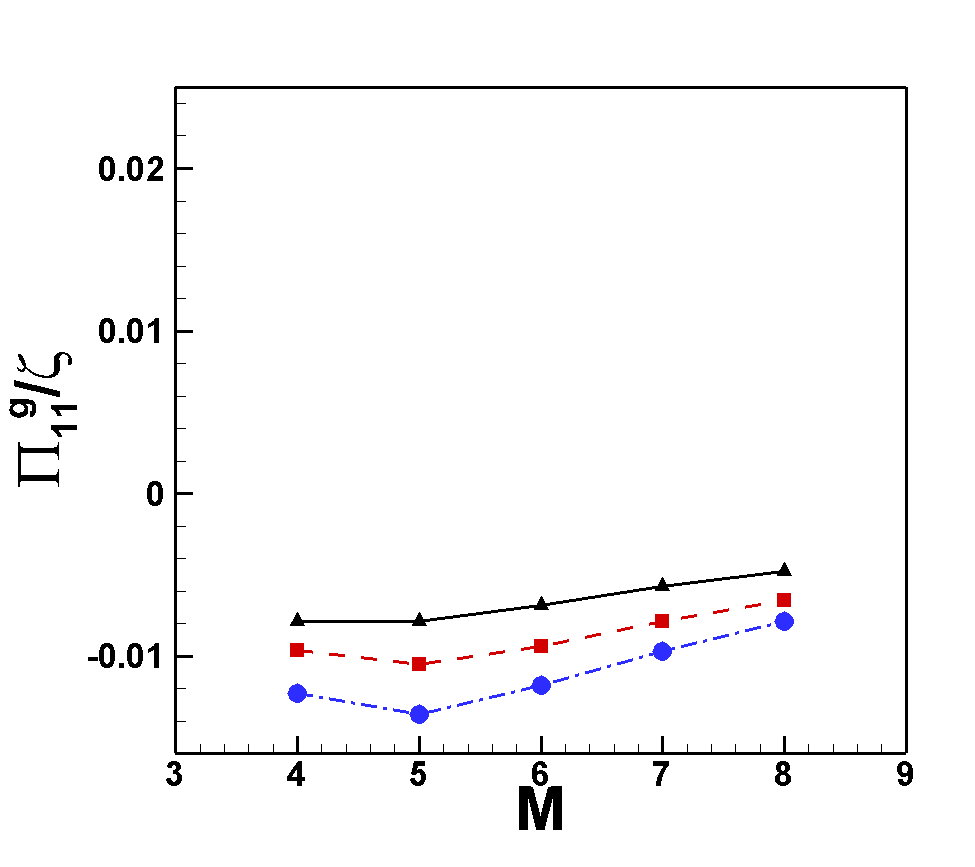}}
     \subfloat[$\epsilon_{11}^g$]{\includegraphics[width=0.33\textwidth, keepaspectratio]{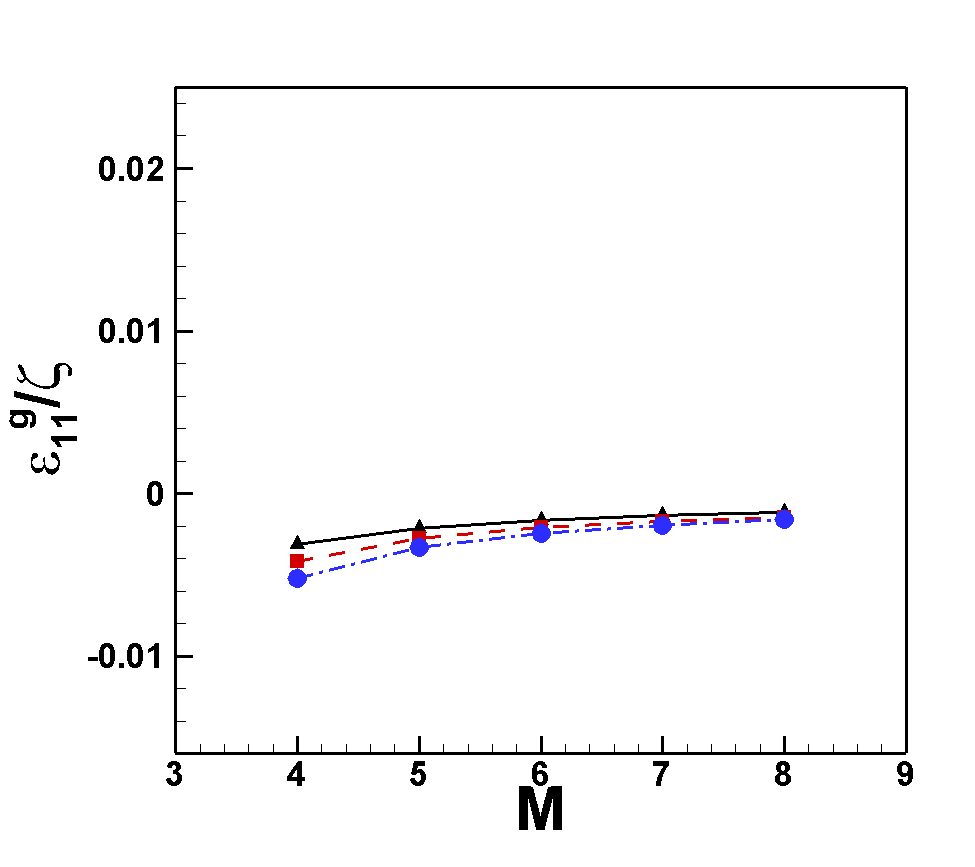}}
    \caption{Global averaged terms in the streamwise kinetic energy ($R_{11}$) budget: (a) production $P_{11}^g$, (b) pressure-strain correlation $\Pi_{11}^g$ and (c) dissipation $\epsilon_{11}^g$ for the most unstable second mode. All terms in the budget are normalized by $\zeta=2k^g U_\infty/L_r$.}
    \label{fig:SmodeStreamEnergyBudget}
\end{figure*}

The inter-component energy transfer amongst the normal stresses is considered next. Since, the most unstable second mode is always aligned with the streamwise direction, the spanwise modes have no energy. Figures \ref{fig:SmodeStreamEnergyBudget}(a-c) plots the global averaged production of the streamwise kinetic energy ($P_{11}^g$), the streamwise pressure-strain correlation ($\Pi_{11}^g$) and the streamwise component of the dissipation tensor ($\epsilon_{11}^g$), respectively. Similar to the first mode, $P_{11}^g$ equals twice the total production of kinetic energy. Pressure-strain correlation extracts energy from the streamwise mode. The amount of energy extracted increases with Prandtl number. A small portion of the energy gained via production is dissipated, and the effect of dissipation increases slightly with Prandtl number.

\begin{figure*}
    \centering
    \subfloat[$\Pi_{22}^g$]{\includegraphics[width=0.33\textwidth, keepaspectratio]{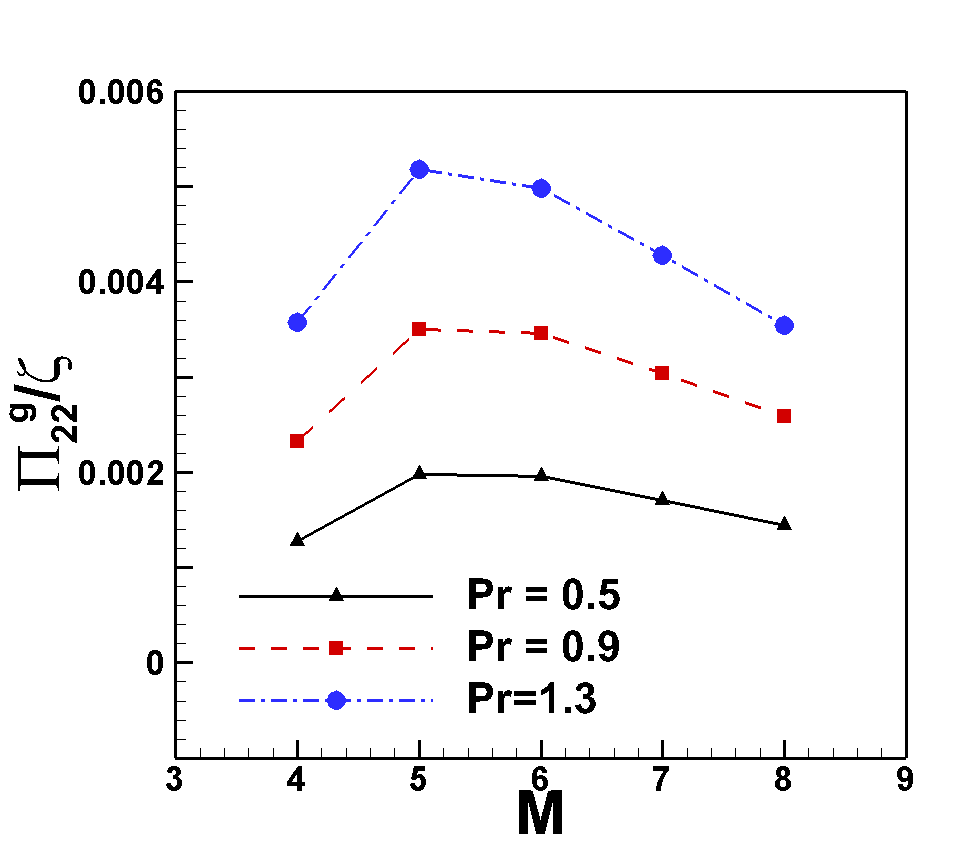}}
    \subfloat[$\epsilon_{22}^g$]{\includegraphics[width=0.33\textwidth, keepaspectratio]{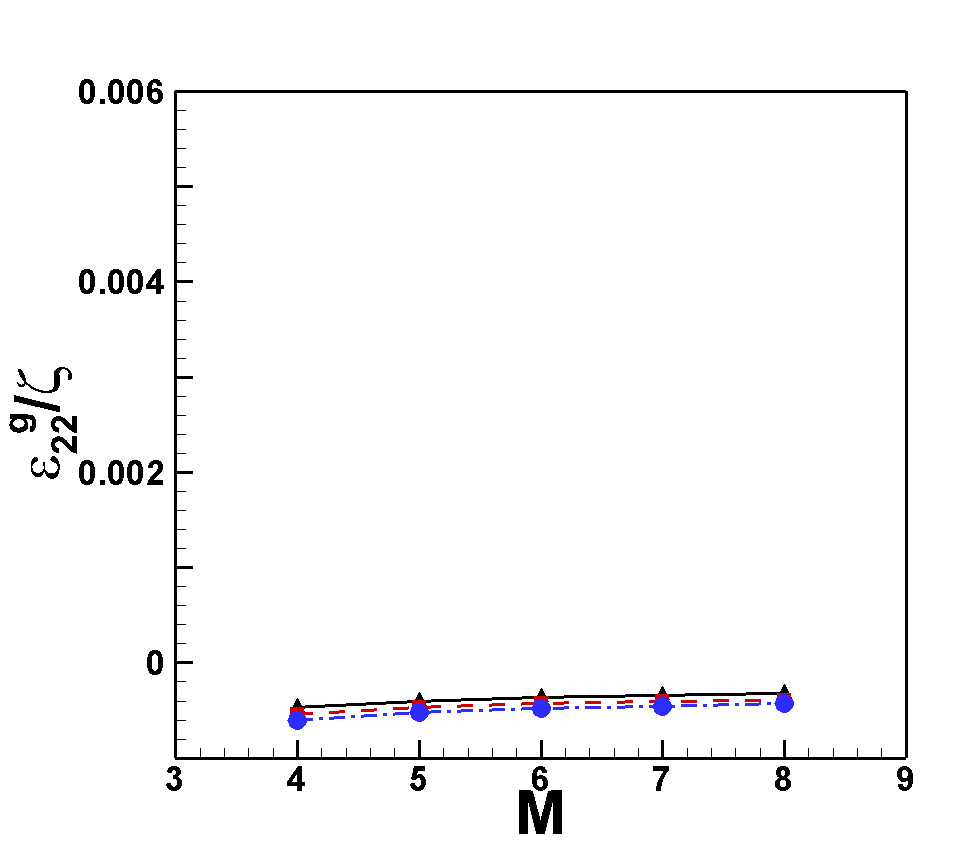}}
     \subfloat[$R_{22}^g$]{\includegraphics[width=0.33\textwidth, keepaspectratio]{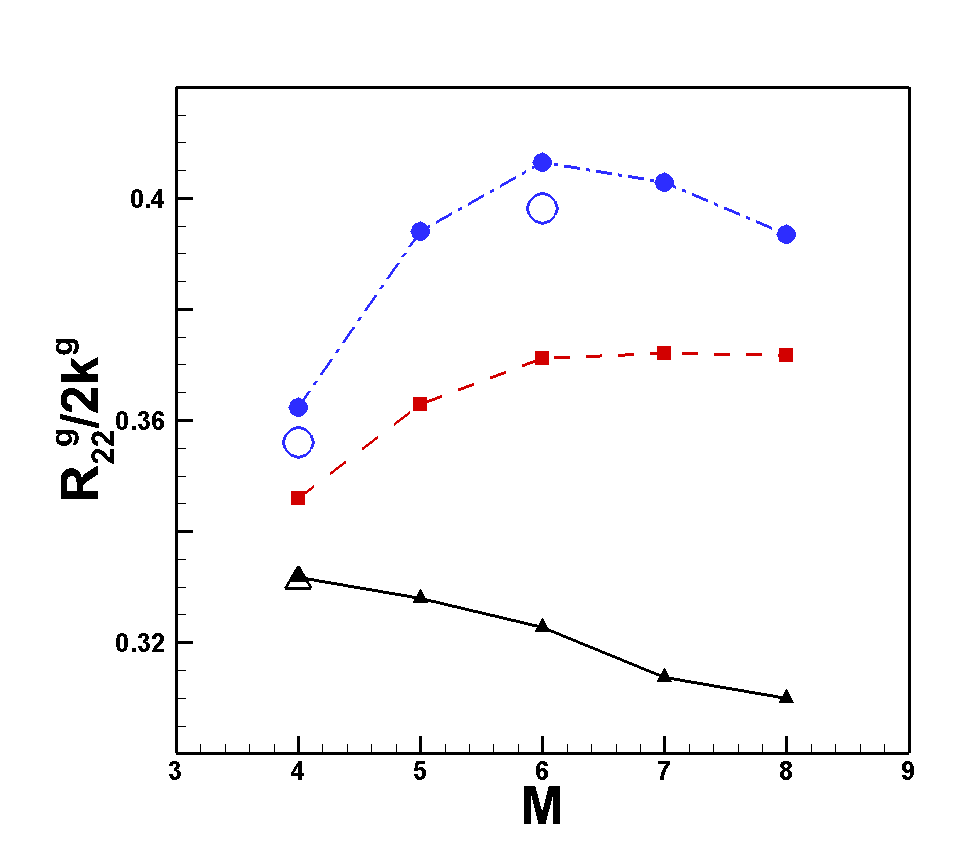}}
    \caption{Global averaged terms in the wall-normal kinetic energy ($R_{22}$) budget: (a) pressure-strain correlation $\Pi_{22}^g$ and (b) dissipation $\epsilon_{22}^g$ for the most unstable second mode. (c) Wall-normal kinetic energy fraction ($R_{22}^g/2k^g$) for the most unstable second mode. All budget terms are normalized by $\zeta=2k^g U_\infty/L_r$. The symbols are same as figure \ref{fig:Grate}.}
    \label{fig:SmodeWallEnergyBudget}
\end{figure*}

Once again, $\Pi_{22}^g$ provides a source of wall-normal energy. This is evident from figures \ref{fig:SmodeWallEnergyBudget}(a-b) wherein the global averaged terms of the wall-normal kinetic energy budget are shown. A small fraction of the energy attained by $\Pi_{22}^g$ is dissipated by the wall-normal perturbations via $\epsilon_{22}$. $\Pi_{22}^g$ increases with Prandtl number whereas $\epsilon_{22}^g$ is fairly constant across all Prandtl numbers. The higher pressure-strain correlation level leads to increased energy content in the wall-normal mode with increasing Prandtl number. This is shown in figure \ref{fig:SmodeWallEnergyBudget}(c) wherein the fraction of wall-normal energy is presented. The DNS results are also presented in the same figure corroborating the linear analysis findings. 

\begin{figure*}
    \centering
        \subfloat[$P_{12}^g$]{\includegraphics[width=0.33\textwidth, keepaspectratio]{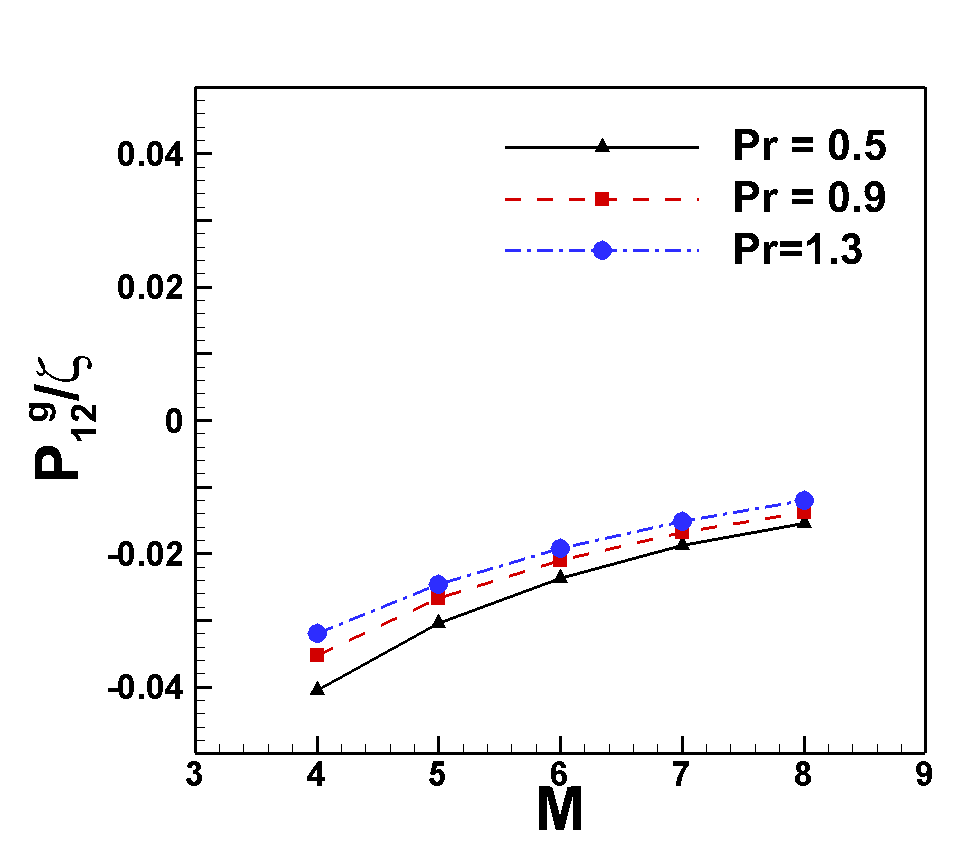}}
    \subfloat[$|\Pi_{12}^g/P_{12}^g|$]{\includegraphics[width=0.33\textwidth, keepaspectratio]{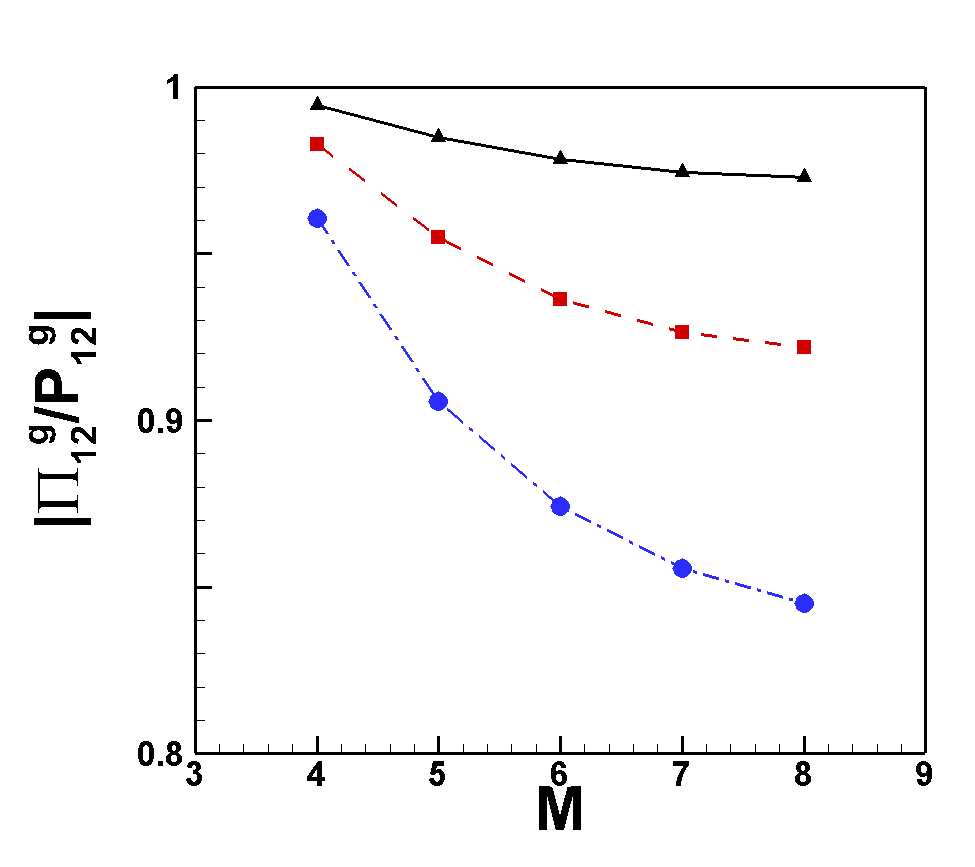}}
     \subfloat[$P_{12}^g+\Pi_{12}^g+\epsilon_{12}^g$]{\includegraphics[width=0.33\textwidth, keepaspectratio]{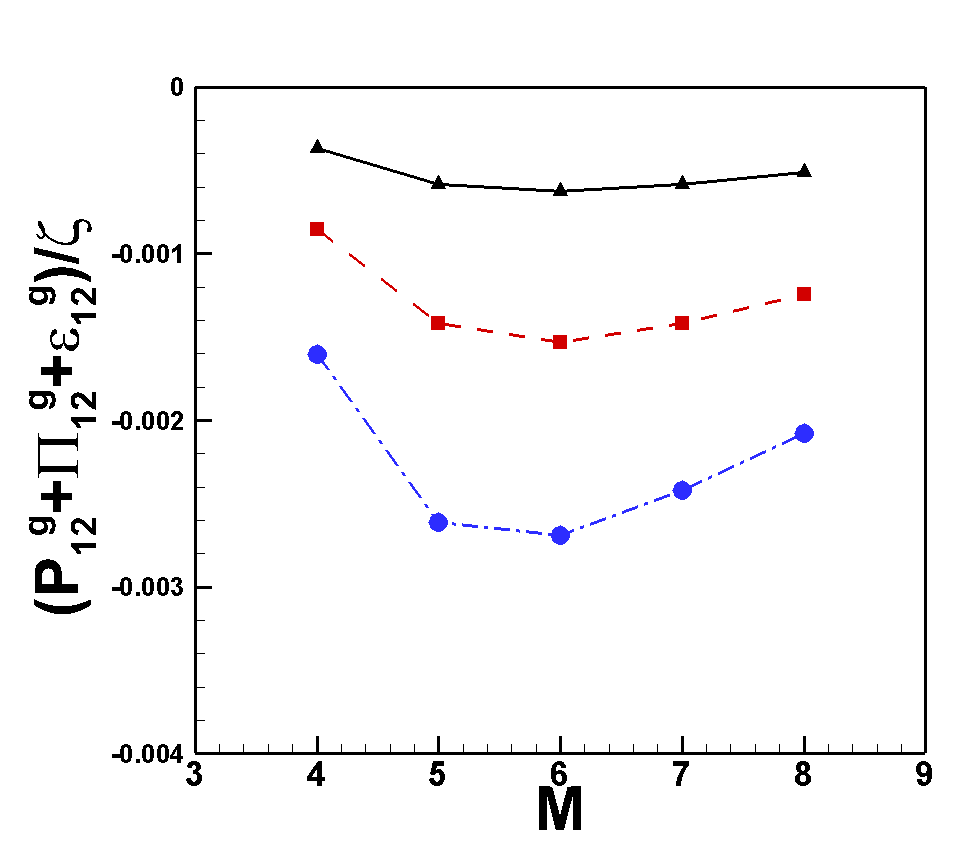}}
    \caption{Global averaged shear stress production $P_{12}^g$, (b) absolute value of ratio of pressure-strain correlation $\Pi_{12}^g$ to shear stress production for the most unstable second mode. (c) Sum of all the terms on the right hand side of the shear stress budget equation \eqref{eq:ReynoldsStressBudget}. All budget terms are normalized by $\zeta=2k^g U_\infty/L_r$.}
    \label{fig:Smodeu1u2EnergyBudget}
\end{figure*}

The wall-normal component of kinetic energy contributes to the production of shear stress ($R_{12}$). The shear stress budget is examined in figures \ref{fig:Smodeu1u2EnergyBudget}(a)-(c). Unlike the first mode, the shear stress production ($P_{12}^g$) decreases in magnitude with Prandtl number. 
The absolute value of the ratio of $\Pi_{12}^g$ to $P_{12}^g$ is of the order of unity. This is consistent with the findings of \cite{bertsch2012rapid} for homogeneous shear flows. For a given Prandtl number, the ratio decreases monotonically with Mach number. $|\Pi_{12}^g/P_{12}^g|$ decreases with Prandtl number at a given $M$. 
 This results in net magnitude increase of the shear stress budget as shown in figure \ref{fig:Smodeu1u2EnergyBudget}(c).    

\begin{figure*}
    \centering
    \subfloat[$R_{12}^g$]{\includegraphics[width=0.45\textwidth, keepaspectratio]{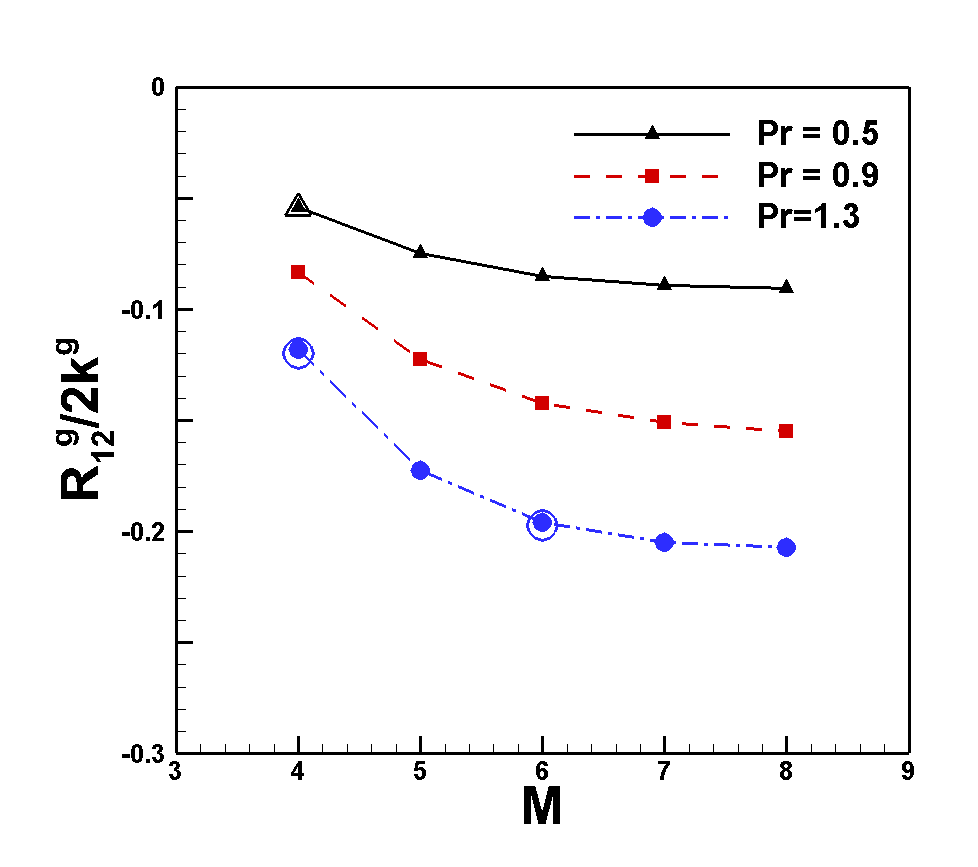}} 
    \subfloat[$d\overline{U}_1/dx_2$]{\includegraphics[width=0.45\textwidth, keepaspectratio]{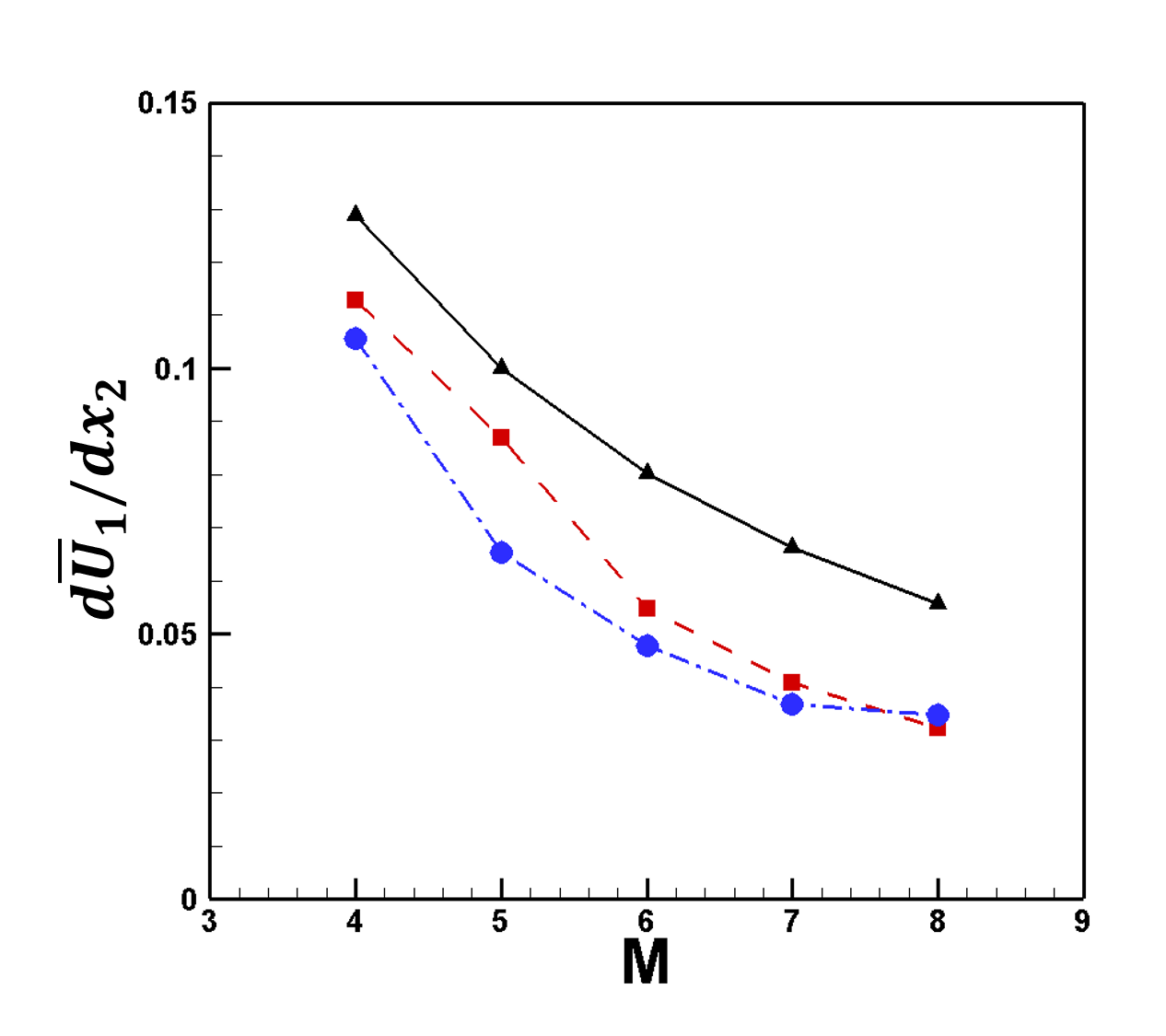}}
    \caption{(a) Shear stress anisotropy ($R_{12}^g/2k^g$) for the most unstable second mode. (b) Mean velocity gradient ($d\overline{U}_1/dx_2$) at peak production for the most unstable second mode. The symbols are same as figure \ref{fig:Grate}.}
    \label{fig:Smodeu1u2}
\end{figure*}

We now examine the kinetic energy production. Figures \ref{fig:Smodeu1u2}(a-b) plots the averaged shear stress anisotropy and the base velocity gradient at peak production respectively. The maximum production is at the the sonic line for low Prandtl number fluids but moves toward the boundary layer edge as the Prandtl number is increased. 
The sonic line ($x_{2a}$) is defined as the location in the flow wherein the relative disturbance Mach number ($M_r$) equals 1. The relative disturbance Mach number \citep{mack1984boundary} is defined as,
\begin{equation}
    \label{eq:Mrel}
    M_r=\frac{(\alpha \overline{U}_1 -\omega_r)M}{[(\alpha^2+\beta^2)\overline{T}]^{1/2}}.
\end{equation}
The magnitude of globally averaged shear stress anisotropy increases with Prandtl number but the base velocity gradient at peak production decreases with Prandtl number. The significant increase in the shear stress magnitude permits the high levels of production observed in figure \ref{fig:SmodeKineticEnergyBudget}(a). For a fixed Prandtl number, the magnitude of $R_{12}^g$ increases with Mach number while the base flow gradient at peak production decreases. This results in a non monotonic dependence of production with respect to Mach number as shown in figure \ref{fig:SmodeKineticEnergyBudget}(a). The mean $R_{12}^g$ obtained from DNS is also shown in figure \ref{fig:Smodeu1u2}(a). The DNS results are in excellent agreement with linear analysis.   

\begin{figure*}
    \centering
    \subfloat[$-\Pi_k^g$]{\includegraphics[width=0.33\textwidth, keepaspectratio]{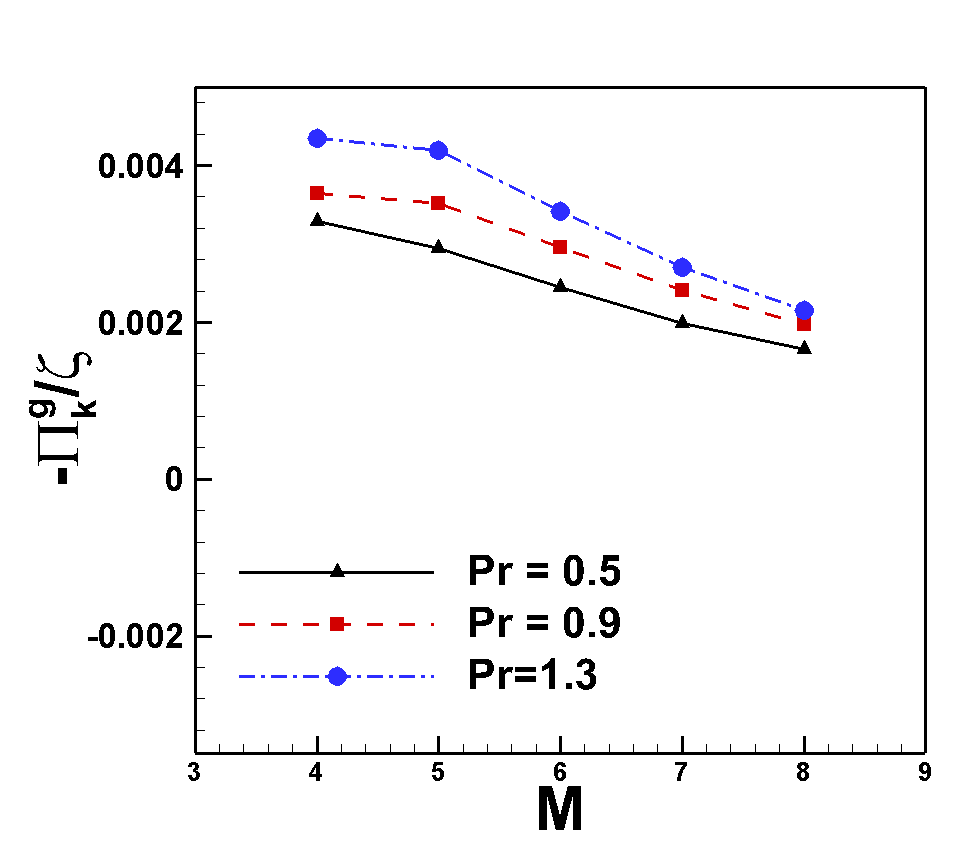}}
    \subfloat[$T_s^g$]{\includegraphics[width=0.33\textwidth, keepaspectratio]{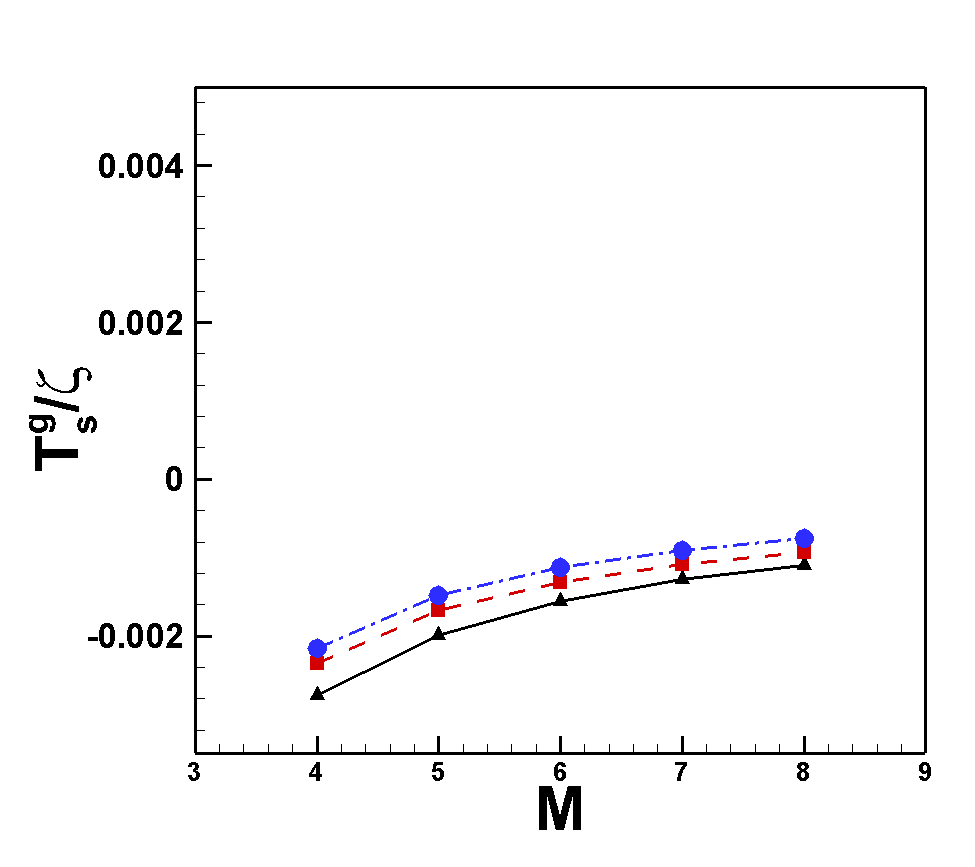}}
     \subfloat[$\epsilon_s^g$]{\includegraphics[width=0.33\textwidth, keepaspectratio]{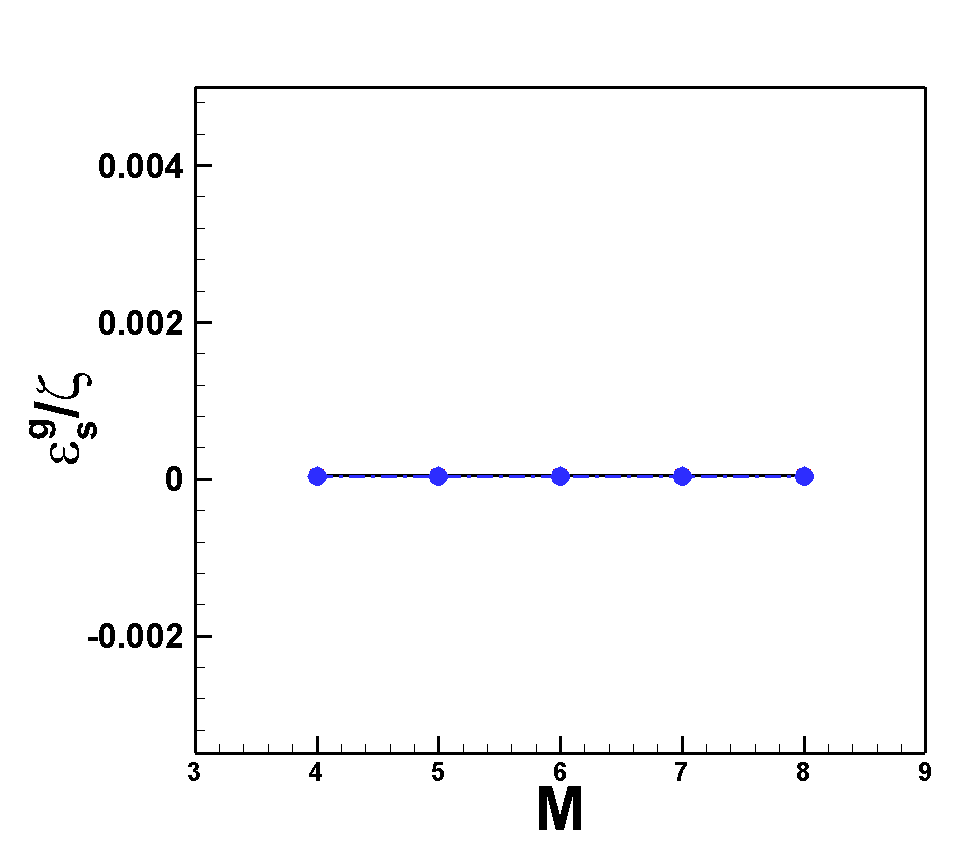}}
    \caption{Global averaged terms in the internal energy budget: (a) negative of pressure-dilatation $-\Pi_{k}^g$, (b) thermal flux $T_{s}^g$ and (c) viscous term $\epsilon_s^g$  for the most unstable second mode. All budget terms are normalized by $\zeta=2k^g U_\infty/L_r$.}
    \label{fig:SmodeInternalEnergyBudget}
\end{figure*}

Finally, we examine the internal energy budget \eqref{eq:InternalEnergyBudget} in figures \ref{fig:SmodeInternalEnergyBudget}(a-c). Internal and kinetic modes exchange energy via pressure-dilatation. The kinetic to internal energy transfer increases with Prandtl number at a given Mach number. The perturbation internal mode also interacts with the mean internal mode via the thermal flux ($T_s^g$) and viscous terms ($\epsilon_s^g$). A significant portion of the energy gained by the perturbation internal mode is transferred to the mean internal mode via the action of thermal flux. The thermal flux action weakens with increasing Prandtl number as it is inversely proportional to Prandtl number.  The viscous action is small compared to $\Pi_k^g$ and $T_s^g$. The increased pressure work combined with reduced thermal flux action leads to higher internal energy content in high Prandtl number fluids (figure \ref{fig:SmodeInternalEnergy}(a)).

\subsection{Flow physics underlying Prandtl number effects} 
\begin{figure*}
    \centering
    \subfloat[$R_{22}$]{\includegraphics[width=0.33\textwidth, keepaspectratio]{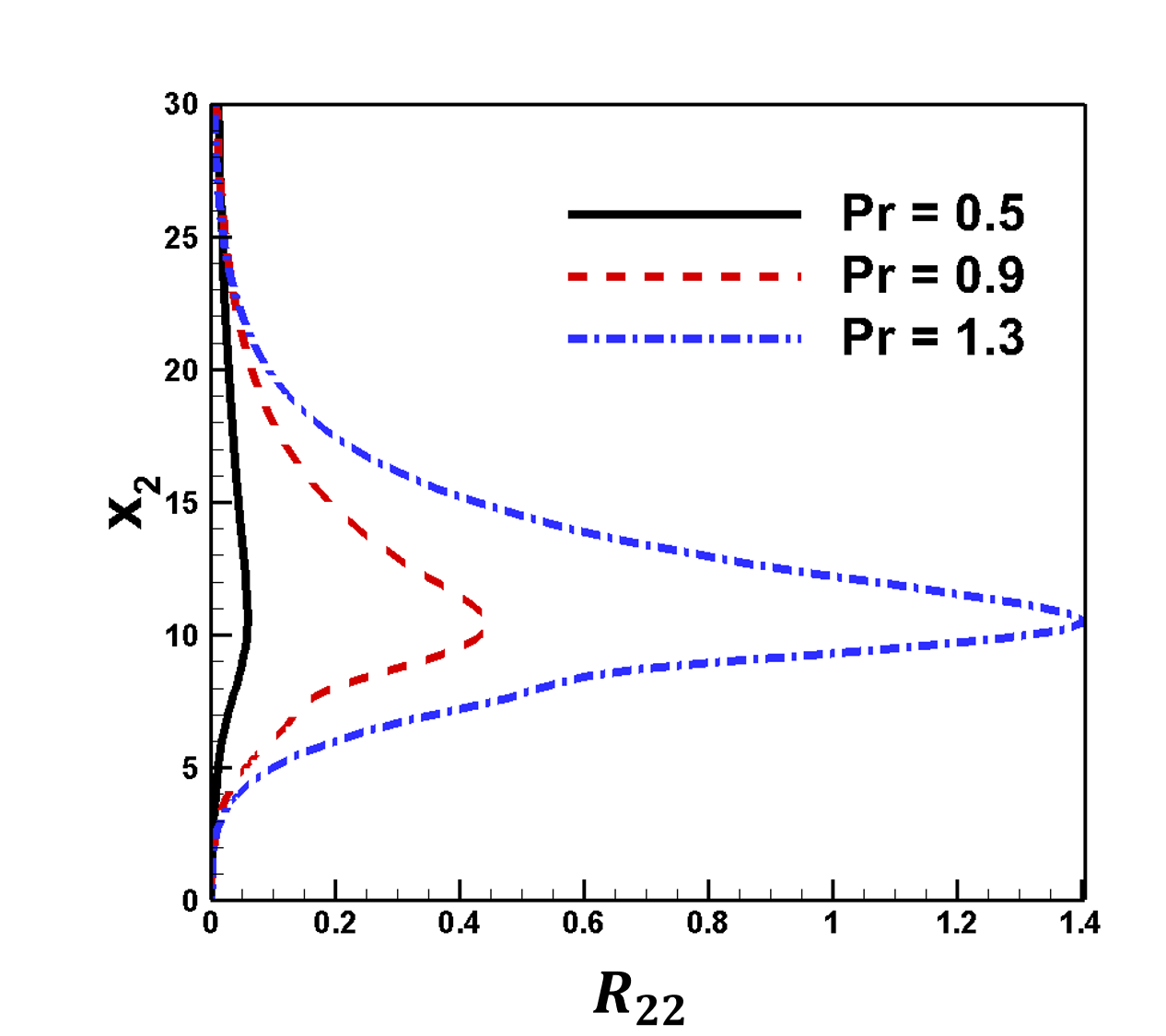}}
    \subfloat[$d\overline{U}_1/dx_2$]{\includegraphics[width=0.33\textwidth, keepaspectratio]{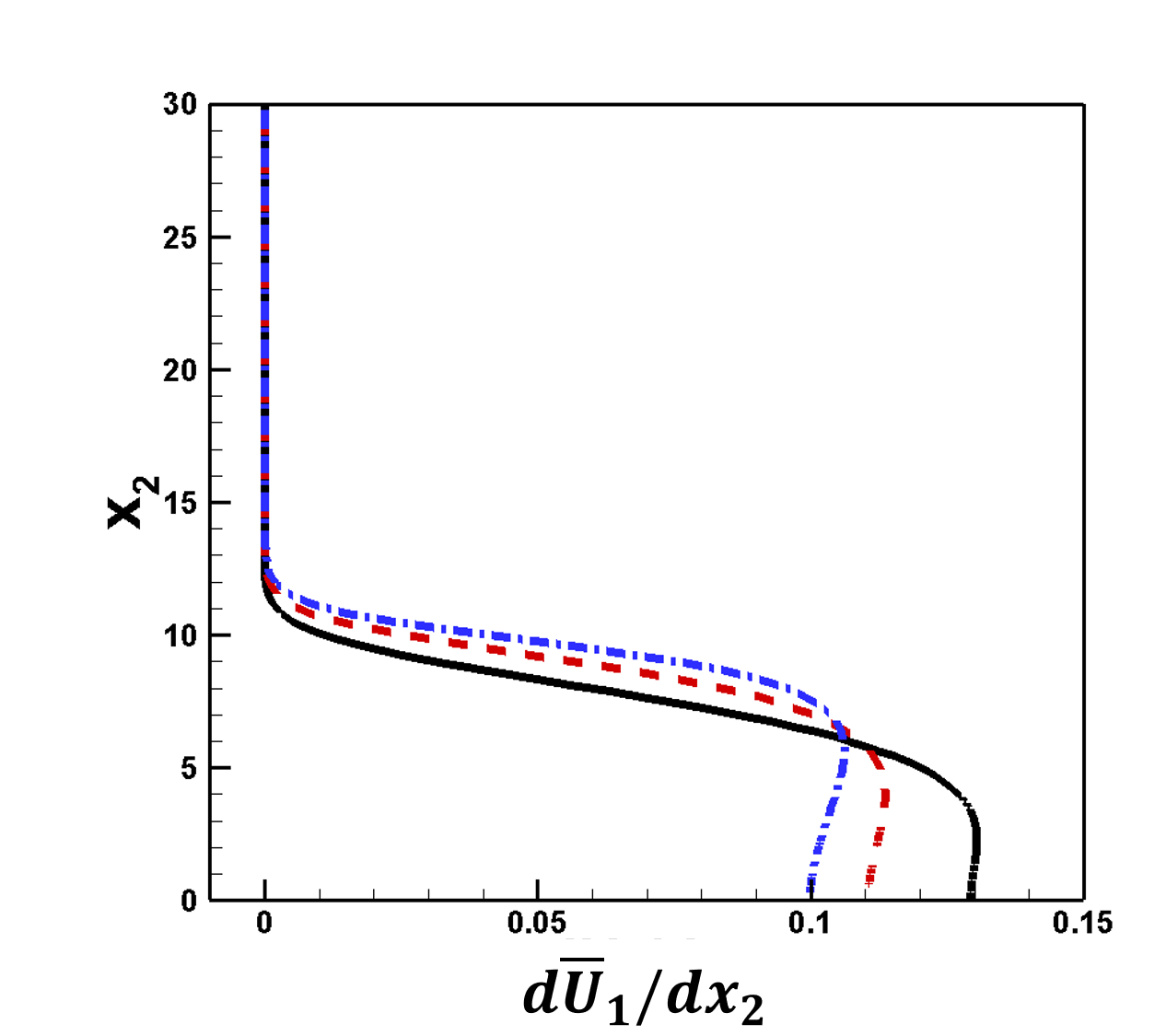}}
    \subfloat[$P_{12}$]{\includegraphics[width=0.33\textwidth, keepaspectratio]{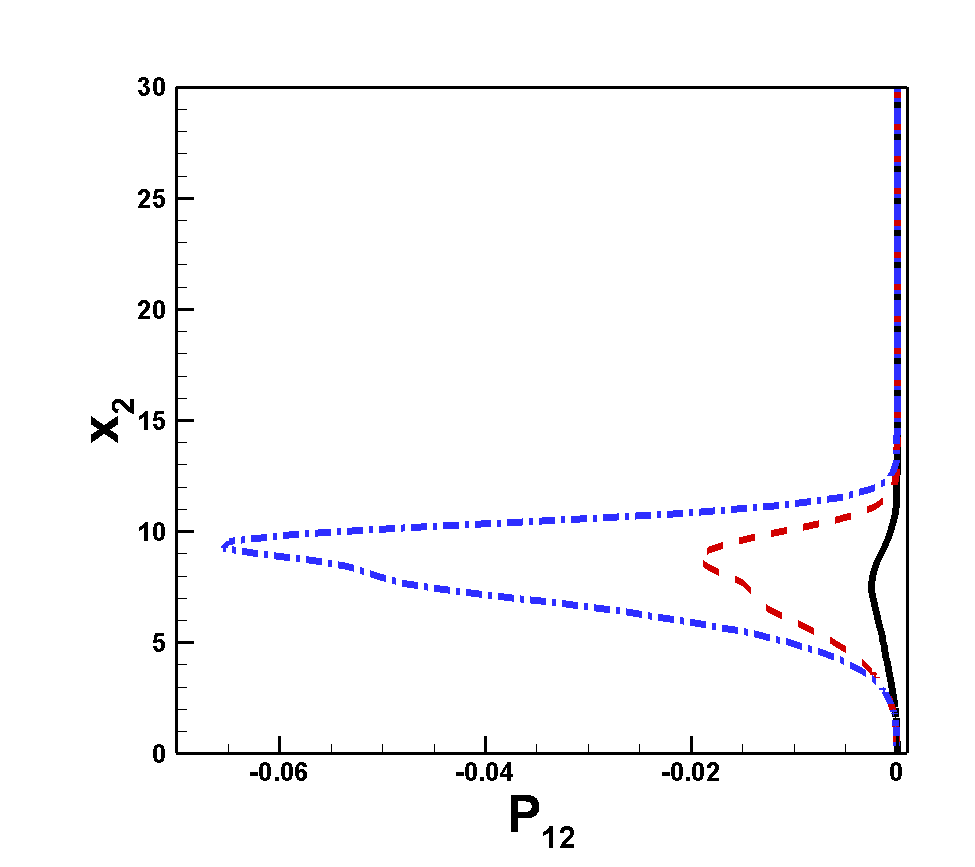}}
    \caption{Profiles of (a) wall-normal kinetic energy ($R_{22}$), (b) mean velocity gradient $d\overline{U}_1/dx_2$ and (c) production of shear stress ($P_{12}$) for the most unstable first mode at $M=4$.}
    \label{fig:FmodeProfile}
\end{figure*}
In this subsection the sequence of flow processes underlying the observed $Pr$ effects on first and second mode will be summarized. It is already established that for both first and second mode the averaged shear stress anisotropy increases with Prandtl number. And yet, $P_{12}^g$ increases with Prandtl number for first mode while it decreases for the second mode. We examine the contrasting behaviour of $P_{12}^g$ by considering the profiles of wall-normal kinetic energy ($R_{22}$), base flow gradient ($d\overline{U}_1/dx_2$) and $P_{12}$. The profiles for first mode at $M=4$ are shown in figures \ref{fig:FmodeProfile}(a)-(c). For the first mode, the wall-normal kinetic energy increases with Prandtl number throughout the boundary layer. The base flow gradient near the $R_{22}$ peak is also higher at high Prandtl numbers. Therefore, $P_{12}^g$ increases with Prandtl number for the first mode.  

\begin{figure}
    \centering
    \includegraphics[width=0.7\textwidth, keepaspectratio]{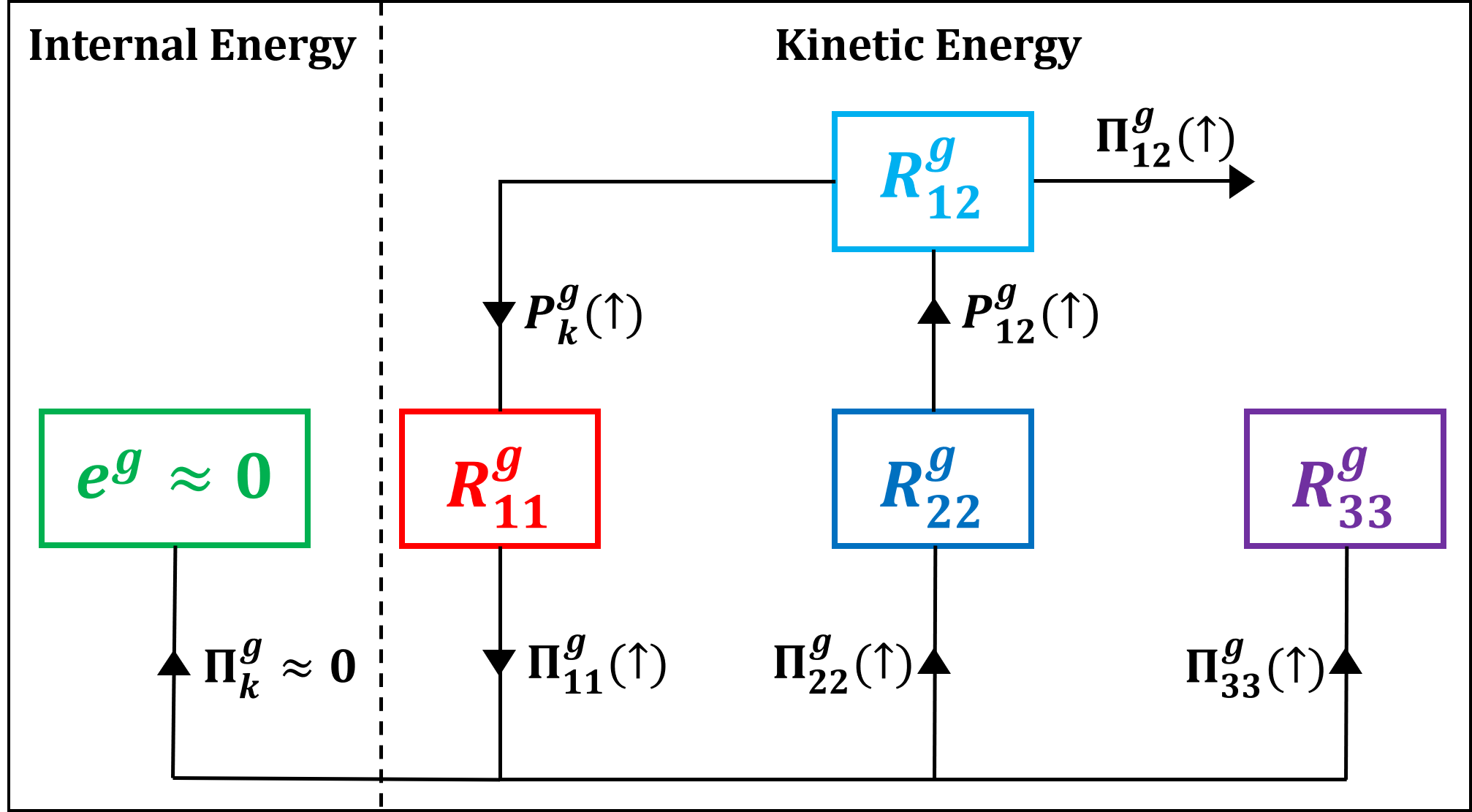}
    \caption{Schematic of energy interactions between internal and kinetic modes for the first mode instability. Up/down arrows indicate increasing/decreasing magnitude of the processes with increasing Prandtl number. }
    \label{fig:FModeFlowChart}
\end{figure}
We now summarize the flow thermodynamic interactions for the first mode. A schematic representing the key interactions is displayed in figure \ref{fig:FModeFlowChart}. Energy is extracted from the mean flow via production and transferred to the streamwise perturbations. Pressure-dilatation (pressure work) is not significant and does not play an important role in the instability dynamics. The effect of Prandtl number manifests through pressure-strain correlation.
 Production and all components of pressure-strain correlation increase with Prandtl number. The more energetic wall-normal mode leads to a rise in the production of shear stress. The ratio of $P_{12}^g$ to $\Pi_{12}^g$ decreases with Prandtl number. As a result the difference of $P_{12}^g$ and $\Pi_{12}^g$ increases with Prandtl number leading to increased magnitude of shear stress anisotropy. The higher shear stress anisotropy ultimately allows for higher production leading to increased destabilization at higher Prandtl number.
\begin{figure*}
    \centering
    \subfloat[$R_{22}$]{\includegraphics[width=0.33\textwidth, keepaspectratio]{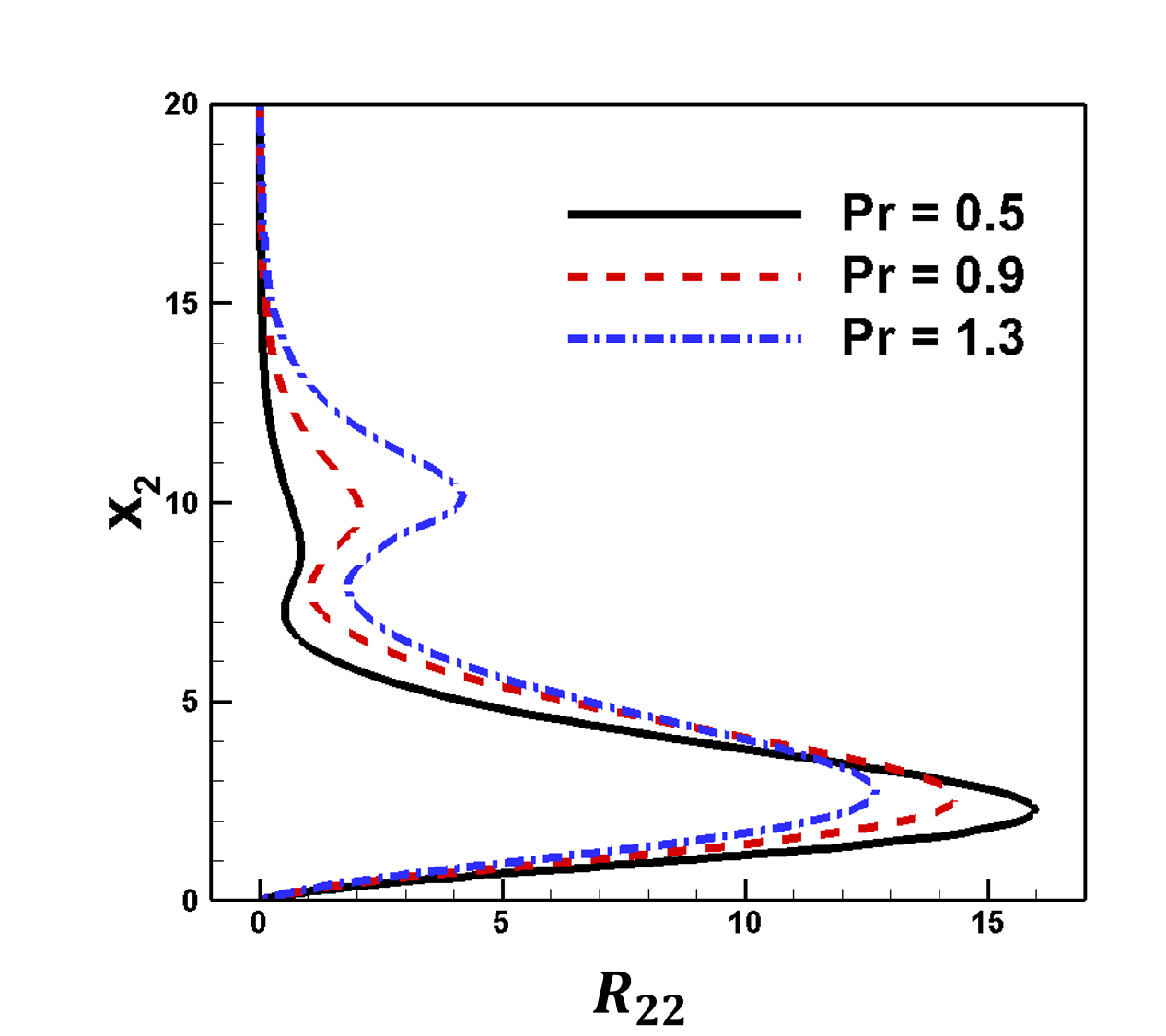}}
    \subfloat[$d\overline{U}_1/dx_2$]{\includegraphics[width=0.33\textwidth, keepaspectratio]{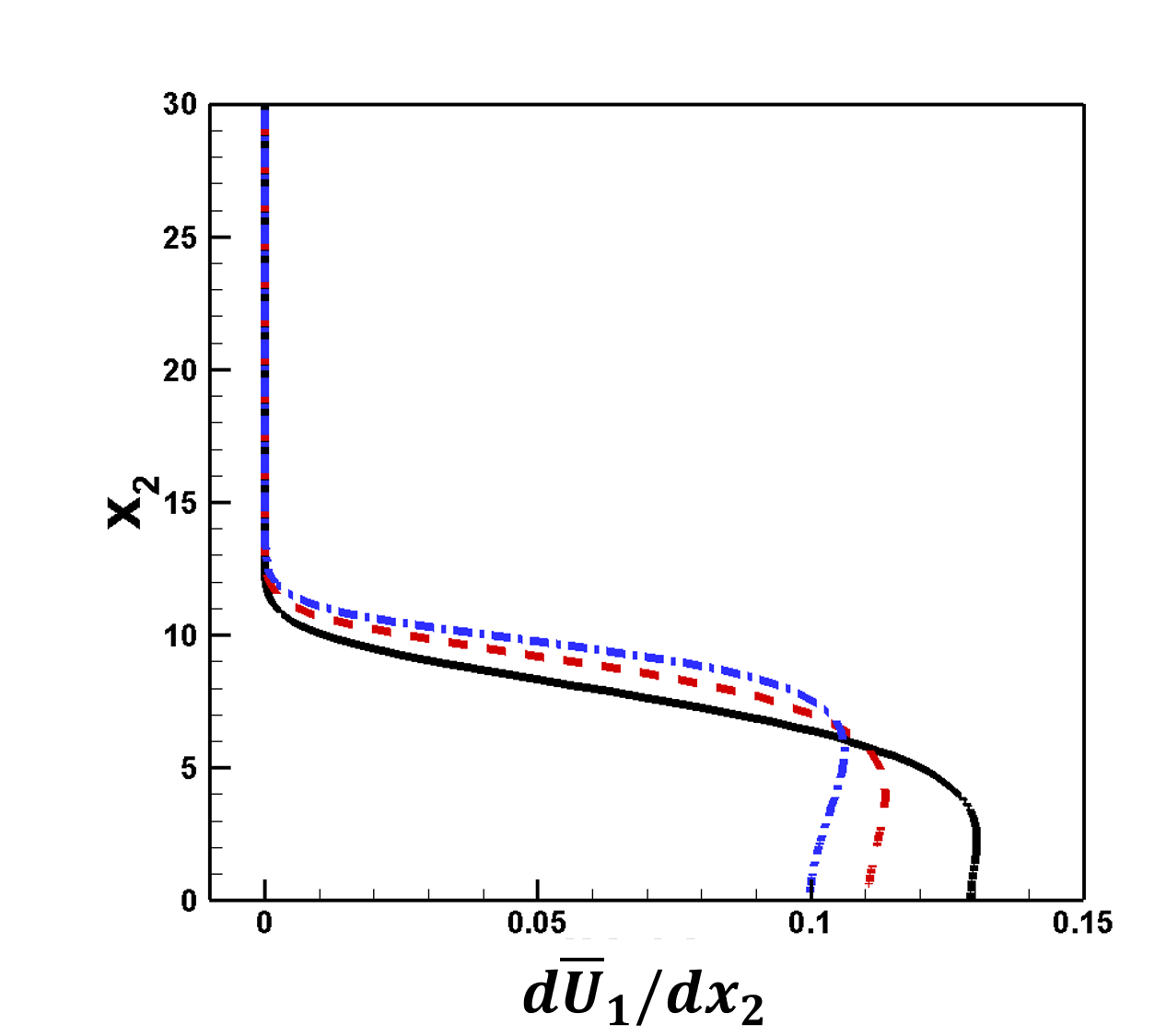}}
    \subfloat[$P_{12}$]{\includegraphics[width=0.33\textwidth, keepaspectratio]{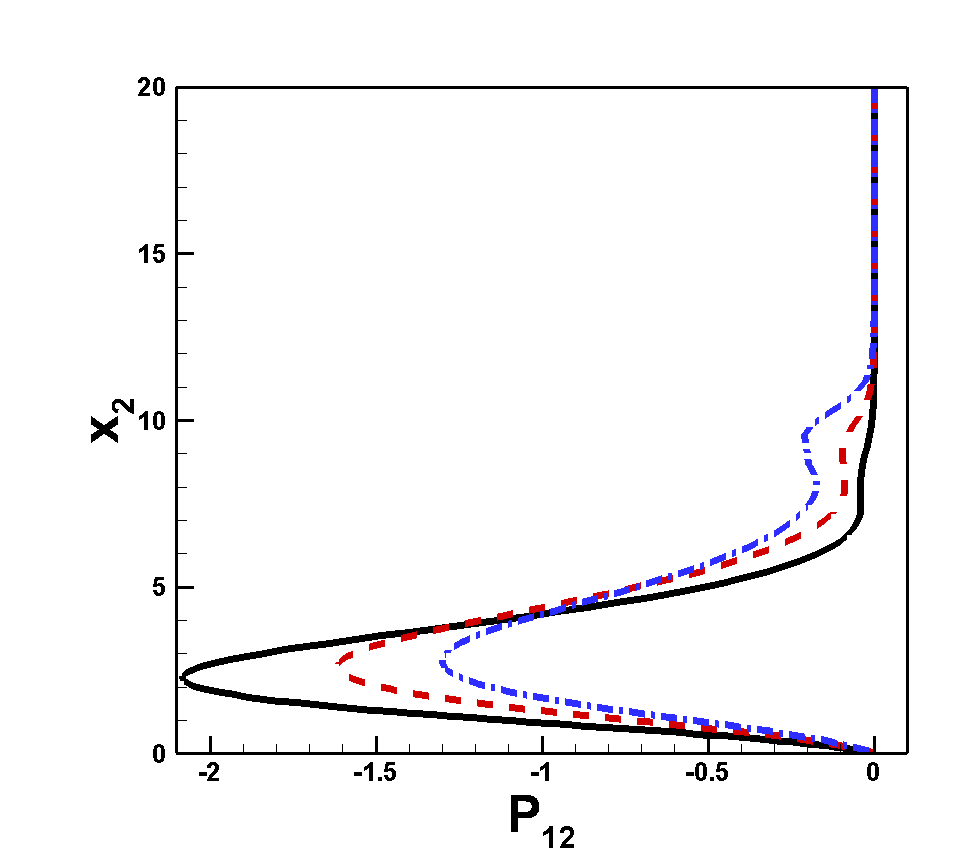}}
    \caption{Profiles of (a) wall-normal kinetic energy ($R_{22}$), (b) mean velocity gradient $d\overline{U}_1/dx_2$ and (c) production of shear stress ($P_{12}$) for the most unstable second mode at $M=4$.}
    \label{fig:SmodeProfile}
\end{figure*}

For the second mode, the global averaged wall-normal energy ($R_{22}^g$) increases with Prandtl number while $P_{12}^g$ decreases in magnitude with Prandtl number. The profiles of $R_{22}$, base flow gradient, and $P_{12}$ for the second mode at $M=4$ are shown in figures \ref{fig:SmodeProfile}(a-c) respectively. Unlike the case of first mode, the $R_{22}$ profile has a global and a local maxima. The global maxima is located near the sonic line ($x_2\approx3$).
 The global maxima of the wall-normal energy decreases with Prandtl number. The mean velocity gradient near the sonic line also decreases with Prandtl number. Hence, the low Prandtl number fluid has a much stronger global minima of $P_{12}$, located near the sonic line. The local $R_{22}$ maxima, near the boundary layer edge ($x_{2}\approx 10$) is stronger for higher Prandtl number fluid. However, the mean flow gradient is negligible near the boundary layer edge. Consequently, the local $P_{12}$ minima near the boundary layer edge is substantially smaller compared to the global minima leading to decrease in $P_{12}^g$ magnitude with increasing Prandtl number. 

\begin{figure}
    \centering
    \includegraphics[width=0.7\textwidth, keepaspectratio]{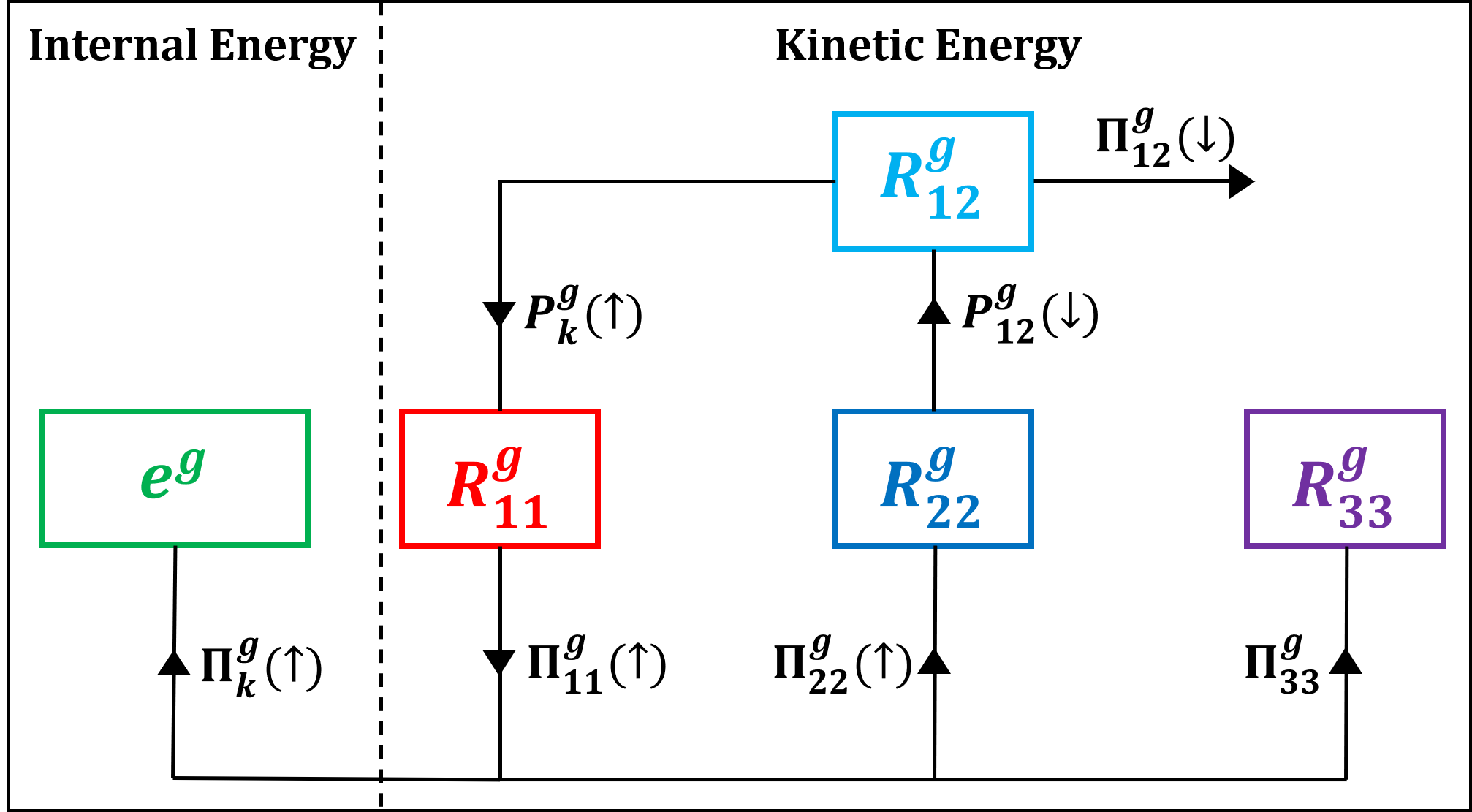}
    \caption{Schematic of energy interactions between internal and kinetic modes for the second mode instability. Up/down arrows indicate increasing/decreasing magnitude of the processes with increasing Prandtl number. }
    \label{fig:SModeFlowChart}
\end{figure}
The flow thermodynamic interactions for the second mode is summarized in the schematic shown in figure \ref{fig:SModeFlowChart}. Energy is transferred from the mean flow to the streamwise mode of perturbation kinetic energy via production.
Unlike the first mode, pressure-dilatation is significant for the second mode. Pressure work transfers energy from the kinetic to the internal mode and pressure-strain correlation redistributes energy amongst the stress components.  Production, pressure-dilatation and the diagonal components of $\Pi_{ij}^g$ increase with Prandtl number. This leads to higher wall-normal and internal energy with increasing Prandtl number. The production of shear stress and the shear component of pressure-strain correlation decrease with Prandtl number. The difference of $P_{12}^g$ and $\Pi_{12}^g$ increases with Prandtl number, leading to high shear stress anisotropy. The increased shear stress anisotropy permits the higher level of production leading to increased instability growth rate at higher Prandtl number. 

\section{Boundary layer response to random pressure forcing at different Prandtl numbers}

The nature and composition of free stream disturbances is not necessarily known for all transition experiments \citep{hader2018towards}. Consequently, \cite{hader2018towards} modelled the free-stream disturbance by random pressure (acoustic) disturbances to simulate natural transition in high speed boundary layers. More recently, \cite{mittal2021linear} employed random acoustic disturbances to study instability evolution in Poiseuille flow. Following previous works, we perform direct numerical simulations of temporally evolving boundary layers with randomly generated pressure forcing at different Mach and Prandtl numbers. The temporal simulations are initialized by laminar basic state superposed with low intensity random pressure forcing  \citep{mittal2021linear,hader2018towards}. The intensity of pressure forcing is $1\%$ of free-stream pressure. 
The simulations are performed at $M=3$ and $M=6$ for $Pr=\{0.5,0.7,1.3\}$. The grid sizes and relevant parameters for the simulations are listed in table \ref{table:ParametersBasic2}.

\begin{figure*}
    \centering
    \subfloat[$k^g$]{\includegraphics[width=0.45\textwidth, keepaspectratio]{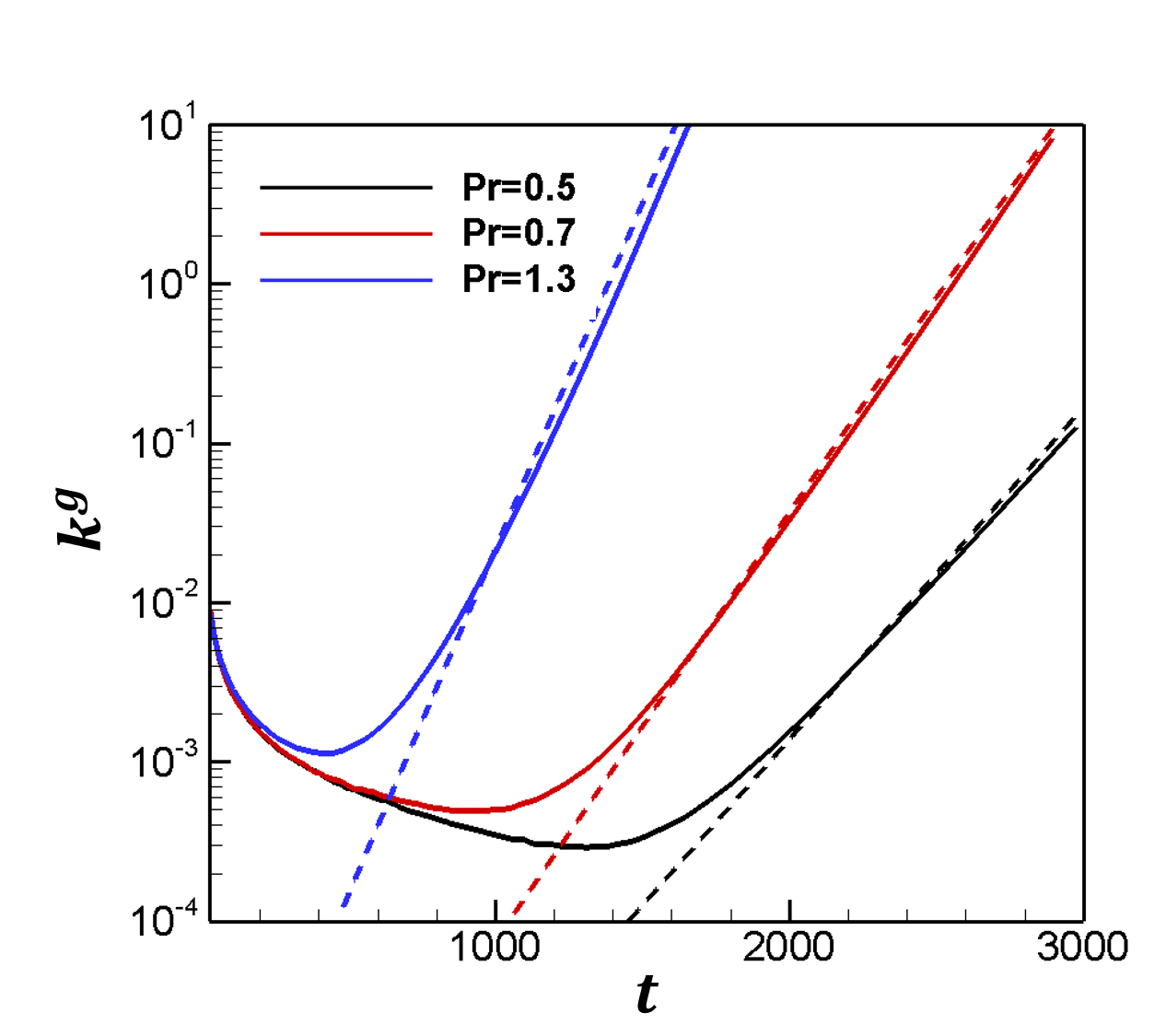}} 
    \subfloat[$e^g$]{\includegraphics[width=0.45\textwidth, keepaspectratio]{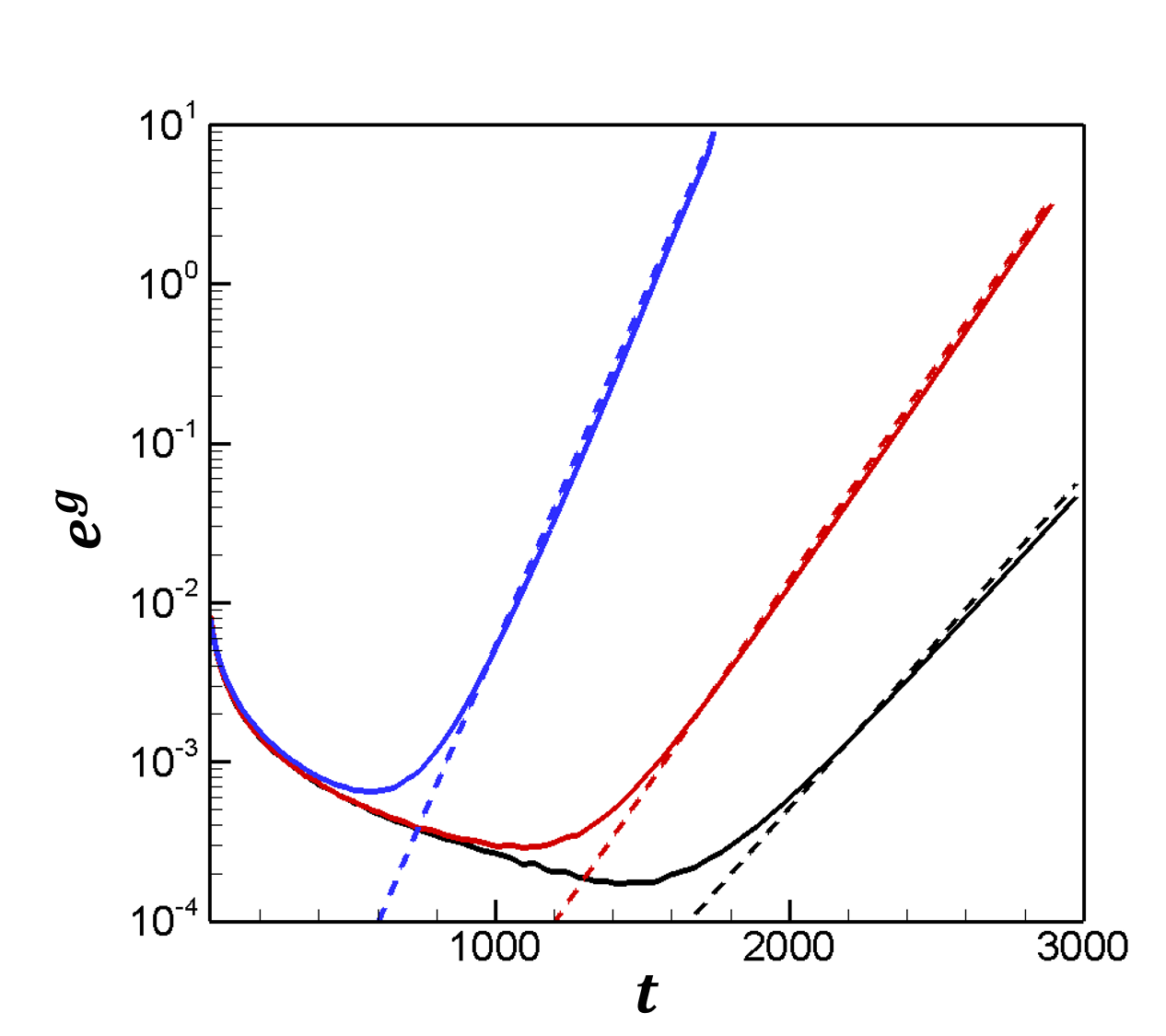}} 
    \caption{Evolution of globally averaged (a) kinetic energy ($k^g$) and (b) internal energy ($e^g$) at $M=6$ and $Re=4000$ for three different Prandtl number. The dashed lines represent the dominant eigenmode growth at the rate predicted by LSA.}
    \label{fig:RandKeM6}
\end{figure*}

\begin{figure*}
    \centering
    \subfloat[$t=30$]{\includegraphics[trim=0 10 10 0, clip,width=0.32\textwidth, keepaspectratio]{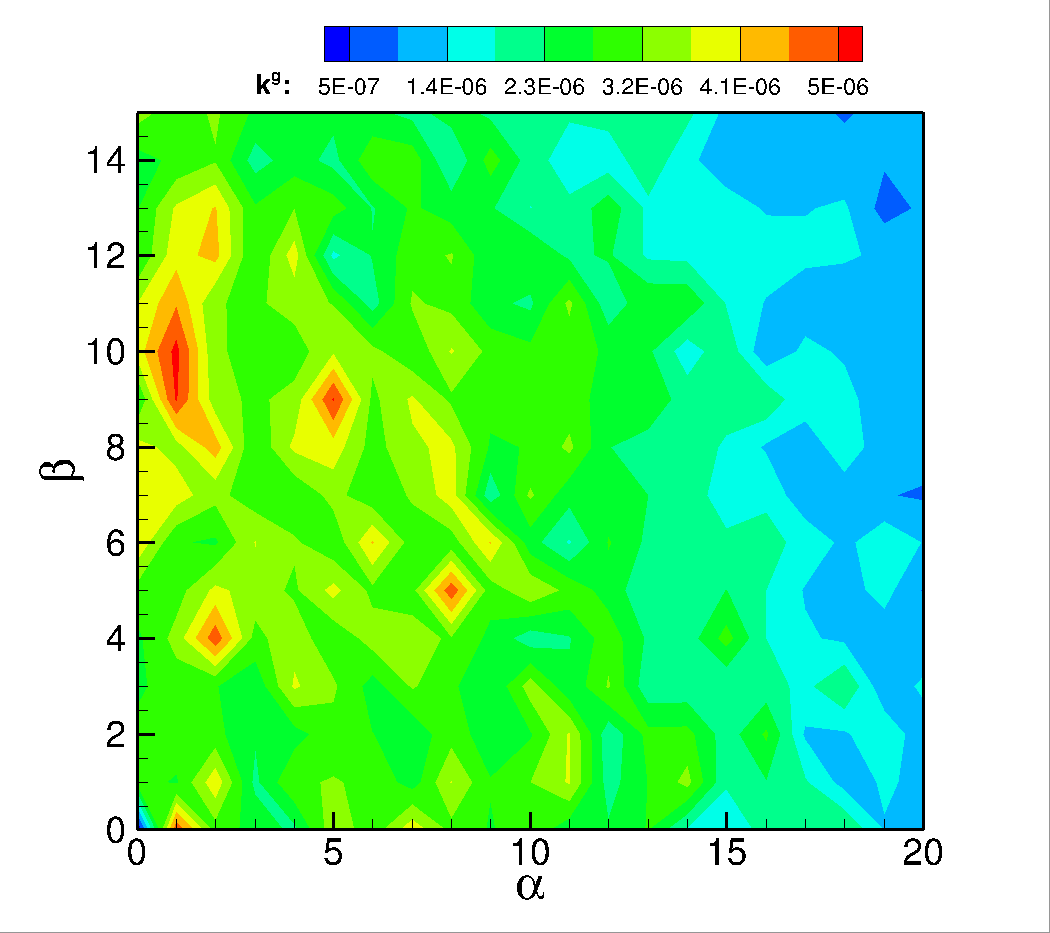}} 
    \subfloat[$t=500$]{\includegraphics[trim=0 10 10 0, clip,width=0.32\textwidth, keepaspectratio]{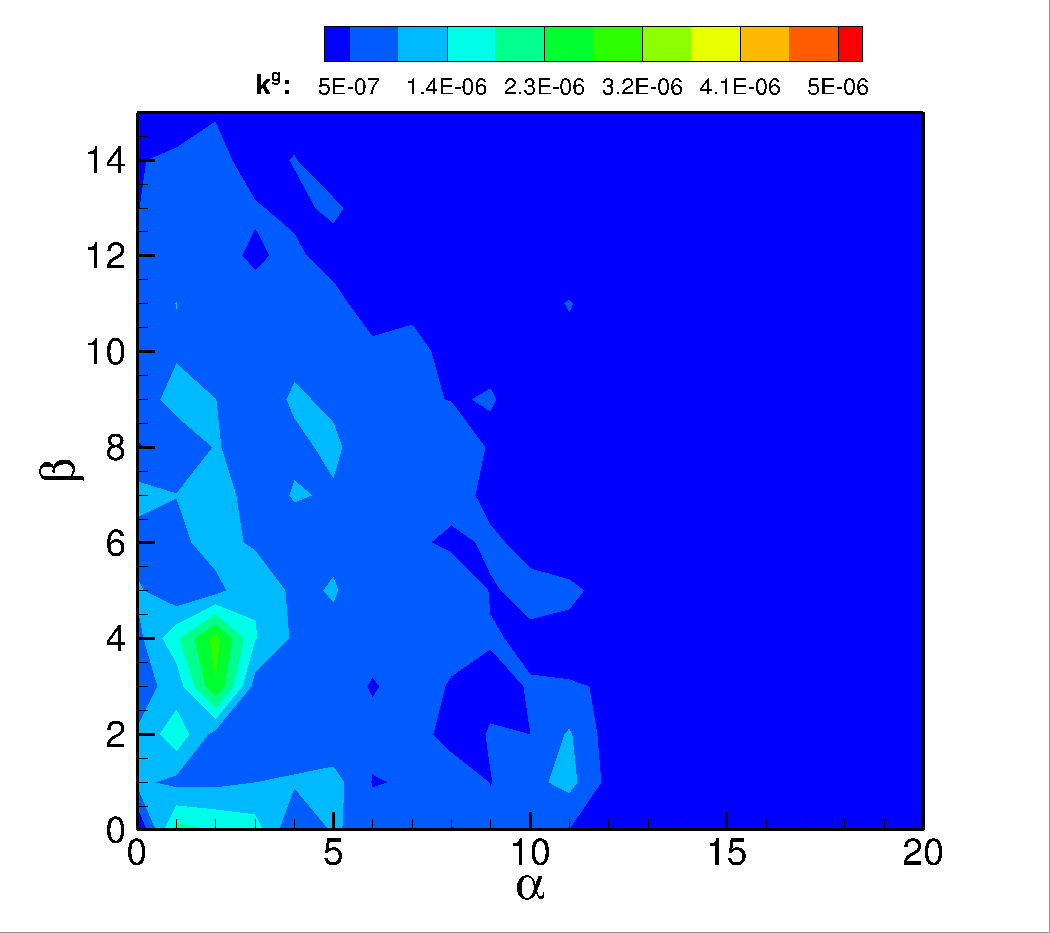}} 
    \subfloat[$t=1500$]{\includegraphics[trim=0 10 10 0, clip,width=0.32\textwidth, keepaspectratio]{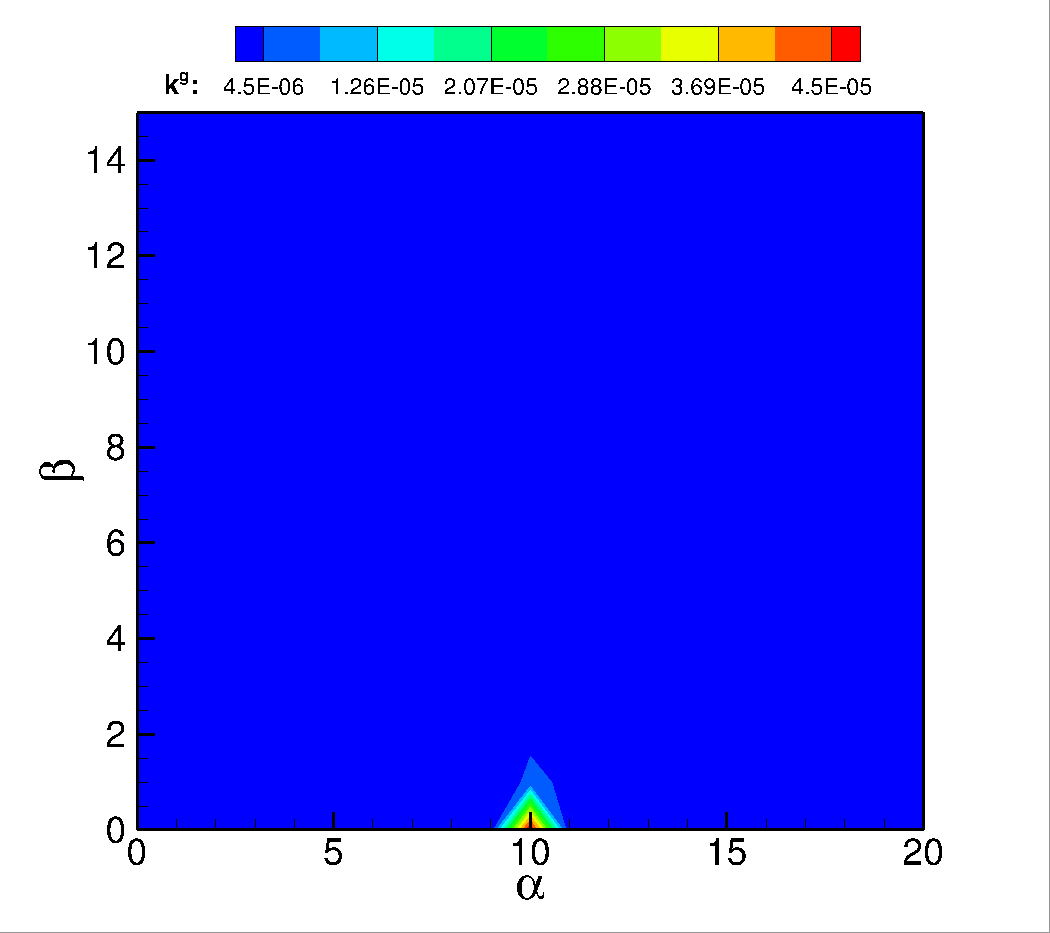}} \\
    \subfloat[$t=30$]{\includegraphics[trim=0 10 10 0, clip,width=0.32\textwidth, keepaspectratio]{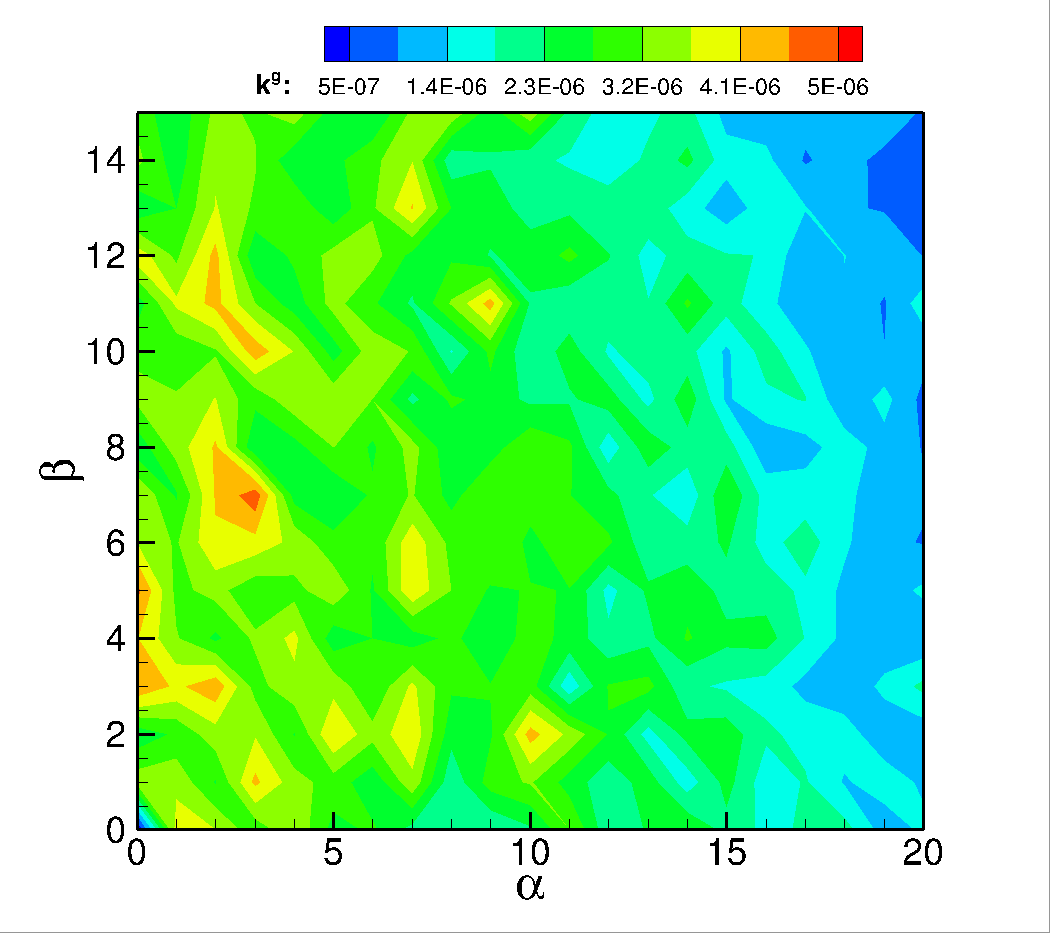}} 
    \subfloat[$t=500$]{\includegraphics[trim=0 10 10 0, clip,width=0.32\textwidth, keepaspectratio]{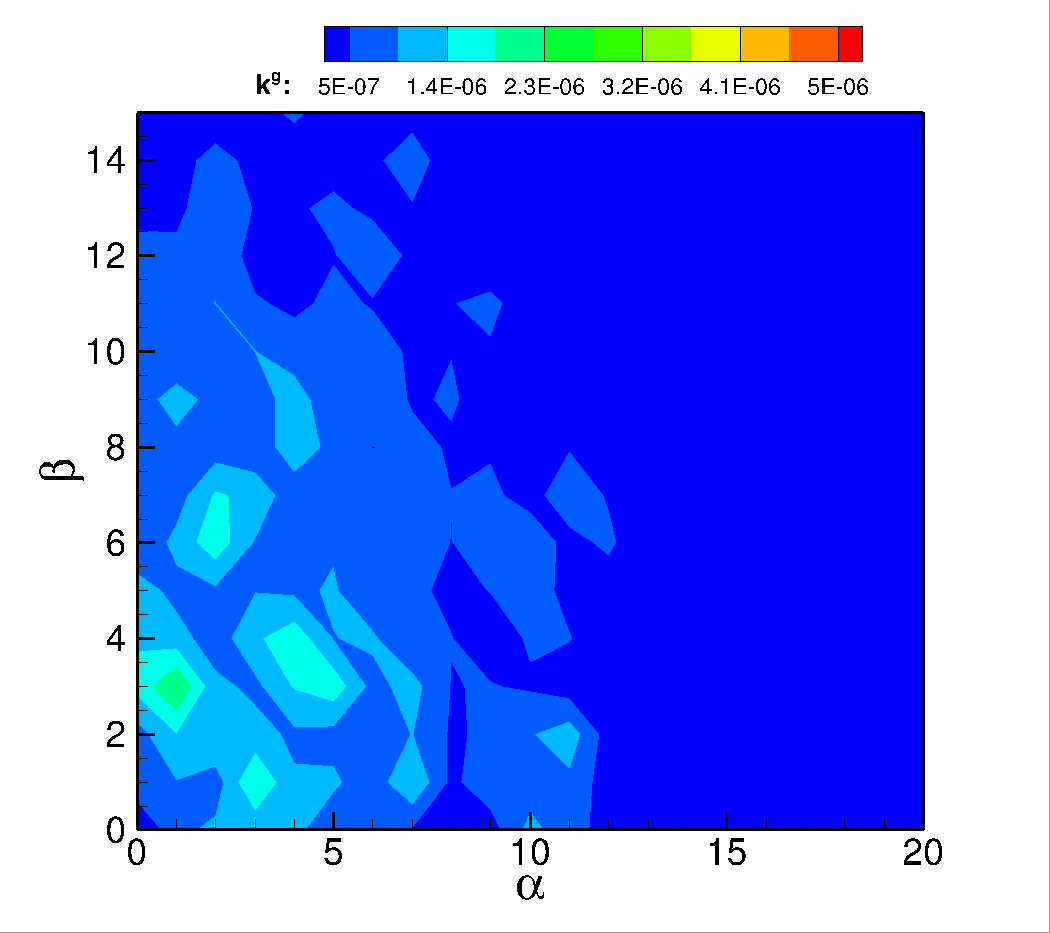}} 
    \subfloat[$t=1500$]{\includegraphics[trim=0 10 10 0, clip,width=0.32\textwidth, keepaspectratio]{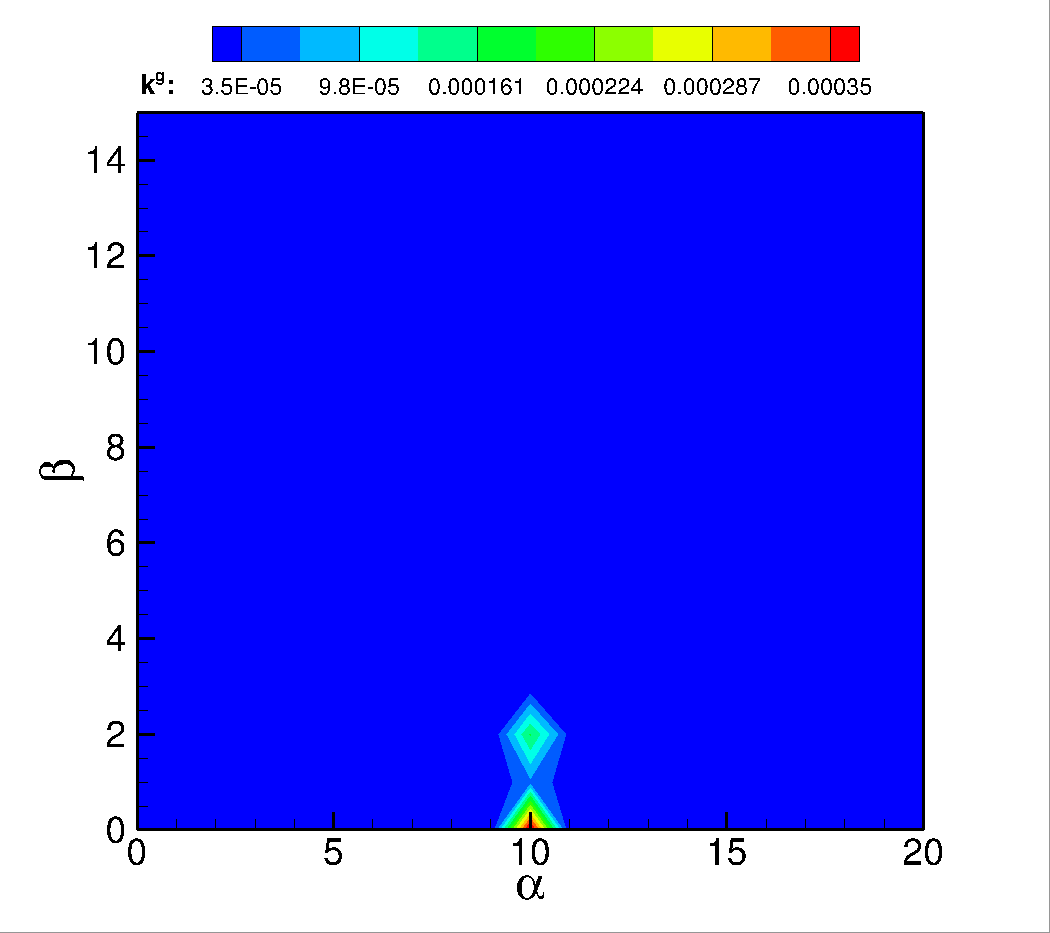}} \\
    \subfloat[$t=30$]{\includegraphics[trim=0 10 10 0, clip,width=0.32\textwidth, keepaspectratio]{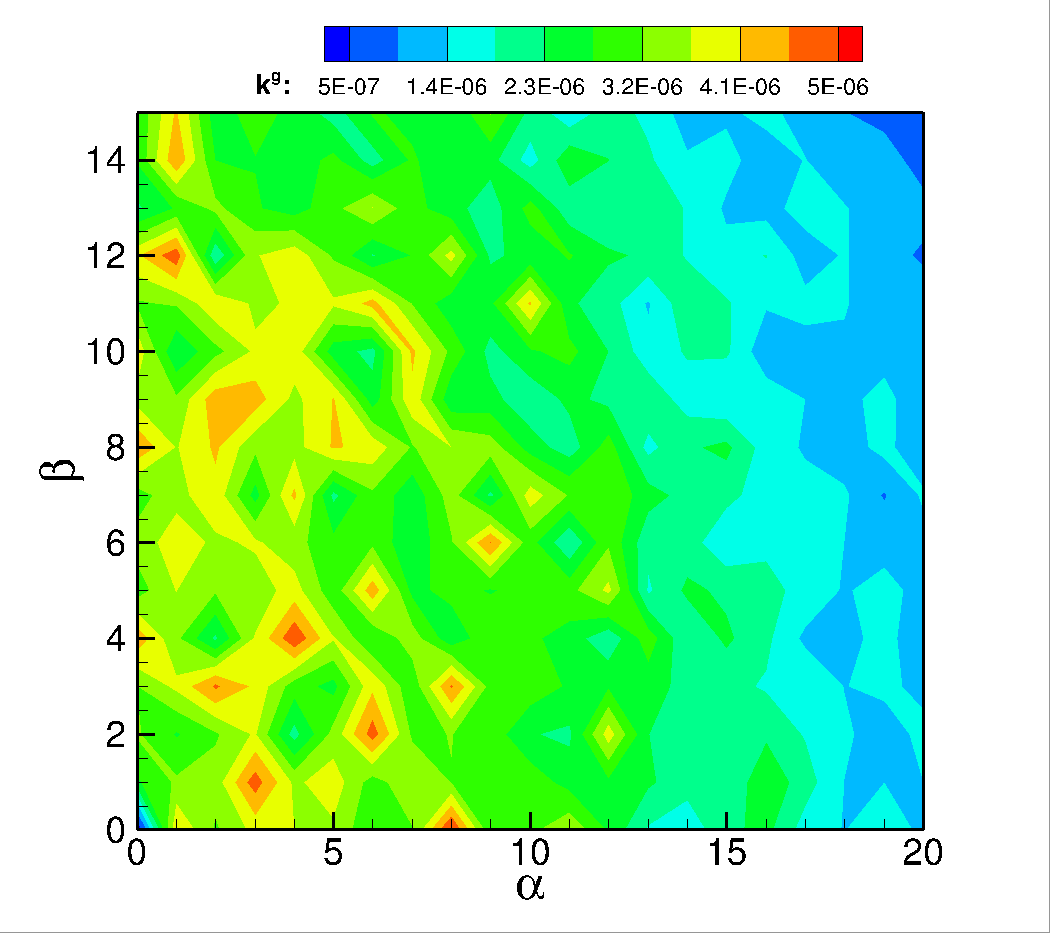}} 
    \subfloat[$t=500$]{\includegraphics[trim=0 10 10 0, clip,width=0.32\textwidth, keepaspectratio]{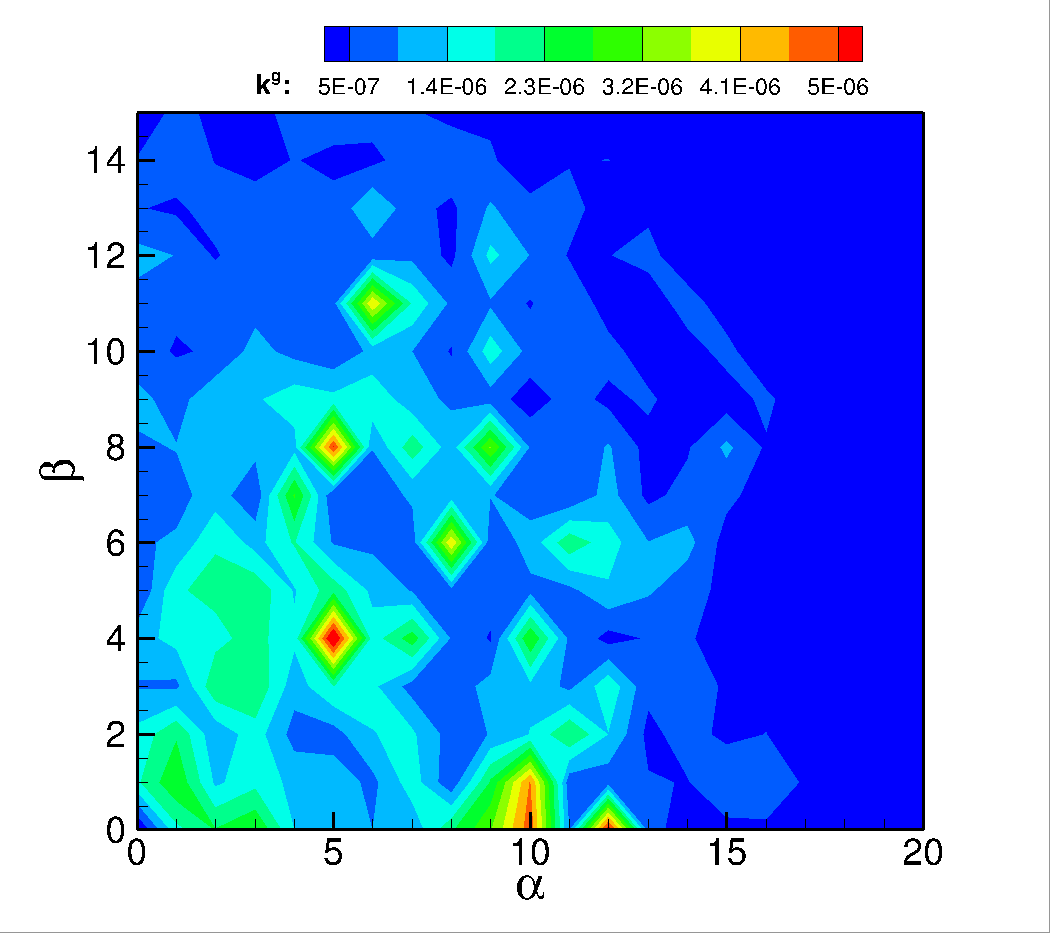}} 
    \subfloat[$t=1500$]{\includegraphics[trim=0 10 10 0, clip,width=0.32\textwidth, keepaspectratio]{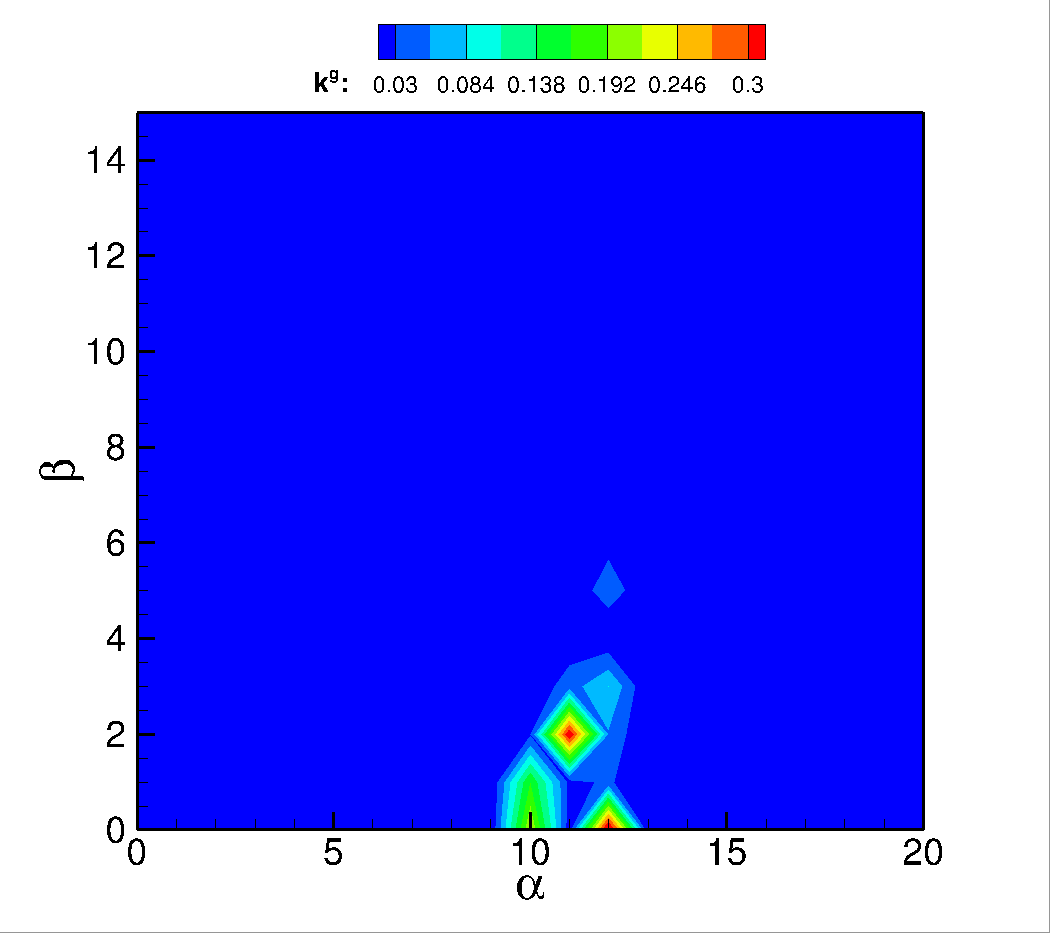}} 
    \caption{Global kinetic energy ($k^g$) spectrum at $M=6$, $Re=4000$ for (a)-(c) $Pr=0.5$, (d)-(f) $Pr=0.7$ and (g)-(i) $Pr=1.3$.}
    \label{fig:RandKeSpecM6}
\end{figure*}

\begin{figure*}
    \centering
    \subfloat[$\hat{u}_1$]{\includegraphics[trim=0 10 10 0, clip,width=0.32\textwidth, keepaspectratio]{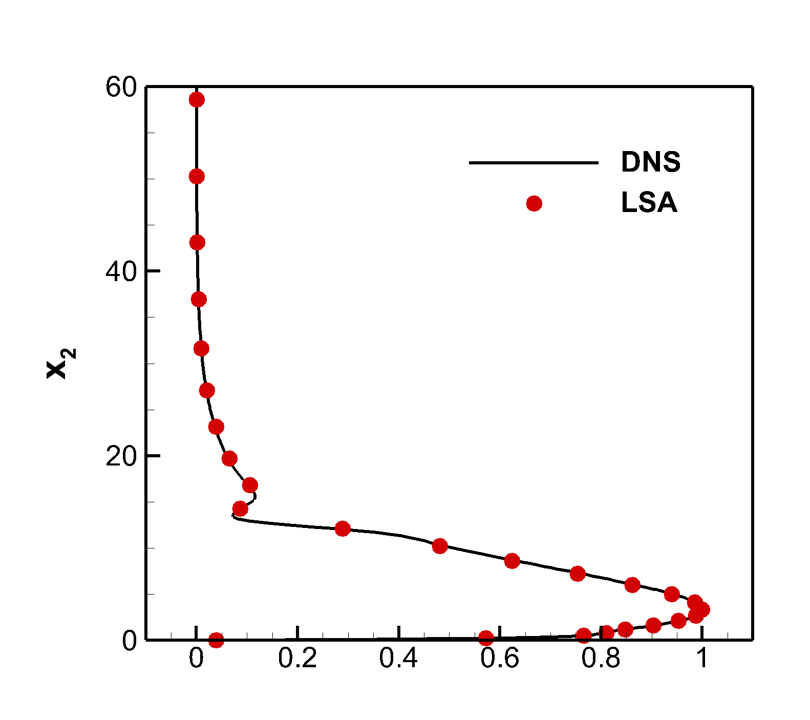}} 
    \subfloat[$\hat{T}$]{\includegraphics[trim=0 10 10 0, clip,width=0.32\textwidth, keepaspectratio]{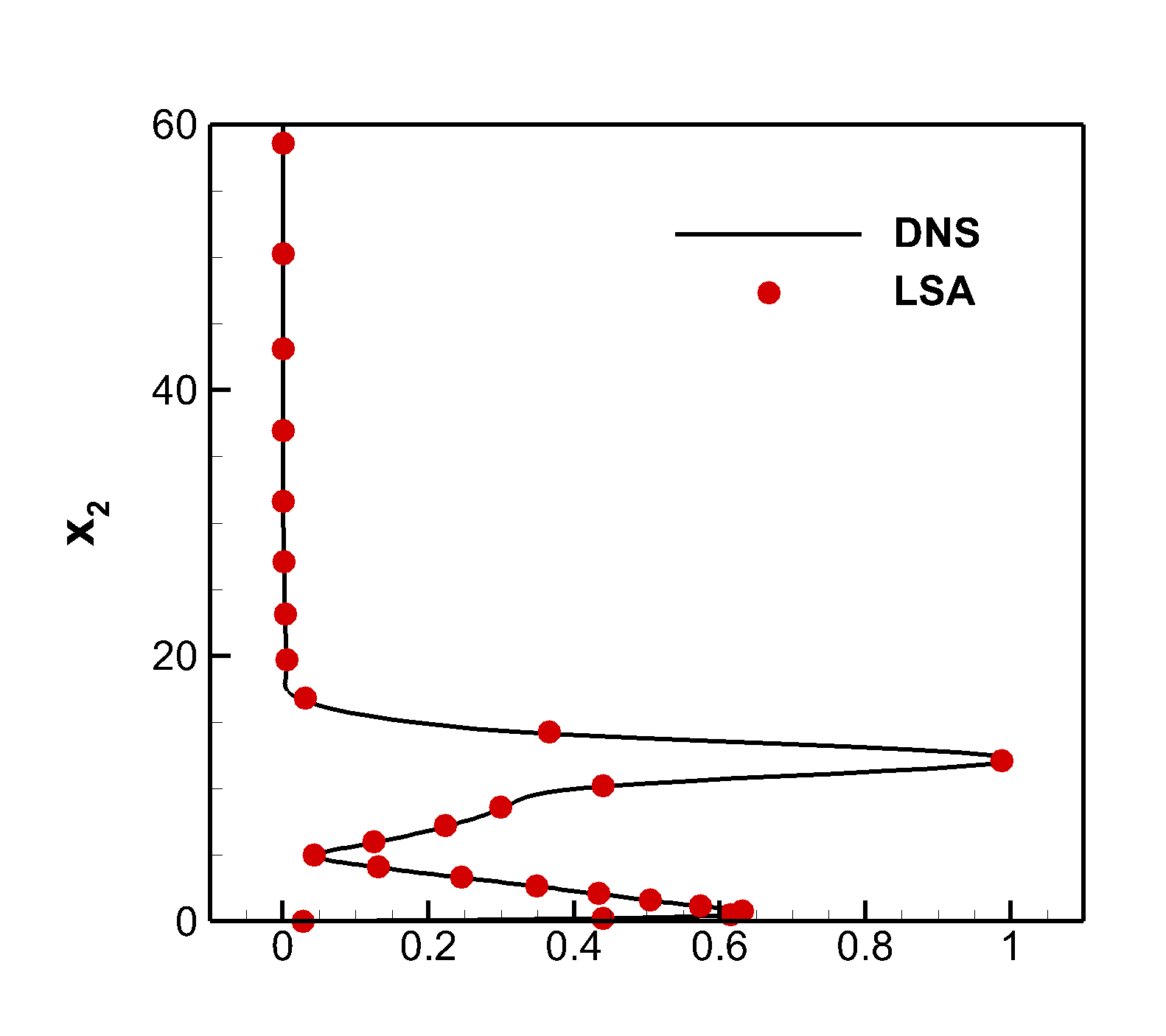}} 
    \subfloat[$\hat{p}$]{\includegraphics[trim=0 10 10 0, clip,width=0.32\textwidth, keepaspectratio]{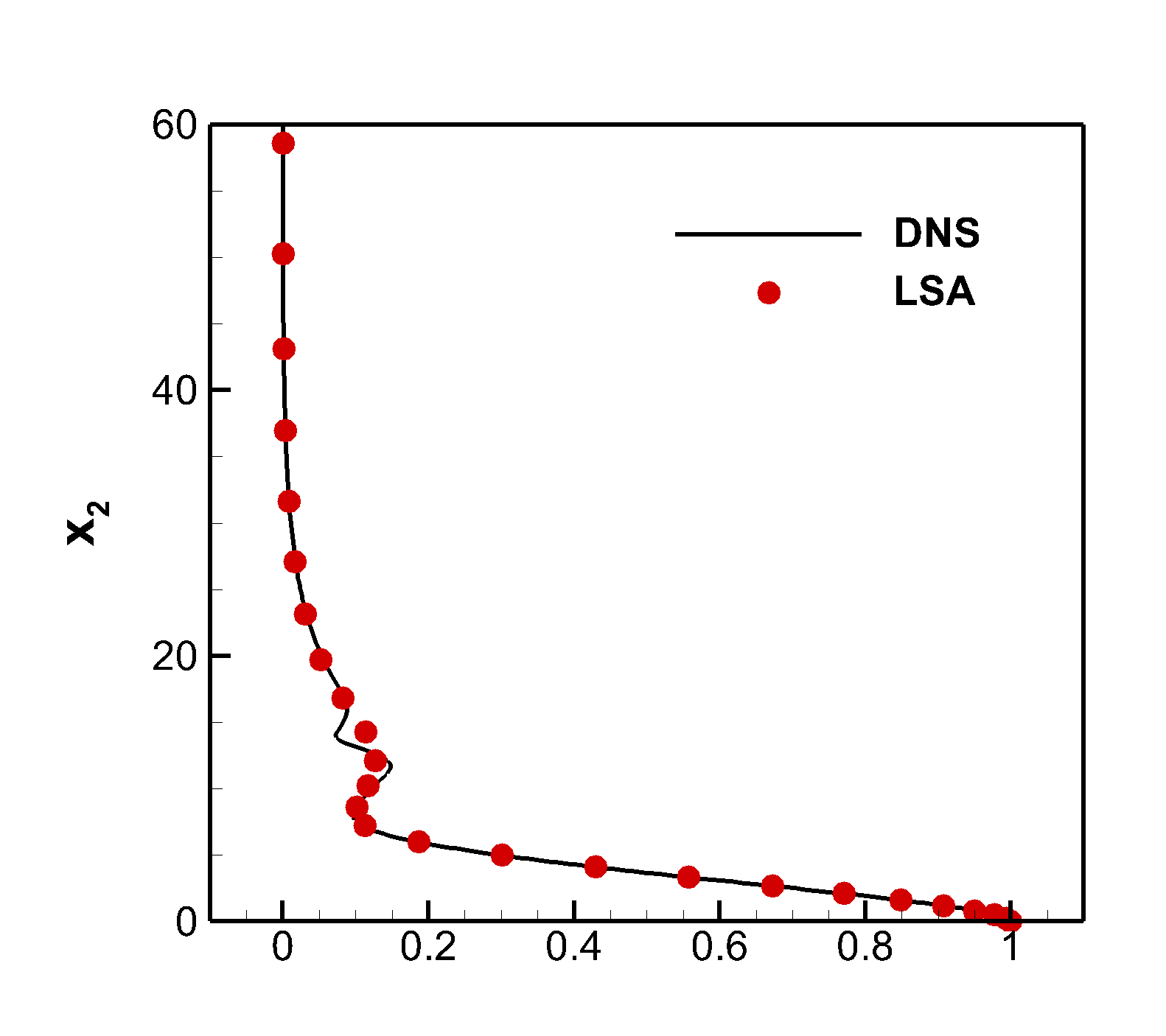}} \\
    \subfloat[$\hat{u}_1$]{\includegraphics[trim=0 10 10 0, clip,width=0.32\textwidth, keepaspectratio]{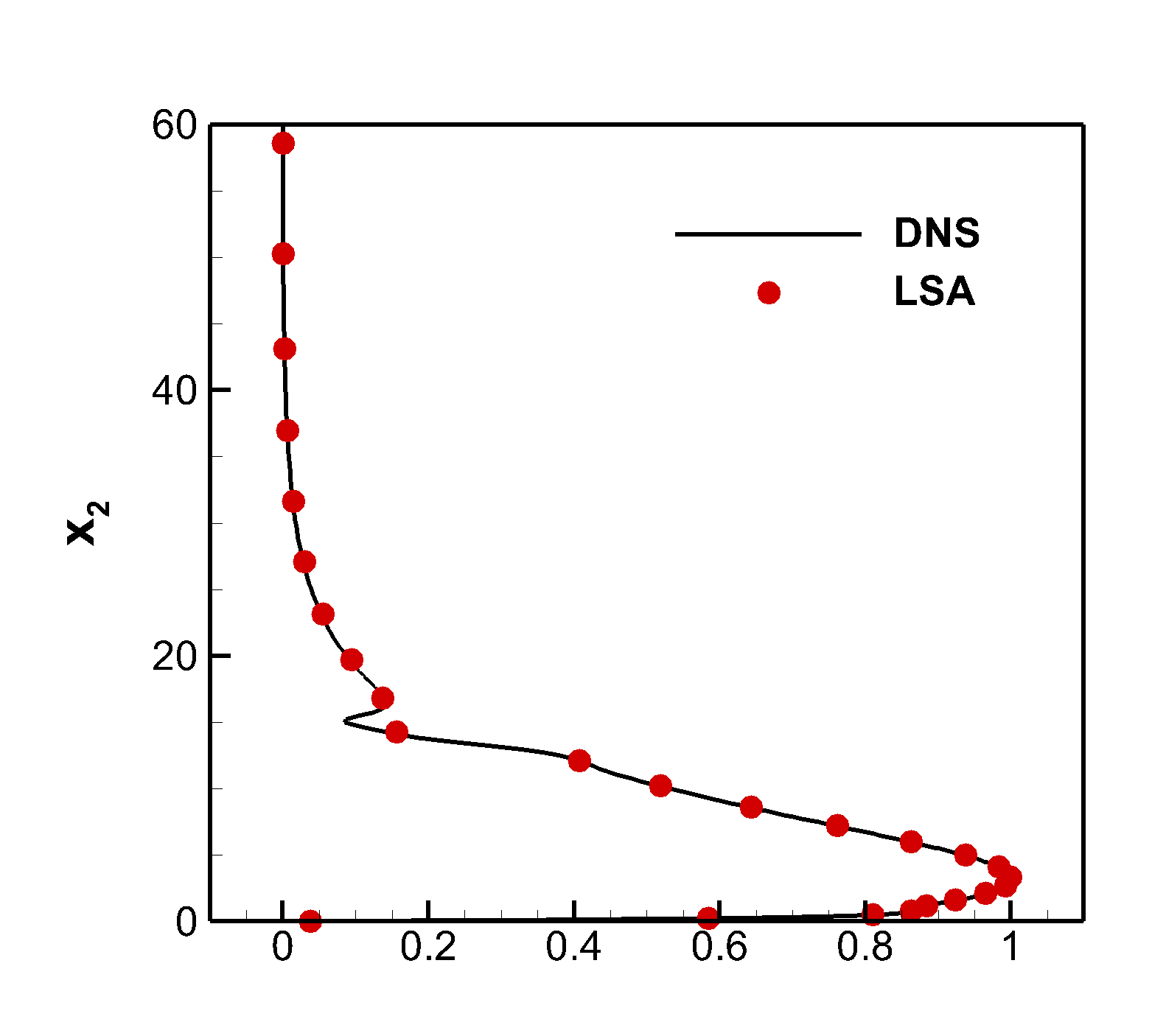}} 
    \subfloat[$\hat{T}$]{\includegraphics[trim=0 10 10 0, clip,width=0.32\textwidth, keepaspectratio]{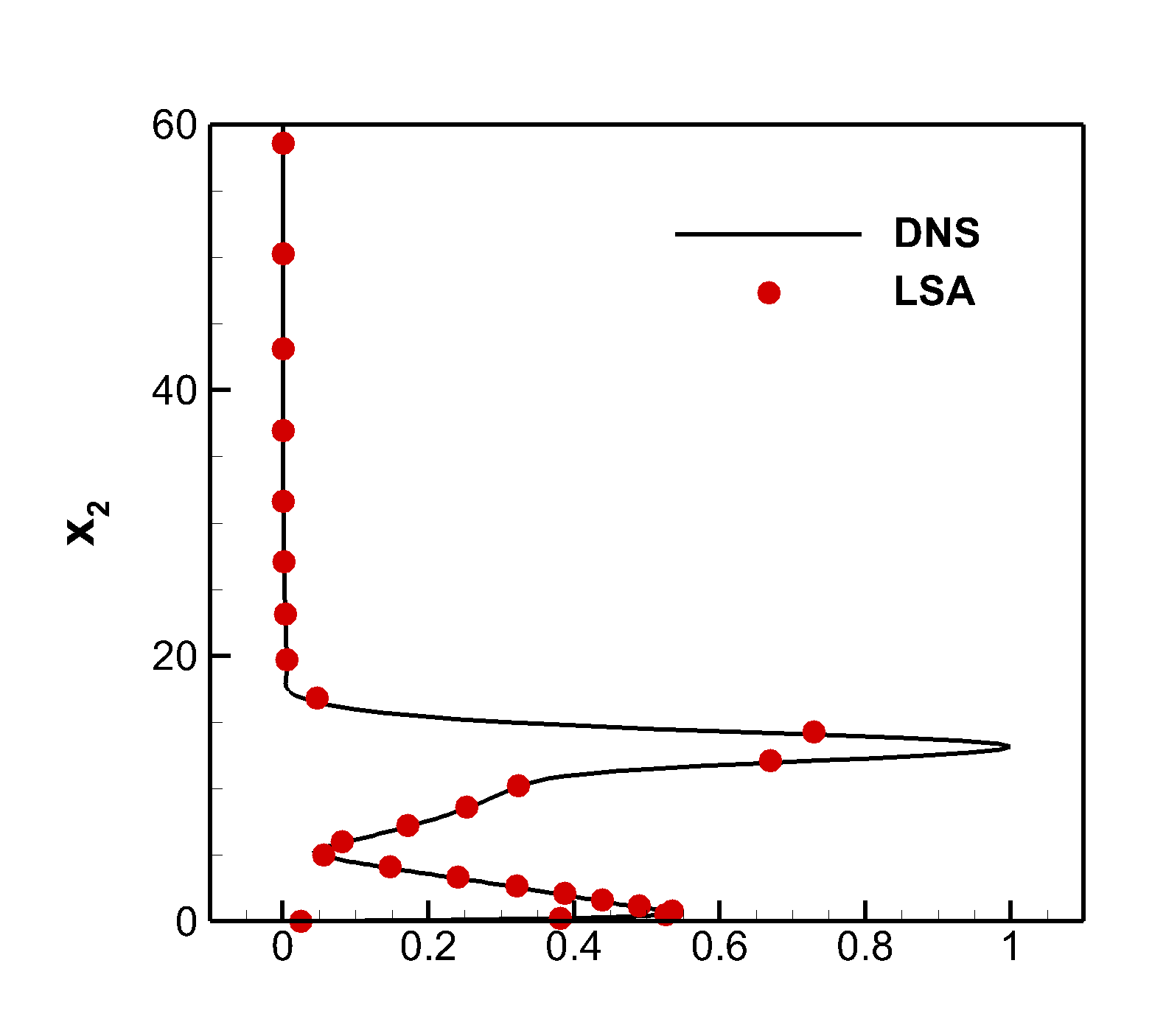}} 
    \subfloat[$\hat{p}$]{\includegraphics[trim=0 10 10 0, clip,width=0.32\textwidth, keepaspectratio]{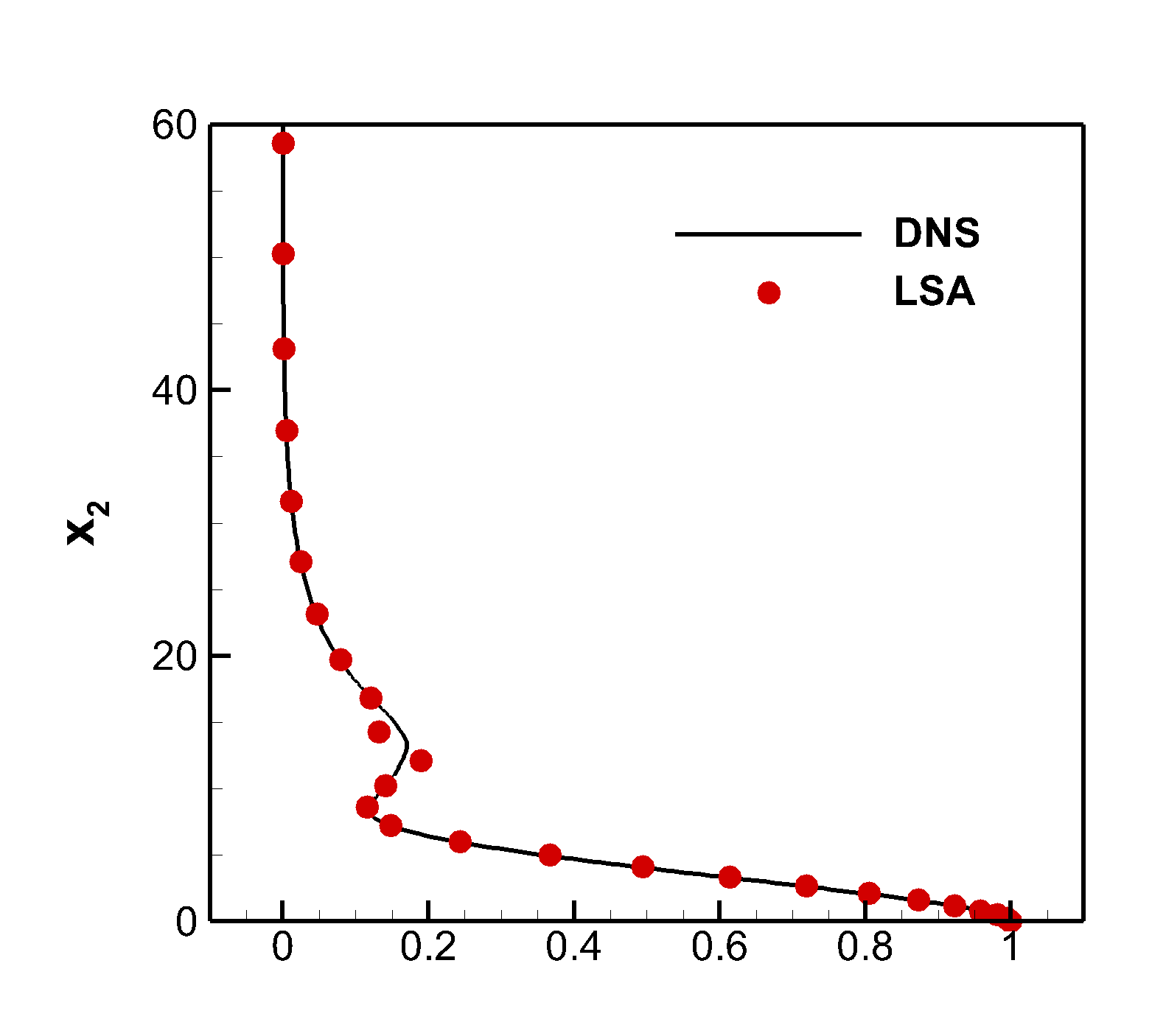}} \\
    \subfloat[$\hat{u}_1$]{\includegraphics[trim=0 10 10 0, clip,width=0.32\textwidth, keepaspectratio]{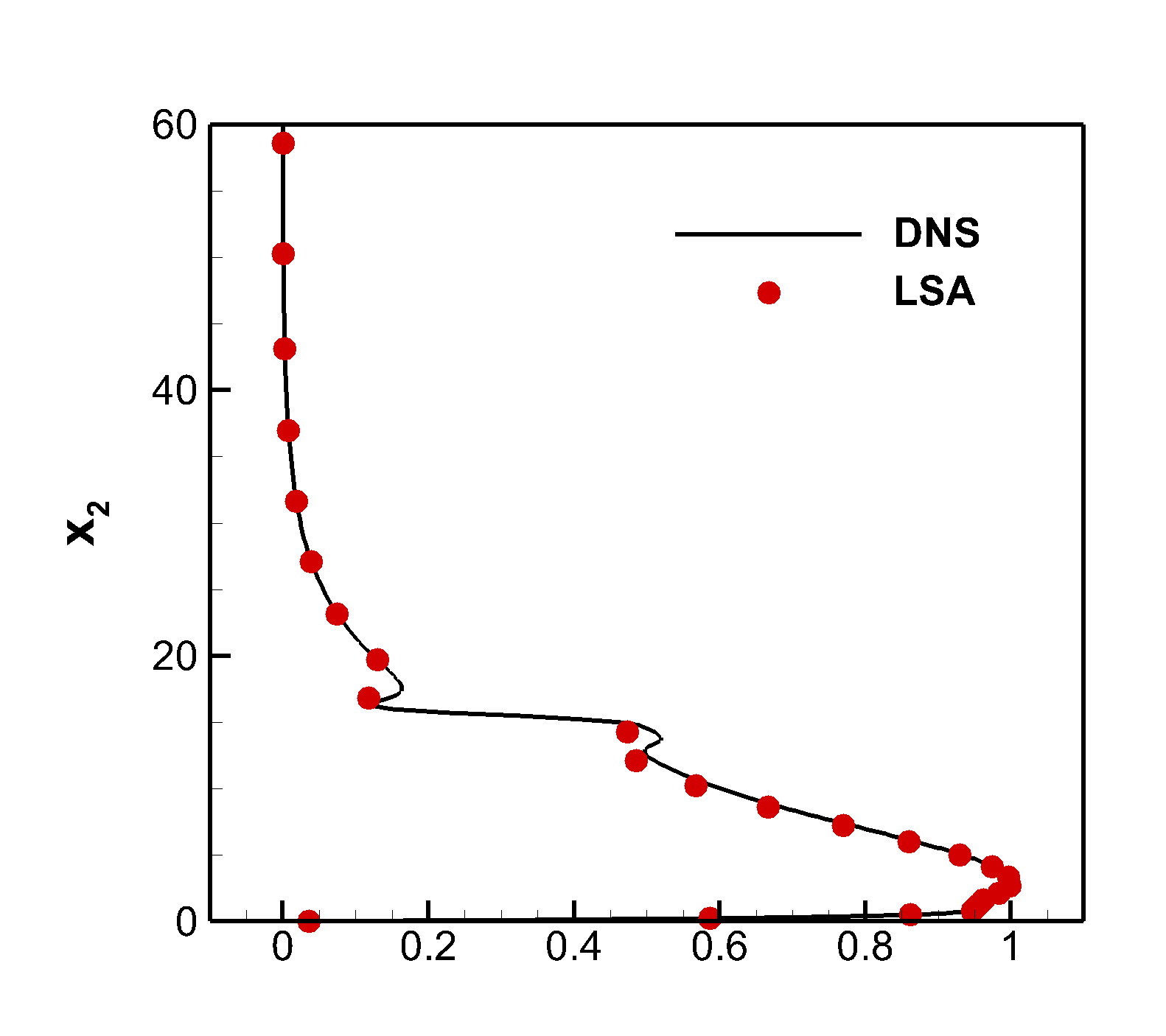}} 
    \subfloat[$\hat{T}$]{\includegraphics[trim=0 10 10 0, clip,width=0.32\textwidth, keepaspectratio]{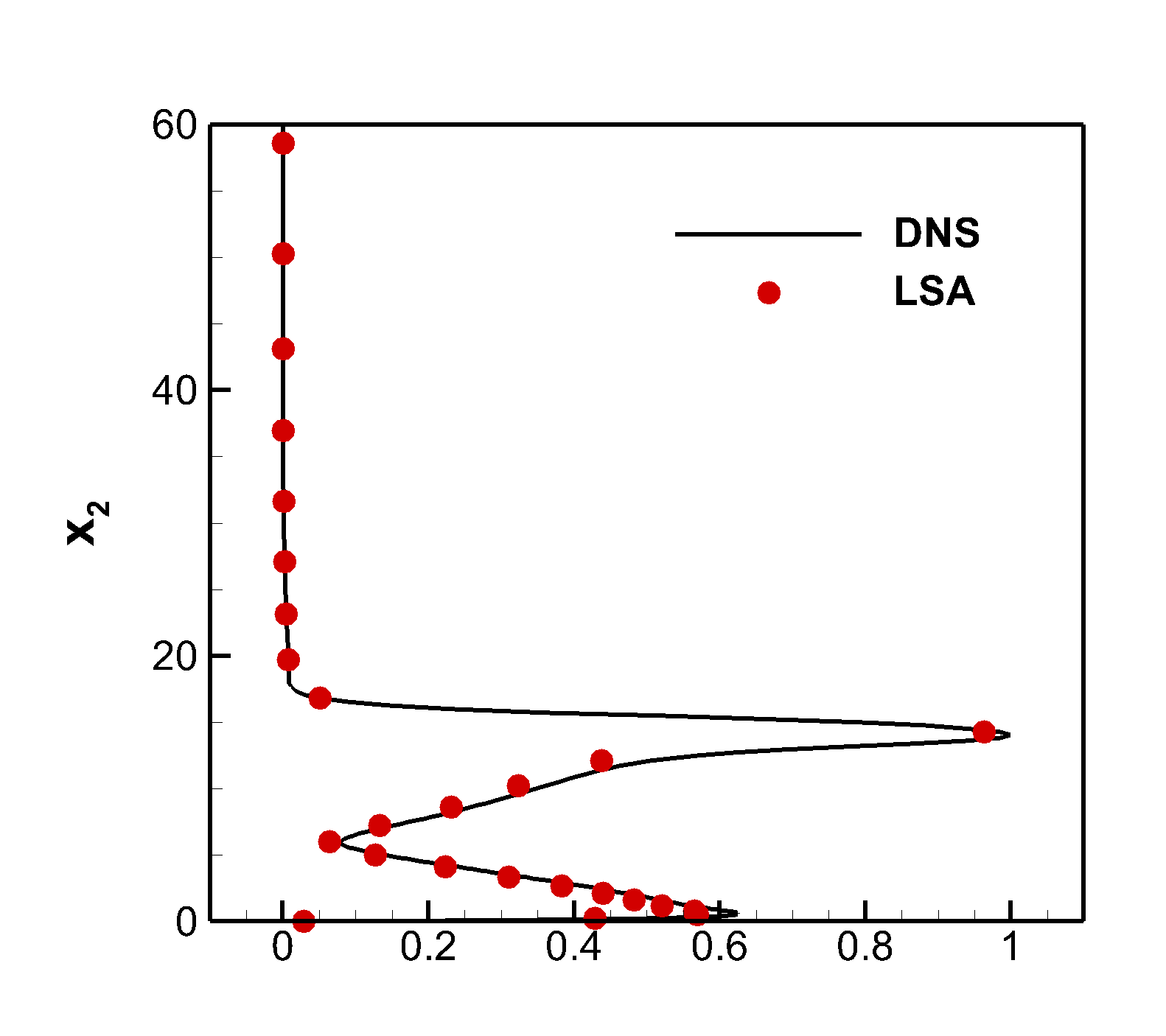}} 
    \subfloat[$\hat{p}$]{\includegraphics[trim=0 10 10 0, clip,width=0.32\textwidth, keepaspectratio]{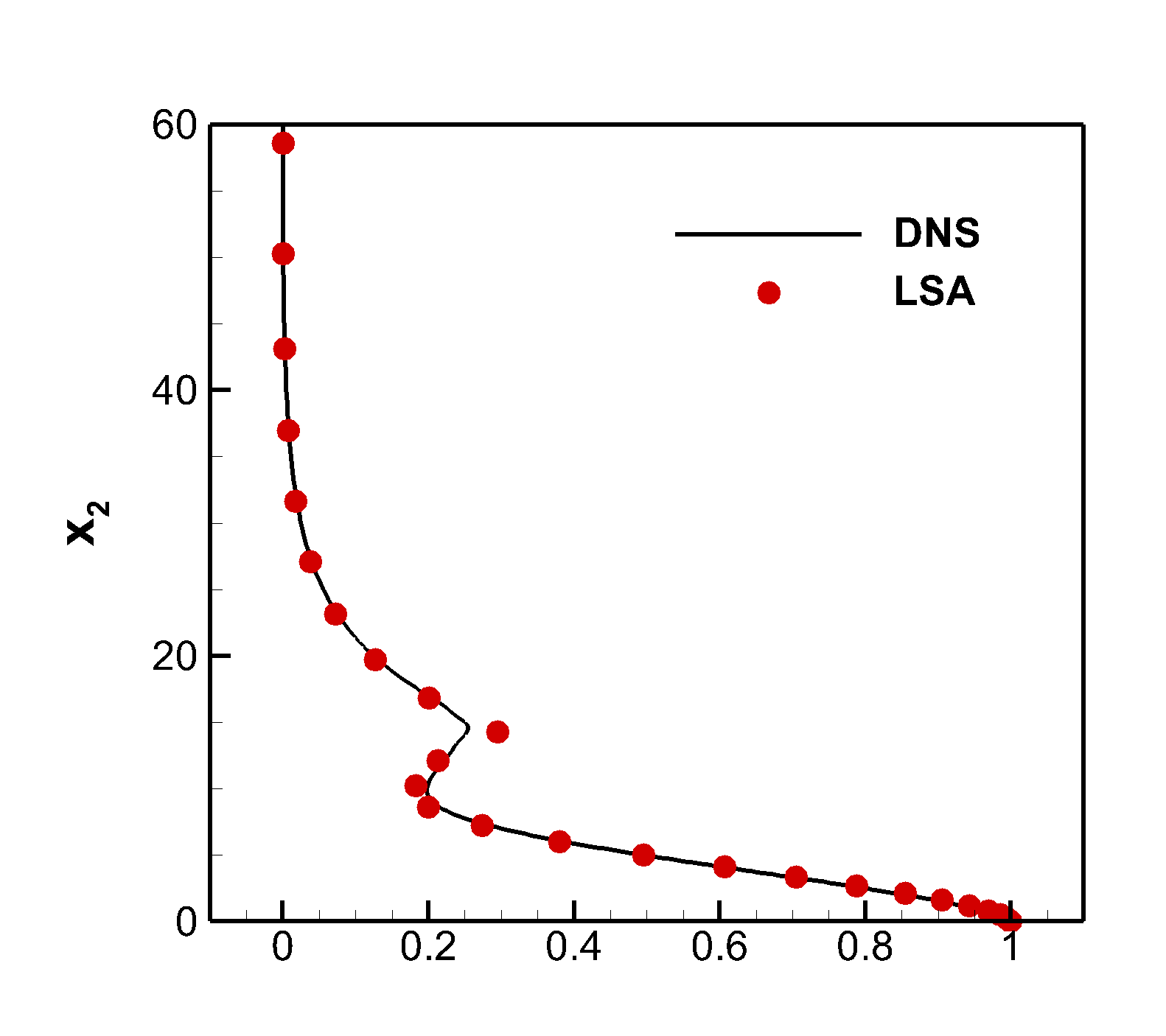}} 
    \caption{Mode shapes corresponding to mode $(10,0)$ at $M=6$, $Re=4000$ for (a)-(c) $Pr=0.5$, (d)-(f) $Pr=0.7$ and (g)-(i) $Pr=1.3$.}
    \label{fig:ModeShapeM6}
\end{figure*}

 \begin{table}
\begin{center}
\begin{tabular}{cccccccccccc}
 	  \textbf{Case} & $\mathbf{Re}$ & $\mathbf{Pr}$ & $\mathbf{M}$ & $\mathbf{\rho_\infty (kg\cdot m^{-3})}$ & $\mathbf{T_\infty (K)}$ & $\mathbf{L_{x1}}$ & $\mathbf{L_{x2}}$ & $\mathbf{L_{x3}}$ & $\mathbf{N_{x1}}$ & $\mathbf{N_{x2}}$ & $\mathbf{N_{x3}}$   \\ \hline 
   	$R_1$ & $4000$ & $0.5$ & $3.0$ & $1.0$ & $353$ & $924$ & $59$ & $924$ & $200$ & $300$ & $200$ \\
   	$R_2$ & $4000$ & $0.7$ & $3.0$ & $1.0$ & $353$ & $924$ & $59$ & $924$ & $200$ & $300$ & $200$ \\
   	$R_3$ & $4000$ & $1.3$ & $3.0$ & $1.0$ & $353$ & $924$ & $59$ & $924$ & $200$ & $300$ & $200$ \\
  	$R_4$ & $4000$ & $0.5$ & $6.0$ & $1.0$ & $353$ & $359$ & $124$ & $359$ & $200$ & $300$ & $200$ \\
  	$R_5$ & $4000$ & $0.7$ & $6.0$ & $1.0$ & $353$ & $359$ & $124$ & $359$ & $200$ & $300$ & $200$  \\
  	$R_6$ & $4000$ & $1.3$ & $6.0$ & $1.0$ & $353$ & $359$ & $124$ & $359$ & $200$ & $300$ & $200$ \\
\end{tabular}
\end{center}
	\caption{Non-dimensional parameters, freestream properties and grid sizes for DNS with random pressure forcing.}
	\label{table:ParametersBasic2}
\end{table}
The evolution of globally averaged kinetic and internal energy at $M=6$ for three different Prandtl numbers is shown in figure \ref{fig:RandKeM6}. After the lapse of an initial transience, kinetic energy grows exponentially at the dominant eigenmode growth rate predicted by LSA. The dominant eigenmode corresponds to the streamwise-spanwise wavenumber pair $(10,0)$ for $Pr=\{0.5,0.7\}$, while the mode $(12,0)$ dominates for $Pr=1.3$. Here $(m,n)$ denotes the $m^{th}$ and $n^{th}$ harmonic of the fundamental streamwise and spanwise wavenumber resolved by the simulation. The asymptotic growth rate of $k^g$ increases with increasing Prandtl number. Since the dominant mode at $M=6$ is a second mode the internal energy also grows beyond the transient region and the growth rate of $e^g$ agrees well with LSA. The initial forcing generates a broadband spectrum with energy randomly distributed in all wavenumbers. This is evident from the spectral spread of global kinetic energy shown in figure \ref{fig:RandKeSpecM6}. The high wavenumbers are linearly stable and dissipate quickly resulting in kinetic energy decay in the transient stage. As a result by $t=500$ most of the energy is contained in the linearly unstable wavenumbers. 
During the transient stage non-modal interactions are prevalent and the evolution can't be determined by an eigenmode analysis. 
The length of the transient stage decreases with increasing Prandtl number (figure \ref{fig:RandKeM6}). Consequently, the higher Prandtl number cases attain the asymptotic state faster and are more energetic at all times than the low $Pr$ case. For $Pr=0.5$ the Fourier mode $(10,0)$ becomes the most energetic mode at late times. Linear stability analysis also predicts the mode $(10,0)$ as the fastest growing mode. The kinetic energy spectrum for higher Prandtl number cases has multiple modes with similar energy content. The mode $(10,0)$ and $(10,1)$ are the most energetic modes for $Pr=0.7$. This is perhaps due to the fact that the most unstable eigenmode corresponding to $(10,0)$ and $(10,1)$ have similar growth rates. Similarly the modes $(10,0)$, $(11,1)$ and $(12,0)$ are most energetic for $Pr=1.3$. 
The Fourier mode shapes corresponding to mode $(10,0)$ at late times is presented in figure \ref{fig:ModeShapeM6}. For all three Prandtl numbers the mode shapes obtained from DNS are in excellent agreement with the dominant eigenmode shapes obtained from LSA. This confirms that in the asymptotic limit the boundary layer response to external forcing is completely characterized by the most unstable eigenmode(s).

\begin{figure*}
    \centering
    \subfloat[$P_k^g$]{\includegraphics[trim=0 10 10 0, clip, width=0.32\textwidth, 
    keepaspectratio]{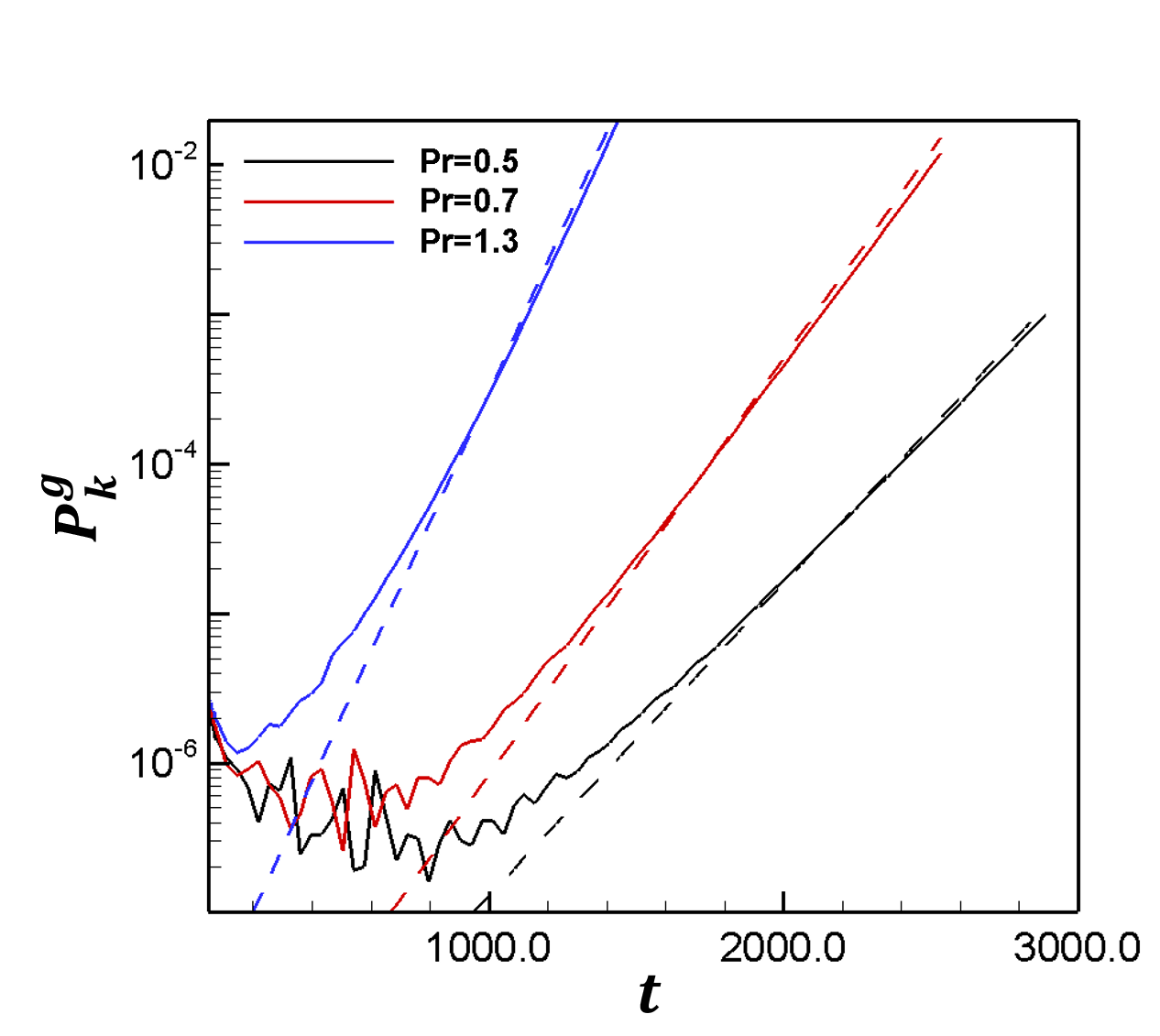}}
    \subfloat[$|\Pi_k^g|$]{\includegraphics[trim=0 10 10 0, clip, width=0.32\textwidth, keepaspectratio]{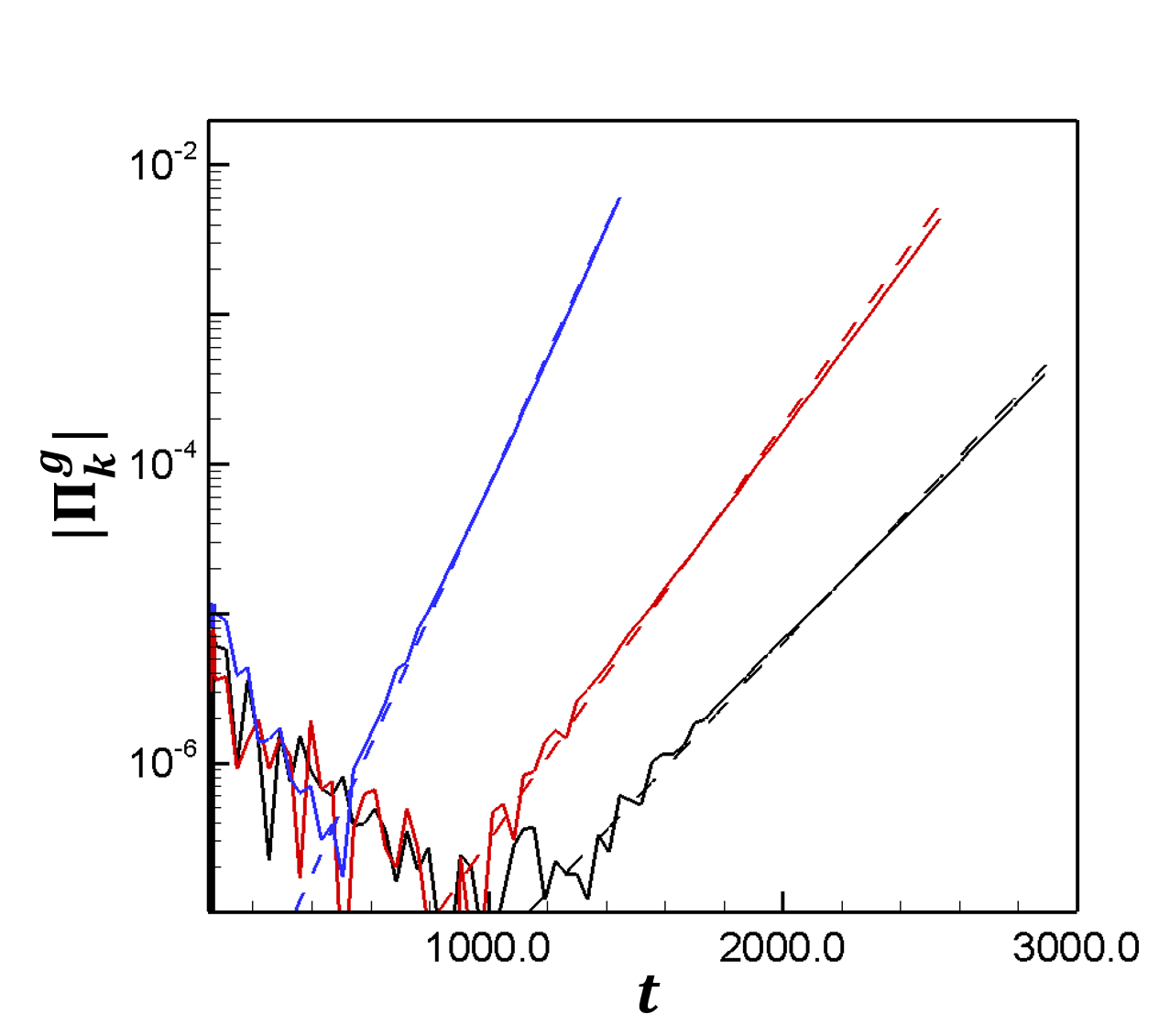}}
    \subfloat[$-\epsilon_k^g$]{\includegraphics[trim=0 10 10 0, clip, width=0.32\textwidth, keepaspectratio]{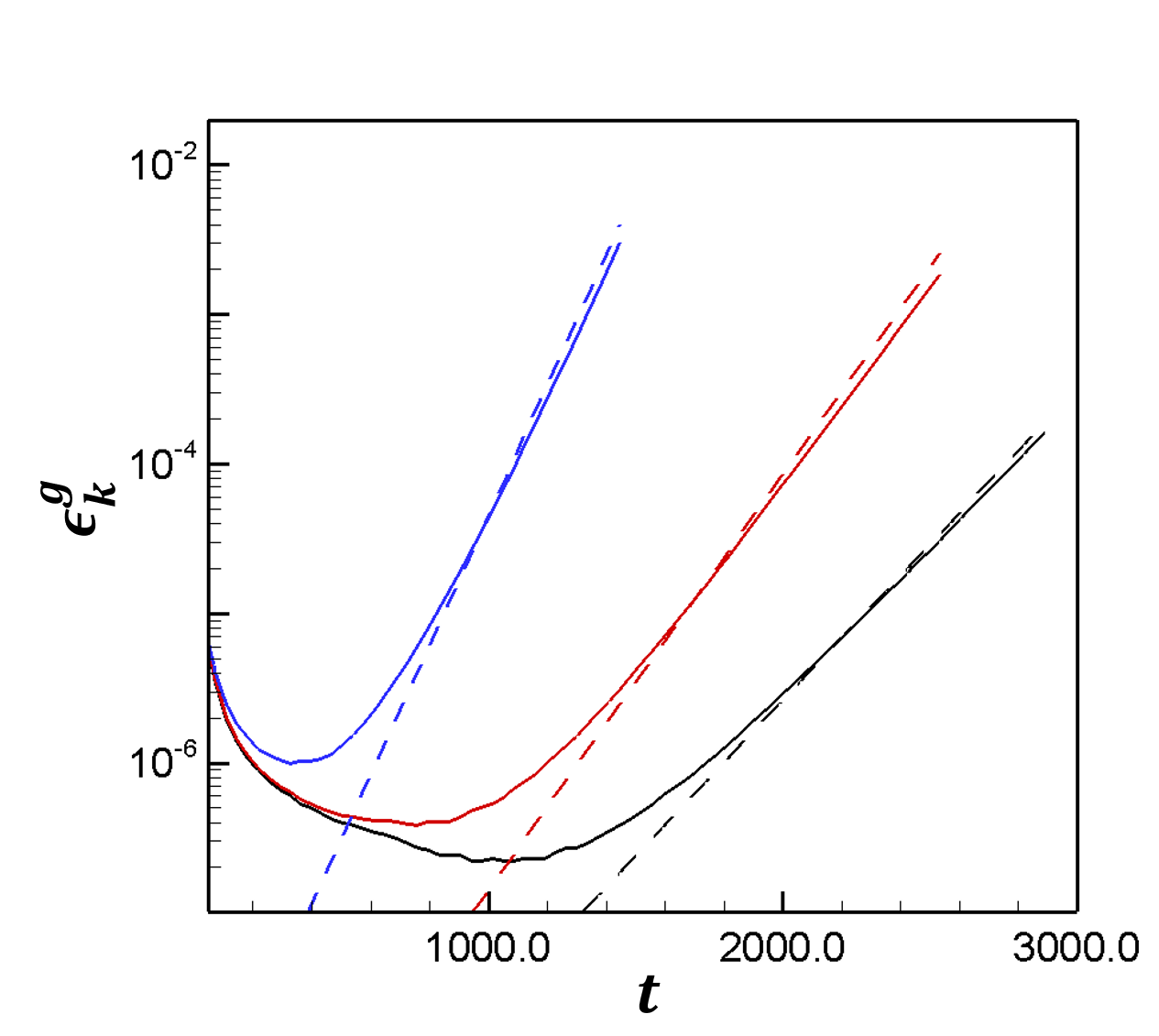}}
    \caption{Evolution of globally averaged terms in the kinetic energy budget: (a) production ($P_k^g$), (b) absolute value of pressure-dilatation ($|\Pi_k^g|$) and (c) dissipation ($-\epsilon_k^g$) at $M=6$ and $Re=4000$ for three different Prandtl number. The dashed lines represent dominant eigenmode growth at the rate predicted by LSA.}
    \label{fig:RandKeBudget}
\end{figure*}

\begin{figure*}
    \centering
    \subfloat[$Pr=0.5$]{\includegraphics[width=0.32\textwidth, keepaspectratio]{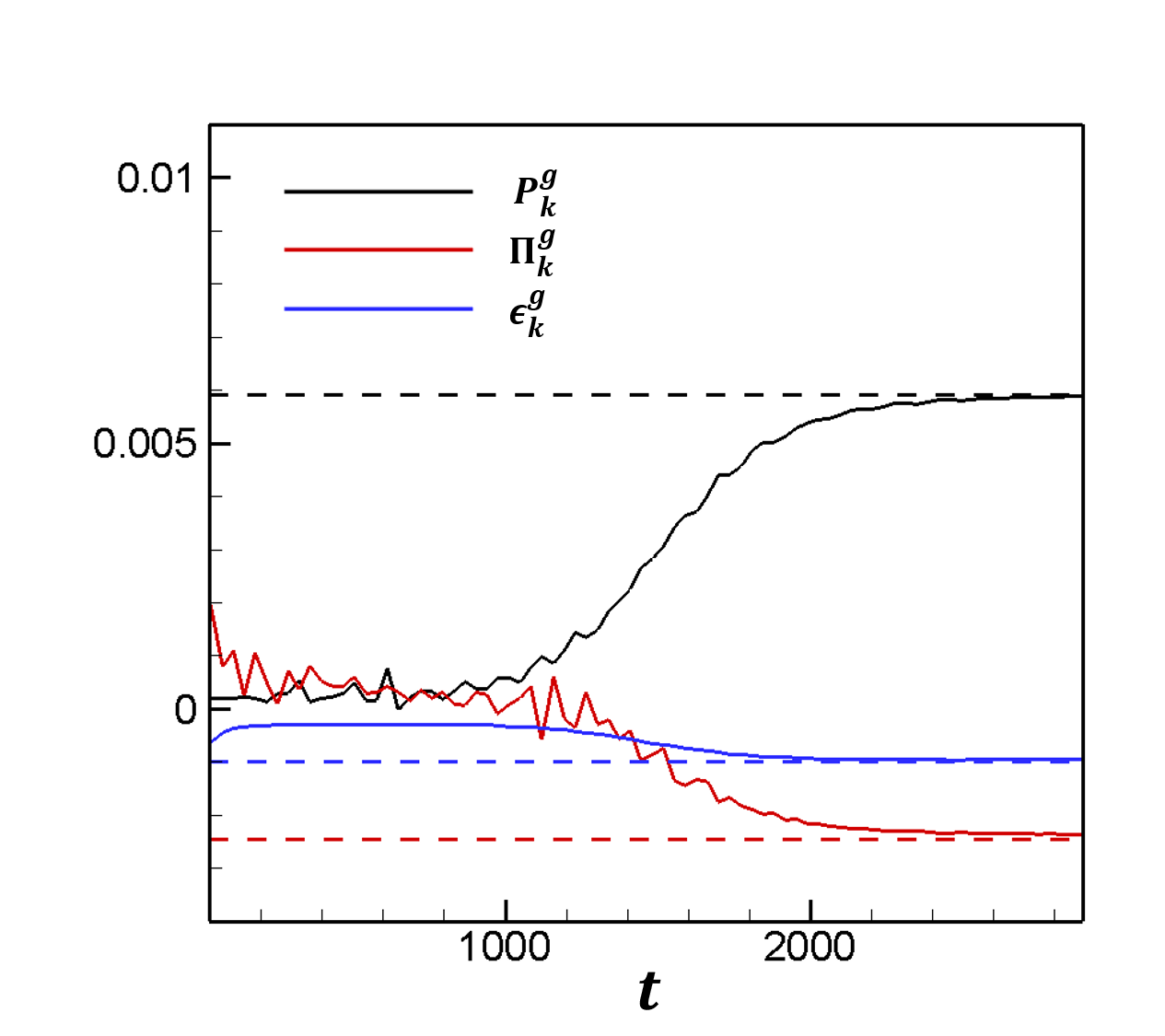}} 
    \subfloat[$Pr=0.7$]{\includegraphics[width=0.32\textwidth, keepaspectratio]{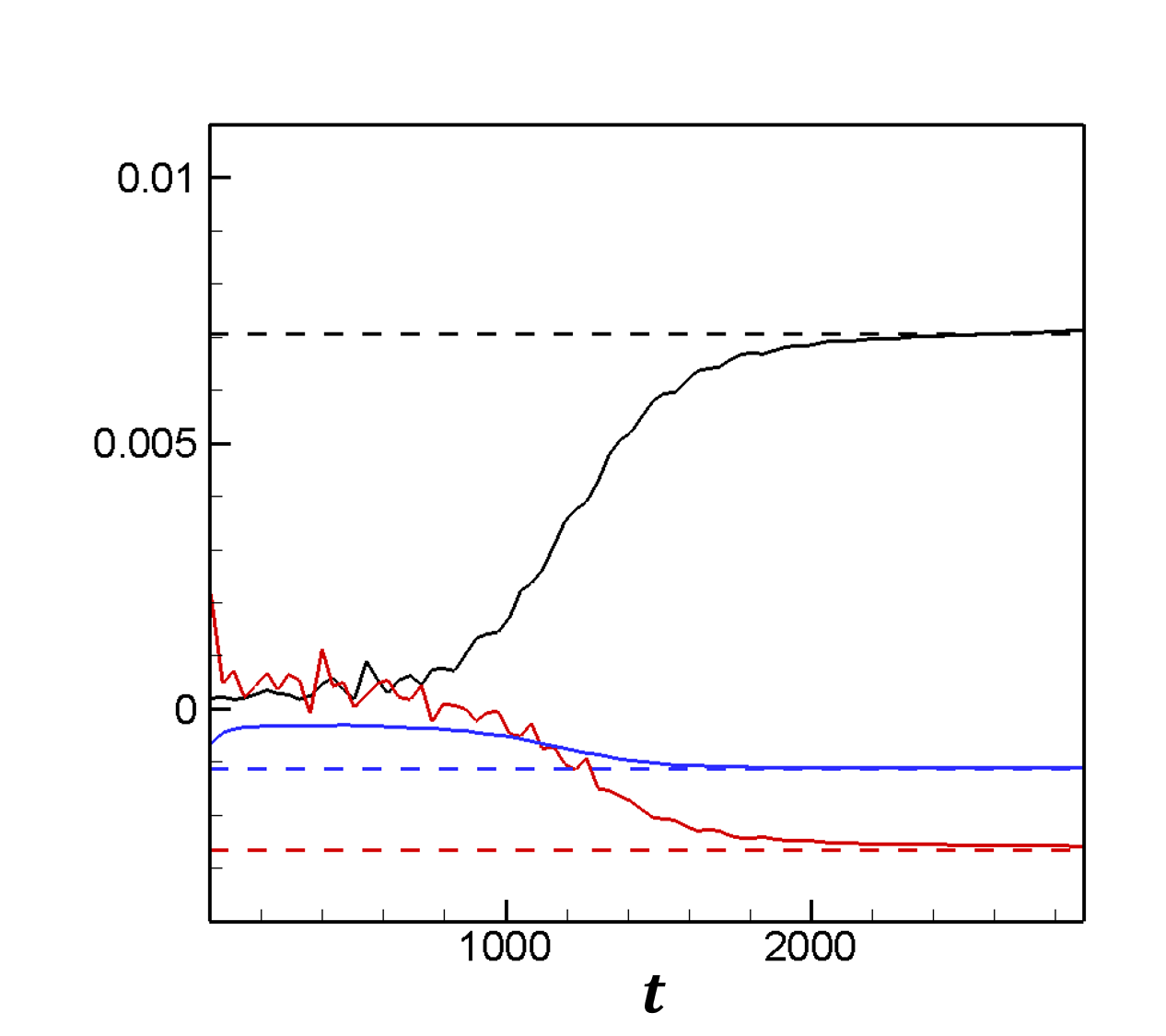}}
    \subfloat[$Pr=1.3$]{\includegraphics[width=0.32\textwidth, keepaspectratio]{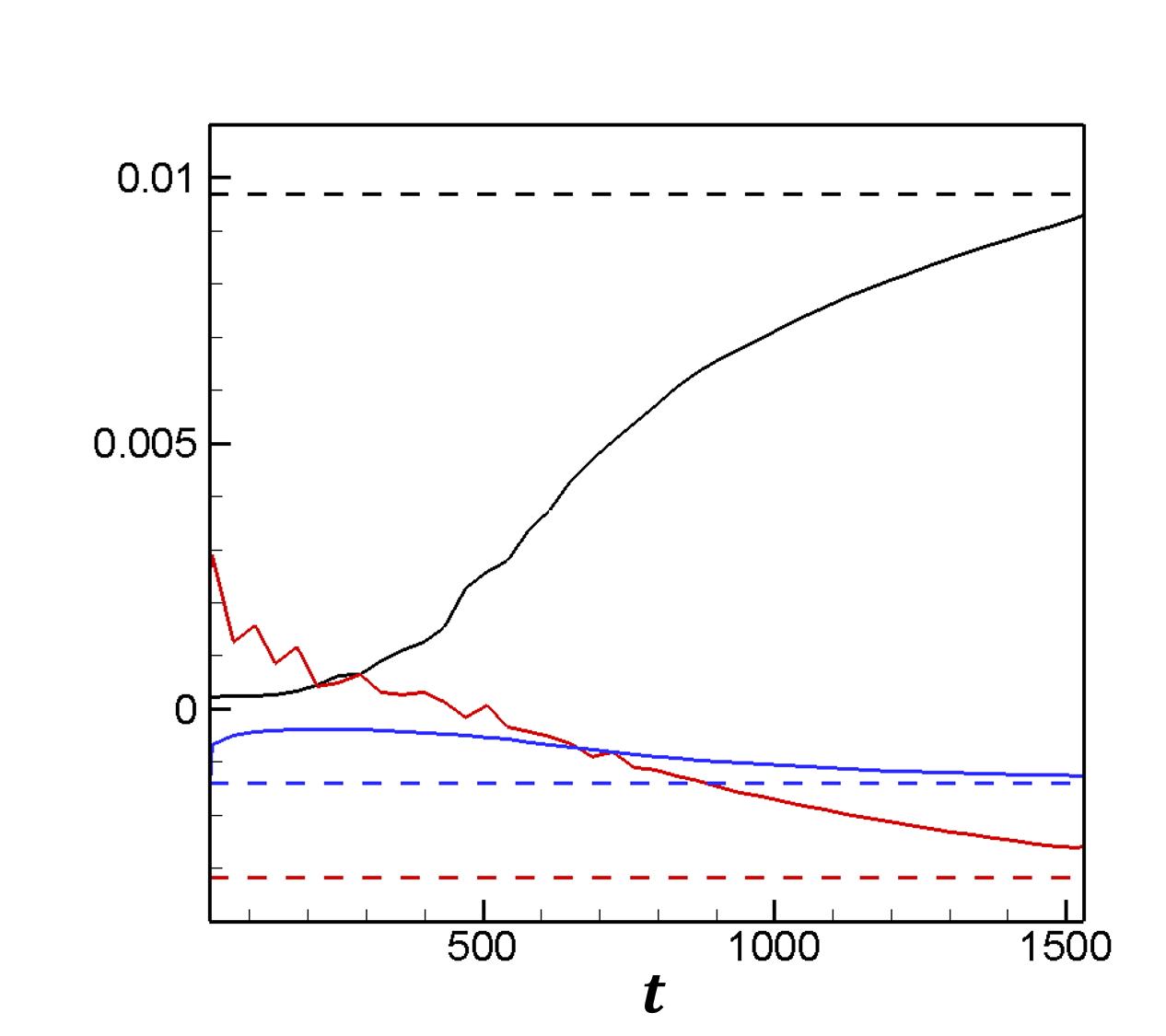}}
    \caption{Evolution of globally averaged terms in the kinetic energy budget normalized by instantaneous kinetic energy for (a) $Pr=0.5$, (b) $Pr=0.7$ and (c) $Pr=1.3$ at $M=6$ and $Re=4000$. All budget terms are normalized by $\zeta=2k^g U_\infty/L_r$. The dashed lines represent the budget contributions of the most unstable eigenmode predicted by LSA.}
    \label{fig:RandKeBudgetInstant}
\end{figure*}

The evolution of globally averaged terms in the kinetic energy budget are shown in figure \ref{fig:RandKeBudget}(a)-(c). Production increases exponentially after the lapse of transient stage for all $Pr$. In the transient regime, $P_k^g$ exhibits a oscillatory behaviour and is always positive. The production level increases with Prandtl number at all times. Dissipation decays with time initially and grows exponentially once the transient stage ends. Much like production, dissipation is higher at high Prandtl numbers for all times. The evolution of pressure-dilatation magnitude shown in figure \ref{fig:RandKeBudget}(b), suggests that $\Pi_k^g$ also has an exponential variation in the asymptotic limit. At late times, pressure-dilatation is negative as energy is transferred from the kinetic to internal mode. However, in the transient stage $\Pi_k^g$ is dominantly positive as energy is transferred from the internal to the kinetic mode. This is not surprising as initially all the perturbation energy is in the internal mode and pressure does work on the velocity field to facilitate the energy transfer. Similar to $P_k^g$ and $\epsilon_k^g$, at late times the magnitude of $\Pi_k^g$ is also higher for high Prandtl number. During the transient stages however pressure-dilatation is similar in magnitude for all three Prandtl number. Figure \ref{fig:RandKeBudgetInstant} presents the kinetic energy budget terms normalized by the instantaneous kinetic energy at $M=6$. The normalization ensures that the budget term asymptotes to a constant at late times. The asymptotic value is approximately equal to the budget contributions of the most unstable eigenmode predicted by linear theory. This is also not surprising as at late times the evolution is determined by the most unstable eigenmode.

\begin{figure*}
    \centering
    \includegraphics[trim=0 10 10 0, clip, width=0.6\textwidth, keepaspectratio]{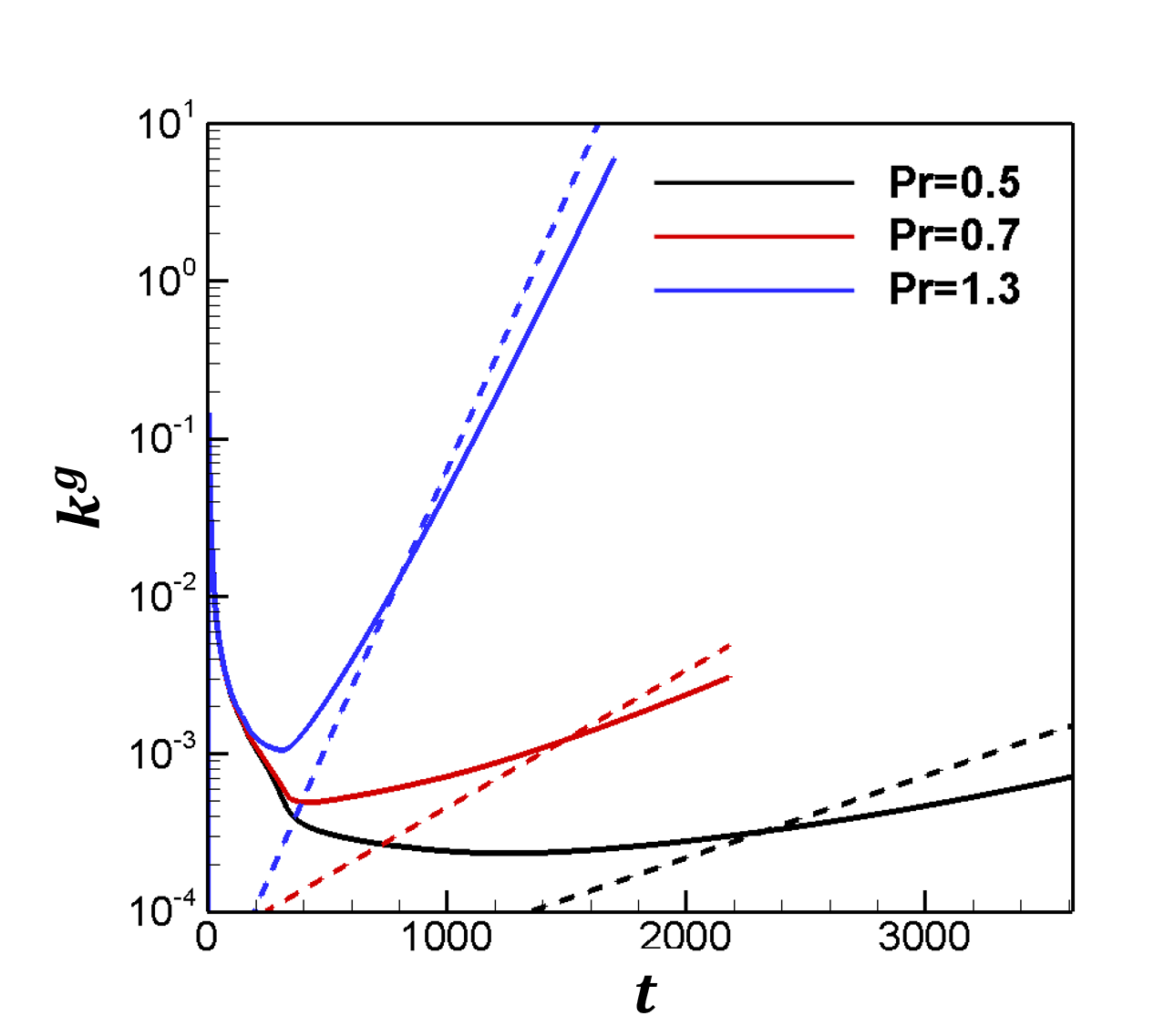} 
    \caption{Evolution of globally averaged kinetic energy ($k^g$) at $M=3$ and $Re=4000$ for three different Prandtl number. The dashed lines represent the dominant eigenmode growth at the rate predicted by LSA.}
    \label{fig:RandKeM3}
\end{figure*}

\begin{figure*}
    \centering
    \subfloat[$Pr=0.5$]{\includegraphics[width=0.32\textwidth, keepaspectratio]{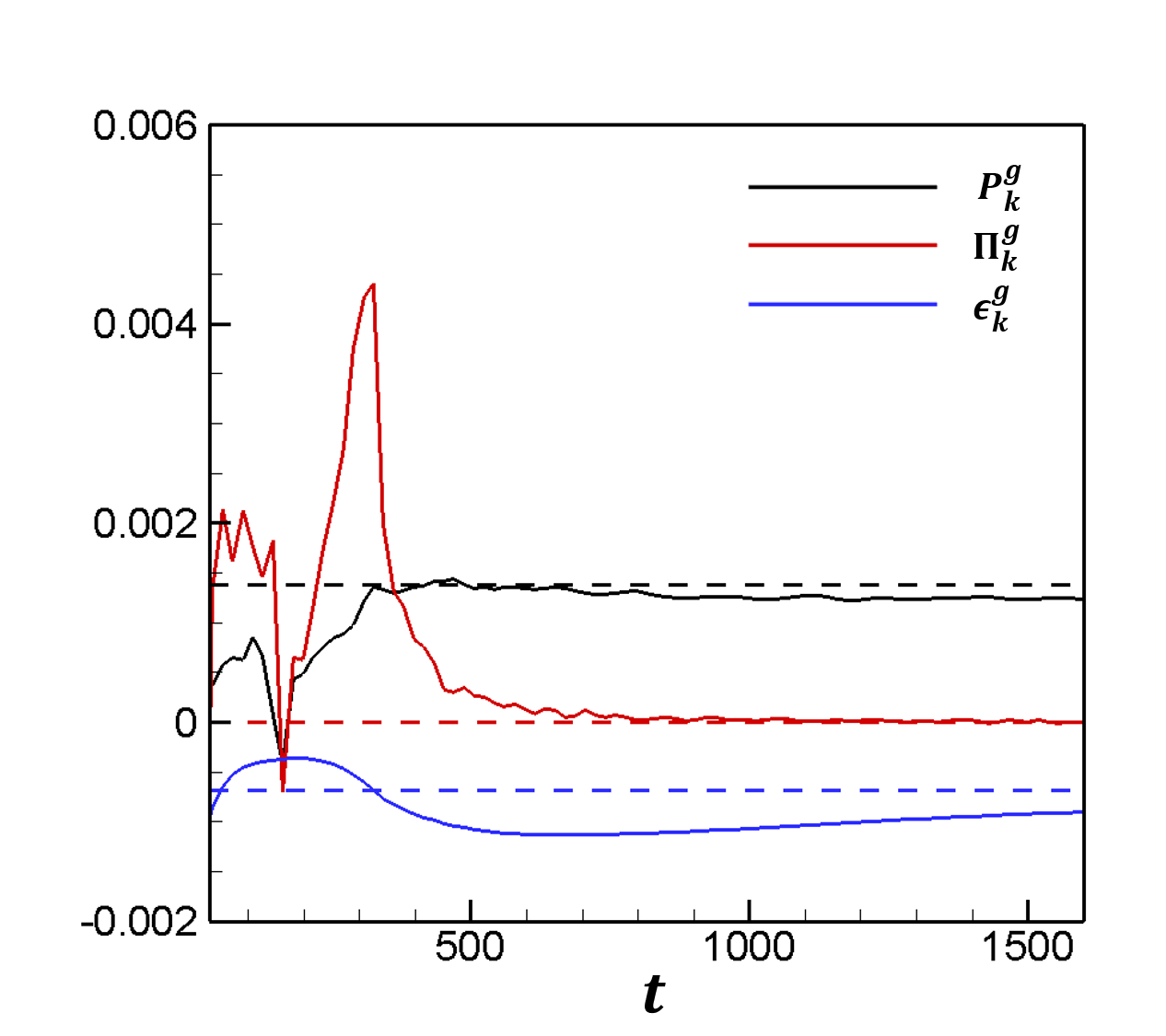}} 
    \subfloat[$Pr=0.7$]{\includegraphics[width=0.32\textwidth, keepaspectratio]{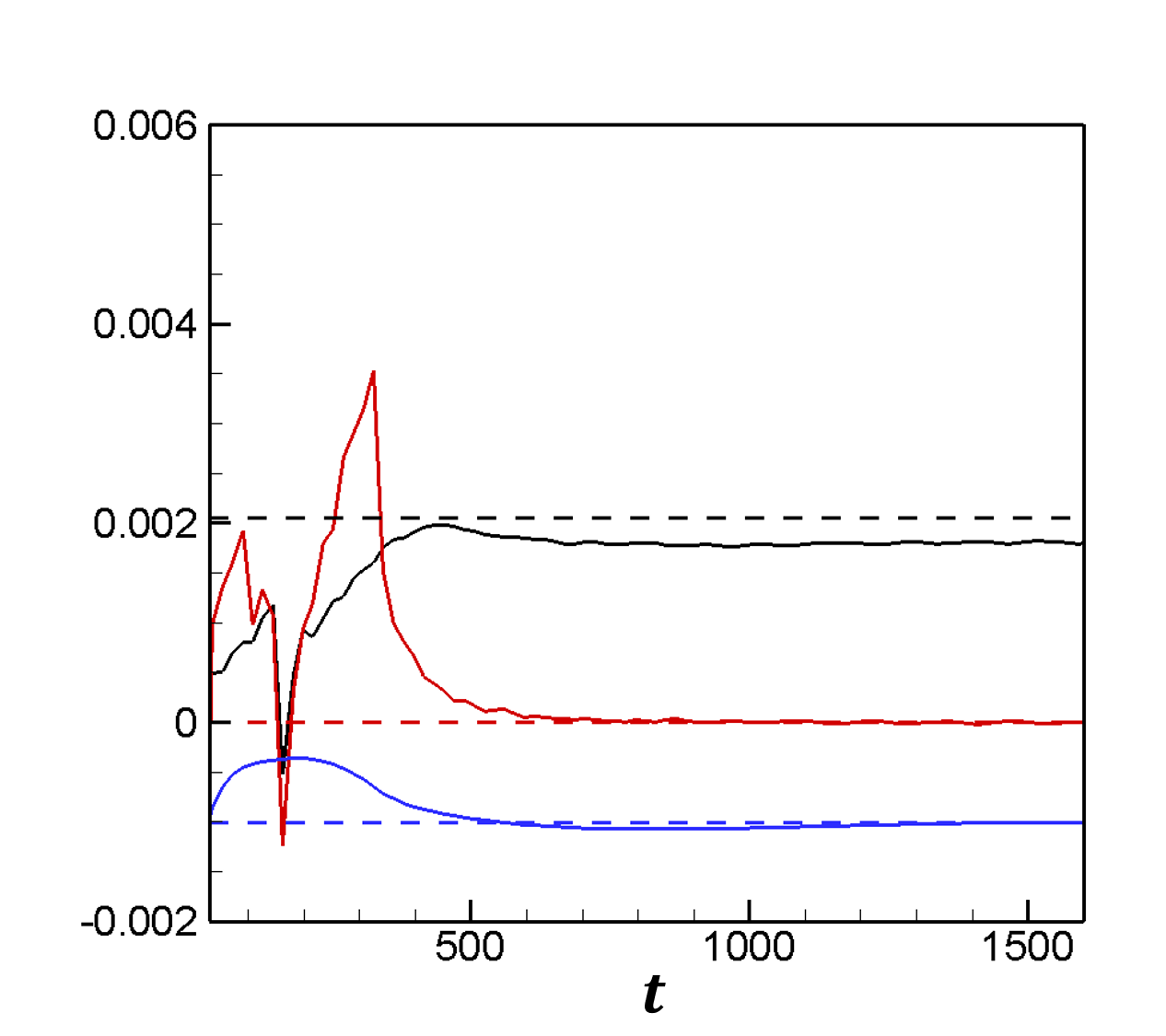}}
    \subfloat[$Pr=1.3$]{\includegraphics[width=0.32\textwidth, keepaspectratio]{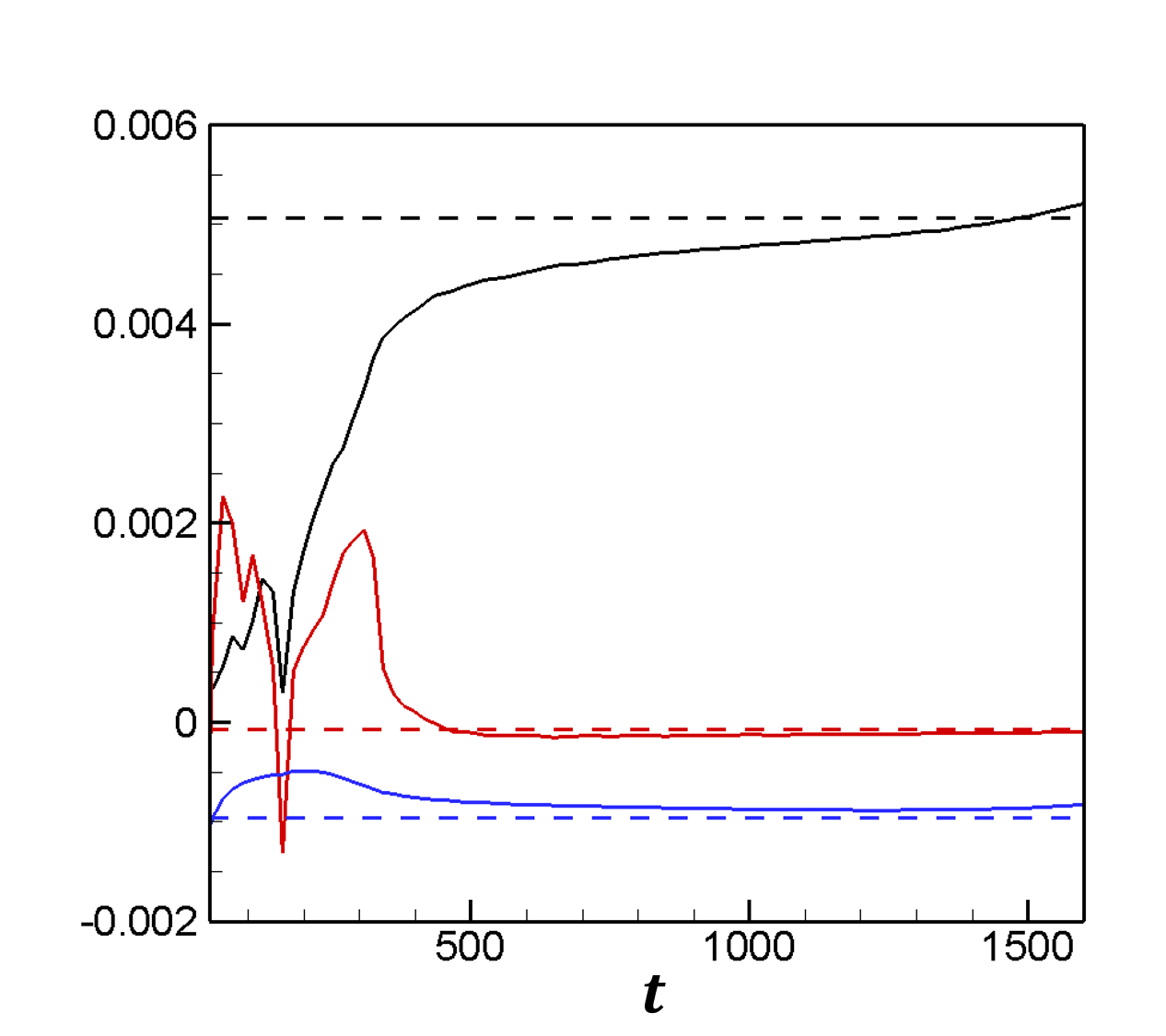}}
    \caption{Evolution of globally averaged terms in the kinetic energy budget normalized by instantaneous kinetic energy for (a) $Pr=0.5$, (b) $Pr=0.7$ and (c) $Pr=1.3$ at $M=3$ and $Re=4000$. All budget terms are normalized by $\zeta=2k^g U_\infty/L_r$. The dashed lines represent the budget contributions of the most unstable eigenmode predicted by LSA.}
    \label{fig:RandKeBudgetInstantM3}
\end{figure*}

The kinetic energy evolution at $M=3$ and $Re=4000$ for different $Pr$ is shown in figure \ref{fig:RandKeM3}. Similar to the $M=6$ case, $k^g$ at $M=3$ also increases exponentially after a transient decay stage. The length of transient stage also decreases with increasing $Pr$. The kinetic energy growth at late times is slower than the most unstable eigenmode growth rate predicted by LSA. At $M=3$ for all Prandtl numbers, there exist several modes with growth rates nearly identical to the most unstable mode. As a result, there are multiple Fourier modes with similar energy content at late times. However, the mode shapes for these modes are qualitatively similar. 
The kinetic energy budget terms normalized by instantaneous kinetic energy is presented in figure \ref{fig:RandKeBudgetInstantM3}. The budget terms at $M=3$ also asymptote to the contributions of the most unstable eigenmode.        

Overall, the response of boundary layer to random pressure forcing can be completely determined by the most unstable eigenmode(s) at late times. 
Since the growth rate of most unstable eigenmode increases with $Pr$, random pressure forcing leads to substantial increase in perturbation energy growth for high Prandtl numbers.   
There exist a transient region where non-modal interactions are significant. 
During the transient stage total energy decays as dissipation dominates production. 
The length of transient regime increases with decreasing $Pr$.

\section{Conclusions}
In this work, the influence of Prandtl number on the flow thermodynamic interaction for compressible boundary layer instability is examined. A linear stability analysis and DNS of compressible boundary layers are performed in the parameter regime: $M\in[0.5,8]$ and $Pr\in[0.5,1.3]$. The most unstable first and second mode are identified and the influence of both Mach and Prandtl number is studied. Both first and second mode are destabilized by increasing Prandtl number. The underlying flow-thermodynamic interactions are investigated.
 It is shown that the internal energy content is negligible compared to kinetic energy for the first mode, while, the second mode has considerable internal energy content. Pressure-dilatation is not significant for the first mode and doesn't play an important role in the instability dynamics. For the second mode pressure work transfers considerable energy from the perturbation kinetic to the internal mode. 

Despite the marked difference in pressure-dilatation both modes exhibit increased production with Prandtl number. In the case of the first mode all components of the pressure-strain correlation increase with Prandtl number. Higher energy in the wall-normal mode leads to increased production of the shear stress at high Prandtl number. The shear pressure-strain correlation also increases with Prandtl number but not as rapidly as shear production. Thus, the net difference between $P_{12}^g$ and $\Pi_{12}^g$ increases with Prandtl number. The net shear stress anisotropy increase permits higher production leading to increased destabilization of the first mode at higher Prandtl numbers. In the case of second mode, both $P_{12}^g$ and $\Pi_{12}^g$ decrease with Prandtl number. However, the net difference of $P_{12}^g$ and $\Pi_{12}^g$ increases with Prandtl number. The resulting higher shear stress anisotropy allows for increased production levels at higher Prandtl numbers leading to increased growth rates at high Prandtl numbers.
These findings not only enhance our understanding of Prandtl number effects, they also provide insight needed for closure model development.

\vspace{0.4cm}
\noindent \textbf{Acknowledgements}
\vspace{0.1cm}

Portions of this research were conducted with the advanced computing resources provided by Texas A\&M High Performance Research Computing

\vspace{0.4cm}
\noindent
Declaration of Interests. The authors report no conflict of interest.

\appendix

\section{}\label{app:A}
\subsection{Linearized Perturbation Equations}\label{app:A1}
The linearized perturbation equations \citep{malik1989prediction} for a compressible parallel boundary layer flow are detailed in this subsection.
The linearized continuity equation is given by the following relation.
\begin{equation}
    \label{eq:PerturbationContinuity}
     \frac{\partial \rho'}{\partial t} + \overline{U}_i\frac{\partial \rho'}{\partial x_i} +\frac{\partial  \overline{\rho}}{\partial x_i}u_i' + \overline{\rho}\frac{\partial u_i'}{\partial x_i}  = 0
\end{equation}
The momentum equations are obtained as,
\begin{equation}
\begin{split}
        \overline{\rho}\frac{\partial u_i'}{\partial t} + \overline{\rho}\overline{U}_k\frac{\partial u_i'}{\partial x_k} +\overline{\rho}\frac{\partial \overline{U}_i}{\partial x_k}u_k' =-\frac{\partial p'}{\partial x_i}  +\frac{1}{Re}\frac{\partial \tau_{ik}'}{\partial x_k},  
\end{split}
\label{eq:PerturbationMomentum}
\end{equation}
where, $\tau_{ik}'$ is the linearized viscous stress tensor. The components of the linearized  viscous stress tensor $\tau_{ik}'$ are given by:
\begin{equation}
    \begin{split}
        \tau_{ik}'=\overline{\mu}\left(\frac{\partial u_i'}{\partial x_k}+\frac{\partial u_k'}{\partial x_i}\right)+\mu'\left(\frac{\partial \overline{U_i}}{\partial x_k}+\frac{\partial \overline{U_k}}{\partial x_i}\right)+\overline{\lambda}\frac{\partial u_k'}{\partial x_k}\delta_{ik}
    \end{split}
    \label{eq:ViscousStressTensor}
\end{equation}
Here, $\overline{\lambda}$ is the bulk viscosity and is set to $\overline{\lambda}=-2\overline{\mu}/3$. 

The energy equation can be expressed in multiple formulations. In the present work, energy equation in both enthalpy and pressure formulation is utilized. The energy equation expressed in the enthalpy formulation is employed for linear stability computations whereas the pressure formulation is used to describe the flow-thermodynamic interactions. The linearized energy equation in the enthalpy formulation is expressed as  
\begin{equation}
    \label{eq:PerturbationEnergy}
    \begin{split}
     \overline{\rho}\frac{\partial T'}{\partial t} + \overline{\rho}\overline{U}_i\frac{\partial T'}{\partial x_i} + \overline{\rho}u_i'\frac{\partial \overline{T}}{\partial x_i} = &  (\gamma-1)M^2\left[\frac{\partial p'}{\partial t}+\overline{U}_i\frac{\partial p'}{\partial x_i}\right] - \frac{1}{RePr}\frac{\partial q_k'}{\partial x_k} \\
      & + \frac{(\gamma-1)M^2}{Re}\left[\tau_{ij}'\frac{\partial \overline{U_i}}{\partial x_j} 
       + \overline{\tau}_{ij}\frac{\partial u_i'}{\partial x_j}\right].
     \end{split}
\end{equation}
Here, $q_k'$ is the perturbation thermal conductivity as defined below.
\begin{equation}
    \label{eq:ThermalConductivity}
    q_k'=-\overline{\kappa}\frac{\partial T'}{\partial x_k}-\kappa'\frac{d\overline{T}}{dx_k}
\end{equation}
The energy equation expressed in the pressure formulation is given by the following relation.
 \begin{equation}
     \label{eq:PerturbationEnergyPressure}
     \begin{split}
     \frac{\partial p'}{\partial t} + \overline{U}_i\frac{\partial p'}{\partial x_i} =  -\gamma\overline{P}\frac{\partial u_k'}{\partial x_k} - \frac{1}{RePrM^2}\frac{\partial q_k'}{\partial x_k} 
      + \frac{\gamma-1}{Re}\left[\tau_{ij}'\frac{\partial \overline{U}_i}{\partial x_j} + \overline{\tau}_{ij}\frac{\partial u_i'}{\partial x_j}\right].
     \end{split}
\end{equation}
The density perturbations are related to pressure and temperature perturbations by the following state equation. 
\begin{equation}
    \label{eq:linear_state}
    \rho'=\gamma M^2\frac{p'}{\overline{T}}-\frac{\overline{\rho}}{\overline{T}}T'
\end{equation}

\subsection{Components of Matrix $\mathbf{A}$ and $\mathbf{B}$}\label{app:A2}
 The components of the coefficient matrices $\mathbf{A}$ and $\mathbf{B}$ in equation \eqref{eq:linear_eig} are detailed in this appendix. It must be noted that the components of the coefficient matrices for the same eigenvalue problem expressed in a marginally different form are presented in Appendix \rom{1} of \cite{malik1990numerical}. For the sake of clarification the components of the coefficient matrices are presented here as well. The non-zero elements of matrix $\mathbf{A}$ are as follows.
\begin{equation}
    \begin{split}
        A_{11}= & \thinspace-1;\quad A_{22}=\thinspace-1;\quad A_{33}=\thinspace-1;\quad
        A_{44}= \thinspace\frac{1}{\overline{T}}; \\A_{45}= & \thinspace-\gamma M^2;\quad  A_{54}=-\thinspace1;\quad A_{55}=  \thinspace(\gamma-1)M^2\overline{T}.
    \end{split}
\end{equation}

The elements of matrix $\mathbf{B}$ are listed below.
\begingroup
\allowdisplaybreaks
    \begin{align*}
        B_{11} = & \thinspace-\alpha\overline{U}_1 + \frac{\iota(\beta^2+2\alpha^2)\overline{\mu}\overline{T}}{Re} + \frac{\iota\alpha^2\overline{\lambda}\overline{T}}{Re} - \frac{\iota\overline{T}}{Re}\frac{d\overline{\mu}}{d\overline{T}}\frac{d\overline{T}}{dx_2}D 
         - \frac{\iota\overline{\mu}\overline{T}}{Re}D^2; \\
        B_{12} = & \thinspace\iota\frac{d\overline{U}_1}{dx_2} + \frac{\alpha\overline{T}}{Re}\frac{d\overline{\mu}}{d\overline{T}}\frac{d\overline{T}}{dx_2} + \frac{\alpha\overline{T}}{Re}(\overline{\mu}+\overline{\lambda})D; \\
        B_{13} =  & \thinspace\frac{\iota\alpha\beta\overline{T}}{Re}(\overline{\mu}+\overline{\lambda});\\
        B_{14} = &\thinspace -\frac{\iota\overline{T}}{Re}\frac{d\overline{\mu}}{d\overline{T}}\frac{d^2\overline{U}_1}{dx_2} - \frac{\iota\overline{T}}{Re}\frac{d^2\overline{\mu}}{d\overline{T}^2}\frac{d\overline{T}}{dx_2}\frac{d\overline{U}_1}{dx_2}-\frac{\iota\overline{T}}{Re}\frac{d\overline{\mu}}{d\overline{T}}\frac{d\overline{U}_1}{dx_2}D ; \\
        B_{15} =  & \thinspace -\alpha\overline{T}; \\
        B_{21} = & \thinspace \frac{\alpha\overline{T}}{Re}\frac{d\overline{\lambda}}{d\overline{T}}\frac{d\overline{T}}{dx_2} + \frac{\alpha\overline{T}}{Re}(\overline{\mu}+\overline{\lambda})D; \\
        B_{22} = & \thinspace-\alpha\overline{U} + \frac{\iota\overline{\mu}\overline{T}}{Re} - \frac{\iota(\beta^2+\alpha^2)\overline{T}}{Re}\left(2\frac{d\overline{\mu}}{d\overline{T}}+\frac{d\overline{\lambda}}{d\overline{T}}\right)\frac{d\overline{T}}{dx_2}D 
        \thinspace- \frac{\iota\overline{T}}{Re}(2\overline{\mu}+\overline{\lambda})D^2; \\
        B_{23} = & \thinspace\frac{\beta\overline{T}}{Re}\frac{d\overline{\lambda}}{d\overline{T}}\frac{d\overline{T}}{dx_2}+\frac{\beta\overline{T}}{Re}(\overline{\mu}+\overline{\lambda})D ;\\
        B_{24} =  & \thinspace \frac{\alpha\overline{T}}{Re}\frac{d\overline{\mu}}{d\overline{T}}\frac{d\overline{U}_1}{dx_2} ;\\
        B_{25} =  & \thinspace\iota\overline{T}D; \\
        B_{31} = & \thinspace\frac{\iota\alpha\beta\overline{T}}{Re}(\overline{\mu}+\overline{\lambda}); \\
        B_{32} = & \thinspace\frac{\beta\overline{T}}{Re}\frac{d\overline{\mu}}{d\overline{T}}\frac{d\overline{T}}{dx_2} + \frac{\beta\overline{T}}{Re}(\overline{\mu}+\overline{\lambda})D; \\
        B_{33} = & \thinspace-\alpha\overline{U}_1+\frac{\iota(\alpha^2+2\beta^2)\overline{\mu}\overline{T}}{Re} + \frac{\iota\beta^2\overline{\lambda}\overline{T}}{Re}  
         - \frac{\iota\overline{T}}{Re}\frac{d\overline{\mu}}{d\overline{T}}\frac{d\overline{T}}{d x_2}D  -\frac{\iota\overline{\mu}\overline{T}}{Re}D^2; \tag{\stepcounter{equation}\theequation} \\
        B_{34} = & \thinspace0; \\
        B_{35} = & \thinspace -\beta\overline{T}; \\
        B_{41} = & \thinspace -\alpha;\quad B_{42} =   \thinspace -\frac{\iota}{\overline{T}}\frac{d\overline{T}}{dx_2} + \iota D ; \quad
        B_{43} = \thinspace-\beta; \\
        B_{44} = & \thinspace\frac{\alpha\overline{U}_1}{\overline{T}} ;\quad
        B_{45} =   \thinspace-\alpha\gamma M^2\overline{U}_1; \\
        B_{51} = & \thinspace\frac{-2\iota(\gamma-1)M^2\overline{\mu}\overline{T}}{Re}\frac{d\overline{U}_1}{dx_2}D; \\
        B_{52} = & \thinspace \iota\frac{d\overline{T}}{dx_2} + \frac{2\alpha(\gamma-1)M^2\overline{\mu}\overline{T}}{Re}\frac{d\overline{U}_1}{dx_2};\\
        B_{53} = & \thinspace 0; \\
        B_{54} = & \thinspace -\alpha\overline{U}-\frac{\iota(\gamma-1)M^2\overline{T}}{Re}\frac{d\overline{\mu}}{d\overline{T}}{\left(\frac{d^2\overline{U}_1}{dx_2^2}\right)}^2 -\frac{\iota\overline{T}}{PrRe}\frac{d\overline{\kappa}}{d\overline{T}}\frac{d^2\overline{T}}{dx_2^2}
         -\frac{\iota\overline{T}}{PrRe}\frac{d^2\overline{\kappa}}{d\overline{T}^2}{\left(\frac{d\overline{T}}{dx_2}\right)}^2 \\ 
         \thinspace & +\frac{\iota\overline{\kappa}\overline{T}}{PrRe}(\alpha^2+\beta^2)  -\frac{2\iota\overline{T}}{PrRe}\frac{d\overline{\kappa}}{d\overline{T}}\frac{d\overline{T}}{dx_2}D  -\frac{\iota\overline{\kappa}\overline{T}}{PrRe}D^2; \\
        B_{55} = & \thinspace (\gamma-1)M^2\alpha\overline{U}\overline{T} \\
    \end{align*}
\endgroup
Here the symbols $D$ and $D^2$ denote the first and second order derivative in the wall-normal direction respectively.
\section{Validation of GKM-DNS}\label{app:B}
\begin{figure*}
    \centering
    \subfloat[$C_3$]{\includegraphics[trim=0 10 10 0, clip, width=0.45\textwidth, keepaspectratio]{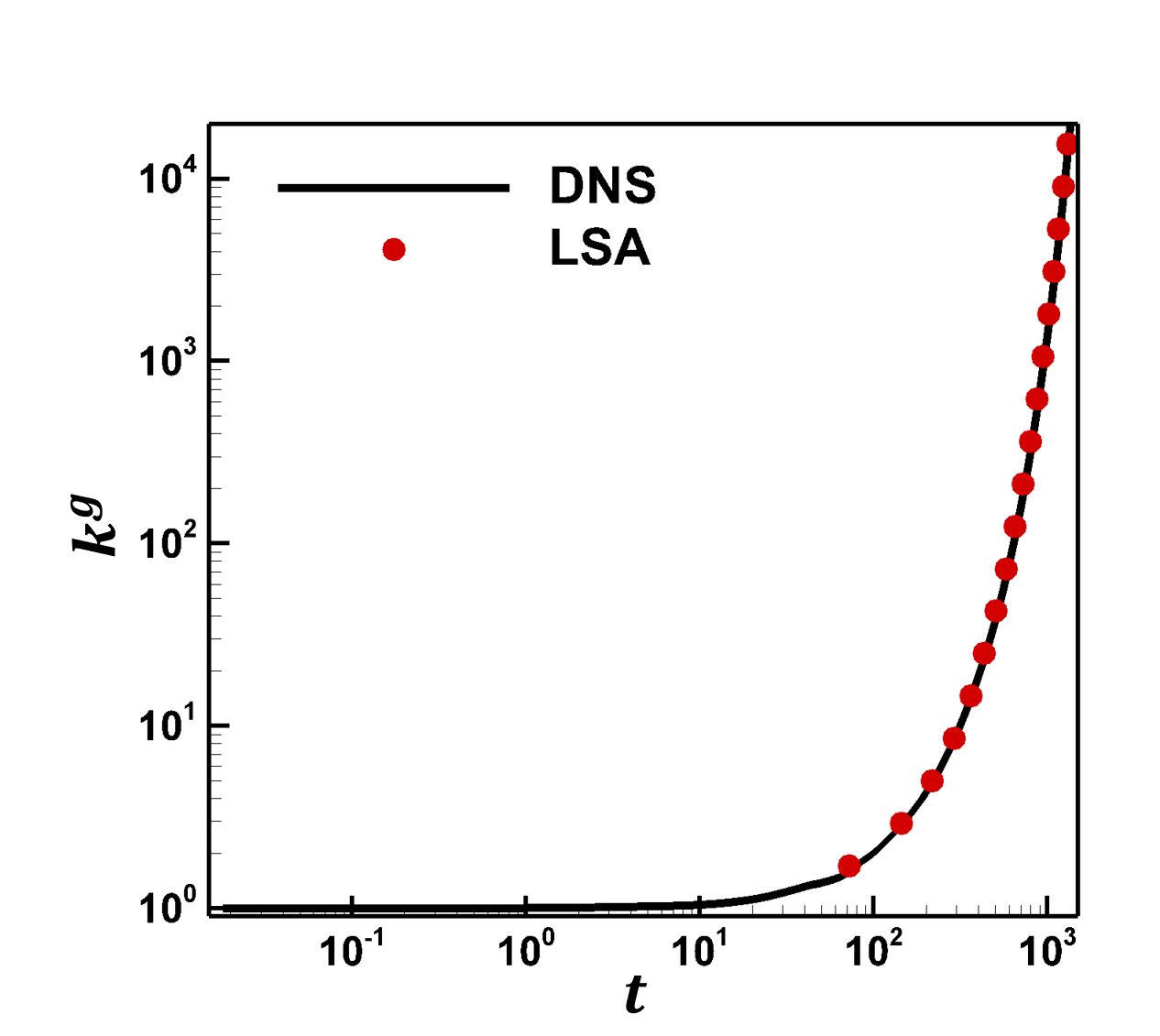}} 
    \subfloat[$C_6$]{\includegraphics[trim=0 10 10 0, clip, width=0.45\textwidth, keepaspectratio]{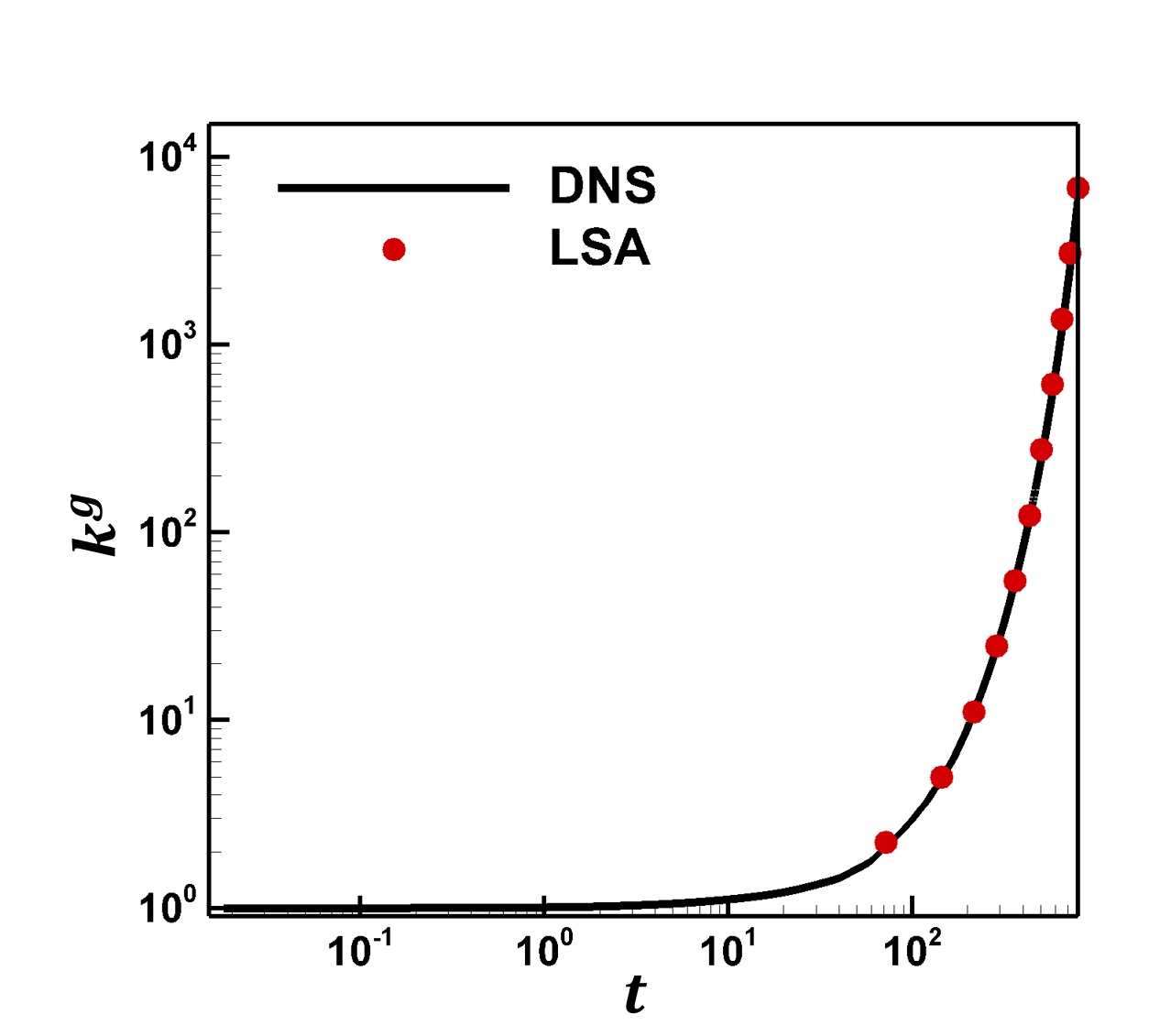}}
    \caption{Evolution of globally averaged kinetic energy ($k^g$) for (a) case $C_3$ and (b) case $C_6$. Solid black line corresponds to DNS results and, kinetic energy growth based on LSA are marked with red symbols.}
    \label{fig:DNSKinteicEnergy}
\end{figure*}
The DNS results are validated by comparing against linear stability analysis. The evolution for volume averaged perturbation kinetic energy $k$ is shown in figure \ref{fig:DNSKinteicEnergy} for cases $C_3$ and $C_6$. The kinetic energy is normalized by the initial perturbation kinetic energy. For both cases, the kinetic energy grows exponentially. The growth rate of kinetic energy agrees very well with the rate predicted by linear stability analysis.

\begin{figure*}
    \centering
  	\subfloat[$\hat{u}_1$]{\includegraphics[trim=0 115 10 0, clip, width=0.245\linewidth]{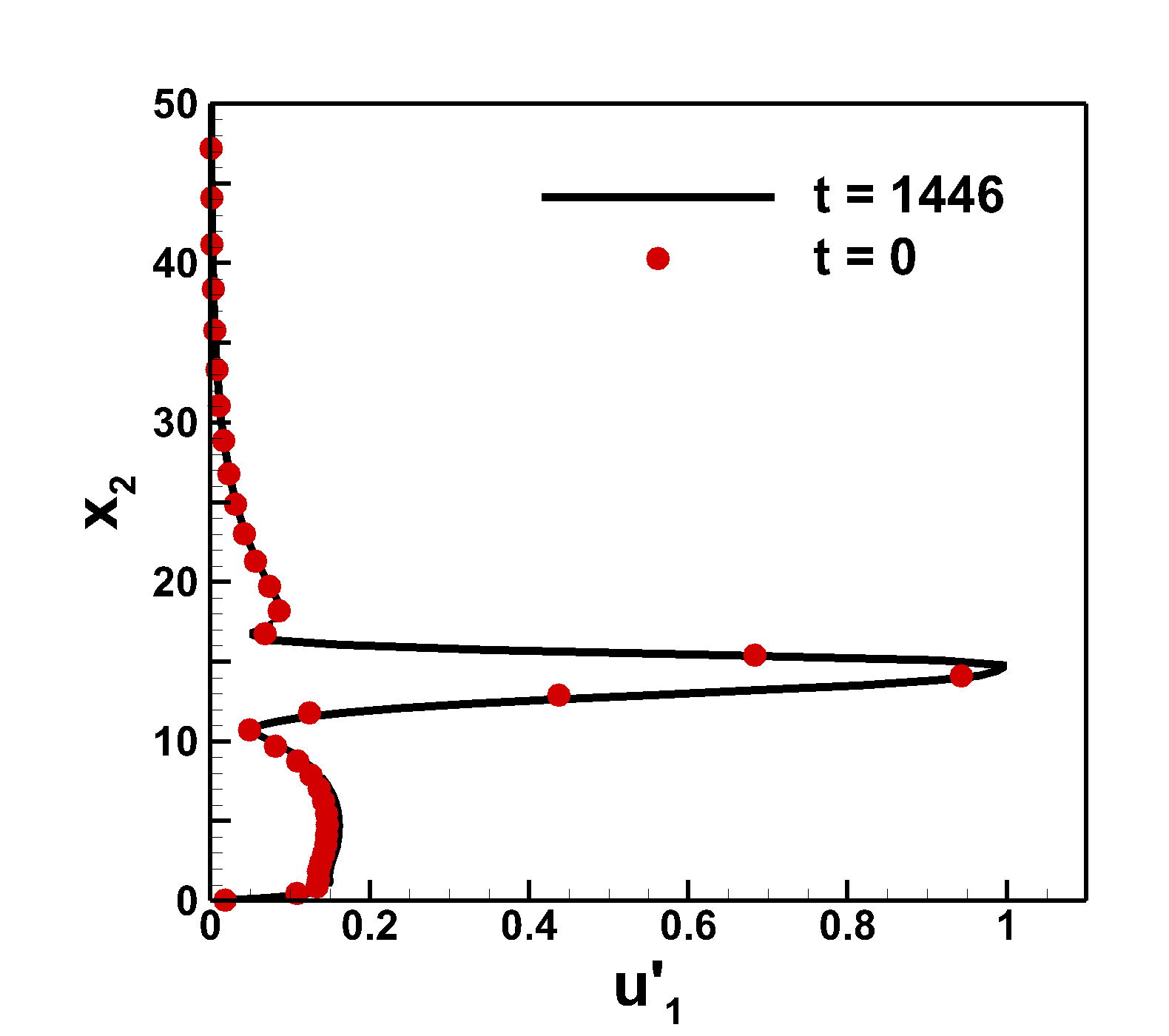}}
  	\subfloat[$\hat{u}_2$]{\includegraphics[trim=0 115 10 0,clip,  width=0.245\linewidth]{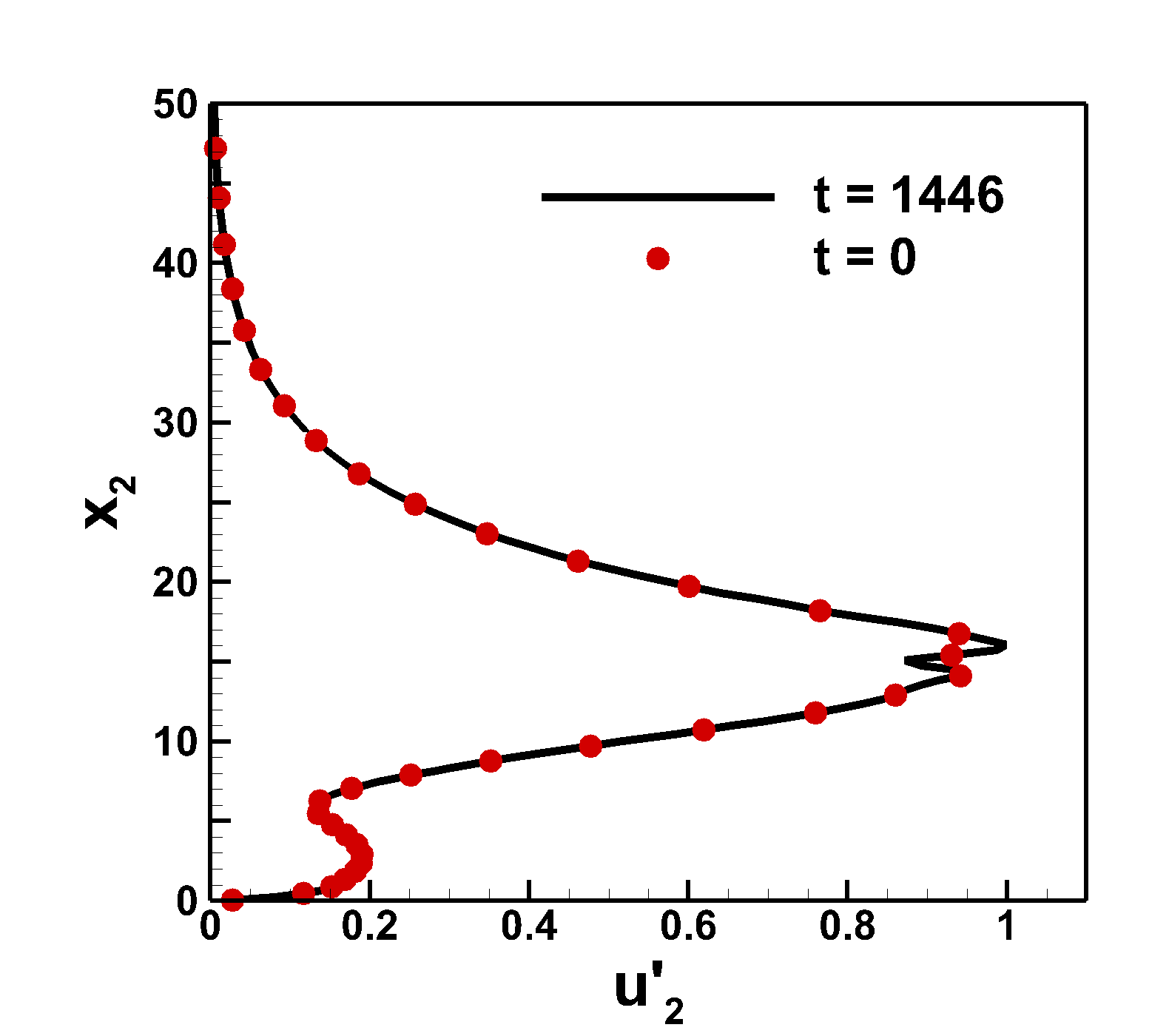}}
  	\subfloat[$\hat{T}$]{\includegraphics[trim=0 115 10 0, clip, width=0.245\linewidth]{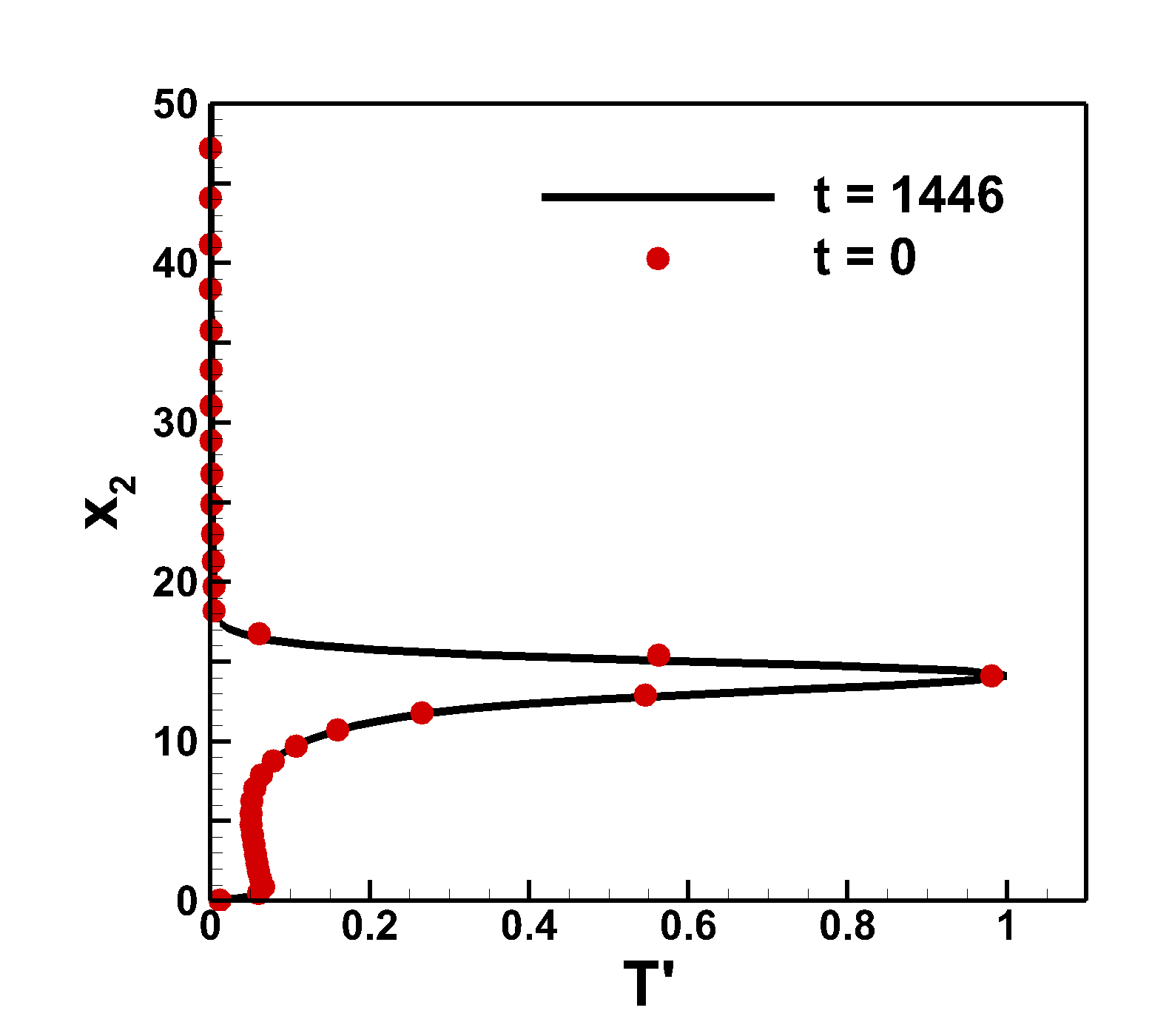}}
  	\subfloat[$\hat{\rho}$]{\includegraphics[trim=0 115 10 0, clip, width=0.245\linewidth]{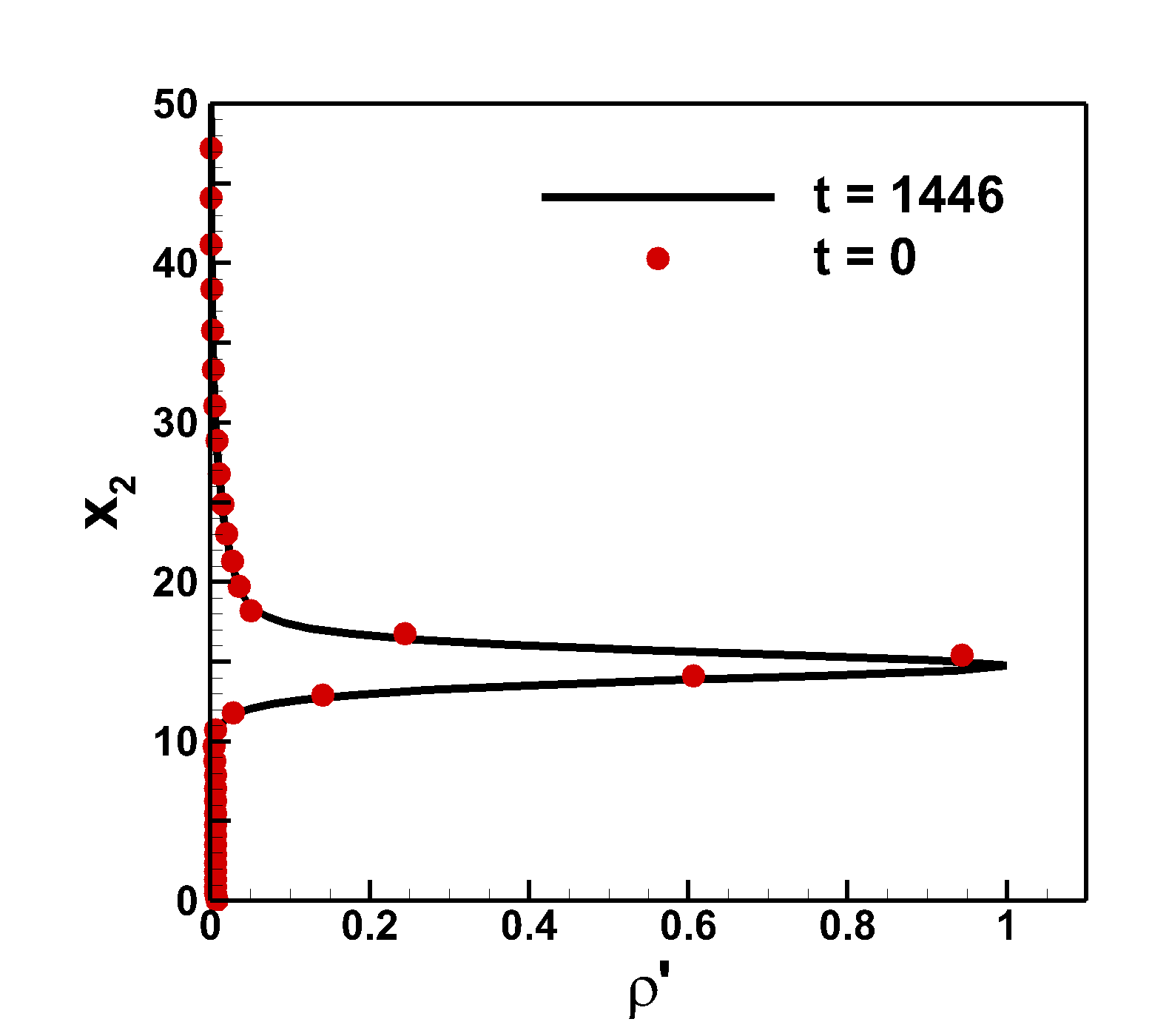}}
	\caption{Mode shapes of (a) streamwise velocity, (b) wall-normal velocity, (c) temperature and (d) density for case $C_3$ at two different times.}
	\label{fig:DNSModeShapeC1}
\end{figure*}

\begin{figure*}
    \centering
  	\subfloat[$\hat{u}_1$]{\includegraphics[trim=0 115 10 0, clip, width=0.245\linewidth]{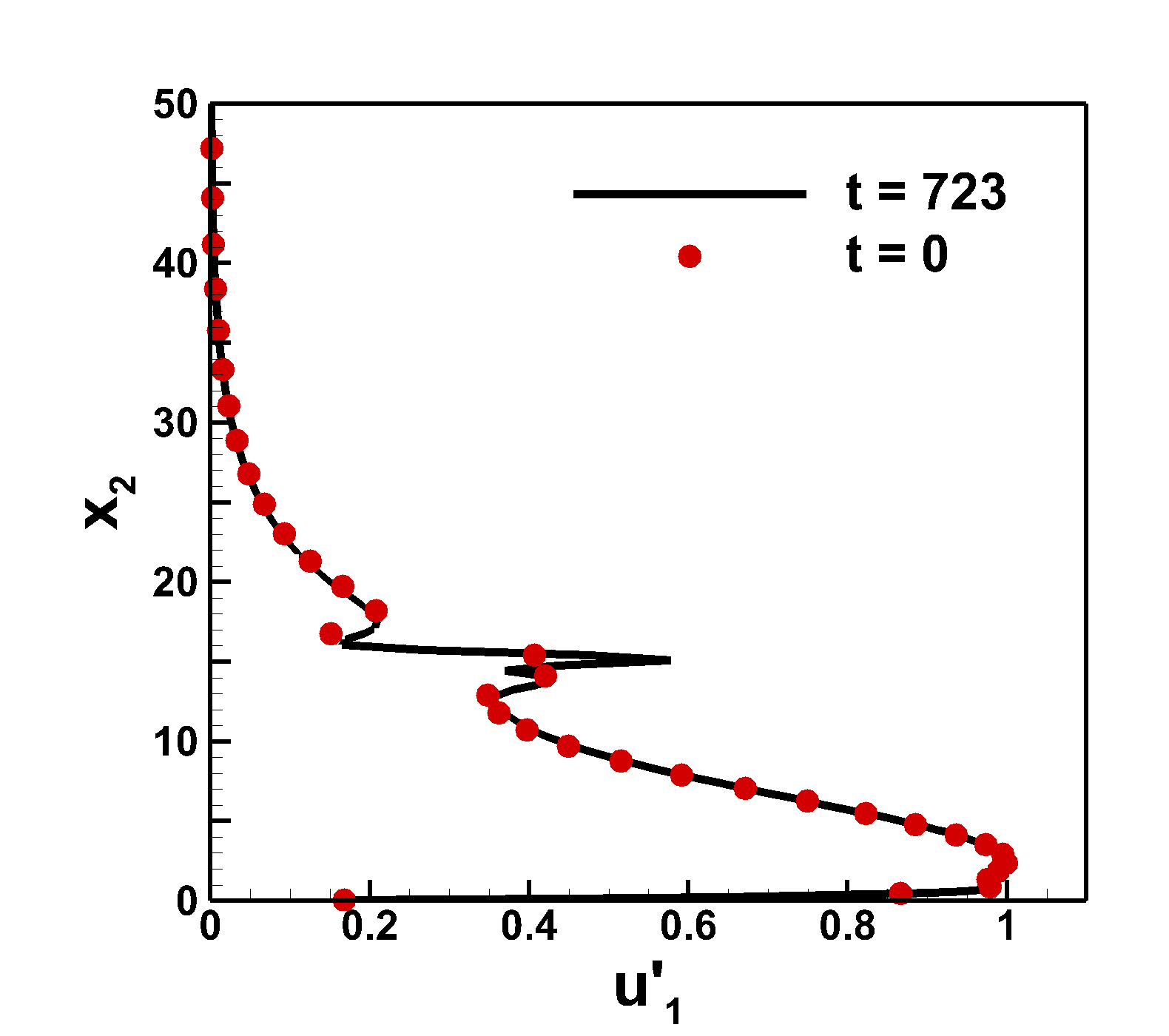}}
  	\subfloat[$\hat{u}_2$]{\includegraphics[trim=0 115 10 0,clip,  width=0.245\linewidth]{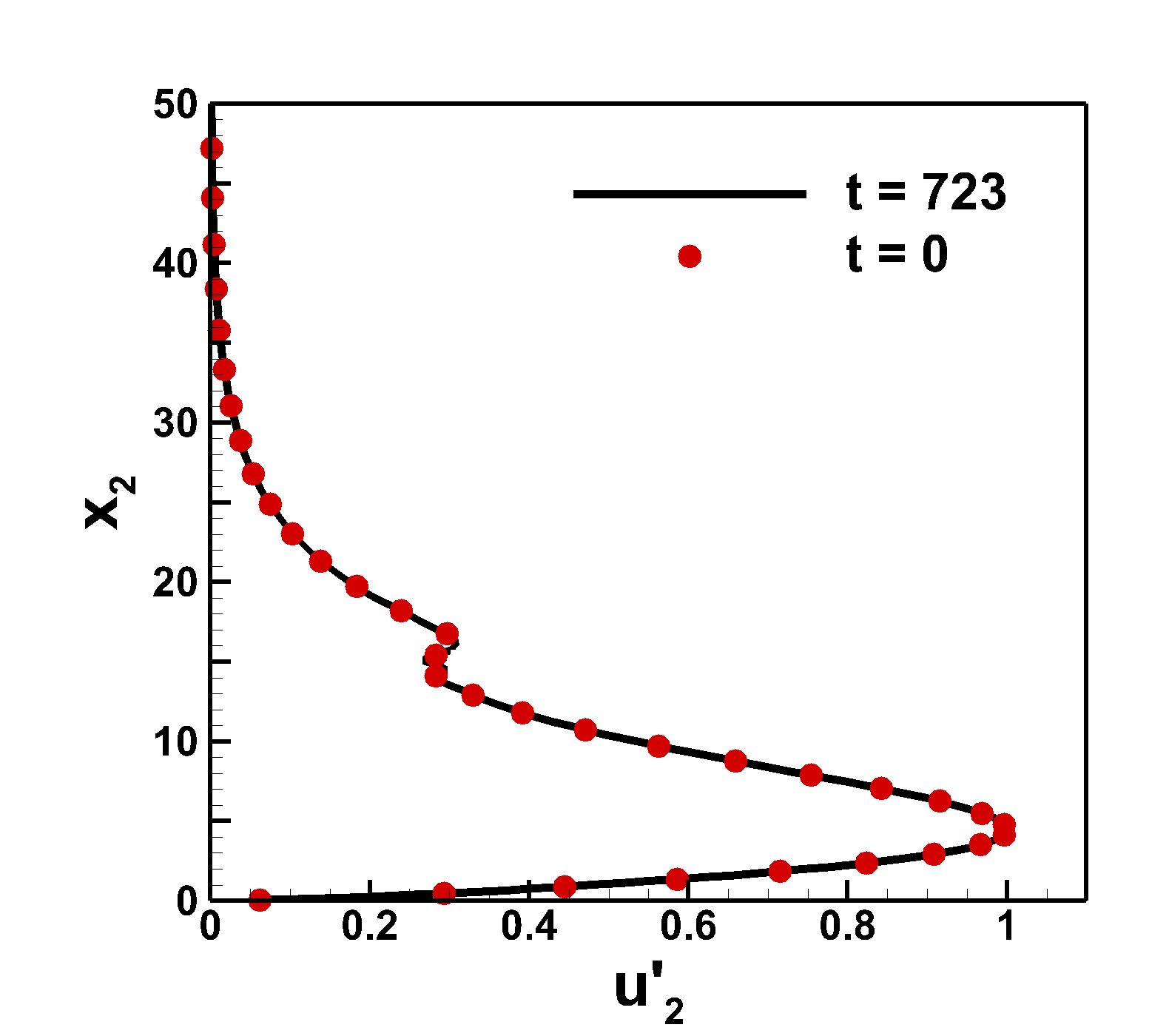}}
  	\subfloat[$\hat{T}$]{\includegraphics[trim=0 115 10 0, clip, width=0.245\linewidth]{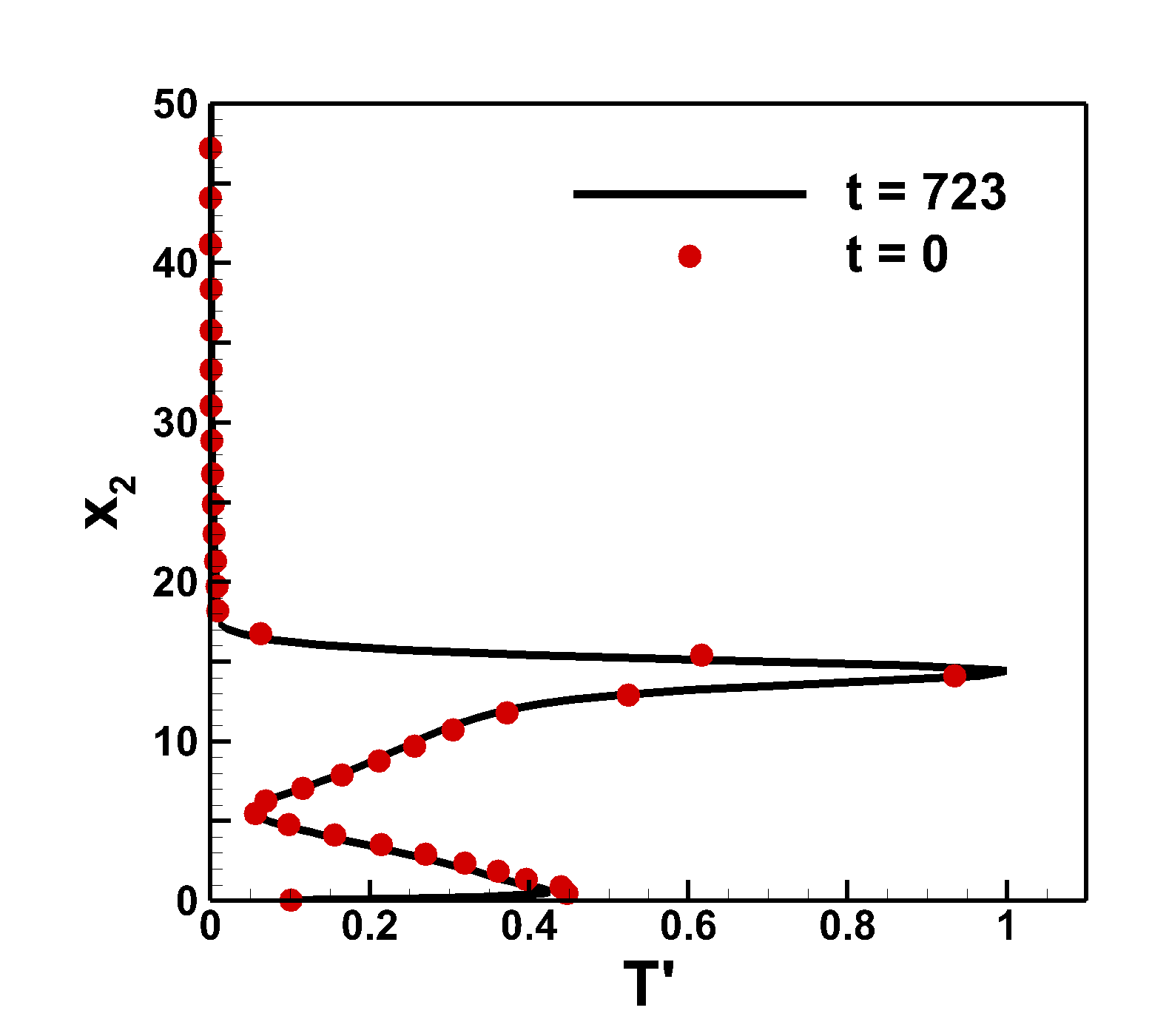}}
  	\subfloat[$\hat{\rho}$]{\includegraphics[trim=0 115 10 0, clip, width=0.245\linewidth]{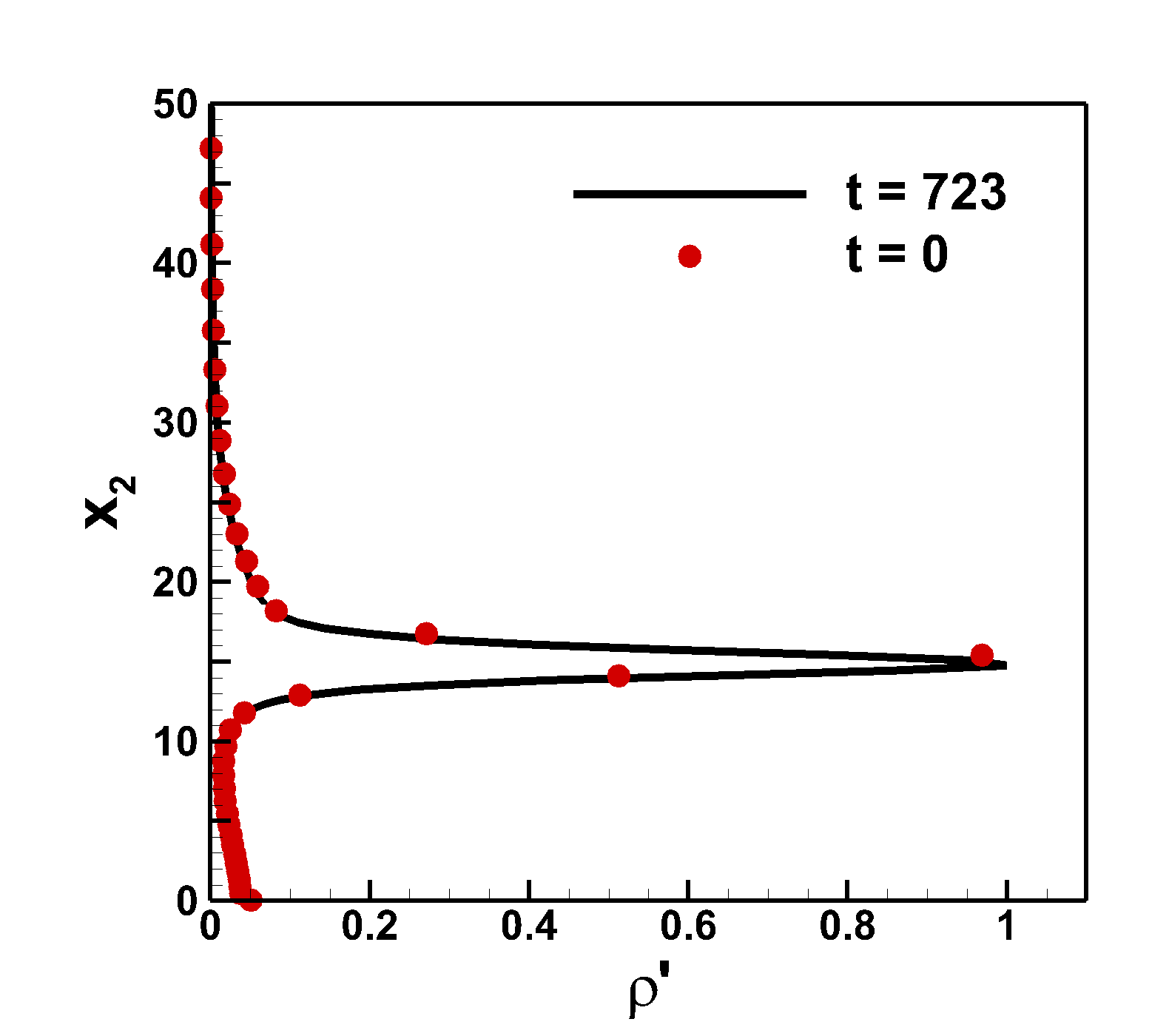}}
	\caption{Mode shapes of (a) streamwise velocity, (b) wall-normal velocity, (c) temperature and (d) density for case $C_6$ at two different times.}
	\label{fig:DNSModeShapeC2}
\end{figure*}

The mode shapes profiles for the case $C_3$ at $t=1446$ and $t=0$ are shown in figure \ref{fig:DNSModeShapeC1}. Specifically, the mode shapes for streamwise velocity, wall-normal velocity, temperature and density are presented. The mode shapes are initialized with the profiles based on LSA. The mode shape profiles of all the disturbances are retained throughout the duration of the simulation. A similar plot of the mode shapes for the case $C_6$ is displayed in figure \ref{fig:DNSModeShapeC2}. The mode shapes for the second mode case are also retained throughout the simulation. Overall, the GKM-DNS results are in excellent agreement with LSA results.   

\bibliographystyle{jfm}
\bibliography{main1}

\end{document}